CAROLINE VAN BORM

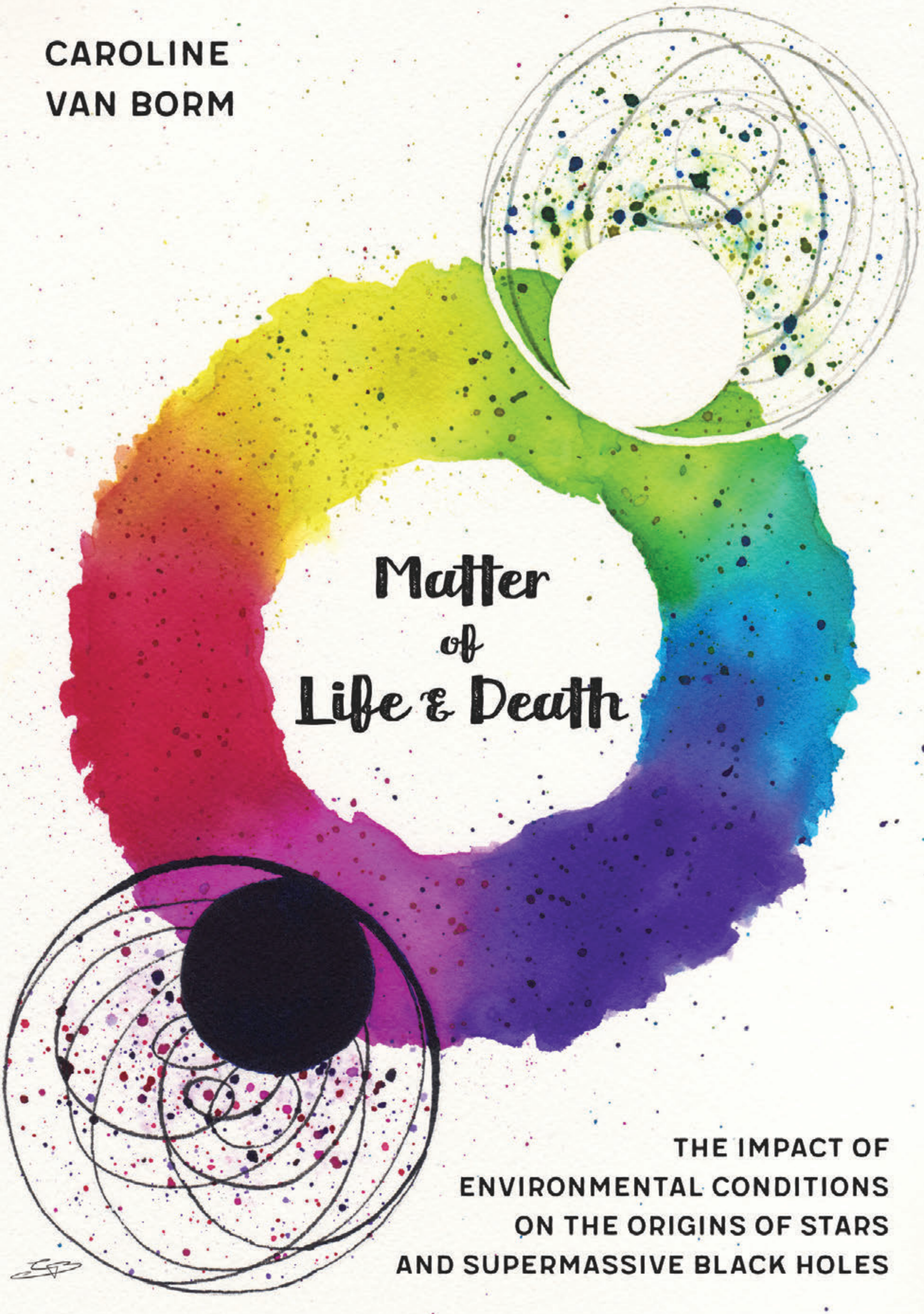

# Matter of Life & Death

THE IMPACT OF ENVIRONMENTAL CONDITIONS ON THE ORIGINS OF STARS AND SUPERMASSIVE BLACK HOLES

## IN THIS THESIS:

I OBSERVATIONAL EVIDENCE SUGGESTS THAT SOME VERY LARGE SUPERMASSIVE BLACK HOLES (SMBHs) ALREADY EXISTED LESS THAN 1 GYR AFTER THE BIG BANG. IT IS QUITE A CHALLENGE TO EXPLAIN THE FORMATION AND GROWTH OF THESE SMBHs. HERE, WE EXPLORE THE FORMATION OF THEIR 'SEEDS' IN THE DIRECT COLLAPSE SCENARIO, USING BOTH 3D HYDRODYNAMICAL SIMULATIONS AND A ONE-ZONE MODEL TO INVESTIGATE THE IMPACT OF TURBULENCE, ROTATION, UV RADIATION FIELD AND MAGNETIC FIELDS.

II FEEDBACK PROCESSES FROM STARS AND BLACK HOLES SHAPE THE INTERSTELLAR MEDIUM (ISM) OUT OF WHICH NEW GENERATIONS OF LUMINOUS OBJECTS FORM. UNDERSTANDING THE PROPERTIES OF THESE OBJECTS, E.G. THE STELLAR INITIAL MASS FUNCTION, REQUIRES KNOWLEDGE OF THE CHEMICAL AND THERMODYNAMICAL PROPERTIES OF THE FEEDBACK-REGULATED ISM. TO GAIN A BETTER UNDERSTANDING OF THE CHEMO-THERMAL STATE AND FRAGMENTATION BEHAVIOR OF GAS IN HIGH-REDSHIFT GALAXIES, WE DEVELOP A COMPUTATIONAL CODE, PDR-Zz, AND USE IT TO SYSTEMATICALLY EXPLORE THE OVERALL IMPACT OF VARIOUS FEEDBACK EFFECTS, BOTH RADIATIVE AND CHEMICAL, ON GAS IN DIFFERENT PHYSICAL REGIMES.

# Matter of Life & Death

## THE IMPACT OF ENVIRONMENTAL CONDITIONS ON THE ORIGINS OF STARS AND SUPERMASSIVE BLACK HOLES

CAROLINE VAN BORM

# Matter of Life & Death

The impact of environmental conditions
on the origins of stars and supermassive black holes

**PhD thesis**

to obtain the degree of PhD at the
University of Groningen
on the authority of the
Rector Magnificus Prof. E. Sterken
and in accordance with
the decision by the College of Deans.

This thesis will be defended in public on

Monday 17 October 2016 at 14.30 hours

by

**Caroline Van Borm**

born on 7 December 1989
in Hagen, Germany

**Supervisors**
Prof. M.C. Spaans
Prof. J.C. Niemeyer

**Assessment Committee**
Prof. P.D. Barthel
Prof. S. Dreizler
Prof. P. Caselli

GEORG-AUGUST-UNIVERSITÄT
GÖTTINGEN

# Matter of Life & Death

## The impact of environmental conditions on the origins of stars and supermassive black holes

Dissertation

to acquire the doctoral degree in mathematics and natural science

'Doctor of Philosophy' (Ph.D.)

at the Georg-August-Universität Göttingen

in the doctoral degree programme PROPHYS (Physics)

at the Georg-August University School of Science (GAUSS)

submitted by

**Caroline Van Borm**

from Hagen, Germany

Göttingen, 2016

**Thesis Committee**

Prof. dr. M. C. Spaans, Kapteyn Institute, University of Groningen

Prof. dr. J. C. Niemeyer, Institut für Astrophysik, Georg-August-Universität Göttingen

**Members of the Examination Board**

Reviewer: Prof. dr. S. Dreizler, Institut für Astrophysik, Georg-August-Universität Göttingen

Second Reviewer: Prof. dr. P. D. Barthel, Kapteyn Institute, University of Groningen

Additional Reviewer: Prof. dr. P. Caselli, Center for Astrochemical Studies, Max Planck Institute for Extraterrestrial Physics, Garching

**Date of the oral examination:** Monday 17 October 2016

# CONTENTS

*A UNIVERSITY IS VERY MUCH LIKE A CORAL REEF. IT PROVIDES CALM WATERS AND FOOD PARTICLES FOR DELICATE YET MARVELLOUSLY CONSTRUCTED ORGANISMS THAT COULD NOT POSSIBLY SURVIVE IN THE POUNDING SURF OF REALITY, WHERE PEOPLE ASK QUESTIONS LIKE 'IS WHAT YOU DO OF ANY USE?' AND OTHER NONSENSE.*

– TERRY PRATCHETT
(THE SCIENCE OF DISCWORLD)

CHAPTER

# 1

# INTRODUCTION

## 1.1 COSMOLOGY

### 1.1.1 FROM THE BIG BANG TO THE COSMIC WEB

FAR-LONG AGO, OUR UNIVERSE BEGAN
WITH A BANG, RIPPLING INTO THE
HERE-NOW WHERE YOU AND I EXIST

The currently most widely accepted cosmological model, the Big Bang theory, says that about 13.8 billion years ago, the Universe started out as almost inconceivably hot and dense. And ever since then, it has expanded and cooled, eventually reaching the cold, tenuous state we see today. The simplest model that describes our observations of the Universe reasonably well is the ΛCDM (Lambda Cold Dark Matter) model, also known as the standard model of Big Bang cosmology. In this model, the Universe contains a cosmological constant Λ, associated with dark energy (an unknown form of energy which is hypothesized to permeate all of space, tending to accelerate the expansion of the Universe), and dark matter which is 'cold' (an unknown type of massive particle that does not interact with photons, and which moves slowly compared to the speed of light). Assuming the standard model is correct, the energy density in the Universe today is comprised of approximately 5% baryons, 25% cold dark matter, and 70% dark energy (Hinshaw et al. 2013; Planck Collaboration et al. 2015).

1

For a fraction of a second after the Big Bang, the Universe underwent a brief period of extremely fast expansion, known as inflation. During inflation, inhomogeneities, anisotropies and the curvature of space were all smoothed out, resulting in a Universe which is homogeneous (appears the same in all locations), isotropic (appears the same in all directions) and flat (its large scale geometry is the usual Euclidean geometry) on large spatial scales. During this process, tiny quantum mechanical fluctuations were enlarged to macroscopic scales and imprinted on the Universe as density fluctuations (of magnitude $\delta\rho/\rho \sim 10^{-5}$), which acted as seeds for the formation of structure.

After the end of inflation, the Universe consisted of a soup of elementary particles. Within the first second after the Big Bang, many interactions took place between the particles and what was eventually left were photons, neutrinos, protons, neutrons, electrons, and dark matter particles. About three

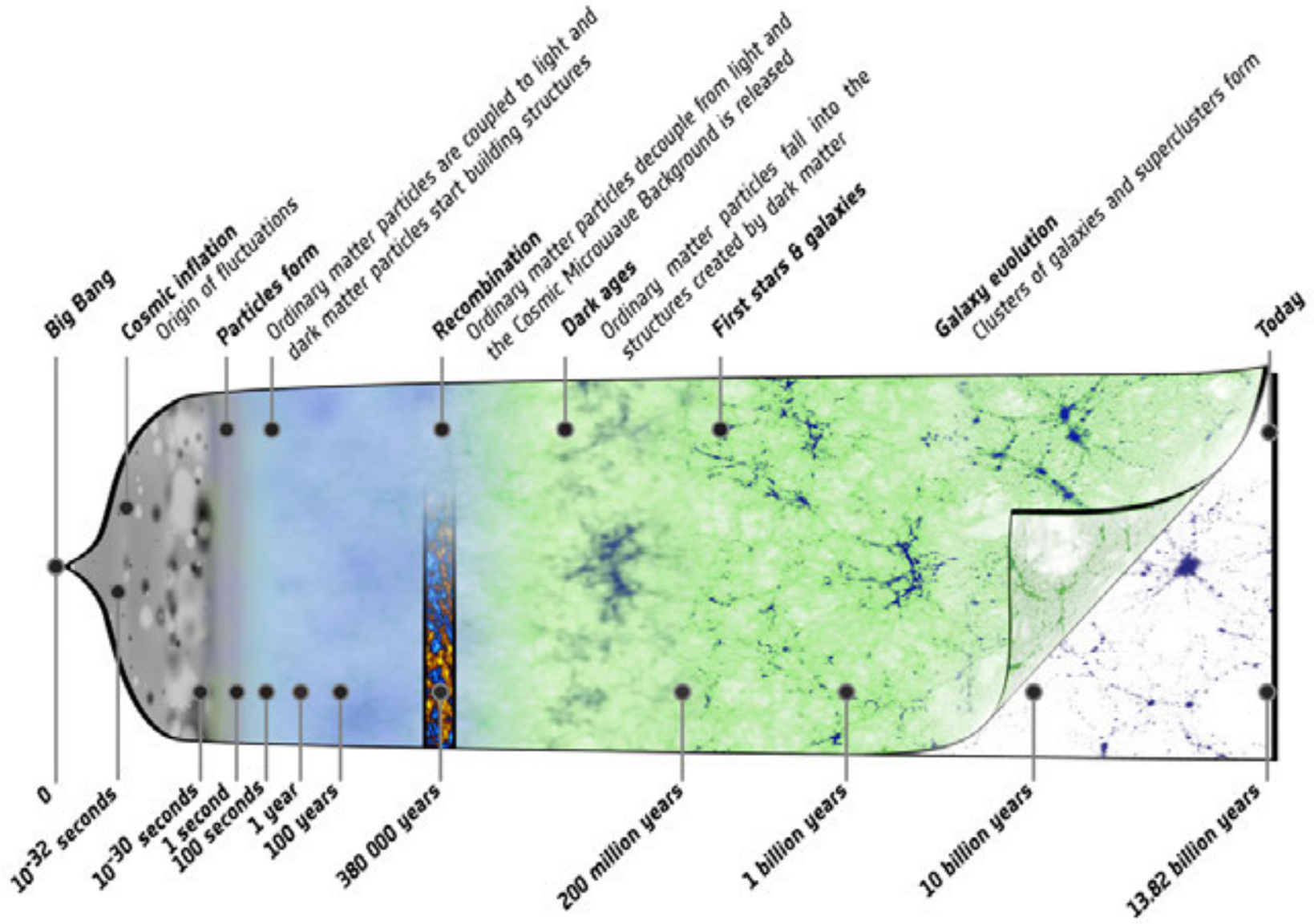

**Figure 1.1** – Illustration of the history of the Universe, showing the main events that occurred between the initial phase of the cosmos, where its properties were almost uniform and punctuated only by tiny fluctuations, to the rich variety of cosmic structure that we observe today, from stars and planets to galaxies and galaxy clusters. (Credit: ESA – C. Carreau; inverted from original)

minutes after the Big Bang, protons and neutrons had combined to form the nuclei of hydrogen and helium.

While the Universe was still very hot and dense, and many collisions occurred between photons and ordinary matter, density fluctuations in the matter could not grow, as the radiation pressure pushed away any matter concentration that may have been created under the effect of gravity. However, the dark matter was not coupled to the photons, and thus positive fluctuations in the dark matter distribution were able to grow denser and more massive.

About 380 000 years after the Big Bang (corresponding to redshift $z \sim 1100$), the Universe had expanded and cooled down enough for protons and electrons to combine into neutral hydrogen atoms, known as the epoch of recombination, and matter and radiation became decoupled. Photons were free to propagate freely through space, no longer constantly scattered by electrons and protons; this radiation is observed as the cosmic microwave background (CMB). Like the dark matter before it, ordinary matter was now also susceptible to the influence of gravity. At this point, the dark matter particles had already created a network of denser structures, and the densest of such concentrations gravitationally attract the ordinary matter. Contrary to dark matter, ordinary matter can dissipate energy quite effectively through radiative cooling, allowing it to collapse even further into the potential wells created by the dark matter. These processes gave rise to a highly sub-structured network of filaments, sheets and denser clusters of ordinary and dark matter, separated by immense voids. This foam-like structure is also known as the cosmic web, which constitutes the skeleton supporting the later emergence of stars and galaxies.

A few hundred million years after the Big Bang, the density in some of the nodes in the cosmic web had become high enough for the formation of stars and galaxies to become possible. The formation of the first star brought the first source of light to the Universe and ended the previous Dark Ages.

For a more in-depth discussion of cosmology, including the Big Bang theory, inflation, nucleosynthesis, structure formation and more, see the classic textbooks by e.g. Kolb & Turner (1990); Peebles (1993); Peacock (1999).

### 1.1.2 THE BEGINNING OF COMPLEXITY

The ΛCDM model predicts that the formation of structure in the Universe happened in a 'bottom-up' scenario, also called 'hierarchical' structure formation. Small overdensities are able to overcome the cosmological expansion and collapse first, and the resulting dark matter 'halos' (perhaps no more massive than globular clusters) then merge to form increasingly larger halos, which become the sites of galaxy formation. Further conglomeration gave rise to groups, clusters and superclusters of galaxies. The merger process continues to this day, as exemplified by our own Milky Way, which is undergoing at least two minor mergers with dwarf galaxies in the Local Group (e.g. Ibata et al. 1994). In this picture of hierarchical structure formation, the first (proto)galaxies were the basic building blocks for galaxy formation.

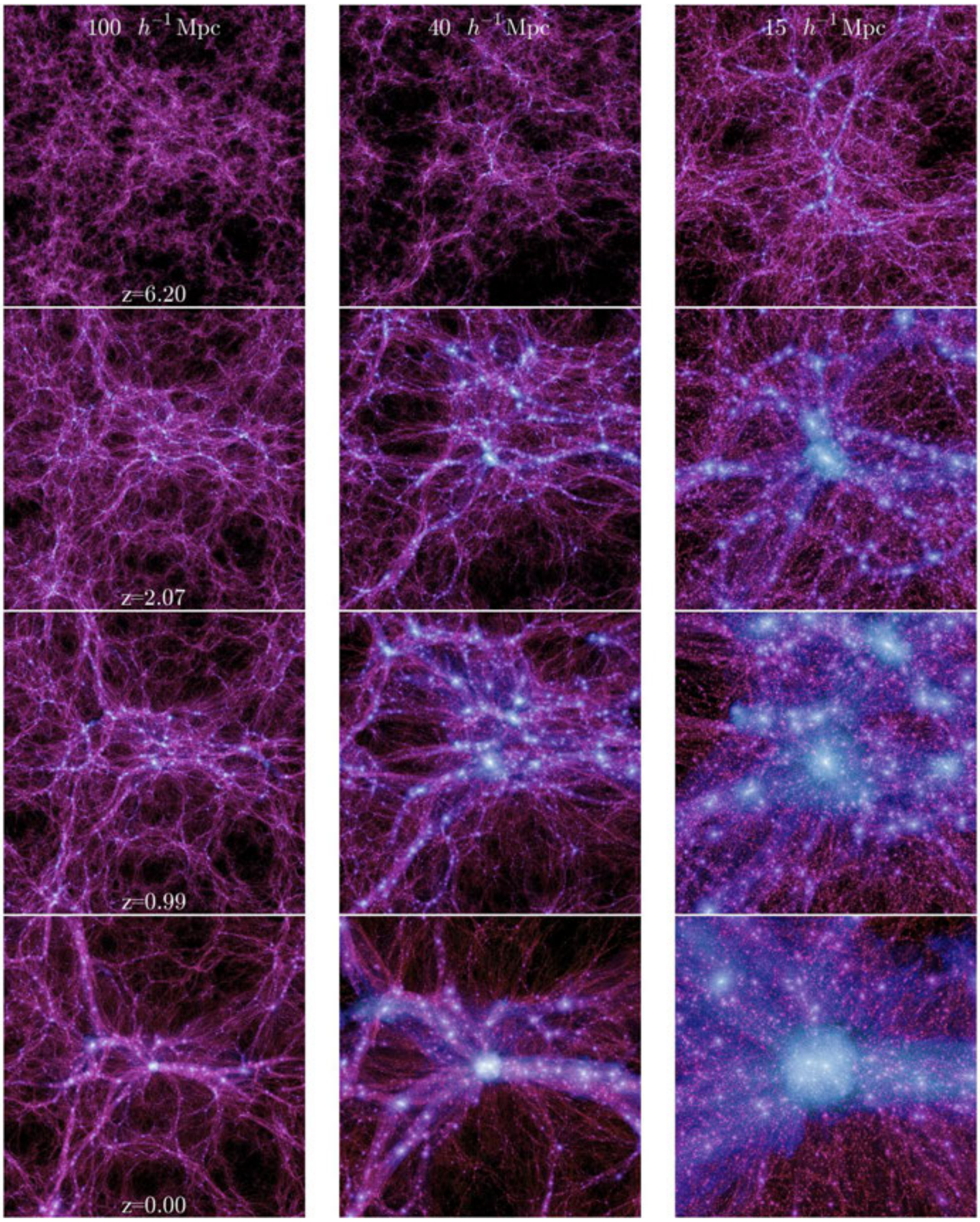

**Figure 1.2** – Time evolution of the largest halo in the Millenium-II cosmological simulation. The halo is shown at three comoving scales (from left to right: 100, 40 and 15 $h^{-1}$ Mpc) and at four different redshifts (from top to bottom: $z = 6.2$, 2.07, 0.99 and 0) (Boylan-Kolchin et al. 2009).

The complex physics associated with galaxy formation and evolution is still not well understood. Major ingredients to the recipe for galaxy assembly are the feedback processes from stars and black holes formed inside these galaxies, or their smaller progenitor systems (e.g. Rees 1993; Barkana & Loeb 2001; Cattaneo et al. 2009). In general, the concept of 'feedback' involves a back reaction of a process on itself, or on the source of the process. Feedback can be either negative or positive, intimately linked to the possibility that a system can become self-regulated. Some types of feedback processes can be disruptive, while others can drive a system towards some sort of steady state. The first stars fundamentally transformed the early Universe through their radiative, chemical, and mechanical feedback (for a review, see Ciardi & Ferrara 2005). They were the first sources of hydrogen-ionizing photons, beginning the process of reionizing the Universe, and of photons capable of dissociating molecular hydrogen (for a review, see Meiksin 2009; Zaroubi 2013). They produced the first metals, or chemical elements that are heavier than hydrogen and helium (aside from the trace amounts of lithium and deuterium resulting from the Big Bang) and the first dust (for a review, see Karlsson et al. 2013). Large amounts of energy were injected into the environment through stellar winds or when ending their lives as supernovae (SNe), creating shocks and turbulence, and possibly expelling gas out of the host halo (e.g. Wise & Abel 2008). Additionally, the first stars may have been the sites where dynamically significant magnetic fields were created for the first time (e.g. Tan & Blackman 2004; Schleicher et al. 2010a; Sur et al. 2010; Widrow et al. 2012). Feedback from black holes also includes radiative feedback, in particular their X-ray emission, and mechanical feedback in the form of winds and jets (e.g. Haiman et al. 2000; Aykutalp et al. 2014). The formation of black holes and their connection to galaxies will be discussed in more detail further on.

## 1.2 FIRST STARS

### 1.2.1 THE FIRST GENERATION OF STARS

The very first generation of stars, formed out of the pristine primordial gas, are often called Population III, or Pop III, stars. Their feedback radically changed the medium in their environment, resulting in a shift in star forma-tion mode for the next generation of stars to one more resembling what we see today (Population II/I). For a comprehensive review on the first stars, see e.g. Bromm (2013); Glover (2013).

The likely sites of the very first star formation were small dark matter halos, with masses of $\sim 10^6\,\mathrm{M}_\odot$ and forming at redshifts $z \sim 20-30$, which are often called minihalos or molecular cooling halos (e.g. Tegmark et al. 1997). For stars to form inside a minihalo, the halo mass must fulfill the Rees-Ostriker-Silk criterion (Rees & Ostriker 1977; Silk 1977), which requires that the cooling timescale, the timescale on which gas dissipates its internal energy, is shorter than the dynamical or free-fall timescale, which represents the fastest timescale on which matter can collapse. For a system with uniform density $\rho$, the dynamical time scales as $t_{\rm dyn} \propto 1/(G\rho)^{0.5}$. If the cooling time is much longer than the dynamical time, a state of equilibrium can be reached, otherwise, the system will collapse in a dynamical time. The sites of first star formation were predicted by theory, by considering the minimum halo mass that fulfills this collapse criterion at a given redshift, and comparing this to the likelihood that a halo of such mass arises from the gravitational collapse of a peak in the primordial density distribution. This picture has also been confirmed by 3D numerical simulations within a cosmological framework (Abel et al. 1998; Bromm et al. 2002; Yoshida et al. 2003).

The dark and baryonic matter inside a halo exist in a state of 'virial equilibrium', where the kinetic and gravitational potential energy are of the same order of magnitude. When the gas collapses in the halo, it will heat up, due to either adiabatic compression or shock heating. The gas temperature is given by the virial temperature, which is related to the virial velocity of the dark matter particles (Barkana & Loeb 2001),

$$T_{\rm vir} \simeq 2\times 10^3\,\mathrm{K}\left(\frac{M_{\rm halo}}{10^6\,\mathrm{M}_\odot}\right)^{2/3}\left(\frac{1+z}{20}\right) \tag{1.1}$$

It is apparent from this expression that minihalos typically have temperatures below $10^4$ K, which is the threshold for gas cooling through atomic hydrogen.

In order for star formation to occur, the gas has to contract and become denser. This means that gravity must be able to overcome the pressure inside the gas. Starting from the virial theorem, the condition for equilibrium of a stable, gravitationally bound system, a criterion can be obtained for the minimum mass susceptible to collapse; this is the Jeans mass. Assuming a spherical, homogeneous cloud of atomic gas, the thermal Jeans mass is given by

$$M_J = \left(\frac{5k_B T}{G\mu m_H}\right)^{3/2} \left(\frac{3}{4\pi\rho}\right)^{1/2}, \quad (1.2)$$

$$\propto \left(\frac{T^3}{\mu^3 \rho}\right)^{1/2}, \quad (1.3)$$

with $\rho$ the density, $T$ the gas temperature, $\mu$ the mean molecular weight, $G$ the gravitational constant, $k_B$ the Boltzmann constant, and $m_H$ the hydrogen mass. The same criterion for gravitational collapse can also be expressed in terms of cloud radius $R$, using the Jeans length $\lambda_J$:

$$\frac{\lambda_J}{2} = R_J = \left(\frac{15 k_B T}{4\pi G \mu m_H \rho}\right)^{1/2}. \quad (1.4)$$

Note that this is a simplification of the real situation, as it ignores rotation, possible internal macroscopic velocity gradients and magnetic fields, and any external pressure on the cloud. A cloud with a mass larger than the Jeans mass will start to collapse. Contraction will proceed if it is not opposed by internal pressure; however, the release of gravitational energy tends to increase the temperature and thus the pressure. Therefore, the gas must be able to dissipate energy, which it can do through radiative cooling.

Without atomic hydrogen as a viable coolant, the primordial gas in minihalos will instead have to cool through molecular hydrogen (Saslaw & Zipoy 1967). Formation of molecular hydrogen in the gas phase through the collision of two hydrogen atoms is difficult, due to the symmetry of the molecule and hence the absence of a permanent electrical dipole moment (Gould & Salpeter 1963). This means the excess energy cannot be radiated away via rapid (allowed) electric dipole radiation; instead, it must occur via slow (forbidden) magnetic quadrupole radiation. Before the excess kinetic energy can be radiated away, the compound system of two hydrogen atoms will decay and separate again. In the presence of dust, molecular hydrogen can form efficiently on the surface of dust grains, as the grain can absorb the excess energy (Hollenbach & Salpeter 1971). However, in the early Universe, before star formation had a chance to enrich the gas with metals and dust, this pathway is not available. It is still possible to form a significant amount of molecular hydrogen in the gas phase, but through a two-step process instead of direct formation. The most efficient pathway turns out to be the $H^-$-route, which relies on the presence of free electrons (McDowell 1961):

$$H + e^- \longrightarrow H^- + \gamma, \tag{1.5}$$

followed by

$$H^- + H \longrightarrow H_2 + e^-. \tag{1.6}$$

During the cloud collapse, in most cases the Jeans mass will decrease with increasing density. Any initial density fluctuations which may have been present within the cloud will cause individual regions to cross the instability threshold and collapse locally, leading to the formation of multiple smaller objects. Fragmentation stops when the fragments become opaque to their cooling radiation, leading to heating and an increase in internal pressure. Collapse will become close to adiabatic, and the Jeans mass now increases with increasing density, setting a minimum fragment mass (e.g. Omukai & Nishi 1998), and thus a minimum mass for the protostar, of $\sim 0.01\,M_\odot$ (Yoshida et al. 2008).

At present, there is broad agreement on the details of the Pop III protostar formation process. However, the further evolution of the protostar until it reaches the main sequence is much less clear, meaning there is still much uncertainty about the final mass of a Pop III star. Because the cooling in low-metallicity gas is so different from (near-)solar metallicity gas, the fragmentation behavior, and thus the star formation mode, is expected to be significantly different from Pop II/I stars. As molecular hydrogen cooling is less efficient than cooling in the presence of metals and dust, the gas temperature was likely higher. Higher temperatures mean a higher Jeans mass, which would suggest that the typical mass of a Pop III star is larger than the typical mass of a star at present, $\sim 0.1 - 1\,M_\odot$. Initially, it was thought that the initial mass function (IMF; describes the distribution of initial masses for a population of stars) would indeed be quite top-heavy, with typical stellar masses of $\sim 100\,M_\odot$ (Abel et al. 2002; Bromm et al. 2002). However, more recent simulations that follow the gas collapse beyond the formation of the first core find that fragmentation may be effective after all, and thus the first stars may tend to form in clusters with much lower individual masses than initially expected, $\lesssim 10\,M_\odot$ (Turk et al. 2009; Stacy et al. 2010, 2012; Clark et al. 2011b; Greif et al. 2011; Bovino et al. 2014b; Susa et al. 2014).

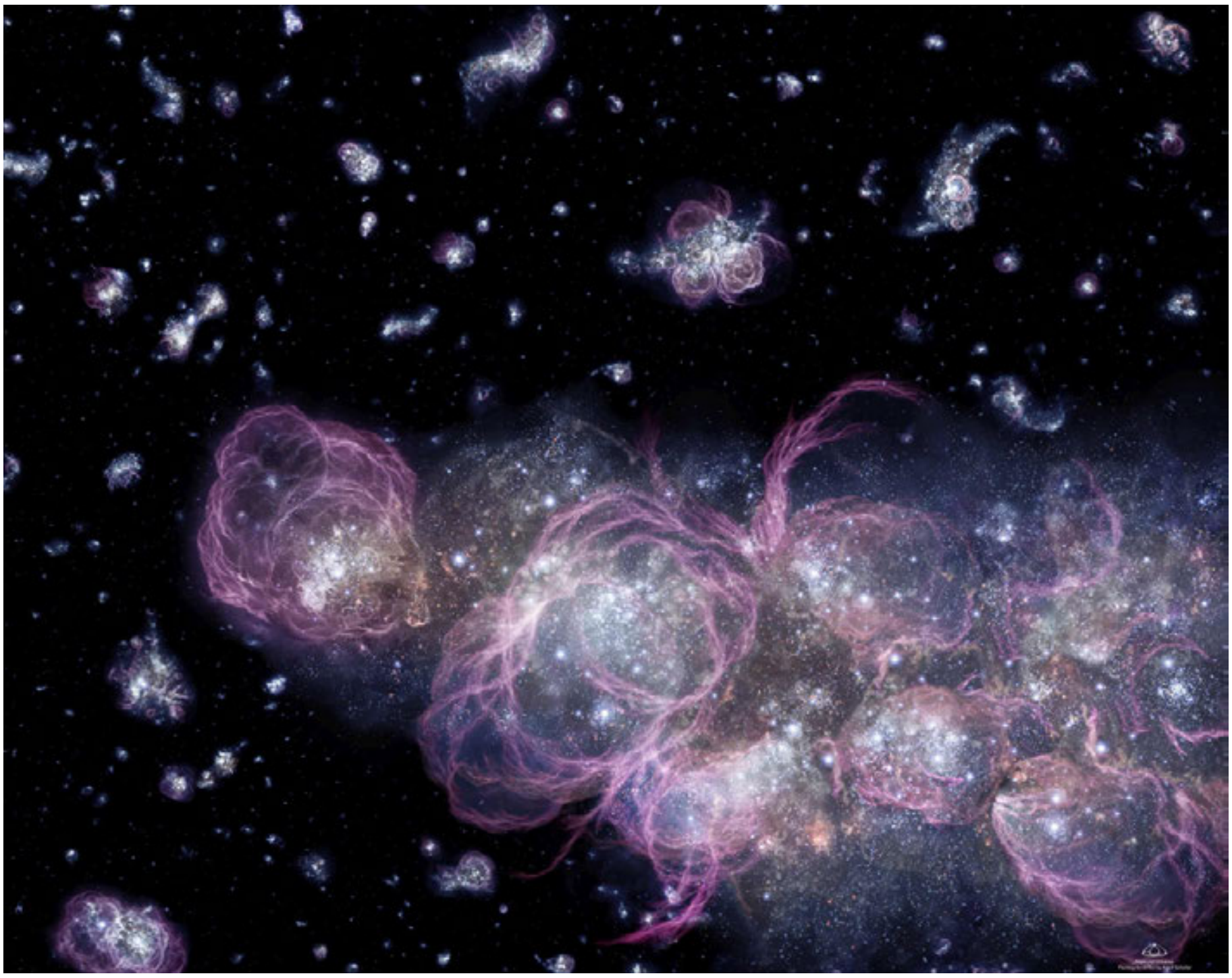

**Figure 1.3** – Artist's impression of how the very early Universe might have looked when going through intense bursts of star formation. (Credit: Adolf Schaller for STScI)

Another important, and as yet unsolved, problem, which is related to the IMF of the first stars, is to understand how and when cosmic star formation transitioned from the early star formation mode to the present mode.

### 1.2.2 THE SECOND GENERATION OF STARS

So far, we have discussed the very first stars, formed from pristine gas and unaffected by feedback from other luminous objects, sometimes referred to as Pop III.1. It may be conceivable that a next generation of stars, Pop III.2, formed from gas that has been altered significantly by radiative and/or mechanical feedback, but not by metal enrichment (Tan & McKee 2008). If the metal-free gas is irradiated by ultraviolet radiation that is efficient at dissociating $H_2$ molecules, the temperature will be higher, which may result in higher accretion rates, since $\dot{M} \propto T^{3/2}$, and a larger final mass (e.g. Hirano et al. (2015); see also Section 1.4). On the other hand, radiative feedback from the first generation of stars may also act to increase the ionization fraction. The larger abundance of free electrons boosts $H_2$ formation and cooling, possibly lowering the temperature to the point where HD cooling can become effective, which may even enable the gas to cool down to the temperature floor set by the CMB (e.g. Nakamura & Umemura 2002; Tan & McKee 2008). The lower temperatures and change in thermodynamical evolution may result in a lower final mass.

Several other processes could also lead to an enhanced ionization fraction and potentially allow the gas to reach the HD-dominated regime. Collisional ionization due to the strong shocks occurs when more massive halos undergo virialization (Johnson et al. 2008; Greif et al. 2008), though it may not be trivial to prevent the gas in these halos from having already been enriched with metals.

Another possibility is that second generation stars form in relic HII regions, which have previously been ionized by Pop III stars but have recombined, typically resulting in a boosted abundance of $H_2$ and HD (Yoshida et al. 2007). This may be possible as the volume that becomes ionized by a single massive Pop III star is much larger than the volume that becomes enriched with metals after the star dies as a supernova (Greif et al. 2010). A hurdle for this scenario may be the low density in such environments, as a consequence of outflows driven by the photoionization heating (Alvarez et al. 2006).

Finally, an enhanced ionization fraction could also be the result of irradiation by X-rays or cosmic rays. While the presence of an X-ray background appeared to be a promising pathway (Haiman et al. 2000), more recent simulations show that when employing realistic models for the X-ray background, star formation is relatively insensitive to the X-rays (Glover & Brand 2003; Machacek et al. 2003; Hummel et al. 2015; Latif et al. 2015). Cosmic ray ionization may therefore be the more important effect, though the magnitude of the cosmic ray ionization rate in the early Universe is still largely unknown (Jasche et al. 2007; Stacy & Bromm 2007; Inayoshi & Omukai 2011; Nakauchi et al. 2014).

### 1.2.3 TRANSITION FROM POP III TO POP II

As soon as the first metals and dust have been created and dispersed by energetic Pop III supernovae, the physics and chemistry of subsequent star formation will be fundamentally changed. Efficient cooling through atomic lines and dust grains becomes possible, and conditions will start to resemble the present-day interstellar medium in the Milky Way. It has been argued that there may be a threshold level of enrichment, a 'critical metallicity', that characterizes the transition from Pop III star formation mode to the more normal Pop II mode (Omukai 2000; Bromm et al. 2001a).

Currently, there are two main scenarios. One of them assumes that the main cooling channels are CII and OI finestructure lines, setting critical abundances on carbon and oxygen of $[C/H]_{\rm crit} \sim -3.5$ and $[O/H]_{\rm crit} \sim -3.0$ (Bromm & Loeb 2003b; Santoro & Shull 2006; Safranek-Shrader et al. 2014). The other assumes dust is the main coolant, and predicts critical abundances that are typically smaller by a factor of 10-100 (Schneider et al. 2003, 2012a; Omukai et al. 2005; Omukai 2012; Clark et al. 2008; Dopcke et al. 2011). On the other hand, it has also been suggested that there may not be a critical metallicity or dust fraction below which fragmentation does not occur at all (Jappsen et al. 2009a,b; Dopcke et al. 2013). There may also be other factors at work, like the redshift, which sets the CMB temperature floor, or magnetic fields. Clearly, the situation is very complex, and it seems that observational evidence is needed to truly understand the important physics and chemistry.

### 1.2.4 ORIGIN OF THE IMF

Observations have provided us with a pretty good idea of what the initial mass function of present-day star formation looks like. The IMF above $\sim 1\,{\rm M}_\odot$ has a power-law slope, ${\rm d}N_*/{\rm d}M_* \propto M_*^\alpha$, with $\alpha = 2.3(5)$, originally identified by Salpeter (1955). The IMF peaks at a few tenths of a solar mass, which sets the typical stellar mass. Below a solar mass, the distribution flattens (Miller & Scalo 1979; Kroupa 2001; Chabrier 2003), and there appears to be an upper mass limit of $\sim 150\,{\rm M}_\odot$ (Elmegreen 2000; Figer 2005).

Current evidence suggests that the IMF is near-universal over a wide range of star-forming environments, especially throughout the Milky Way and local galaxies. However, systematic variations of the IMF have been reported for extreme starburst environments and massive elliptical galaxies, potentially due to their unique formation environment. For a review of the observational evidence regarding the universality of the IMF and the potential variations in extreme environments, see Kroupa et al. (2013); Offner et al. (2014). If the features of the IMF as described above are indeed nearly universal, it means the star formation process must be relatively independent of the mean density, the amount of turbulence, the magnetic field strength, and, but only to a limited extent, metallicity.

For a comprehensive review on the theory of present-day star formation, see e.g. McKee & Ostriker (2007). Stars form out of cold, molecular gas, which is found in molecular clouds in the interstellar medium (ISM), the gas and (in the non-primordial case) dust present in a (proto)galaxy which is not incorporated in stellar systems. In the Milky Way, molecular clouds occupy a small fraction of the ISM volume, however, they comprise a significant fraction of the mass. Molecular clouds are embedded in regions of more tenuous and predominantly neutral, atomic gas, with a gradual transition in between. The presence of far-ultraviolet radiation, with photon energies in the range 6-13.6 eV (e.g. from stars), will significantly affect the chemistry and thermal balance of this gas, and such regions are then called photodissociation or photon-dominated regions (PDRs) (e.g. Spitzer 1978; Draine 2011). In terms of physical and chemical processes, essentially most of the ISM is in PDRs. Far-UV radiation traveling into a PDR becomes gradually attenuated, and ionized species first give way to neutral atomic species, while deeper into the region molecules become stable, transitioning into a molecular cloud.

Molecular clouds have a complex, hierarchical structure that extends from the scale of the cloud (~10-100 pc) down to the thermal Jeans mass in the case of gravitationally bound clouds, and down to much smaller masses for unbound structures. Two relevant structures that have been identified are clumps (~1 pc), which are likely the birthplaces of stellar clusters, and cores (~0.1 pc), which are the birthplaces of individual or binary stars.

Sufficient numbers of cores (dense condensations, often containing embedded protostars) have been observed within molecular clouds to derive core mass functions (CMF), analogous to the IMF (e.g. André et al. 2010; Könyves et al. 2010). The distribution of core masses is remarkably similar to the IMF, suggesting that they serve as gas reservoirs from which stars accrete most of their mass. If stellar masses are indeed largely determined by the amount of gas in their parent cores, it is important to understand the origin and distribution of the core masses in order to understand the IMF. The shapes of the CMF and IMF are similar, though the CMF is shifted to masses of about a factor 3 higher. This shift implies that stellar feedback and other processes in the collapse or post-collapse stage will affect stellar masses.

The presence of supersonic turbulence plays an essential role in defining the CMF. Turbulence is ubiquitous in the interstellar medium, and has a dual effect: it can both locally compress the gas, and provide support against gravity (e.g. Larson 1981; Mac Low & Klessen 2004; McKee & Ostriker 2007; Federrath & Klessen 2012). Because density fluctuations grow exponentially while velocities grow linearly, the presence of supersonic turbulence will always produce local dense regions that are less stable against collapse (the cores), even if turbulent motions balance gravity on a global scale. The most robust features of theoretical models describing the turbulent origins of the CMF are the Salpeter-like slope (similar to the IMF slope above a solar mass), and the log-normal-like turnover towards lower masses. The location of the turnover may be closely tied to the scale below which thermal (or magnetic) support dominates over turbulence. As a result, the location of the peak of the CMF, and by extension also of the IMF, will depend strongly on the gas thermodynamics (Larson 2005; Jappsen et al. 2005; Bonnell et al. 2006; Klessen et al. 2007). In any case, star formation regions vary so widely in their physical properties, including their turbulence and magnetic fields, that these do not really provide a compelling explanation for a universal typical mass scale.

While turbulent and magnetic support may delay gravitational collapse, such support will eventually dissipate or diffuse, leaving only thermal pressure to counteract gravity (e.g. Mac Low & Klessen 2004). The thermal Jeans mass, which depends strongly on temperature and thus on the chemical and thermodynamical properties of the gas, must therefore play a key role in determining the typical stellar mass. The relation between temperature and density can be described using a polytropic equation of state (EOS), $P \propto \rho^{\gamma_p}$, and assuming the pressure-temperature relation of an ideal gas, $P = nk_B T$. The polytropic exponent $\gamma_p$ then encompasses most of the thermodynamical properties, and is a measure for the compressibility of the gas (Vazquez-Semadeni et al. 1996; Scalo et al. 1998; Passot & Vázquez-Semadeni 1998; Spaans & Silk 2000; Scalo & Biswas 2002). The Jeans mass can be expressed in terms of the polytropic exponent, giving $M_J \propto \rho^{\frac{3}{2}(\gamma_p - \frac{4}{3})}$. When $\gamma_p < 1$, and thus temperature decreases with increasing density, the Jeans mass decreases, favoring the occurrence of fragmentation. When $1 \lesssim \gamma_p < 4/3$, the temperature increases with density, and the Jeans mass decreases much more slowly, suggesting fragmentation is much less likely. Finally, if $\gamma_p > 4/3$, fragmentation and collapse will cease entirely. Thus, the density at which the temperature reaches its minimum value and begins to increase may be the highest density at which fragmentation is probable, setting the typical mass scale (provided there is only a single temperature minimum).

Numerical simulations have also shown that the amount of fragmentation is very sensitive to the exact temperature-density relation in collapsing clouds (Li et al. 2003; Jappsen et al. 2005; Bonnell et al. 2006; Klessen et al. 2007; Federrath & Banerjee 2015). The thermodynamical properties, and thus the polytropic exponent, depend strongly on e.g. metallicity and the background radiation field, which is related to the redshift through the CMB. Therefore, in order to determine the IMF of the first, as well as later generations of stars, it is vital to understand the chemical and thermodynamical properties of the gas under the relevant conditions.

## 1.3 FIRST GALAXIES

While minihalos are regarded as the birthplaces of the first stars, they may not be considered true galaxies, if one condition for a system to be called a galaxy is that the potential well of the halo is deep enough to retain photoheated gas and gas heated and accelerated by supernova (SN) explosions. If the IMF of the first stars was not too different from the present IMF, minihalos would not be severely affected by feedback and could have hosted the first galaxies. However, if many more massive stars were formed, strong negative feedback would shut off star formation altogether by blowing out the gas. Halos with a deeper potential well would be required to revirialize the gas affected by stellar feedback, which is the case for halos with a mass of $\sim 10^8\,M_\odot$, collapsing at $z \sim 10$. Such systems may be better candidates for being the first galaxies. They have a virial temperature of $\gtrsim 10^4$ K, which is above the threshold for atomic hydrogen cooling; therefore, such halos are often referred to as atomic cooling halos. These systems may also be viable formation sites for the seeds of the first supermassive black holes, which will be discussed in more detail further on. For a comprehensive review on the first galaxies, see e.g. Bromm & Yoshida (2011).

## 1.4 FIRST MASSIVE BLACK HOLES

### 1.4.1 BLACK HOLES 101

As mentioned before, our current picture of galaxy formation and evolution is not complete without black holes. Black holes are some of the strangest and most fascinating objects that exist in our Universe. They are objects of extreme density, with such strong gravitational attraction that nothing, not even light, can escape from its pull once it gets close enough. As such, black holes do not directly emit light and therefore cannot be seen in the traditional sense, which presents a challenge to observers. However, we can still 'observe' them in other ways, for example by looking for the effects of the black hole's strong gravity on its surroundings, e.g. by observing the motions of nearby stars, and by looking for the X-ray radiation that originates from material being accreted by the black hole. A pair of black holes in the process of merging can also be observed through the detection of their gravitational wave emission (Abbott et al. 2016).

#### BLACK HOLE TYPES

A black hole has only three independent physical properties: mass, angular momentum, and electric charge. The size of a black hole, $r_S$, given by the radius of the event horizon or Schwarzschild radius, is roughly proportional to the mass $M$ through

$$r_S = \frac{2GM}{c^2} \approx 2.95 \frac{M}{M_\odot} \text{km} \tag{1.7}$$

This relation is exact only for black holes with zero charge and angular momentum; for more general black holes it can differ by up to a factor of 2.

Black holes are commonly classified according to their mass, independent of angular momentum or charge.

- Micro or mini black holes are hypothetical tiny black holes for which quantum mechanics effects play an important role, with sizes smaller than an atom, up to ~0.1 mm with a mass of $\sim M_{Moon}$. Shortly after the Big Bang, when the Universe was very dense, such small primordial black holes could have formed.
- Currently, the most common type of black holes are stellar black holes, which result from the gravitational collapse at the end of the life of a massive star. They typically have masses ranging from about 5 $M_\odot$ up to several tens of solar masses. It is estimated that the Milky Way contains a few hundred million stellar black holes.
- More massive black holes, with masses in the range of ~100 to $10^5$ $M_\odot$, are called intermediate mass black holes (IMBHs). So far, there has not yet been an unambiguous detection of a black hole of such a mass, though there have been hints that they exist, for example, in globular clusters.
- Finally, the very largest type of black holes are the supermassive black holes, with masses exceeding $10^5$ $M_\odot$. Evidence suggest that almost all massive galaxies contain a supermassive black hole at their center. In the Milky Way, the supermassive black hole corresponds to the location of the compact radio source Sagittarius A$^*$, and it has a mass of $\sim 4 \times 10^6$ $M_\odot$ while being about the size of the Sun. Dynamical estimates suggest that, across a wide range, the central black hole mass equals about 0.1% of the mass of the spheroidal component of the host galaxy (Magorrian relation; Magorrian et al. (1998)). This and other correlations between the black hole mass and the properties of the host galaxy point to a common root or co-evolution between galaxies and their central black hole.

In the case where there is enough matter present near the supermassive black hole (SMBH) in the center of a galaxy, an accretion disk will form, feeding gas and dust into the black hole. Dissipative processes in the accretion disk transport material inwards and angular momentum outwards, while causing the accretion disk to heat up. This results in broadband emission with a significant X-ray component (about 80% UV and 10% X-ray emission, in terms of bolometric luminosity). An accreting supermassive black hole would be visible as a compact region at the center of a galaxy, with a much higher than normal luminosity over at least some portion (and possibly all) of the electromagnetic spectrum. Observations of such objects are called active galactic nuclei or AGN, and the most likely explanation for their source is indeed an accreting supermassive black hole. The most energetic and distant AGNs are also called quasars, or quasi-stellar radio sources.

### 1.4.2 FORMATION OF SMBH SEEDS

Several very bright quasars, with bolometric luminosities $\gtrsim 10^{47}\,\mathrm{erg\,s^{-1}}$, have been detected already at $z > 6$, when the Universe was less than a tenth of its current age. These high-redshift quasars are very rare, with a space density of the order of $\sim 1\,\mathrm{Gpc^{-3}}$, and have only been found in large surveys, such as the Sloan Digital Sky Survey (SDSS), or the smaller-area but deeper CFHQS and UKIDSS surveys. This suggests that some SMBHs with masses of $\simeq 10^9\,\mathrm{M_\odot}$ already existed less than 1 Gyr after the Big Bang (Fan 2006; Willott et al. 2010; Mortlock et al. 2011; Venemans et al. 2013). It is possible that these bright quasars represent only the tail of the mass distribution, which would imply that large numbers of massive black holes might have existed at that time.

Explaining how SMBHs with such large masses could have assembled so soon after the Big Bang presents quite a challenge. The main questions concern how and when the ‘seeds’ of these SMBHs formed and how their subsequent growth proceeded. Several pathways leading to the formation of seed black holes (SBHs) have been proposed. A diagram which summarizes

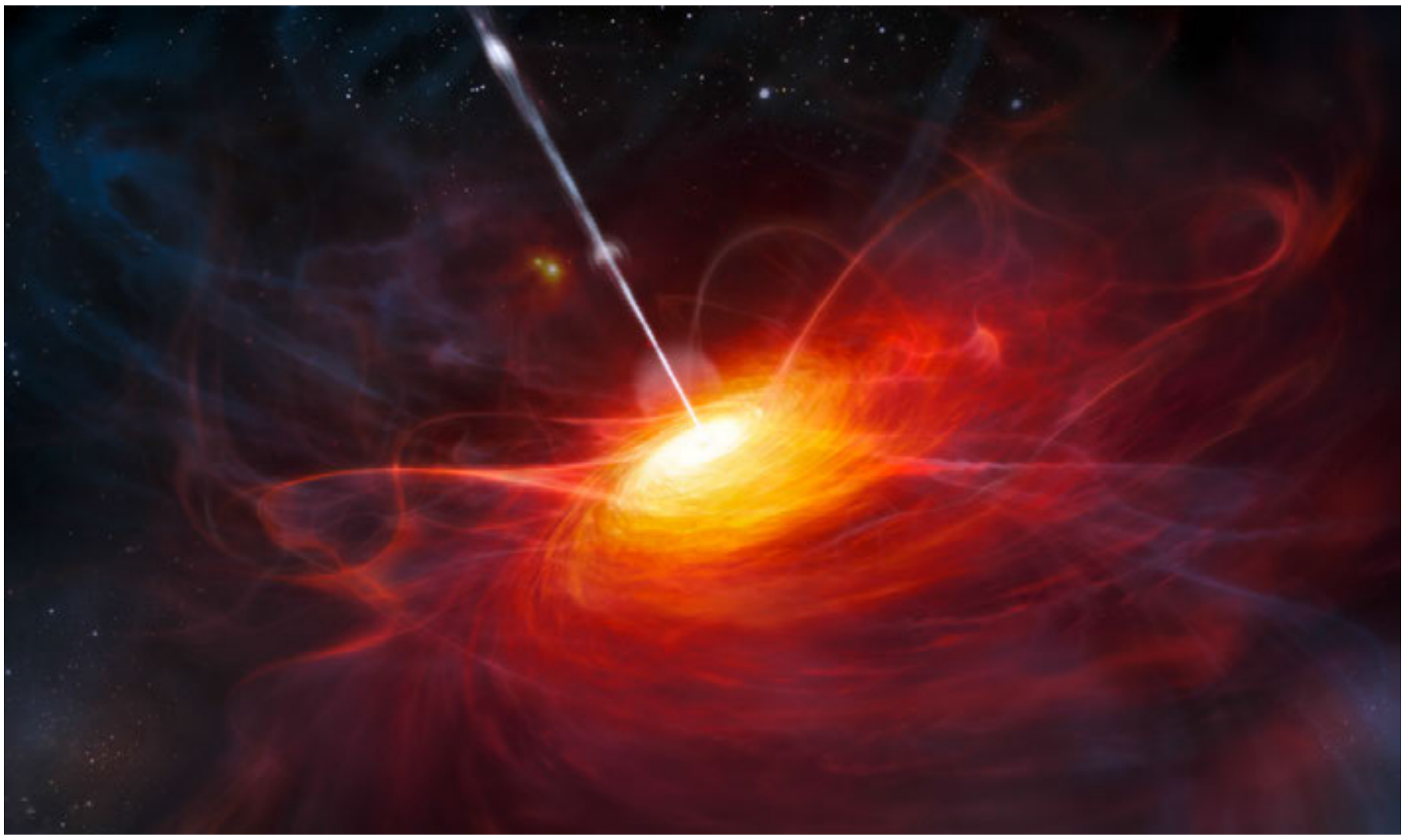

**Figure 1.4** – Artist's impression of how ULAS J1120+0641, a very distant quasar powered by a black hole with a mass of $\sim 2 \times 10^9\,M_\odot$, may have looked, just 770 million years after the Big Bang (Mortlock et al. 2011). (Credit: ESO/M. Kornmesser)

the main possibilities for the formation of such seed black holes in high-redshift galaxies is shown in Figure 1.5 (for a comprehensive review, see Volonteri 2010; Haiman 2013; Latif & Ferrara 2016).

### A REMNANTS OF POP III STARS

Perhaps the most obvious scenario assumes that SMBHs grow from the remnants of the first stars. Whether or not a Pop III star will end its life as a black hole depends on its mass and angular momentum. The fate of non-rotating stars as a function of mass and metallicity was investigated by (Heger & Woosley 2002; Heger et al. 2003). Low-metallicity stars with masses in the range ~25-140 $M_\odot$ are expected to form black holes, either by fallback after a core collapse supernova or directly, with $M_{BH} \sim 10-50\,M_\odot$ (Zhang et al. 2008). The problem with these light black holes is that they may not be dynamically stable within the center of their host: they may move around due to interactions with stars of similar mass, rather than settling at the center of the potential well. Stars with masses in the range ~140-260 $M_\odot$ are predicted to explode as pair-instability supernovae, and leave no remnant behind. And still more massive stars, with masses in excess of ~260 $M_\odot$, are also expected to form black holes, with masses at least half of the initial stellar mass, $M_{BH} \sim 100 - 600\,M_\odot$ (Bond et al. 1984; Fryer et al. 2001). These would be good seed black hole candidates. However, the shape of the initial mass function of Pop III stars is still an unsolved problem, and it is not known if there are enough, if any, Pop III stars with masses above the threshold (~260 $M_\odot$) for the formation of these intermediate mass black holes.

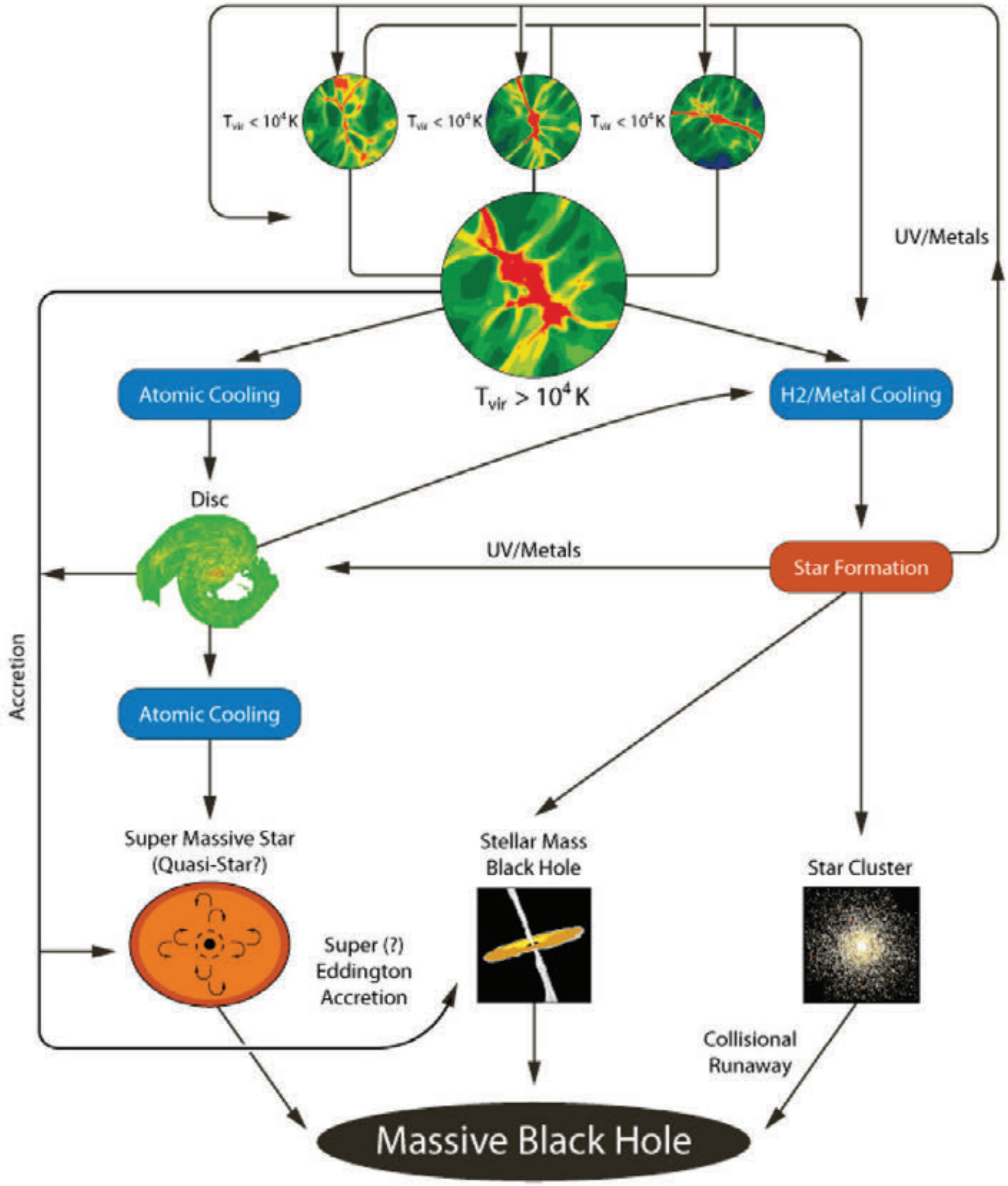

**Figure 1.5** – Summary of the possible pathways to massive black hole formation in DM halos with $T_{\rm vir} \gtrsim 10^4$ K: via a stellar seed black hole, via a very massive or quasi star (also called the direct collapse scenario), or via runaway collisions in a nuclear star cluster (Regan & Haehnelt 2009a).

Even if very massive stars were able to form and collapse into seed black holes, it would be difficult for these relatively light seeds to accrete sufficient mass and grow into a supermassive black hole in the available time. It has been suggested that super-Eddington accretion may be necessary to accomplish this (e.g. Madau et al. 2014). However, the ionized HII region formed around the seed black hole significantly reduces the accretion rate onto the seed (Milosavljević et al. 2009b,a; Alvarez et al. 2009; Johnson et al. 2011; Aykutalp et al. 2014).

### B STELLAR CLUSTER COLLISIONS

Another scenario predicts the formation of seed black holes from very compact nuclear star clusters, which may form at redshifts of $\sim 10 - 15$, after some metal enrichment has occurred so that metal line-cooling becomes effective, and in the presence of trace amounts of dust (Schneider et al. 2003, 2012a,b; Omukai et al. 2005; Omukai 2012; Clark et al. 2008; Klessen et al. 2012; Dopcke et al. 2011, 2013; Safranek-Shrader et al. 2014; Bovino et al. 2014a). In such a cluster, stellar collisions can occur in a runaway fashion and lead to the formation of a very massive star, finally resulting in a seed black hole with a mass up to ~3000 $M_\odot$ (Begelman & Rees 1978; Portegies Zwart et al. 2004; Devecchi & Volonteri 2009; Devecchi et al. 2010, 2012; Lupi et al. 2014). Although this is significantly more massive than what is expected for the first generation of stars, it will still be difficult for such seeds to grow into the observed SMBHs in the available time.

## C DIRECT COLLAPSE

In this thesis, we focus on a third group of scenarios, in which the primordial gas in a halo would collapse directly into a single central object, without fragmenting, resulting in seed black holes with masses $\sim 10^4$-$10^6\,M_\odot$ (e.g. Bromm & Loeb 2003a; Koushiappas et al. 2004; Begelman et al. 2006; Lodato & Natarajan 2006; Spaans & Silk 2006; Schleicher et al. 2010b; Latif et al. 2013a; Regan et al. 2014). The most likely host candidates are atomic cooling halos with virial temperatures $\gtrsim 10^4$ K at redshifts $\sim 10-15$. For such a direct collapse to occur, it is necessary that fragmentation is limited, which is possible if the gas in the halo is kept hot (and thus the Jeans mass high). Hence, the formation of molecular hydrogen must be inhibited so that cooling can occur only through atomic hydrogen, as otherwise $H_2$ cooling will lower temperatures to $\sim$200 K and fragmentation may occur. In the absence of $H_2$ cooling, self-gravitating gas will collapse nearly isothermally until it becomes optically thick to radiative cooling, at which point the (nearly) adiabatic phase sets in.

One plausible mechanism for suppressing the formation of sufficient $H_2$ is the presence of a supercritical ultraviolet (UV) radiation background. Such a radiation field could be present in a small subset of all halos due to the presence of a stellar population in a neighboring halo (Dijkstra et al. 2008; Agarwal et al. 2014; Dijkstra et al. 2014; Visbal et al. 2014). If massive, the first generation of stars (Pop III) is expected to have a stellar spectrum with a characteristic temperature of $\sim 10^5$ K (T5 spectrum; Tumlinson & Shull 2000; Bromm et al. 2001b; Schaerer 2002), while the lower mass second generation of stars (Pop II) has a softer spectrum with several $10^4$ K (T4 spectrum). UV radiation in the Lyman-Werner bands (11.2-13.6 eV) with an intensity above a certain threshold is able to photodissociate $H_2$ and $H^-$, which is an important intermediary species for the formation of $H_2$, and keep their abundance very low. A T5 spectrum will mainly directly photodissociate $H_2$, while a T4 spectrum is better at photodetaching $H^-$.

Alternative mechanisms for inhibiting $H_2$ cooling comprise dissipation of a sufficiently strong magnetic field (Schleicher et al. 2009; Sethi et al. 2010; Van Borm & Spaans 2013), trapping of Ly$\alpha$ photons (Spaans & Silk 2006), or the presence of strong shocks (Inayoshi & Omukai 2012), all of which result in collisional dissociation of $H_2$.

If fragmentation can be suppressed in the collapsing gas, it will contract until rotational support halts the collapse. Usually, this will happen before a black hole has been created; instead a disk will form. Additional mechanisms for angular momentum transport are required to further condense the gas and eventually form a black hole. These disks may again be susceptible to fragmentation Bromm & Loeb (2003a); Regan & Haehnelt (2009b), in part depending on the amount of rotation, with stronger rotation inducing more fragmentation (Bromm et al. 2002; Machida 2008; Hocuk & Spaans 2010), and the presence of turbulence, which plays an important role in angular momentum transport and determining the fragmentation properties of collapsing gas clouds, since it can both locally compress the gas as well as provide additional support against collapse on larger scales (e.g. Larson 1981; Mac Low & Klessen 2004; McKee & Ostriker 2007; Federrath & Klessen 2012). However, the occurrence of mild fragmentation at this point is not a major hurdle, as simulations show that this does not prevent the growth of a central massive object through accretion and mergers (Latif et al. 2013a; Shlosman et al. 2016).

Although the detailed evolutionary pathways are still not well understood, a possible outcome of the direct collapse scenario is the continued collapse of some gas to smaller scales in the galactic nucleus. As the gas flows in, it becomes optically thick and cooling photons can no longer escape; further evolution will be close to adiabatic. If the radiation pressure is strong enough to temporarily balance gravity, the resulting object will be a 'protostar'. Once the protostar has formed, it will accrete and evolve into either a supermassive star or a quasistar, depending on the accretion rate. The work by Schleicher et al. (2013) suggests that for accretion rates $>0.14\,M_\odot/\mathrm{yr}$, a quasistar will be the result, while lower accretion rates lead to the formation of a supermassive star (SMS). A supermassive star (with a mass in the range $10^3$-$10^6\,M_\odot$) of fixed mass, supported by radiation pressure, is thought to evolve as an $n = 3$ polytrope and finally collapse into a black hole containing most of the stellar mass (Johnson et al. 2011; Whalen et al. 2013; Hosokawa et al. 2012a, 2013). However, if the mass accretion rate is high enough, the outer layers of the SMS cannot thermally relax. In this case, it is not well-described by an $n = 3$ polytrope, but will have a more complex structure with a convective core surrounded by a convectively stable envelope that contains most of the mass. The core will burn up its hydrogen, and subsequently collapse into a black hole with a mass of a few $M_\odot$. The resulting structure, where the black hole accretes material from the massive, radiation-pressure-supported envelope, is called a 'quasistar' (Begelman et al. 2006, 2008; Begelman 2010; Volonteri & Begelman 2010; Ball et al. 2011, 2012).

### 1.4.3 FROM SEED TO SMBH

Once a seed black hole is formed, it must grow rapidly within a short timespan to explain the observed high-redshift quasars. Mass accretion at the Eddington rate causes a black hole to increase in mass over time as

$$M_{\rm BH}(t) = M_{\rm BH}(0)\exp\left(\frac{1-\epsilon}{\epsilon}\frac{t}{t_{\rm Edd}}\right), \tag{1.8}$$

where $t_{\rm Edd} = 0.45\,$Gyr and $\epsilon$ is the radiative efficiency. This means that, for a 'standard' efficiency of ~0.1, it takes a $10^2\,M_\odot$ seed at least ~0.81 Gyr and a $10^5\,M_\odot$ seed at least ~0.46 Gyr to grow into a $10^9\,M_\odot$ black hole. A larger radiative efficiency of 0.2 increases the growing time to ~1.81 Gyr for a $10^2\,M_\odot$ seed and to ~1.03 Gyr for a $10^5\,M_\odot$ seed. However, the black hole might not accrete at the Eddington rate the whole time, since the accretion rate could be limited by several different factors, both 'external' and 'internal' effects. On one hand, the external conditions relate to the amount of gas available for accretion. The constant availability of gas in the halo during the accretion period could require halos to merge, since episodes of star formation and feedback from supernovae might deplete the gas. On the other hand, the internal effects relate to feedback from the radiative output produced by the accreting black hole itself (see e.g. Pelupessy et al. 2007; Johnson & Bromm 2007; Milosavljević et al. 2009b,a; Park & Ricotti 2011; Spaans et al. 2012). The UV and X-ray photons produced by the accretion disk around the black hole may drive an HII region that expels gas from the central regions and strongly affects the thermodynamics of (star-forming) interstellar gas. Radiative pressure may play a role within a few pc from the black hole (Aykutalp et al. 2013). Another internal effect is the mechanical feedback from strong outflows (e.g. jets) that may be driven by the accreting black hole. These outflows can also expel gas, as well as create turbulence in the young galaxy.

## 1.5 THIS THESIS

### PART I: SUPERMASSIVE BLACK HOLE SEEDS

In the first part of this thesis, we investigate the formation of seeds of the first supermassive black holes through the direct collapse scenario.

Chapter 2 of the thesis explores the formation of a protostar resulting from the collapse of primordial gas in the presence of a strong UV radiation background, which is one way of keeping the gas hot. Particularly, we investigate the impact of turbulence and rotation on the fragmentation behavior of the gas cloud. In order to do this, we perform numerical 3D adaptive mesh refinement (AMR) simulations using the ENZO code, with the addition of the KROME package to improve modeling of the chemical and thermal processes. This enables us to simulate the formation of a protostar up to unprecedentedly high central densities of $10^{21}$ cm$^{-3}$ and spatial scales of a few solar radii.

Other ways in which the gas may be able to stay hot and avoid fragmentation are explored in Appendix 2.B, using an analytical one-zone model. Particularly, we examine the interplay between magnetic fields, turbulence, and a UV radiation background during the gravitational collapse of primordial gas in a halo. We follow the evolution of a cloud of primordial gas from its initial cosmic expansion through turnaround, virialization, and collapse up to a density of $10^{7}$ cm$^{-3}$.

### PART II: ISM THERMODYNAMICS AND FRAGMENTATION

In the second part of the thesis, we focus on the equilibrium state of interstellar material (gas and dust) under various physical conditions, to better understand the chemo-thermal state and fragmentation behavior of gas in high-redshift galaxies.

In Chapter 3 of the thesis, we describe a photodissociation region (PDR) code. Our code, PDR-Zz, is based on the PDR code by Meijerink & Spaans (2005), which we updated and significantly improved, and also extended for use in a wider range of physical conditions (e.g. much higher densities, metallicities from essentially 0 to solar, redshifts >0). This computational code can be used to study traditional PDRs, and can additionally be applied to many other situations where one is interested in steady-state gas at different optical depths, under various conditions; from diffuse and molecular clouds in the Milky Way to clouds near AGN and material in galaxies in the high-redshift Universe. The input parameters to the code are the total hydrogen density, the metallicity, the incident UV radiation field strength, the cosmic ray ionization rate, and the redshift; the combination of which defines the coupled chemical and thermal solution for the simulated region. In this chapter, we describe in detail which chemical species, reactions, and rates are used, as well as which cooling and heating processes are included in the model. Additionally, we describe our improvements to the convergence algorithms for both the chemical and thermal balance, which result in a large speedup and enable us to run many models in a relative short time.

In Chapter 4, we explore in a systematic way the overall impact of various feedback effects, both radiative and chemical, on the chemical and thermal balance of the gas in different regimes. A grid of models is run using our numerical code PDR-Zz, which provides 1D models using a steady-state approach. The grid covers a sizable range in total hydrogen density, metallicity, redshift, UV radiation field scaling parameter, and cosmic ray ionization rate. We provide insight in the most important processes and dominant chemical species in different regimes. We show the effects of radiative and chemical feedback on the effective gas temperature both in the outer layers and in the bulk of the cloud, and describe the differences. Finally, we examine the equations of state in our parameter space, identify several regions of interest, and relate these to the IMF.

Finally, in Chapter 5 we conclude with an outlook, where we consider the possibilities for future work, both theoretical and observational, related to the first black holes, first stars, and the IMF.

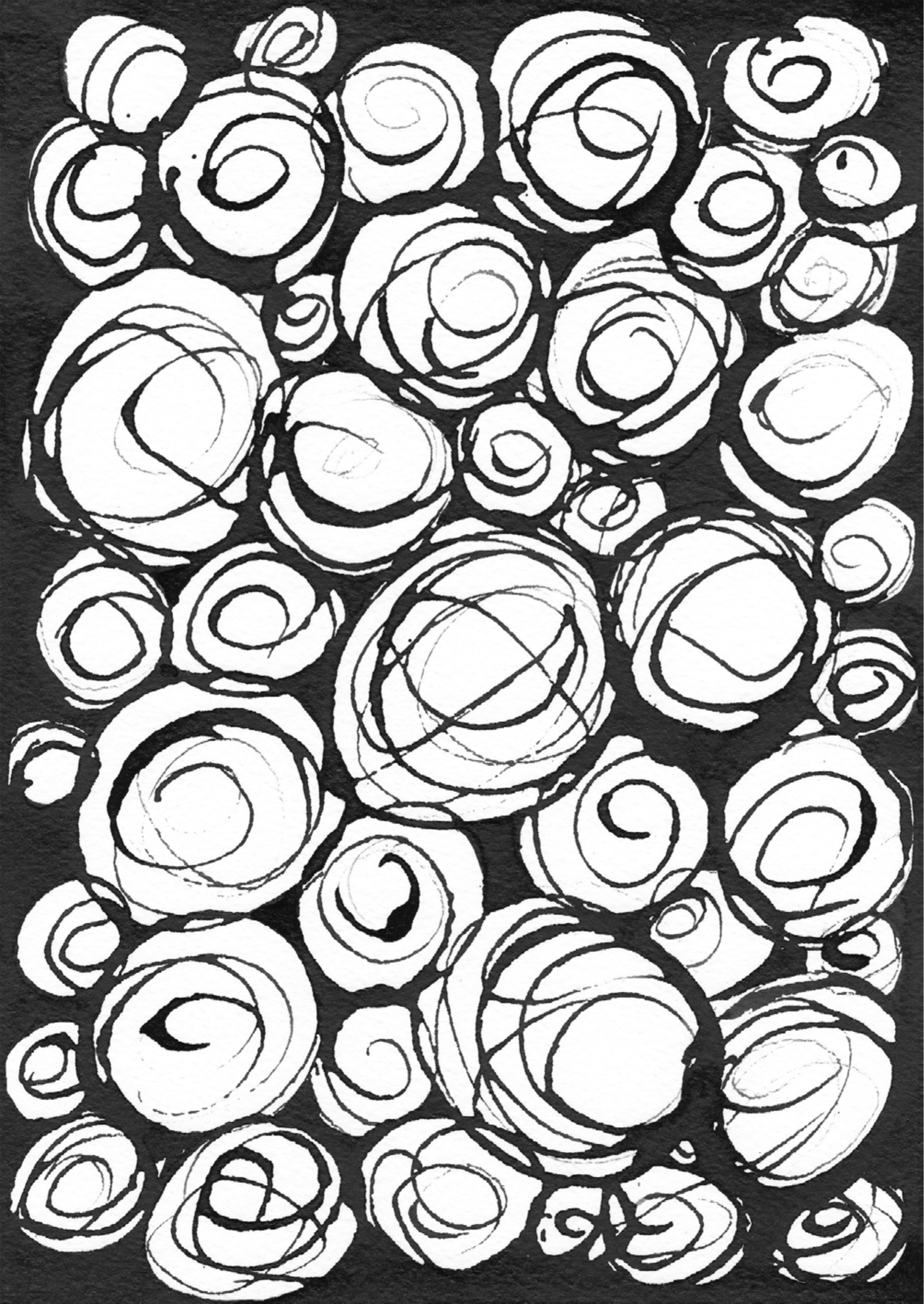

# Part I

# Supermassive Black Hole Seeds

*THE ONLY TRUE WISDOM IS IN KNOWING YOU KNOW NOTHING.*

— SOCRATES

CHAPTER

# 2

# EFFECTS OF TURBULENCE AND ROTATION ON PROTOSTAR FORMATION AS A PRECURSOR OF MASSIVE BLACK HOLES

## ABSTRACT

**Context** The seeds of the first supermassive black holes may have resulted from the direct collapse of hot primordial gas in $\gtrsim 10^4$ K halos, forming a supermassive or quasistar as an intermediate stage.
**Aims** We explore the formation of a protostar resulting from the collapse of primordial gas in the presence of a strong Lyman-Werner radiation background. Particularly, we investigate the impact of turbulence and rotation on the fragmentation behavior of the gas cloud. We accomplish this goal by varying the initial turbulent and rotational velocities.
**Methods** We performed 3D adaptive mesh refinement simulations with a resolution of 64 cells per Jeans length using the ENZO code, simulating the formation of a protostar up to unprecedentedly high central densities of $10^{21}$ cm$^{-3}$ and spatial scales of a few solar radii. To achieve this goal, we employed the KROME package to improve modeling of the chemical and thermal processes.
**Results** We find that the physical properties of the simulated gas clouds become similar on small scales, irrespective of the initial amount of turbulence and rotation. After the highest level of refinement was reached, the simulations have been evolved for an additional ~5 freefall times. A single bound clump with a radius of $2 \times 10^{-2}$ AU and a mass of $\sim 7 \times 10^{-2}$ $M_\odot$ is formed at the end of each simulation, marking the onset of protostar formation. No strong fragmentation is observed by the end of the simulations, regardless of the initial amount of turbulence or rotation, and high accretion rates of a few solar masses per year are found.
**Conclusion** Given such high accretion rates, a quasistar of $10^5$ $M_\odot$ is expected to form within $10^5$ years.

## 2.1 INTRODUCTION

Several very bright quasars have been detected at $z > 6$, which suggests that supermassive black holes (SMBHs) with masses of $\sim 10^9$ $M_\odot$ already existed when the Universe was less than 1 Gyr old (Fan 2006; Willott et al. 2010; Mortlock et al. 2011; Venemans et al. 2013). It is challenging to explain how such SMBHs could have assembled so soon after the Big Bang, in particular

2

how and when the 'seeds' of these SMBHs formed and how their subsequent growth proceeded. Various scenarios for the formation of seed black holes in the early Universe have been proposed and are briefly discussed below (for a detailed review, see Volonteri (2010); Haiman (2013)).

Perhaps the most obvious scenario assumes that SMBHs grow from the first stellar remnants. The first stars are thought to form at redshifts of $\sim 20-50$ in minihalos of $\sim 10^6\,M_\odot$, cooled by molecular hydrogen (Tegmark et al. 1997; Abel et al. 2002; Bromm et al. 2002; Latif et al. 2013c; Bovino et al. 2013; Hirano et al. 2014). The first generation of stars was expected to have a more top-heavy initial mass function (typical stellar masses $\sim 100\,M_\odot$) than the current mode of star formation, resulting from the inefficient cooling in these minihalos. However, more recent simulations that follow the collapse beyond the formation of the first core find that fragmentation may be effective after all, and thus the first stars may tend to form in clusters with much lower individual masses than initially expected, $\lesssim 10\,M_\odot$ (Turk et al. 2009; Stacy et al. 2010, 2012; Clark et al. 2011b; Greif et al. 2011; Bovino et al. 2014b; Susa et al. 2014). Accretion luminosity does not seem to have much influence on the fragmentation behavior (Smith et al. 2011, 2012). On the other hand, stellar UV feedback appears to inhibit accretion onto the protostar, which would result in an upper limit on the stellar mass of $\sim 50-100\,M_\odot$ (Hosokawa et al. 2011, 2012b; Susa 2013). Even if very massive stars were able to form and collapse into seed black holes, it would be difficult for them to accrete sufficient mass in the available time. It has been suggested that super-Eddington accretion may be necessary to accomplish this (e.g. Madau et al. 2014). However, the HII region formed around the seed black hole significantly reduces the accretion rate onto the seed (Milosavljević et al. 2009b,a; Alvarez et al. 2009; Johnson et al. 2011). In addition, it has been found that stars with masses $\sim$100-10 000 $M_\odot$ may form in massive primordial halos irradiated by a moderate UV background, with a strong correlation between the strength of the UV flux and the mass of a protostar (Latif et al. 2014b).

Another scenario predicts the formation of seed black holes from very compact nuclear star clusters, which may form at redshifts of $\sim 10-15$, after some metal enrichment has occurred so that metal line-cooling becomes effective, and in the presence of trace amounts of dust (Schneider et al. 2003, 2012a,b; Omukai et al. 2005; Omukai 2012; Clark et al. 2008; Klessen et al. 2012; Dopcke et al. 2011, 2013; Safranek-Shrader et al. 2014; Bovino et al.

2014a). In such a cluster, stellar collisions can occur in a runaway fashion and lead to the formation of a very massive star, finally resulting in a seed black hole with a mass up to $\sim 3000\,M_\odot$ (Begelman & Rees 1978; Portegies Zwart et al. 2004; Devecchi & Volonteri 2009; Devecchi et al. 2010, 2012; Lupi et al. 2014). Although this is significantly more massive than what is expected for the first generation of stars, it will still be difficult for such seeds to grow into the observed SMBHs in the available time.

In this work, we focus on a third pathway: the so-called direct collapse scenario. In this case, the primordial gas in a halo would collapse directly into a single central object, without fragmenting (e.g. Bromm & Loeb 2003a; Koushiappas et al. 2004; Begelman et al. 2006; Lodato & Natarajan 2006; Spaans & Silk 2006; Schleicher et al. 2010b; Latif et al. 2013a; Regan et al. 2014). The most likely host candidates are halos with virial temperatures $\gtrsim 10^4$ K at redshifts $\sim 10-15$. For a direct collapse to occur, it is important that fragmentation is suppressed, which is possible if the gas in the halo is kept hot (and thus the Jeans mass high). Hence, the formation of $H_2$ must be inhibited so cooling can occur only through atomic hydrogen, because otherwise molecular hydrogen cooling will lower temperatures to $\sim 200$ K and fragmentation may occur. In the absence of $H_2$ cooling, self-gravitating gas will collapse nearly isothermally until it becomes optically thick and the adiabatic phase sets in.

One plausible mechanism for suppressing the formation of sufficient $H_2$ is the presence of a UV radiation background. If massive, the first generation of stars (Pop III) is expected to have a stellar spectrum with a characteristic temperature of $\sim 10^5$ K (T5 spectrum; Tumlinson & Shull 2000; Bromm et al. 2001b; Schaerer 2002), while the lower mass second generation of stars (Pop II) has a softer spectrum with several $10^4$ K (T4 spectrum). These two spectral types have been used in several studies (e.g. Omukai 2001; Bromm & Loeb 2003a; Shang et al. 2010). Lyman-Werner radiation (11.2 – 13.6 eV) with an intensity above a certain threshold is able to photodissociate $H_2$ and $H^-$ (important for the formation of $H_2$) and keep their abundance very low. A T5 spectrum will mainly directly photodissociate $H_2$, while a T4 spectrum will be better at photodetaching $H^-$. The critical intensity required to suppress $H_2$ formation in massive halos where direct gas collapse can occur has been estimated at $J_{21}^{\rm crit} \gtrsim 10^2 - 10^3$, where $J_{21}$ denotes the specific intensity just below the Lyman limit (13.6 eV), in units of $10^{-21}\,\mathrm{erg\,cm^{-2}\,sr^{-1}\,s^{-1}\,Hz^{-1}}$ (e.g.

2

Omukai 2001; Bromm & Loeb 2003a; Schleicher et al. 2010b; Shang et al. 2010; Van Borm & Spaans 2013; Latif et al. 2014a,b). This is relatively high compared to the expected cosmic UV background at the relevant redshifts ($J_{21}^{\rm bg} \sim 10$ at $z \sim 10$) (e.g. Ahn et al. 2009; Holzbauer & Furlanetto 2012; Dijkstra et al. 2014). However, the UV background distribution has a long bright-end tail, owing to the presence of close (about 10 kpc) luminous neighbors, which means that there is a small but significant subset of halos that is exposed to supercritical intensities (Dijkstra et al. 2008; Agarwal et al. 2014; Dijkstra et al. 2014; Visbal et al. 2014). Recently, though, it has been shown that it is important to consider spectra generated from realistic stellar populations, taking the mode of star formation (continuous or bursty) and the age, metallicity, and mass of the stars into account (Agarwal & Khochfar 2015; Sugimura et al. 2014). This has implications for the $H_2$ photodissociation rate and the $H^-$ photodetachment rate, and thus affects the value of the critical intensity, $J_{21}^{\rm crit}$. (Agarwal & Khochfar 2015) computed the reaction rate coefficients for $H_2$ photodissociation and $H^-$ photodetachment using realistic spectra resulting from a stellar synthesis code, and found that these depend on the age and metallicity of the stars, in contrast to the findings of (Sugimura et al. 2014). The latter used a one-zone model and realistic stellar spectra to also calculate $J_{21}^{\rm crit}$, finding values in the range between 1000 and 1400. Latif et al. (2015) have studied the impact of varying the temperature of a blackbody spectrum in 3D cosmological simulations, to more closely resemble a realistic spectrum generated by Pop II stars. They found an even higher value for $J_{21}^{\rm crit}$, a few times $10^4$, due to additional 3D effects. This value depends only weakly on the adopted radiation spectra in the range between $T_{\rm rad} = 2 \times 10^4$ K and $10^5$ K.

Alternative mechanisms for inhibiting $H_2$ cooling comprise dissipation of a sufficiently strong magnetic field (Schleicher et al. 2009; Sethi et al. 2010; Van Borm & Spaans 2013) or the presence of strong shocks (Inayoshi & Omukai 2012), both of which result in collisional dissociation of $H_2$.

Numerical 3D simulations have found fragmentation to be inhibited and thus show the feasibility of the direct collapse scenario. In some simulations, bar-like instabilities (Wise et al. 2008) or self-gravitating disks on parsec scales (Regan & Haehnelt 2009a) were found, though these employed a resolution of 16 cells per Jeans length. More recently, it was demonstrated that at least 32 cells, and preferably more, are required to properly resolve turbulence (Sur et al. 2010; Federrath et al. 2011; Turk et al. 2012; Latif et al. 2013b). New simulations employing a higher resolution find that it is likely that $\sim 10^5\,M_\odot$ objects will form (Latif et al. 2013b,a,e), though the peak density in these studies is not much higher than $10^{15}\,cm^{-3}$. In the simulations pursued here, we aim to complement these studies exploring collapse and fragmentation on smaller scales.

While simulating the formation of the protostar in 3D was not yet possible, various one-zone models employing detailed chemical models show the expected thermal pathway (Omukai 2000, 2001; Omukai et al. 2005, 2008). For a strong UV background, Omukai (2001) showed that clouds collapse nearly isothermally, cooled successively by Lyman-alpha emission of atomic hydrogen, two-photon emission of atomic hydrogen from the 2$s$ state, and $H^-$ free-bound emission. Afterwards, the adiabatic phase sets in at $\sim 10^{20}\,cm^{-3}$, at which point the minimum Jeans mass, and thus the characteristic mass of the protostar, has been reduced to $0.03\,M_\odot$.

Once the protostar has formed, it will accrete and evolve into either a supermassive star or a quasistar, depending on the accretion rate. The work by Schleicher et al. (2013) suggests that for accretion rates $>0.14\,M_\odot/yr$, a quasistar will be the result, while lower accretion rates lead to the formation of a supermassive star. A supermassive star (SMS, with a mass in the range $10^3$-$10^6\,M_\odot$) of fixed mass, supported by radiation pressure, is thought to evolve as an $n = 3$ polytrope and finally collapse into a black hole containing most of the stellar mass (Johnson et al. 2011; Whalen et al. 2013; Hosokawa et al. 2012a, 2013). However, if the mass accretion rate is high enough, the outer layers of the SMS cannot thermally relax. In this case, it is not well-described by an $n = 3$ polytrope, but will have a more complex structure with a convective core surrounded by a convectively stable envelope that contains most of the mass. The core will burn up its hydrogen, and subsequently collapse into a black hole with a mass of a few $M_\odot$. The resulting structure, where the black hole accretes material from the massive, radiation-pressure-supported envelope, is termed a 'quasistar' (Begelman et al. 2006, 2008; Begelman 2010; Volonteri & Begelman 2010; Ball et al. 2011, 2012).

As is known from present-day star formation, turbulence plays an important role in angular momentum transport and determining the fragmentation properties of collapsing gas clouds, since it can both locally compress the gas as well as provide additional support against collapse on larger scales (e.g. Larson 1981; Mac Low & Klessen 2004; McKee & Ostriker 2007; Federrath & Klessen 2012). Similar effects have been found at high redshifts in simulations of minihalos, where turbulence plays a role in distributing angular momentum (Abel et al. 2002), and affects the fragmentation behavior (Clark et al. 2011a; Turk et al. 2012; Latif et al. 2013c). Also in simulations of more massive, atomic cooling halos, the importance of turbulence has been recognized (Greif et al. 2008; Wise et al. 2008). However, many of these older studies do not employ a sufficient Jeans resolution, as its impact was only recognized later. Latif et al. (2013b) found that the amount of turbulent structure increases significantly with increasing resolution, and in the study by Latif et al. (2013a) it was found that fragmentation occurs occasionally, but that this does not prevent the growth of a central massive object resulting from turbulent accretion and mergers.

Numerical simulations of collapsing gas in minihalos show that fragmentation also depends on the amount of rotation, with stronger rotation inducing more fragmentation (Bromm et al. 2002; Machida 2008; Hocuk & Spaans 2010). The study by Clark et al. (2008) shows that massive disk-like structures are assembled, fragmenting to form protostars. In atomic cooling halos the effects of rotation have not yet been studied in detail, though Bromm & Loeb (2003a) found that a single black hole is formed in low-spin galaxies, while higher spin galaxies tend to form binary black holes. In their simulations of atomic cooling halos, Regan & Haehnelt (2009b) observed the formation of massive compact self-gravitating disks, and found mild fragmentation in one of the three simulated halos.

In this paper we present the first study in which the formation of a massive protostar is simulated in 3D up to unprecedented high central densities ($10^{21}$ cm$^{-3}$), owing to improved modeling of the chemistry. A high spatial resolution is obtained as well; starting from pc scales, we are able to resolve scales down to a few solar radii. In addition, we investigate how the fragmentation behavior of collapsing primordial gas in the presence of a strong Lyman-Werner radiation background is affected by varying amounts of turbulence and rotation. For each case the formation of clumps and their accretion rates are studied.

In Section 2.2 some details are given on the methods and setup of the numerical simulations that have been performed. In Section 2.3 the results for both the one-zone calculations and the 3D simulations are presented and discussed, and we conclude with a summary of the results in Section 2.4.

## 2.2 NUMERICAL METHODOLOGY AND SIMULATION SETUP

ENZO is an open-source adaptive mesh refinement (AMR) simulation code, which provides high spatial and temporal resolution for the modeling of astrophysical fluid flows (Bryan et al. 2014). It contains a wide variety of physics modules, making it suitable for many different astrophysical applications. We use a modified version of ENZO 2.3, replacing the chemistry implementation by a customized build of the KROME chemistry package (Grassi et al. 2014), as discussed in the following subsections. The hydrodynamical equations are solved using the MUSCL scheme, which is a second-order accurate extension of Godunov's method. The implementation in ENZO uses second-order Runge-Kutta time integration, and the Riemann solver employed is the HLLC solver (Harten-Lax-van Leer with Contact), with a fallback to the more diffusive HLL solver (Harten-Lax-van Leer) in case negative energies or densities are computed. The choice of this solver is due to the strong shocks which occur once the central core becomes adiabatic and the central protostars forms. Self-gravity is computed by solving the Poisson equation using a multigrid technique.

### 2.2.1 INITIAL CONDITIONS

We follow the gravitational collapse of an isolated spherical gas cloud with a radius of 15 pc and a top-hat density profile, embedded in a 100 pc simulation box. The Jeans length is resolved by at least 64 cells at all times. Additionally, a refinement criterion based on overdensity is used. These combined criteria result in the simulations using 29 refinement levels, at which point an adiabatic core is formed and no further refinement is necessary. The collapse is followed for another $1.67 \times 10^{-2}$ years, corresponding to ~5 freefall times, after the highest refinement level is reached. To ensure pressure equilibrium between the sphere and its surroundings, we set the initial sphere density to $1000\,\mathrm{cm}^{-3}$ and its temperature to 500 K, while the surrounding gas is initialized with a density of $100\,\mathrm{cm}^{-3}$ and a temperature of 5000 K. The above combination of parameters also ensures that the mass of the cloud (~$3.5 \times 10^5$ $M_\odot$) is greater than the local Jeans mass (~$3 \times 10^4$ $M_\odot$), and thus the cloud will collapse. The total mass contained in the box is ~$2.8 \times 10^6$ $M_\odot$. This setup has been chosen in order to be able to explore the formation of protostars up to very high densities. The cloud is irradiated by a UV background with a T5 spectrum (see Section 2.2.2, under heading g) of intensity $10^5$ in units of $J_{21}$, so that the abundance of $H_2$ is kept low and cooling will occur mainly through atomic hydrogen.

Furthermore, we add a certain amount of initial turbulence to the gas, as well as some rotation of the cloud. These parameters are varied to study and quantify their effects on the collapse dynamics and fragmentation properties. An overview of the different simulations can be found in Table 2.1. The turbulent velocities are sampled from a Maxwellian distribution with a temperature equal to the initial temperature of the gas sphere, and subsequently multiplied by the percentage given in the table. Since the maximum of the Maxwell distribution function is of the order of the sound speed $c_s$, the turbulent velocities are of the order of a given percentage of $c_s$. The amount of rotation is given in percentage of the Keplerian velocity; i.e. 100% rotation means the cloud is rotationally supported.

| **Simulations** | | |
|---|---|---|
| Name | Turbulence (~ % of $c_s$) | Rotation (% of $v_{Kep}$) |
| T40R0 | 40 % | 0 % |
| T40R10 | 40 % | 10 % |
| T40R20 | 40 % | 20 % |
| T20R10 | 20 % | 10 % |
| T80R10 | 80 % | 10 % |

**Table 2.1** – Overview of the different simulations and their initial turbulent and rotational velocities. Turbulent velocities are sampled from a Maxwellian distribution where the temperature is the initial temperature of the gas sphere, and subsequently multiplied by the percentage given in the table. The amount of rotation is given in percentage of the Keplerian velocity.

### 2.2.2 CHEMISTRY, HEATING, AND COOLING

We employ the KROME[1] chemistry package, which has been developed in order to simplify the embedding of the chemistry and the microphysics in numerical simulations. It builds the corresponding rate equations, the solver parameters, and includes a series of thermal processes which are coupled to the chemical evolution. A patch to embed KROME in ENZO is available with the package and has been used within this work. KROME solves the non-equilibrium chemistry together with the temperature equation using the adaptive high-order solver DLSODES, which was shown to be both accurate and efficient for networks that present a corresponding ordinary differential equation system with a sparse Jacobian, and that are typical in astrophysical applications (Bovino et al. 2013; Grassi et al. 2013). We have modified and extended the available package, mainly to obtain the desired cooling processes. The main improvements of our modified version of KROME are the addition of $H^-$ cooling, Rayleigh scattering, and a different evaluation of the critical density used for the chemical heating, following Glover & Abel (2008).

2

#### A CHEMICAL NETWORK

Our chemical network consists of 31 reactions, including 9 species: H, e, $H^+$, $H^-$, $H_2$, $H_2{}^+$, He, $He^+$, and $He^{++}$. All the reactions and their associated rates can be found in Appendix 2.A.

#### B MOLECULAR HYDROGEN COOLING

The molecular hydrogen cooling rates were taken from Glover & Abel (2008), with an opacity correction from Ripamonti & Abel (2004), as implemented in KROME. However, we modified the opacity correction to use the molecular hydrogen density instead of the total density, rendering it usable for cases with a non-zero UV background. Hence, the $H_2$ cooling rate is multiplied by a factor $\min\left[1, \left(n_{H2} / \left(8 \times 10^9\,\mathrm{cm}^{-3}\right)\right)^{-0.45}\right]$, where $n_{H2}$ is the $H_2$ number density. Recent studies by Greif (2014) and Hartwig et al. (2015) calculate the escape fraction of cooling photons using a multi-line, multi-frequency ray-tracing scheme, and an approach based on the TreeCol algorithm, respectively. Greif (2014) find that the radially averaged escape fraction agrees well with the analytical fit from Ripamonti & Abel (2004), while the results of Hartwig et al. (2015) suggest that this fit underestimates the escape fraction after the initial stage of collapse. Presently, it has not yet been investigated which of these two methods yields the most accurate results. However, additional one-zone calculations suggest that even a significantly larger escape fraction does not influence our results, as the ineffectiveness of the cooling is mainly the result of the low $H_2$ abundance. Of course, opacity effects would become more important when considering a case where $H_2$ is the dominant coolant.

[1] Publicly available at http://kromepackage.org/

### C COLLISION-INDUCED EMISSION COOLING

When a collision takes place between an $H_2$ molecule and another $H_2$ molecule, a He molecule, or a H atom, the interacting pair briefly acts as a 'supermolecule' with a non-zero electric dipole, and there is a high probability of a photon being emitted. Collision-induced emission (CIE) may become important at high densities, depending on the gas temperature. We use the fit provided in KROME for the optically thin rate, but modified to ensure it is 0 if $f_{H_2} < 0.1$ and does not become important before $f_{H_2} \sim 0.5$, where $f_{H_2}$ is the $H_2$ mass fraction relative to H, as it is uncertain whether the fit is still valid for extremely dissociated media. The opacity correction at high densities has been adopted from Ripamonti & Abel (2004),

$$\tau_{\rm CIE} = \max\left[10^{-5}, \left(\frac{n}{2\times10^{16}\,{\rm cm^{-3}}}\right)^{2.8}\right], \tag{2.1}$$

where $n$ is the total number density. The CIE cooling rate is then multiplied by $\min\left[1, \left(1-\exp\left(-\tau_{\rm CIE}\right)\right)/\tau_{\rm CIE}\right]$, where $\left(1-\exp\left(-\tau\right)\right)/\tau$ is the usual spherical escape probability.

### D ATOMIC COOLING

KROME employs the atomic cooling rates from Cen (1992). These include the collisional ionization of H, He, $He^+$, and He(2s) by electrons, the recombination of $H^+$, $He^+$, and $He^{++}$, the dielectronic recombination of $He^+$, the collisional excitation of H (all $n$), He ($n = 2,3,4$ triplets), and $He^+$ ($n = 2$), and bremsstrahlung for all ions. The main cooling channel relevant here is the collisional excitation of H. We have added an optical depth approximation for the Rayleigh scattering by H atoms, which will suppress this main channel, as

$$\tau_{\rm Rl} = \sigma_{\rm H,Rl} n_{\rm HI} \frac{\lambda_{\rm J}}{2}, \tag{2.2}$$

where $\lambda_{\rm J}$ is the Jeans length, $n_{\rm HI}$ is the number density of atomic hydrogen, and

$$\begin{aligned}\sigma_{\rm H,Rl} = 5.799 \times 10^{-29} \lambda^{-4} + 1.422 \times 10^{-30} \lambda^{-6} \\ + 2.784 \times 10^{-32} \lambda^{-8} {\rm cm}^2\end{aligned} \tag{2.3}$$

is the Rayleigh scattering cross section of H for radiation with wavelength $\lambda$ (in μm) (Kurucz 1970). The cooling rate is then multiplied by $\exp(-\tau_{\rm Rl})$. Additionally, we have added two fudge factors to mimic optical depth effects and thus reduce cooling at high densities ($n \gtrsim 10^{17}\,{\rm cm}^{-3}$), in accordance with the findings of Omukai (2001). The first factor, $f_1$, represents that the gas should be optically thick to atomic hydrogen line cooling around $\sim 10^{17}\,{\rm cm}^{-3}$, and H ionization becomes the main atomic cooling channel. The second factor, $f_2$, ensures that the gas becomes almost completely optically thick to radiative cooling around $\sim 10^{20}\,{\rm cm}^{-3}$, so that afterwards the evolution is nearly adiabatic. The fudge factors are calculated as

$$f_{\rm i} = \min\left[1, \frac{1 - \exp\left(-\tau_{\rm f_i}\right)}{\tau_{\rm f_i}}\right], \tag{2.4}$$

for $i = 1,2$, using the functional form of the spherical escape probability. The opacities $\tau_{\rm f_1}$ and $\tau_{\rm f_2}$ are given by [2]

$$\tau_{\rm f_1} = \max\left[10^{-5}, \left(\frac{n}{10^{17}\,{\rm cm}^{-3}}\right)^5\right], \tag{2.5}$$

$$\tau_{\rm f_2} = \max\left[10^{-5}, \left(\frac{n}{10^{20}\,{\rm cm}^{-3}}\right)^8\right]. \tag{2.6}$$

[2] The exponents of 5 and 8 do not have a specific physical meaning, but are instead intended to provide a sharp enough cutoff, as in this regime the atomic cooling functions increase steeply with both density and temperature.

### E H⁻ COOLING

Through radiative association of H and e, $H^-$ is formed and a photon is emitted. There will be net cooling if this photon can escape (Omukai 2001; Schleicher et al. 2008b). The cooling rate can then be approximated as

$$\Lambda_{H^-} \approx k_{H^-} n_{HI} n_e E_\gamma, \tag{2.7}$$

where $E_\gamma$ is the approximate energy of the emitted photon. A typical electron undergoing radiative attachment has an energy of the order of $k_B T$, so the average outgoing photon energy can be estimated as $E_\gamma \sim E_0 + k_B T$, where the binding energy $E_0$ of $H^-$ is 0.755 eV. Rayleigh scattering (see Section d), as well as $H^-$ bound-free absorption, will suppress this cooling channel, so optical depth approximations for these processes have been taken into account. The cross section for $H^-$ bound-free absorption is (John 1988)

$$\sigma_{H^-,\mathrm{bf}} = 10^{-18} \lambda^3 \left( \frac{1}{\lambda} - \frac{1}{\lambda_0} \right)^{1.5} f(\lambda), \tag{2.8}$$

where $\lambda$ is the wavelength of the scattered radiation in μm, $\lambda_0 = 1.6419$ μm, and $f(\lambda)$ is given by equation 5 in John (1988).

2

### F CHEMICAL COOLING AND HEATING

Various chemical reactions can result in net cooling or heating of the gas (Omukai 2000). In our case, the most important ones are the three-body formation of $H_2$ (Forrey (2013), see Bovino et al. (2014c) for a comparison of different rates) and collisional dissociation of $H_2$ (Shapiro & Kang (1987); Martin et al. (1996, 1998)). The collisional dissociation process releases 4.48 eV per dissociated $H_2$ molecule (its binding energy), cooling the gas, while the heat deposited by three-body formation is $4.48(1 + n_{cr}/n)^{-1}$ eV per $H_2$ molecule. Here, $n_{cr}$ is the critical density, calculated as (Glover & Abel 2008)

$$n_{cr} = \left( \frac{x_{HI}}{n_{cr,HI}} + \frac{x_{H_2}}{n_{cr,H_2}} \right)^{-1}, \tag{2.9}$$

where $x_{\rm HI}$ and $x_{\rm H_2}$ are the number fractions of HI and $H_2$, respectively, and $n_{\rm cr,HI}$ and $n_{\rm cr,H_2}$ are their respective critical densities, given by

$$n_{\rm cr,HI} = {\rm dex}\left[3 - 0.416\log T_4 - 0.327\left(\log T_4\right)^2\right], \tag{2.10}$$

$$n_{\rm cr,H_2} = {\rm dex}\left[4.845 - 1.3\log T_4 + 1.62\left(\log T_4\right)^2\right], \tag{2.11}$$

where $T_4 = \frac{T}{10^4\,{\rm K}}$.

## G RADIATION BACKGROUND

In our calculations we have used a constant UV background flux with a T5 spectrum below the Lyman limit, which will photodissociate $H_2$ and photodetach $H^-$. The main difference with a T4 spectrum is that lower values of the intensity, $J_{21}$, are required for the gas to collapse isothermally. We do not expect the choice of spectrum or the specific strength of the UV background to matter, as long as the $H_2$ abundance is kept low so that $H_2$ cooling is unimportant. The difference between the spectra is expressed in the photodissociation rate of $H_2$ and photodetachment rate of $H^-$ (see $k_{24}$ and $k_{25}$ in Appendix 2.A). We also include $H_2$ self-shielding, using the improved fit described in Wolcott-Green et al. (2011),

$$f_{\rm sh} = \frac{0.965}{\left(1 + x_{N_{\rm H_2}}/b_5\right)^{1.1}} + \frac{0.035}{\left(1 + x_{N_{\rm H_2}}\right)^{0.5}} \exp\left(-8.5\times10^{-4}\left(1 + x_{N_{\rm H_2}}\right)^{0.5}\right), \tag{2.12}$$

where $x_{N_{\rm H_2}}$ is given by

$$x_{N_{\rm H_2}} = \frac{N_{\rm H_2}}{5\times10^{14}\,{\rm cm}^{-2}}, \tag{2.13}$$

with $N_{\rm H_2}$ the column density in ${\rm cm}^{-2}$, calculated as $N_{\rm H_2} = n_{\rm H_2}\lambda_{\rm J}/2$. The Doppler broadening parameter for $H_2$, $b_5$, is given by

$$b_5 = 10^{-5}\left(\frac{2k_{\rm B}T}{2m_{\rm H}}\right)^{0.5}, \tag{2.14}$$

in units of $10^5$ cm/s. The photodissociation rate of $H_2$ is multiplied by the self-shielding factor $f_{\rm sh}$.

## 2.3 RESULTS

### 2.3.1 ONE-ZONE CALCULATIONS

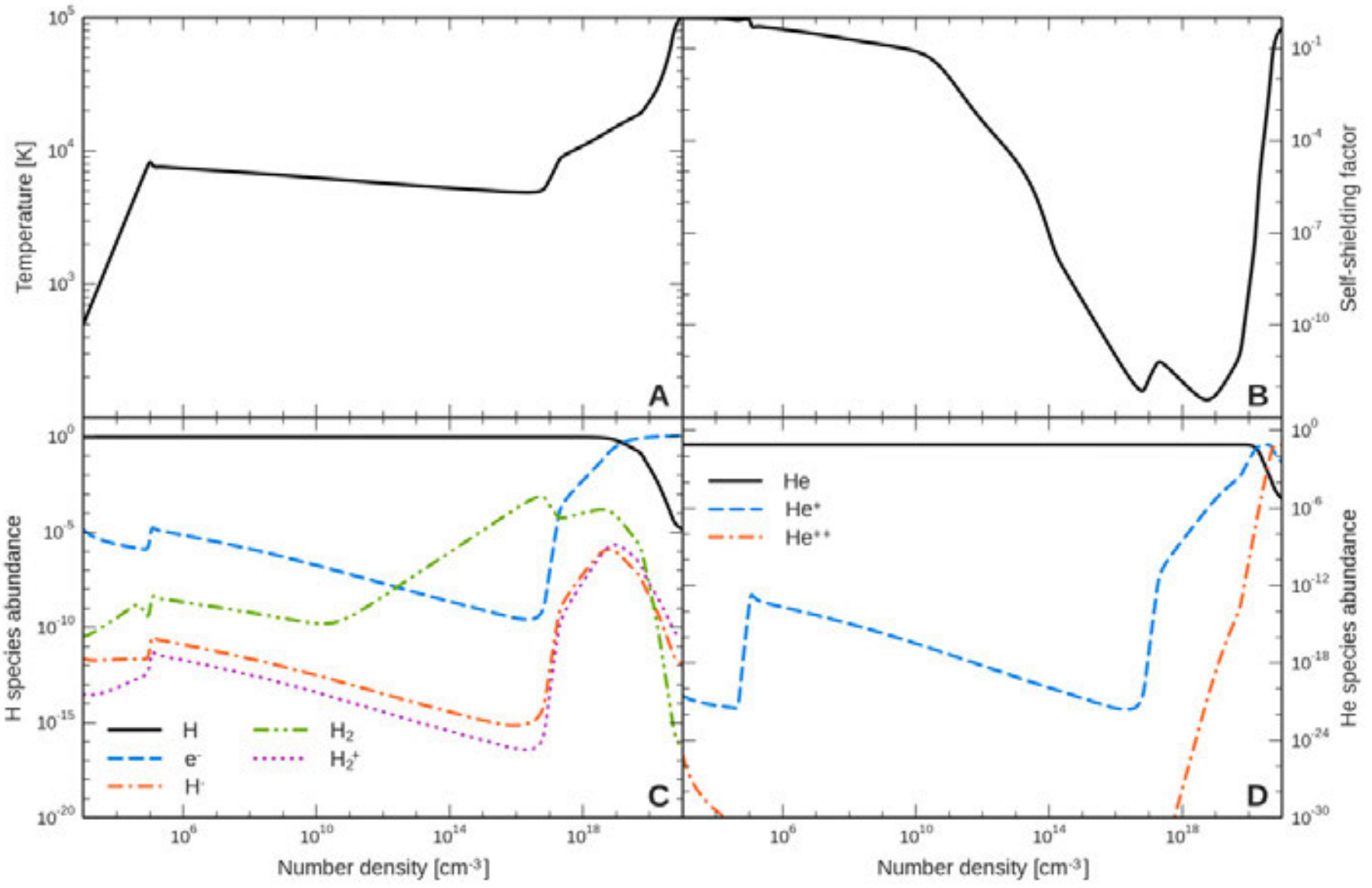

**Figure 2.1** – Physical quantities as a function of number density in a one-zone calculation, irradiated by a strong T5 background, using our modification of KROME. A: temperature; B: self-shielding factor for $H_2$, $f_{sh}$ ($f_{sh}=1$ means no shielding, and the smaller $f_{sh}$, the stronger the shielding); C: number fractions of H species (and electrons); D: number fractions of He species.

We have performed a one-zone freefall collapse test, already included with the KROME package, to verify our chemical model. The results for a T5 background with $J_{21}=10^5$ and initial conditions similar to those of the 3D simulations can be seen in Figure 2.1, calculated up to a number density of $10^{21}$ cm$^{-3}$. Panel A shows the temperature evolution, panel B shows the self-shielding factor for molecular hydrogen, panel C shows the number fractions of the different H species (and electrons), and panel D shows the number fractions of the different He species. We note here that at densities above $10^{17}$ cm$^{-3}$ an equilibrium approximation could be adopted and might speed up the calculations. Nevertheless, we preferred to follow a complete non-equilibrium evolution.

Initially, the temperature increases adiabatically, due to strong compressional heating. Because the molecular hydrogen is strongly dissociated by the UV background, the gas cannot cool through $H_2$ and instead cools via other processes. During the initial adiabatic phase, $H^-$ cooling is the dominant cooling process, but it is not efficient enough to counter the strong heating. When the temperature reaches ~8000 K, around ~$10^5$ cm$^{-3}$, Ly$\alpha$ cooling starts to become dominant and the temperature slope flattens off, now evolving nearly isothermally, though still decreasing slowly. Both chemical cooling and $H^-$ cooling are also important during this phase. Around a number density of ~$10^8$ cm$^{-3}$, both of these rates become higher than the atomic cooling. The $H^-$ cooling channel becomes strongly suppressed around ~$10^{16}$ cm$^{-3}$ as the cloud becomes optically thick to both Rayleigh scattering and $H^-$ bound-free absorption. Chemical cooling still maintains the near-isothermal evolution briefly, but then chemical heating cancels out the cooling and the temperature starts rising. Collisional ionization of H starts at ~$10^{17}$ cm$^{-3}$, resulting in a slowdown of the temperature rise up to ~$10^{19}$ cm$^{-3}$. From this point on, the cloud collapses adiabatically, and after sufficient contraction a protostar is expected to form in the center.

During the whole collapse, the molecular hydrogen fraction never becomes larger than $10^{-3}$, and as a result $H_2$ cooling is not important (except for densities between $10^4$-$10^5$ cm$^{-3}$). Starting from ~$10^{10}$ cm$^{-3}$, three-body formation increases the $H_2$ abundance, peaking just before the rise in temperature at ~$10^{17}$ cm$^{-3}$, after which strong collisional dissociation drastically decreases the abundance again. At low densities the self-shielding is too weak to prevent $H_2$ from being photodissociated (the smaller the factor, the stronger the shielding). At densities above $10^{10}$ cm$^{-3}$ the gas starts to become well-shielded, however, due to the high temperature, collisional dissociation of $H_2$ becomes effective. Additionally, $H_2$ cooling starts to become optically thick at these densities.

### 2.3.2 3D SIMULATION RESULTS

An overview of the different 3D simulations and their abbreviations are listed in Table 2.1. We have performed one simulation for each set of initial conditions, five in total. After reaching the highest refinement level of 29,

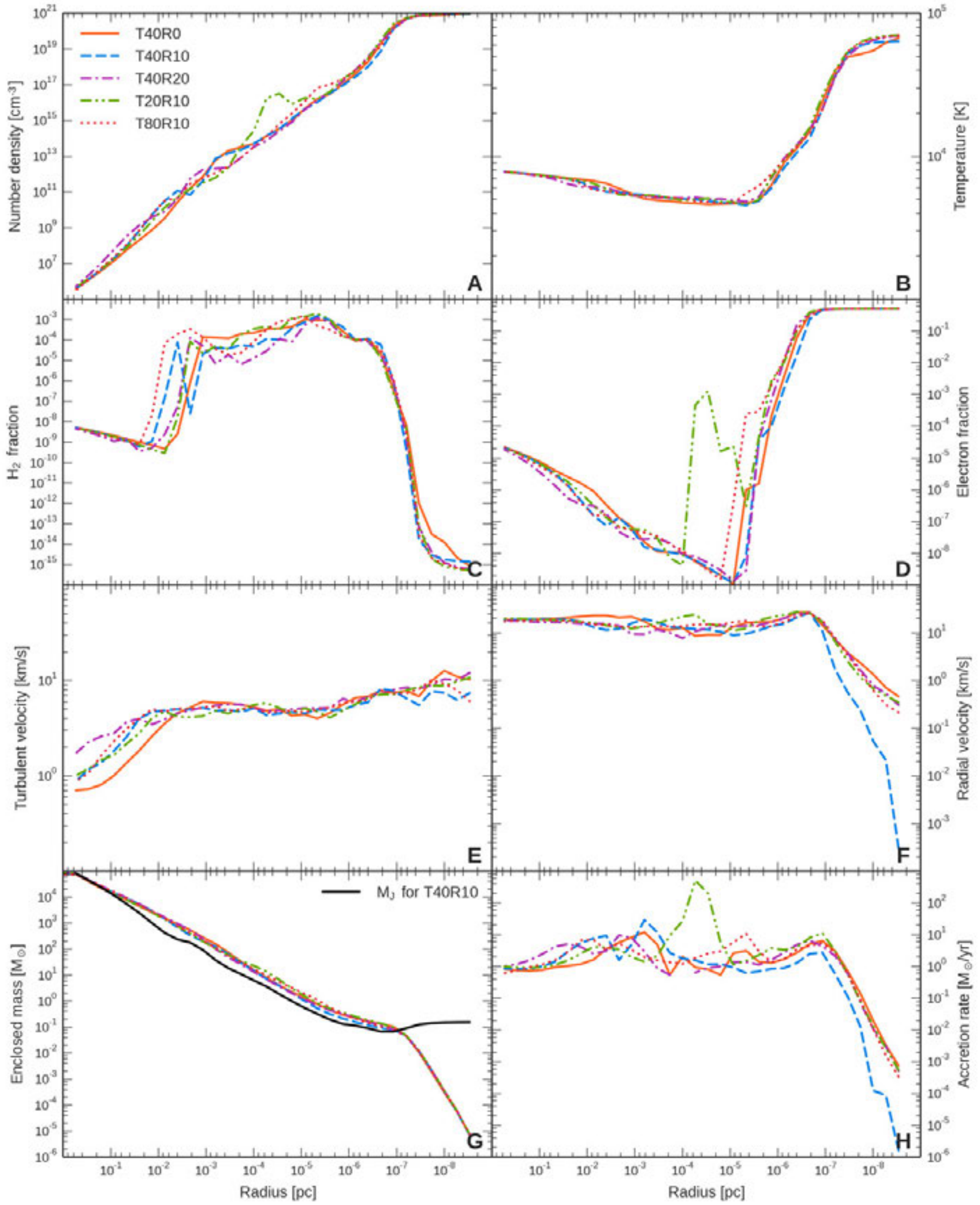

**Figure 2.2** – Physical quantities, weighted by mass, spherically averaged and radially binned, as a function of radius at the peak density output for the different simulations. A: number density; B: temperature; C: $H_2$ number fraction; D: electron number fraction; E: turbulent velocity; F: radial velocity, plotted as $-v_{\rm rad}$; G: enclosed mass, and the Jeans mass for T40R10 (it is very similar for other runs); H: radial mass infall rate, calculated from the density and the radial velocity. The simulation details and abbreviations are listed in Table 2.1.

2

the simulations evolved for another $\sim 1.67 \times 10^{-2}$ years, corresponding to ~5 freefall times, with the freefall time ($\sim 3.5 \times 10^{-3}$ yr) calculated at the moment when the highest refinement level is first reached.

Figure 2.2 shows several spherically averaged, radially binned profiles of various quantities for all simulations, centered on the peak density location (hereafter referred to as the central clump). The data shown has been obtained at the end of each simulation, when a peak density of $10^{21}$ cm$^{-3}$ was reached. From the density profile (shown in panel A) it can be seen that in general, the density increases with decreasing radius, so that overall the evolution of quantities with decreasing radius corresponds to an evolution with increasing density. Specifically, the density increases approximately as $\propto r^{-2}$, as is typical of an isothermal collapse. Deviations from this behavior are caused by local over- or underdensities, resulting from the turbulent nature of the gas. In the very center of the cloud, inside of the radius corresponding to the minimum Jeans mass, the density profile flattens off, indicating the central clump. This clump can also be seen in the enclosed mass profile (shown in panel G), which steeply decreases inside $10^{-7}$ pc, due to enhanced pressure support.

The density profile in simulation T20R10 deviates somewhat from isothermal, with a peak in the density profile around $\sim 4 \times 10^{-5}$ pc. After close inspection of density projections at different scales (see Figure 2.3 and Figure 2.4, particularly at the 50 pc scale), this appears to be due to the presence of a second concentration of mass containing two additional clumps, which have not collapsed as far as the main clump. However, from comparison runs with the same initial conditions, though with a different random seed for the initialization of the turbulent velocity field, we have found that such additional clumps are only sometimes present for the T20R10 initial conditions. Additionally, a second clump is also sometimes found for the other initial conditions discussed, though always with a lower peak density than the main clump. Hence, this fragmentation is likely not related to the amount of initial turbulence or rotation. It is not yet clear whether these additional clumps will continue to collapse, or instead accrete onto the main clump. Based on a simplified 'toy' model of fragmentation in the accretion disk around a protostar, Inayoshi & Haiman (2014) argue that some of the clumps formed in the disk may evolve to zero-age main sequence stars, but that most of these clumps can migrate inward and merge with the central protostar.

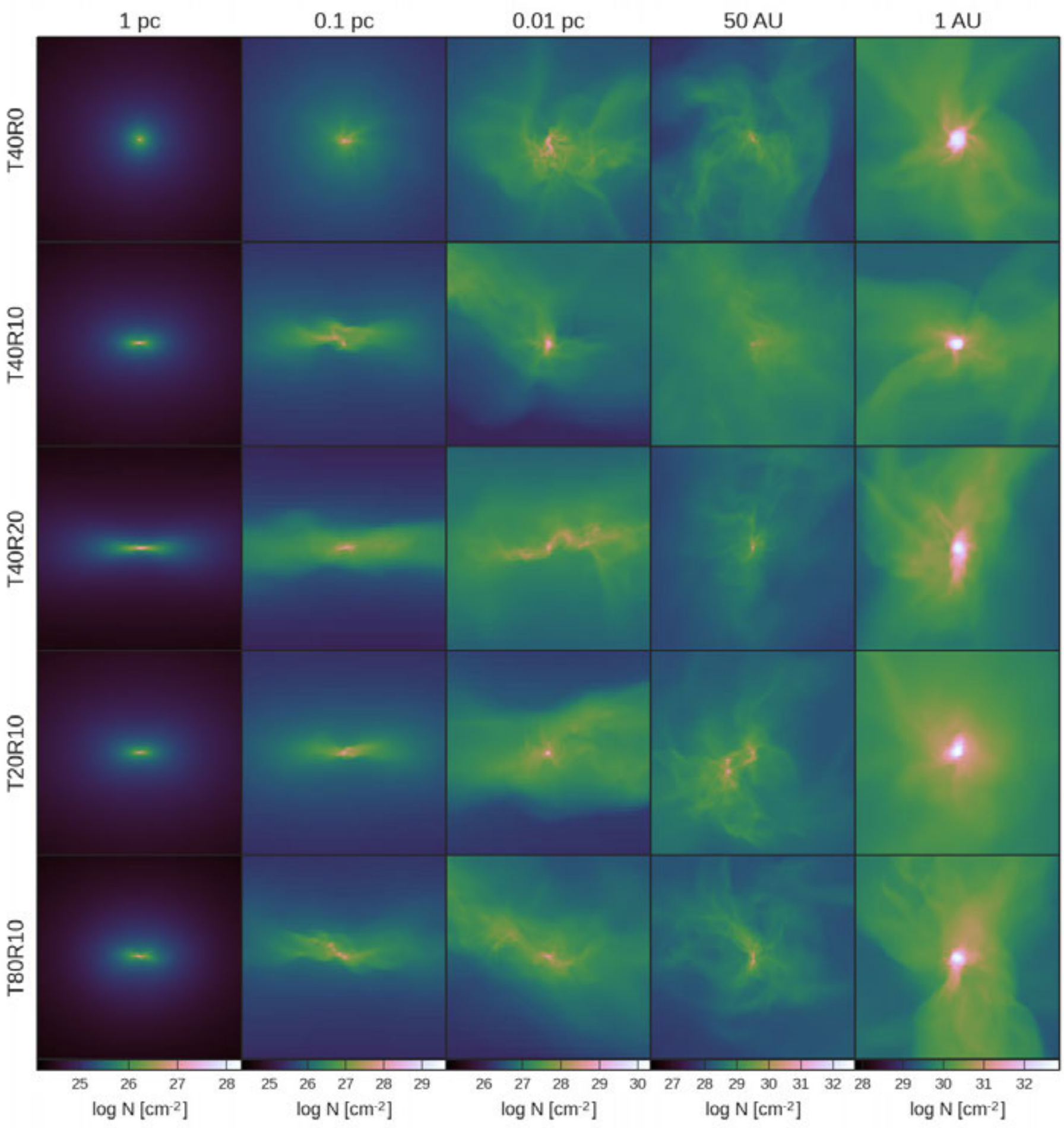

**Figure 2.3** – Density projections along the x-axis for all simulations, showing the integrated number density for various scales at the peak density output.The simulation details and abbreviations are listed in Table 2.1.

The temperature evolution of the gas cloud (shown in panel B of Figure 2.2) is very similar in all simulations. In the final stage displayed in the plots, the outer layers of the cloud are at a temperature of ~8000 K. Further inwards, the temperature evolves nearly isothermally, though still gradually drops to about 4000 K, until reaching a radius of about $3 \times 10^{-6}$ pc. Inside this radius, the evolution proceeds nearly adiabatically and the temperature reaches ~$7 \times 10^4$ K by the time the peak density is reached. This behavior is expected based on the one-zone calculations, and for a more detailed description of the involved processes, see Section 2.3.1.

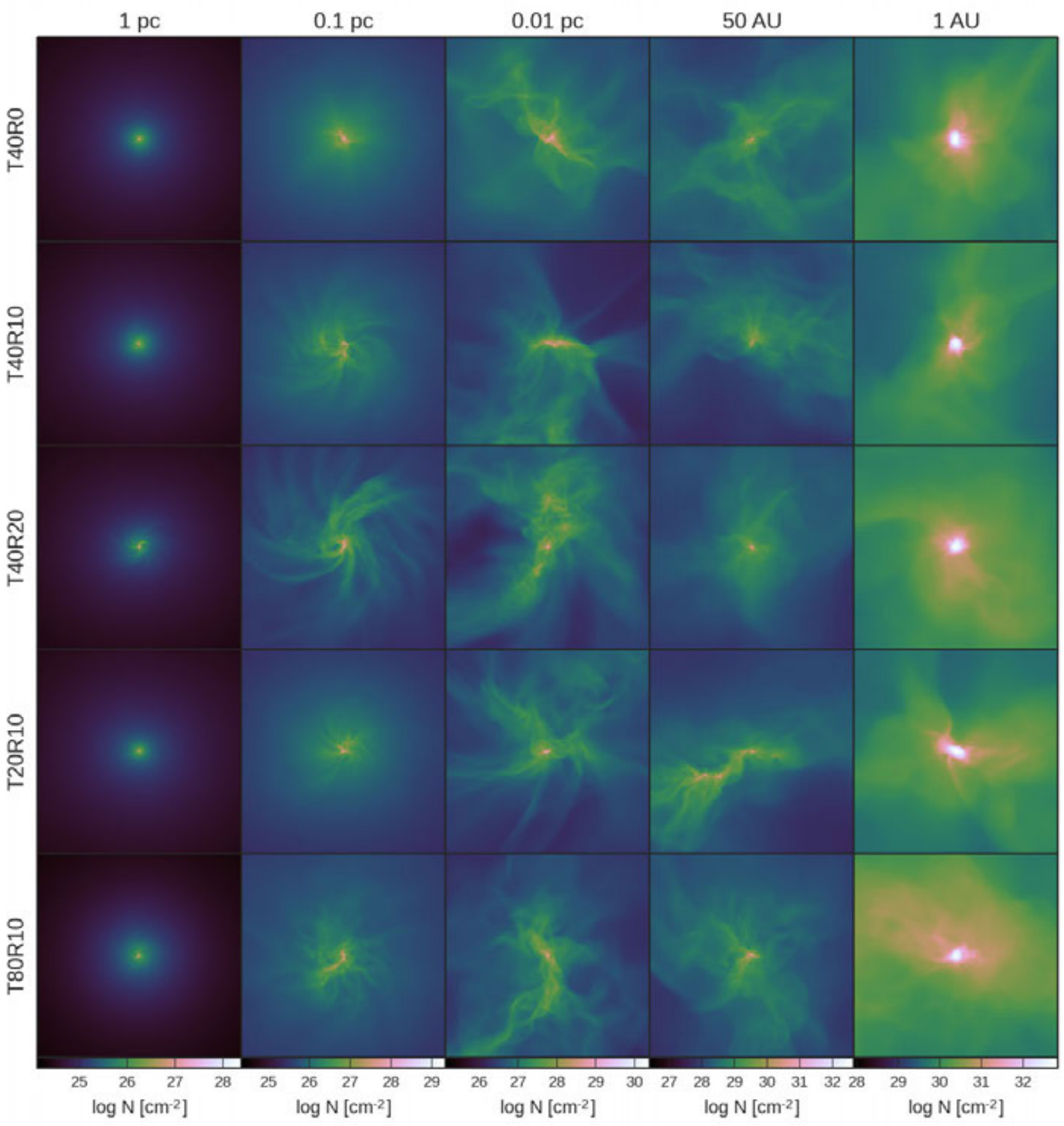

**Figure 2.4** – Density projections along the z-axis for all simulations, showing the integrated number density for various scales at the peak density output.The simulation details and abbreviations are listed in Table 2.1.

The molecular hydrogen number fraction (shown in panel C) exhibits some variation between simulations at larger radii, but converges for small radii, with only the T40R0 case deviating slightly from the others. At large radii, corresponding to low densities, the fraction slowly decreases as $H_2$ is dissociated by the UV background. Next, a steep increase occurs at the radius corresponding to a density of $10^{10}$ cm$^{-3}$, due to 3-body formation becoming efficient, after which formation and dissociation approximately balance each other for a broad density range. For the highest densities, where the gas is heating up, collisional dissociation starts to dominate and the fraction drops drastically. The $H_2$ number fraction never gets much larger than $10^{-3}$, in agreement with the one-zone test, which means that there is never enough $H_2$ for molecular cooling to be important.

The overall evolution of the electron number fraction (shown in panel D) is again quite similar to what is expected from one-zone calculations. The T80R10 case deviates slightly from the others, in that the electron fraction starts to increase already at somewhat larger radii. This is again due to some mass buildup around that radius, reaching slightly higher temperatures than the surrounding matter. The T20R10 case deviates quite strongly around the radius where for the other simulations the minimum occurs, which is due to the aforementioned second mass concentration at that radius.

The RMS turbulent velocity (shown in the panel E) increases slowly with radius from ~1 km/s to ~10 km/s for all simulations. It is interesting that although initially the amount of turbulence is varied, later in the runs this difference is smoothed out and at least in the turbulent velocities there is no longer a clear difference between the high and low turbulence cases. The radial velocity (shown in panel F) is similar for all simulations as well, and stays around ~11 km/s throughout most of the cloud. Only for the smallest radii, inside the minimum Jeans radius, does the radial velocity decrease down to 1-0.1 km/s, due to the pressure support in the clump.

The radial accretion rate (the rate of mass flow towards the central clump; shown in panel H) is calculated as $\mathrm{d}M/\mathrm{d}r = 4\pi r^2 \rho v_{\mathrm{rad}}$, where $\rho$ is the density and $v_{\mathrm{rad}}$ the radial velocity. The rate varies somewhat between different simulations, although there does not seem to be a trend with either turbulence or rotation. The large peak in the accretion rate for the T20R10 run around $\sim 4 \times 10^{-5}$ pc is due to the close connection of the second mass concentration to the central clump, locally boosting the accretion rate. It can be seen that both the density and radial velocity show a peak at the same location, causing the enhanced accretion rate. Similar features in the accretion rate were found by Regan & Haehnelt (2009a), who also attribute them to clumps of high-density gas. Overall accretion rates of a few solar masses per year are observed in all cases. Given such high accretion rates, a supermassive star of $10^5\,M_\odot$ is expected to form within $10^5$ years.

From the density projections (Figure 2.3 and Figure 2.4), it can be seen that for the simulations including rotation a disk has formed. Stronger rotation leads to a flatter, more extended disk, and more pronounced spiral structures. The differences in turbulent structures show on the 0.1 pc scale, with an increased amount of structure for higher initial turbulent velocities, also enhanced by stronger rotation. On smaller scales, the differences are no longer clear, as can also be seen from the turbulent velocity profiles in panel E of Figure 2.2.

Figure 2.5 and Figure 2.6 show temperature slices for two different scales, next to density slices of the same area. It can be seen that there are hot regions of gas surrounding slightly cooler patches. Such warmer and cooler patches result from local compression and expansion of the gas due to turbulent motions.

In Figure 2.7 we explore the properties of the disk, by displaying several disk averaged, radially binned quantities (using the radius in the x-y plane) for all simulations except T40R0 (as there is no disk present), centered on the peak density location (hereafter referred to as the central clump). The data shown has been obtained at the end of each simulation, when a peak density of $10^{21}$ cm$^{-3}$ was reached.

In panel A, the Toomre $Q$ parameter is shown. This parameter is calculated as $Q = \frac{\sigma\Omega}{\pi G\Sigma}$, where $\sigma$ is the RMS of the sound speed and the turbulent velocity (as both thermal and turbulent motions will play a role in stabilizing the disk; plotted in panel C and D), $\Omega$ is the rotation frequency (plotted in panel E), $G$ is the gravitational constant, and $\Sigma$ is the surface density (plotted in panel F). The disk is stable when $Q$ is larger than a critical value, which is of order unity; when $Q$ approaches this threshold, the disk will become gravitationally unstable. It can be seen that the disk is mildly unstable for all simulations. Only for the smallest radii $Q$ becomes decidedly larger than one, which is due to the close-to-adiabatic core in the central region. On the smallest scales, the adiabatic heating thus stabilizes the protostar against further collapse. We note that the adiabatic equation of state is however only an approximation, while real systems may evolve further via Kelvin-Helmholtz contraction.

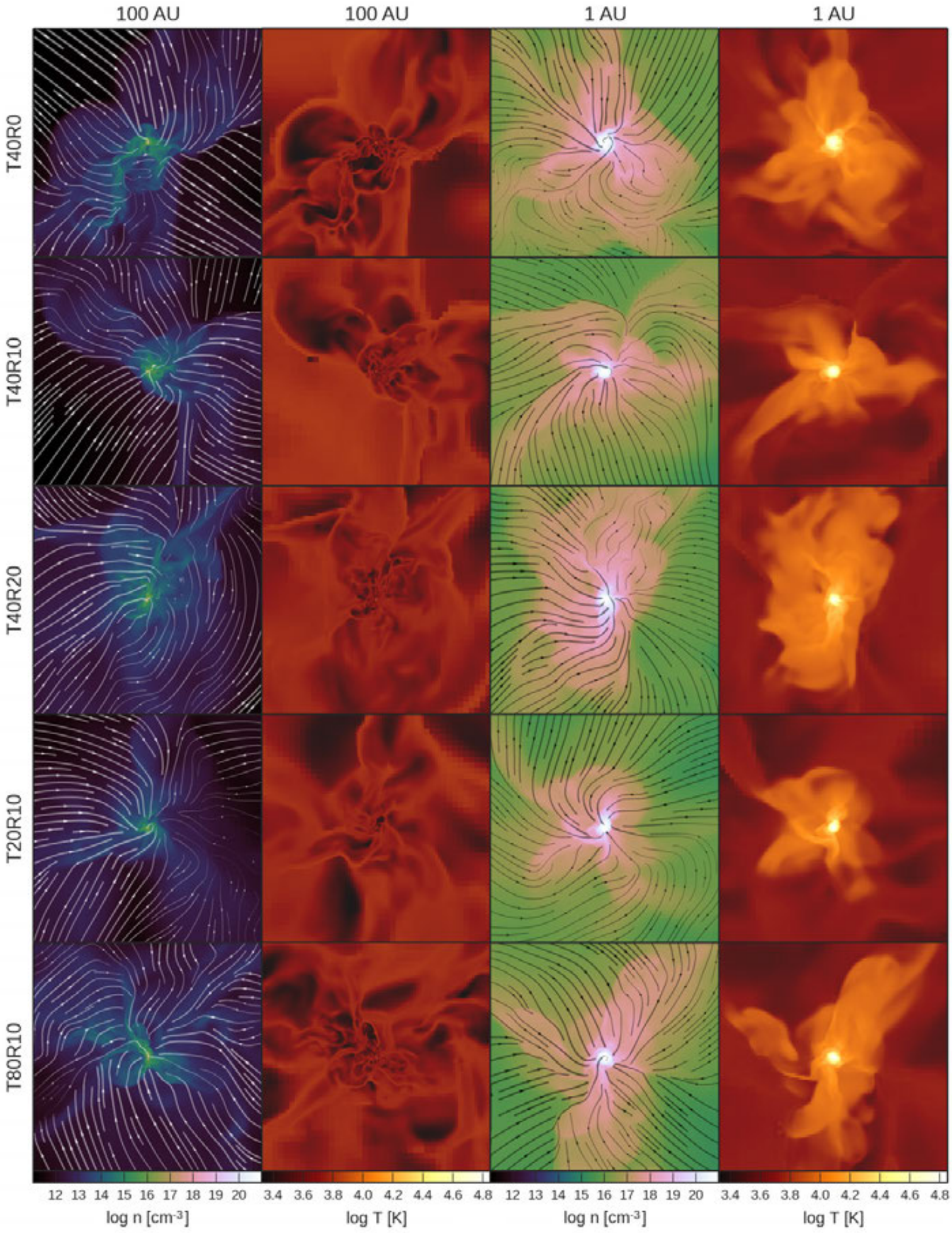

**Figure 2.5** – Density and temperature slices along the x-axis for all simulations and for two different scales at the peak density output. The simulation details and abbreviations are listed in Table 2.1.

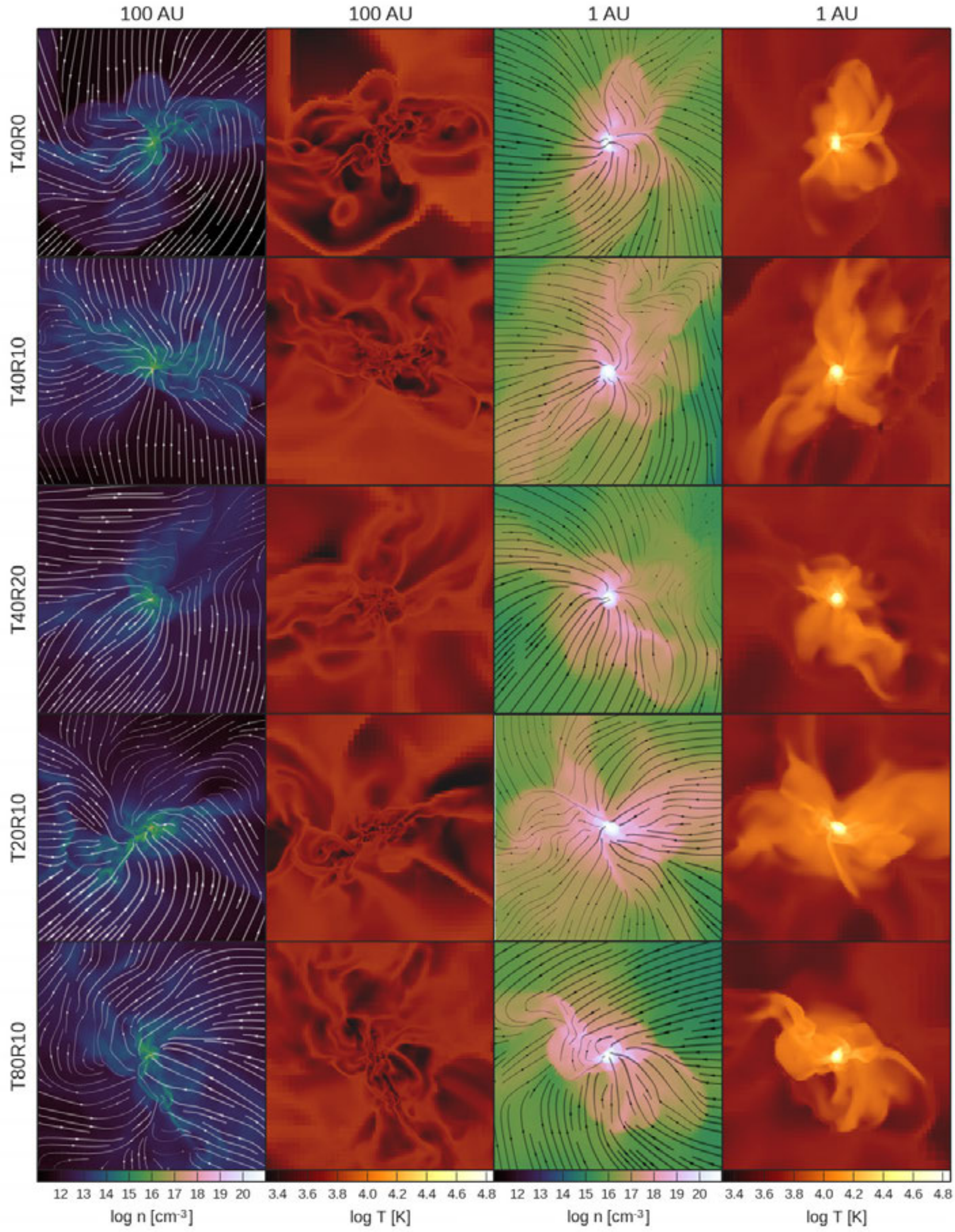

**Figure 2.6** – Density and temperature slices along the z-axis for all simulations and for two different scales at the peak density output. The simulation details and abbreviations are listed in Table 2.1.

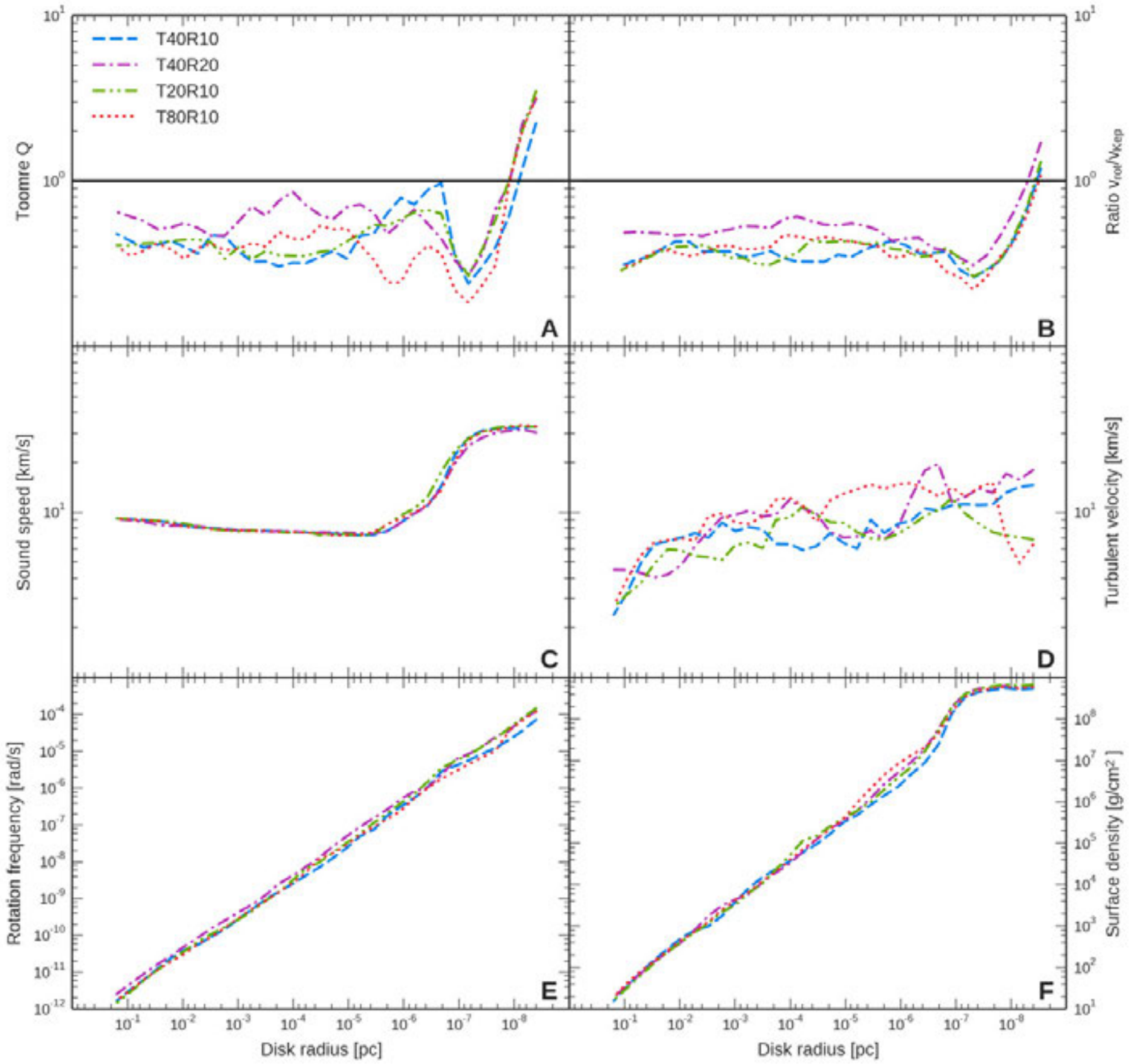

**Figure 2.7** – Physical quantities, weighted by mass, disk averaged and radially binned, as a function of radius at the peak density output for the different simulations. A: Toomre Q parameter; B: ratio of the rotational to the Keplerian velocity; C: sound speed; D: turbulent velocity; E: rotational frequency; F: surface density. The simulation details and abbreviations are listed in Table 2.1.

The ratio of the rotational velocity to the Keplerian velocity (shown in panel B) follows roughly the same behavior as the Toomre $Q$ parameter. The ratio is more or less constant over most radii. Some imprint of the initial amount of rotation remains, as the T40R20 run has the highest ratio over nearly all radii. However, for all runs the ratio has increased compared to the initial value, as a spin-up occurs during collapse.

It is interesting to note that we do not find a clear trend with either turbulence or rotation in any of the measured quantities on the smaller scales. It appears that whatever the initial conditions are, at later stages the initial difference in turbulence and rotation is smoothed out on these scales. Of course, on larger scales the presence and size of a disk does vary according to the initial amount of rotation, and on intermediate scales there are more turbulent structures for an increasing amount of initial turbulence, but this does not affect the overall evolution of density, temperature, accretion rate, and other quantities on scales smaller than 1 pc.

Whether one or more clumps are present does not depend on the initial amount of turbulence or rotation either, as we have concluded from comparison runs with the same initial conditions and a different random seed, in which usually one, sometimes two, and in a single case three of these clumps form. However, we never find more than three clumps, none of which have collapsed as far as the main clump, meaning that there is not much fragmentation, irrespective of turbulence or rotation. As mentioned, the simulations evolved for another $\sim 1.67 \times 10^{-2}$ years after the highest refinement level was reached. Given that no fragmentation occurs in most of our simulations during this time, it can be considered as a lower limit on the fragmentation time scale.

### THE CENTRAL OBJECT

A quantification of the properties of the central clump in each simulation is listed in Table 2.2. We find only one of such collapsed clumps in each simulation. The location of the 'knee' in the enclosed mass profile is taken as the clump radius. The mass enclosed in this radius corresponds approximately to the minimum Jeans mass (see also panel G in Figure 2.2), and thus the clumps are gravitationally bound. This object marks the onset of protostar formation. Due to computational constraints simulations we cannot evolve the simulations further, though we expect the gas in the surroundings to collapse to form a massive protostar.

| Clumps | | |
|---|---|---|
| Run | Radius [pc / AU] | Mass [$M_\odot$] |
| T40R0 | $9\times10^{-8}$ / $2\times10^{-2}$ | $7.7\times10^{-2}$ |
| T40R10 | $9\times10^{-8}$ / $2\times10^{-2}$ | $6.8\times10^{-2}$ |
| T40R20 | $9\times10^{-8}$ / $2\times10^{-2}$ | $6.5\times10^{-2}$ |
| T20R10 | $9\times10^{-8}$ / $2\times10^{-2}$ | $7.5\times10^{-2}$ |
| T80R10 | $9\times10^{-8}$ / $2\times10^{-2}$ | $6.5\times10^{-2}$ |

**Table 2.2** – Properties of the central bound clump found in each simulation. The location of the 'knee' in the enclosed mass function is taken as the minimum clump radius; the mass enclosed in this radius corresponds approximately to the minimum Jeans mass.

Given the radial accretion rates shown in Figure 2.2, a supermassive star of $10^5$ $M_\odot$ is expected to form within $10^5$ years, assuming that the gas reservoir to accrete from is large enough. If the accretion rate remains higher than $10^{-2}$ $M_\odot$/yr, Hosokawa et al. (2012a) found that the star will have a bloated envelope and lower surface temperatures, which inhibits the emission of ionizing radiation. In this case, radiative feedback will not be able to interfere with the accretion process. If accretion rates higher than 0.14 $M_\odot$/yr can be maintained until the core has exhausted its hydrogen content through nuclear burning (after $\sim7\times10^6$ yr), it is likely that the core of the star will collapse into a black hole, resulting in a quasistar (Schleicher et al. 2013).

## 2.4 DISCUSSION & CONCLUSIONS

We have performed 3D adaptive mesh refinement simulations using the ENZO code, simulating the formation of a protostar up to unprecedented high central densities of $10^{21}$ cm$^{-3}$, and spatial scales of a few solar radii. To achieve this goal, we have employed the KROME package to improve the modeling of the chemical and thermal processes. Particularly, we have investigated how the fragmentation behavior of collapsing primordial gas in the presence of a strong Lyman-Werner radiation background is influenced by varying amounts of turbulence and rotation.

We found that in the runs including rotation, a mildly unstable disk forms on scales of ~0.5 pc, with a more extended disk for the stronger rotating case, run T40R20. On somewhat smaller scales, ~0.1 pc, the amount of turbulent structures increases with increasing initial turbulent velocities, as one would expect. However, on even smaller scales, $\lesssim 0.01$ pc, the differences between the runs disappear, and radial profiles of the density, temperature, accretion rate, and other quantities are all very similar, with no dependence on the initial amount of turbulence or rotation. The thermal evolution of all runs is consistent with the one-zone result from Omukai (2001).

In each simulation we have found a single bound clump collapsed to a density of $10^{21}$ cm$^{-3}$, with a radius of $2 \times 10^{-2}$ AU and a mass of $\sim 7 \times 10^{-2}$ M$_\odot$, corresponding to the minimum Jeans mass. This clump marks the onset of protostar formation. Given the observed accretion rates of a few solar masses per year, the protostar is expected to become a quasistar with a mass of $10^5$ M$_\odot$ within $10^5$ years, assuming a high accretion rate can be maintained. Ferrara et al. (2014) have derived a detailed prediction for the initial mass function (IMF) of the first massive black holes formed in atomic cooling halos, combining the physics of SMS evolution and direct collapse black hole formation and growth with cosmological merger-tree simulations. They have found that in the case that minihalos can form stars and pollute the gas, the IMF is bimodal and spans a broad mass range, $M \approx (0.5-20) \times 10^5$ M$_\odot$; while in the case that they cannot form stars, the IMF spans a narrower range, $M \approx (1-2.8) \times 10^6$ M$_\odot$. However, they predominantly consider larger scales (several kpc) on a longer-term evolution than the study presented in this paper, as their focus is on modeling the dynamics of halo mergers and the implications for accretion.

In a single run presented in this study (T20R10), the gas fragments into three clumps instead of one. From comparison runs with the same initial conditions and a different random seed for the realization of the turbulent velocity field, we have concluded that this fragmentation does not depend on the initial amount of rotation or turbulence, as usually one, sometimes two, and in a single case three clumps are found, though never more than three. Thus, we do not find much fragmentation, irrespective of turbulence or rotation. It is not clear whether the additional clumps will continue to collapse, or instead accrete onto the main clump, though based on a simplified model of fragmentation in the accretion disk around a protostar, Inayoshi

& Haiman (2014) argue that most of these clumps can migrate inward and merge with the central protostar. The simulations have been evolved for another $\sim 1.67 \times 10^{-2}$ years ($\sim$5 freefall times) after the highest refinement level was reached. As no fragmentation occurs in most of our simulations during this time, it can be considered as a lower limit on the fragmentation time scale. To quantify the amount of fragmentation with greater certainty, the simulations should be evolved for longer, though our findings at least hint at the robustness of the direct collapse scenario.

For our simulations, we have used a Lyman-Werner background with a T5 spectrum. However, we expect to find the final result to be similarly independent of turbulence or rotation, as long as the intensity of the UV background is above the critical value, regardless of the stellar spectrum.

Recently, Inayoshi et al. (2014) have done a similar simulation to attempt to resolve protostar formation, starting from equally simplified, though somewhat different initial conditions. They start from a marginally supported isothermal sphere with an initial density of $10^4\,\mathrm{cm}^{-3}$ and temperature of 8000 K. The mass and radius of their cloud are slightly smaller than ours, $1.17 \times 10^5\,\mathrm{M}_\odot$ and 10.8 pc, respectively, though they are of the same order of magnitude. They also resolve the Jeans length by at least 64 grid cells, and their limiting resolution is 0.1 AU. There are a few differences between our chemical models. Concerning three-body rates, they do not take reaction $\mathrm{H} + \mathrm{H} + \mathrm{H}_2$ into account, and for reaction $\mathrm{H} + \mathrm{H} + \mathrm{H}$ they use the rate by Shapiro & Kang (1987), while we have used the updated rate by Forrey (2013). Additionally, their opacity corrections at high density are calculated based on the Rosseland mean opacity, while we have used a different treatment for each cooling process, as described in Section 2.2.2. For atomic cooling, we consider opacity from Rayleigh scattering, and a fudge opacity to reduce cooling at high densities, in accordance with findings from previous one-zone studies. For $\mathrm{H}^-$ cooling, opacity from both Rayleigh scattering and $\mathrm{H}^-$ bound-free absorption is taken into account.

They have stopped their simulation when a temperature just in excess of $10^4$ K was reached during the adiabatic phase, and find a hydrostatic core with a mass of $1\,\mathrm{M}_\odot$ and a radius of 2 AU, at a peak density of $\sim 5 \times 10^{16}\,\mathrm{cm}^{-3}$. In our simulations, the onset of the adiabatic phase occurs approximately one order of magnitude in density later, likely due to differences in the chemical model, which results in our central clump being less massive, $\sim 7 \times 10^{-2}\,\mathrm{M}_\odot$, and smaller, $2 \times 10^{-2}$ AU. Findings in agreement with ours include the resulting isothermal profile of the collapsing cloud, the fact that $H_2$ cooling remains inefficient, and the accretion rate. They did not include initial rotation, and do not find a disk, similar to our run without rotation (T40R0).

Presently, there are no simulations that resolve the formation of the protostar starting from cosmological initial conditions. However, our mass infall rates are in agreement with those found in cosmological direct collapse simulations reaching lower peak densities (Latif et al. 2013a,e).

Wise et al. (2008) have performed cosmological simulations following the collapse of two atomic cooling halos up to densities comparable to our peak density, though using a much simpler chemical model, neglecting, for example, $H_2$ and $H^-$ chemistry and cooling. Particularly, they did not consider optical depth effects, overestimating the cooling above column densities of $\sim 10^{13}$ cm$^{-2}$, and thus did not obtain a transition towards an adiabatic equation of state. Therefore, the formation of a quasi-static object, like a protostar, was not observed. This can be clearly seen from their Fig. 5: there is no increase in temperature for the inner regions, nor a significant change in slope in the density or enclosed mass profiles. These important differences make it difficult to directly compare their results to ours. However, we can say that their radial velocity and density profiles are, except in the innermost region ($R < 10^{-6}$ pc), comparable to ours, meaning that the radial accretion rate should be of the same order of magnitude as well.

In this study, we did not take turbulence on subgrid scales into account (for more details on subgrid scale turbulence, see e.g. Schmidt et al. 2006; Schmidt & Federrath 2011). As has been shown by Latif et al. (2013b,a,e), turbulence on unresolved scales affects the morphology of the collapsing gas. Thus, it would be interesting to investigate whether this affects our results. Another caveat is the absence of magnetic fields. Latif et al. (2013d, 2014c) have demonstrated that even a very small seed field can be effectively amplified by the small-scale dynamo mechanism. The resulting strong magnetic field provides additional support against gravity and helps to suppress fragmentation. A further caveat is the simplification of the cooling functions, and more importantly, the opacities. Future work should include a more detailed treatment of optical depth effects, possibly even employing radiative transfer, although this will be computationally more expensive. In the future, it would also be useful to implement equilibrium chemistry, as then it will be possible to follow the evolution for longer, and study the accretion onto the protostar in more detail. Additionally, simulations that resolve protostar formation starting from cosmological initial conditions are needed, to rule out possible effects caused by an idealized setup.

## ACKNOWLEDGMENTS

Computations described in this work were performed using the ENZO code (`http://enzo-project.org`), which is the product of a collaborative effort of scientists at many universities and national laboratories. The simulation results are analysed using yt, a multi-code analysis toolkit for astrophysical simulation data (Turk et al. 2011). CVB, DRGS and MAL acknowledge funding by the Deutsche Forschungsgemeinschaft (DFG) under grant SFB 963/1 (project A12). DRGS, SB, and CVB thank the DFG for funding and computing time via the Schwerpunktprogram SPP 1573 'Physics of the Interstellar Medium' (grant SCHL 1964/1 – 1). TG acknowledges the Centre for Star and Planet Formation funded by the Danish National Research Foundation.

# APPENDICES

## 2.A REACTION RATES

| Reaction | Rate coefficient ($cm^3\,s^{-1}$) | Range | Ref. |
|---|---|---|---|
| $H + e^- \rightarrow H^+ + 2e^-$ | $k_1 = \exp\Big[-32.71396786 + 13.5365560 \ln T_{eV} - 5.73932875 (\ln T_{eV})^2 + 1.56315498 (\ln T_{eV})^3 - 0.28770560 (\ln T_{eV})^4 + 3.48255977 \times 10^{-2} (\ln T_{eV})^5 - 2.63197617 \times 10^{-3} (\ln T_{eV})^6 + 1.11954395 \times 10^{-4} (\ln T_{eV})^7 - 2.03914985 \times 10^{-6} (\ln T_{eV})^8\Big]$ | | 1 |
| $H^+ + e^- \rightarrow H + \gamma$ | $k_2 = 3.925 \times 10^{-13}\, T_{eV}^{-0.6353}$ | $T \leqslant 5500\,K$ | 2 |

2

**Table 2.A.1** – Continued from previous page

| Reaction | Rate coefficient ($cm^3\,s^{-1}$) | Range | Ref. |
|---|---|---|---|
| | $k_2 = \exp\Big[-28.61303380689232$<br>$-0.7241125657826851\ln T_{eV}$<br>$-0.02026044731984691(\ln T_{eV})^2$<br>$-0.002380861877349834(\ln T_{eV})^3$<br>$-0.0003212605213188796(\ln T_{eV})^4$<br>$-0.00001421502914054107(\ln T_{eV})^5$<br>$+4.989108920299513\times10^{-6}(\ln T_{eV})^6$<br>$+5.755614137575758\times10^{-7}(\ln T_{eV})^7$<br>$-1.856767039775261\times10^{-8}(\ln T_{eV})^8$<br>$-3.071135243196595\times10^{-9}(\ln T_{eV})^9\Big]$ | $T > 5500\,K$ | 2 |
| $He + e^- \rightarrow He^+ + 2e^-$ | $k_3 = \exp\Big[-44.09864886$<br>$+23.91596563\ln T_{eV}$<br>$-10.7532302(\ln T_{eV})^2$<br>$+3.05803875(\ln T_{eV})^3$<br>$-0.56851189(\ln T_{eV})^4$<br>$+6.79539123\times10^{-2}(\ln T_{eV})^5$<br>$-5.00905610\times10^{-3}(\ln T_{eV})^6$<br>$+2.06723616\times10^{-4}(\ln T_{eV})^7$<br>$-3.64916141\times10^{-6}(\ln T_{eV})^8\Big]$ | | 1 |
| $He^+ + e^- \rightarrow He + \gamma$ | $k_4 = 3.925\times10^{-13}\,T_{eV}^{-0.6353}$<br>$+1.544\times10^{-9}\,T_{eV}^{-1.5}\Big[0.3\exp\left(-48.596/T_{eV}\right)$<br>$+\exp\left(-40.496/T_{eV}\right)\Big]$ | | 3, 4 |
| $He^+ + e \rightarrow He^{++} + 2e^-$ | $k_5 = \exp\Big[-68.71040990212001$<br>$+43.93347632635\ln T_{eV}$<br>$-18.48066993568(\ln T_{eV})^2$<br>$+4.701626486759002(\ln T_{eV})^3$<br>$-0.7692466334492(\ln T_{eV})^4$<br>$+0.08113042097303(\ln T_{eV})^5$<br>$-0.005324020628287001(\ln T_{eV})^6$<br>$+0.0001975705312221(\ln T_{eV})^7$<br>$-3.165581065665\times10^{-6}(\ln T_{eV})^8\Big]$ | $T > 9280\,K$ | 2 |

**Table 2.A.1** – Continued from previous page

| Reaction | Rate coefficient ($cm^3\,s^{-1}$) | Range | Ref. |
|---|---|---|---|
| $He^{++} + e^- \rightarrow He^+ + \gamma$ | $k_6 = 3.36 \times 10^{-10} T^{-1/2} (T/1000)^{-0.2} (1 + (10^{-6} T)^{0.7})^{-1}$ | | 5 |
| $H + e^- \rightarrow H^- + \gamma$ | $k_7 = 6.775 \times 10^{-15} T_{eV}^{0.8779}$ | | 6 |
| $H^- + H \rightarrow H_2 + e^-$ | $k_8 = 1.43 \times 10^{-9}$ | $T \leqslant 1160\,K$ | 2 |
| | $k_8 = \exp\Big[-20.06913897587003 + 0.2289800603272916 \ln T_{eV} + 0.03599837721023835 (\ln T_{eV})^2 - 0.004555120027032095 (\ln T_{eV})^3 - 0.0003105115447124016 (\ln T_{eV})^4 + 0.0001073294010367247 (\ln T_{eV})^5 - 8.36671960467864 \times 10^{-6} (\ln T_{eV})^6 + 2.238306228891639 \times 10^{-7} (\ln T_{eV})^7\Big]$ | $T > 1160\,K$ | 2 |
| $H + H^+ \rightarrow H_2{}^+ + \gamma$ | $k_9 = 1.85 \times 10^{-23} T^{1.8}$ | $T \leqslant 6700\,K$ | 7 |
| | $k_9 = 5.81 \times 10^{-16} (T/56200)^{(-0.6657 * \log(T/56200))}$ | $T > 6700\,K$ | 7 |
| $H_2{}^+ + H \rightarrow H_2 + H^+$ | $k_{10} = 6.0 \times 10^{-10}$ | | 8 |
| $H_2 + H^+ \rightarrow H_2{}^+ + H$ | $k_{11} = \exp\Big[-24.24914687731536 + 3.400824447095291 \ln T_{eV} - 3.898003964650152 (\ln T_{eV})^2 + 2.045587822403071 (\ln T_{eV})^3 - 0.5416182856220388 (\ln T_{eV})^4 + 0.0841077503763412 (\ln T_{eV})^5 - 0.007879026154483455 (\ln T_{eV})^6 + 0.0004138398421504563 (\ln T_{eV})^7 - 9.36345888928611 \times 10^{-6} (\ln T_{eV})^8\Big]$ | | 2 |
| $H_2 + e^- \rightarrow 2H + e^-$ | $k_{12} = 5.6 \times 10^{-11} \exp(-102124/T) T^{1/2}$ | | 9 |
| $H_2 + e^- \rightarrow H + H^-$ | $k_{13} = 36.7 T^{-2.28} \exp(-47172/T)$ | | 10 |
| $H_2 + H \rightarrow 3H$ | See expression in [11] | | 11 |

**Table 2.A.1** – Continued from previous page

| Reaction | Rate coefficient ($cm^3\,s^{-1}$) | Range | Ref. |
|---|---|---|---|
| $H_2 + H_2 \rightarrow H_2 + 2H$ | $k_{15} = \mathrm{dex}\left[\left(n_H/n_{cr}\,(1+n_H/n_{cr})^{-1}\right)\log k_{15,\mathrm{LTE}} + (1+n_H/n_{cr})^{-1}\log k_{15,\mathrm{v0}}\right]$ <br> $k_{15,\mathrm{v0}} = \left(6.0465\times10^{-30}\,T^{4.1881}\right) / \left(1+6.7606\times10^{-6}\,T\right)^{5.6881} \exp(-54657.4/T)$ <br> $k_{15,\mathrm{LTE}} = 1.3\times10^{-9}\exp(-53300/T)$ <br> See Section f for $n_{cr}$ | | 12, 7 |
| $H^- + e^- \rightarrow H + 2e^-$ | $k_{16} = \exp\left[-18.01849334273 + 2.360852208681\ln T_{eV} - 0.2827443061704(\ln T_{eV})^2 + 0.01623316639567(\ln T_{eV})^3 - 0.03365012031362999(\ln T_{eV})^4 + 0.01178329782711(\ln T_{eV})^5 - 0.001656194699504(\ln T_{eV})^6 + 0.0001068275202678(\ln T_{eV})^7 - 2.631285809207\times10^{-6}(\ln T_{eV})^8\right]$ | | 1 |
| $H^- + H \rightarrow 2H + e^-$ | $k_{17} = 2.5634\times10^{-9}\,T_{eV}^{1.78186}$ | $T \leqslant 1160\,K$ | 2 |
| | $k_{17} = \exp\left[-20.37260896533324 + 1.139449335841631\ln T_{eV} - 0.1421013521554148(\ln T_{eV})^2 + 0.00846445538663(\ln T_{eV})^3 - 0.0014327641212992(\ln T_{eV})^4 + 0.0002012250284791(\ln T_{eV})^5 + 0.0000866396324309(\ln T_{eV})^6 - 0.00002585009680264(\ln T_{eV})^7 + 2.4555011970392\times10^{-6}(\ln T_{eV})^8 - 8.06838246118\times10^{-8}(\ln T_{eV})^9\right]$ | $T > 1160\,K$ | 2 |
| $H^- + H^+ \rightarrow 2H$ | $k_{18} = 6.5\times10^{-9}\,T_{eV}^{-1/2}$ | | 15 |
| $H^- + H^+ \rightarrow H_2^+ + e^-$ | $k_{19} = 4\times10^{-4}\,T^{-1.4}\exp(-15100/T)$ | $T \leqslant 10^4\,K$ | 13 |
| | $k_{19} = 10^{-8}\,T^{-0.4}$ | $T > 10^4\,K$ | 14 |
| $H_2^+ + e^- \rightarrow 2H$ | $k_{20} = 10^{-8}$ | $T \leqslant 617\,K$ | 2 |

**Table 2.A.1** – Continued from previous page

| Reaction | Rate coefficient ($cm^3\,s^{-1}$) | Range | Ref. |
|---|---|---|---|
| | $k_{20} = 1.32 \times 10^{-6}\,T^{-0.76}$ | $T > 617\,K$ | 2 |
| $H_2{}^+ + H^- \rightarrow H + H_2$ | $k_{21} = 5.0 \times 10^{-6}\,T^{-1/2}$ | | 15 |
| $H + H + H \rightarrow H_2 + H$ | $k_{22} = 6 \times 10^{-32}\,T^{-1/4} + 2 \times 10^{-31}\,T^{-1/2}$ | | 16 |
| $H + H + H_2 \rightarrow 2H_2$ | $k_{23} = \left(6 \times 10^{-32}\,T^{-1/4} + 2 \times 10^{-31}\,T^{-1/2}\right)/8$ | | 16, 17 |
| $H^- + \gamma \rightarrow H + e^-$ | $k_{24} = 10^{-10}\,\alpha J_{21}$<br>$\alpha = 2000$ for T4 spectrum<br>$\alpha = 0.1$ for T5 spectrum | | 14 |
| $H_2 + \gamma \rightarrow 2H$ | $k_{25} = 10^{-12}\,\beta J_{21}$<br>$\beta = 3$ for T4 spectrum<br>$\beta = 0.9$ for T5 spectrum | | 14 |

**Table 2.A.1** – Reaction rate coefficients. $T$ and $T_{eV}$ are the gas temperature in units of K and eV, respectively.
References: [1] Janev et al. (1987); [2] Abel et al. (1997); [3] Cen (1992); [4] Aldrovandi & Pequignot (1973); [5] Omukai (2000); [6] de Jong (1972); [7] Shapiro & Kang (1987); [8] Karpas et al. (1979); [9] Donahue & Shull (1991); [10] Capitelli et al. (2007); [11] Martin et al. (1996); [12] Martin et al. (1998); [13] Poulaert et al. (1978); [14] Shang et al. (2010); [15] Dalgarno & Lepp (1987); [16] Forrey (2013); [17] Palla et al. (1983).

2

## 2.B THE INFLUENCE OF MAGNETIC FIELDS, TURBULENCE, AND UV RADIATION ON THE FORMATION OF SUPERMASSIVE BLACK HOLES[3]

### ABSTRACT

**Context** The seeds of the supermassive black holes (SMBHs) with masses of $\sim 10^9\,M_\odot$ observed already at $z \sim 6$ may have formed through the direct collapse of primordial gas in $T_{\rm vir} \gtrsim 10^4\,$K halos, whereby the gas must stay hot ($\sim 10^4$ K) in order to avoid fragmentation.
**Aims** The interplay between magnetic fields, turbulence, and a UV radiation background during the gravitational collapse of primordial gas in a halo is explored; in particular, the possibilities for avoiding fragmentation are examined.
**Methods** Using an analytical one-zone model, the evolution of a cloud of primordial gas is followed from its initial cosmic expansion through turnaround, virialization, and collapse up to a density of $10^7\,{\rm cm}^{-3}$.
**Results** It was found that in halos with no significant turbulence, the critical UV background intensity ($J_{21}^{\rm crit}$) for keeping the gas hot is lower by a factor $\sim 10$ for an initial comoving magnetic field $B_0 \sim 2\,$nG than for the zero-field case, and even lower for stronger fields. In turbulent halos, $J_{21}^{\rm crit}$ is found to be a factor $\sim 10$ lower than for the zero-field-zero-turbulence case, and the stronger the turbulence (more massive halo and/or stronger turbulent heating), the lower $J_{21}^{\rm crit}$.
**Conclusion** The reduction in $J_{21}^{\rm crit}$ is particularly important, since it exponentially increases the number of halos exposed to a supercritical radiation background.

### 2.B.1 INTRODUCTION

Several very bright quasars have been detected already at $z > 6$. This suggests that some supermassive black holes (SMBHs) with masses of $\simeq 10^9\,M_\odot$ already existed when the Universe was less than 1 Gyr old (Fan 2006). Explaining how such massive SMBHs could have assembled so soon after the

[3] Published as Van Borm C., & Spaans, M. 2013, A&A, 553, L9

Big Bang presents quite a challenge. The main questions concern how and when the seeds of these SMBHs formed and how their subsequent growth proceeded.

Several pathways leading to the formation of seed black holes (SBHs) have already been proposed. One group of scenarios suggests that SBHs formed via the direct collapse of metal-free/very metal-poor gas in halos with $T_{vir} \gtrsim 10^4$ K, at redshifts ~5–10, resulting in SBHs with $M \sim 10^4 - 10^5\,M_\odot$ (see, e.g., Bromm & Loeb 2003a; Koushiappas et al. 2004; Begelman et al. 2006; Lodato & Natarajan 2006; Spaans & Silk 2006; Schleicher et al. 2010b; Latif et al. 2013a). For efficient gas collapse to occur, fragmentation must be suppressed, which is possible if the gas in the halo is kept hot (large Jeans mass). Hence, the formation of $H_2$ must be inhibited so cooling can occur only through atomic H, as otherwise $H_2$ cooling will lower temperatures to ~200 K.

Several mechanisms have been suggested that suppress $H_2$ cooling. The most accepted of these mechanisms requires a critical level of Lyman-Werner radiation to photodissociate $H_2$ and keep its abundance very low. The critical intensity needed to suppress $H_2$ in the massive halos where direct gas collapse can occur is large compared to the expected cosmic UV background at the relevant redshifts. However, the distribution has a long bright-end tail, and halos irradiated by supercritical intensities would be a small subset of all halos (Dijkstra et al. 2008). Another mechanism proposes that the dissipation of a sufficiently strong magnetic field can heat the gas in the halo to ~$10^4$ K, which causes $H_2$ to be destroyed by collisional dissociation (Sethi et al. 2010).

A variety of mechanisms exist for generating magnetic fields early in the Universe, both before and after recombination (for a review, see, e.g., Widrow et al. 2012), and also for amplifying an existing field. In the case of a collapsing halo, the most important ones are gravitational compression, the small-scale turbulent dynamo, the large-scale dynamo in protostellar and galactic disks, and the magneto-rotational instability (MRI).

Gravitational compression increases the magnetic field as $B \propto \rho_b^\alpha$ when the field is coupled to the gas. For spherically symmetric collapse, $\alpha = 2/3$, but if the collapse proceeds preferentially along one axis, the scaling is closer to $\alpha = 1/2$. In realistic cases, intermediate values are often found (e.g., Schleicher et al. 2009; Hocuk et al. 2012).

Non-helical turbulent flows can act as small-scale dynamos, which produce disordered, random magnetic fields (Kazantsev 1968). The magnetic field amplification results from the random stretching and folding of the field lines by the turbulent random flow. During gravitational collapse, turbulence is generated by the release of gravitational energy and the infall of accreted gas on the inner, self-gravitating core. This means that, in the context of star and galaxy formation, a strong tangled magnetic field may already be generated during the collapse phase by the small-scale dynamo (Schleicher et al. 2010a). For the formation of SBHs, it implies that the existence of an accretion disk may cause the magnetic field to be further amplified (by a large-scale dynamo and/or the MRI) which provides additional stability and hence reduces fragmentation.

## 2.B.2 THE MODEL

The evolution of a cloud of primordial gas is followed from initial expansion to a high-density core, using a one-zone model, in which the physical variables involved are regarded as those at the cloud center. The model uses standard cosmology, with cosmological parameters given by WMAP7 (Larson et al. 2011).

### A DENSITY EVOLUTION

The spherical collapse model for a top-hat overdensity is used to compute the matter density. At the moment of turnaround, the gas decouples from the dark matter and becomes self-gravitating. Any effects due to rotation are neglected for simplicity. The baryonic matter collapse is expected to proceed like the Larson-Penston similarity solution (for the isothermal case; Larson 1969; Penston 1969), as generalized to polytropic cases by Yahil (1983). According to this solution, the cloud consists of two parts, a central core region and an envelope. The central core region has a flat density distribution, whereas the density in the envelope decreases outwards. The size of the central flat region is roughly given by the local Jeans length, $\lambda_J = c_s\sqrt{\pi/(G\rho_m)}$, with $c_s = \sqrt{\gamma k_B T/(\mu m_H)}$ the sound speed in the central region and $\rho_m$ the total matter density in the central region. The collapse in the core proceeds approximately at the free-fall rate, although additional heat input due to magnetic energy dissipation, for example, may delay gravitational collapse. The mean baryonic density evolution in the central part is described by $\dot{\rho_b} = \rho_b / t_{ff}$, where $t_{ff}$ is the free-fall time $t_{ff} = \sqrt{3\pi/(32G\rho_m)}$.

## B CHEMICAL NETWORK

The species that are included in the chemical network of this model are H, $H^+$, $H^-$, $H_2$, $H_2{}^+$, and $e^-$; HD or other molecules involving deuterium are not included. Reactions with He are not taken into account, but He is considered in the calculation of the mean molecular mass. The He mass fraction is taken to be ~0.248 (corresponding to an abundance $x_{He} \approx 0.0825$) and stays constant throughout the time integration. The fractional abundances of H, $H_2$, and $e^-$ are explicitly followed during the integration.

The evolution of the fractional abundance of electrons $x_e$ is given by the equation (see, e.g., Peebles 1993, for further details)

$$\frac{dx_e}{dt} = \left[\beta_e x_{HI} \exp\left(-\frac{h\nu_\alpha}{k_B T_{CMB}}\right) - \alpha_e x_e^2 n_H\right] C + \gamma_e(T) x_{HI} x_e n_H. \tag{2.15}$$

The first term represents the recombination and photoionization of the primordial plasma, the second term is the collisional recombination term, and the third term represents collisional ionization ($H + e^- \longrightarrow H^+ + 2\,e^-$).

The evolution of the fractional abundance of molecular hydrogen $x_{H_2}$ is given by the equation (Shang et al. 2010; Sethi et al. 2010)

$$\frac{dx_{H_2}}{dt} = k_m x_e x_{HI} n_H - k_{des} x_{H_2} n_H, \tag{2.16}$$

where

$$k_m = \frac{k_9 k_{10} x_{HI} n_H}{k_{10} x_{HI} n_H + k_\gamma + (k_{13} + k_{21})\, x_e n_H + k_{19} x_e n_H + k_{20} x_{HI} n_H + k_{25}}, \tag{2.17}$$

$$k_{des} = k_{15} x_{HI} + k_{17} x_p + k_{18} x_e + k_{28} f_{sh} / n_H. \tag{2.18}$$

Here, $k_m$ is the net rate of formation of $H_2$ through the $H^-$ channel, $k_{des}$ is the net destruction rate of $H_2$, and $k_\gamma$ is the destruction rate of $H^-$ by CMB photons. The reaction rates can be found in Appendix A of Shang et al. (2010), numbered as above. For sufficiently large column densities, $H_2$ can shield itself from radiation in the Lyman-Werner bands. The self-shielding factor $f_{sh}$ is given by Draine & Bertoldi (1996). The updated $f_{sh}$ from Wolcott-Green et al. (2011) results in decreased shielding and may thus enhance our results.

**RADIATION BACKGROUND** A sufficiently intense UV radiation background can either directly photodissociate $H_2$, or photodissociate the intermediary $H^-$. The relevant criterion for suppressing $H_2$ formation is that the photodissociation timescale must be shorter than the formation timescale, which results in $J^{crit} \propto \rho$. The intensity is written as $J_{21}$, which denotes the specific intensity just below 13.6 eV in the units of $10^{-21}$ erg cm$^{-2}$ sr$^{-1}$ s$^{-1}$ Hz$^{-1}$. Here, two different UV spectra are considered. They are both Planck spectra with a blackbody temperature of either $T_* = 10^4$ or $10^5$ K (T4 and T5, respectively). The T4 spectrum is meant to approximate the mean spectrum of a normal stellar population (Pop II), whereas the T5 spectrum is closer to the harder spectrum expected to be emitted by the first generation of stars (Pop III) (Tumlinson & Shull 2000; Bromm et al. 2001b; Schaerer 2002).

2

## C MAGNETIC FIELD EVOLUTION

Gravitational compression and the small-scale dynamo can amplify the magnetic field strength $B$, while ambipolar diffusion will decrease $B$. If the flux-freezing condition applies, the magnetic field depends on the density as $B \propto \rho_b^\alpha$, where $\alpha$ lies in the range $2/3 - 1/2$. Hence, $B$ will increase during gravitational collapse. It is assumed that gas is continually falling in, so the turbulence generated by accretion will not decay, but is instead constantly replenished. However, this depends on the ambient gas reservoir and may in reality be more complicated. It has been shown that the injection scale of such accretion-driven turbulence is close to the size of the system under consideration (thus, $\lambda_J$) (Klessen & Hennebelle 2010; Federrath et al. 2011). The turbulent velocity on the injection scale is expected to be comparable to the velocity of the infalling gas, and for a roughly isothermal density profile, the free-fall velocity is independent of radius. So, while the injection scale changes during the collapse, the injected velocity $v_{in}$ stays the same and is approximately equal to the virial velocity (Greif et al. 2008; Wise &

Abel 2007; Wise et al. 2008). On scales smaller than the injection scale, the turbulent velocity is expected to scale as $v \propto l^{\beta}$, with $\beta = 1/3$ for Kolmogorov turbulence (incompressible gas) and $\beta = 1/2$ for Burgers turbulence (strongly compressed gas).

The magnetic field on a scale $l$ typically grows exponentially on the eddy turnover time $t_{\rm ed} = l/v$, where $v$ is the turbulent velocity on the scale $l$. However, the magnetic field will saturate when the magnetic energy corresponds to a fraction $Rm_{\rm cr}^{-1}$ of the kinetic energy; $Rm_{\rm cr}$ is the critical magnetic Reynolds number. If $Rm > Rm_{\rm cr}$, the magnetic field grows through dynamo action. The maximum magnetic field strength is thus given by $B_{\rm max} = \sqrt{4\pi\rho_{\rm b} v^2/Rm_{\rm cr}}$ (Subramanian & Barrow 1998). However, the exact value of the saturation field strength is still somewhat uncertain. Once $B > B_{\rm max}$ on a certain scale, it is no longer amplified by the small-scale dynamo; however, it is still amplified by gravitational compression ($\propto \rho^{\alpha}$). It can then in principle increase above the saturation level (which only increases $\propto \rho^{1/2}$), but in this case it is subject to turbulent decay. On a given scale, this decay will probably also occur on $t_{\rm ed}$. As a result, the magnetic field strength tends to stay close to $B_{\rm max}$ on that scale.

The most important contribution to the total magnetic energy comes from the integral scale, the scale on which the magnetic field is largest. Schleicher et al. (2010a) have shown that in an atomic cooling halo, the integral scale increases very rapidly to the maximum scale on which the magnetic field is coherent after the start of the simulation. For this reason, only the evolution of the magnetic field at this scale of maximal coherence is followed here. Since the magnetic field is distorted by the gravitational collapse, the largest possible coherence length is always smaller than the Jeans length by some factor $f_{\rm d}$; its precise value is uncertain. We adopted a fiducial value of 0.1; changing this by a factor of a few does not significantly affect the results.

Ambipolar diffusion (AD) is important in a mostly neutral medium where a tangled magnetic field is present. The AD heating rate can be estimated as $L_{\rm AD} \approx \eta_{\rm AD} B^2/\left(4\pi l_{\rm B}^2\right)$ (Shang et al. 2002; Schleicher et al. 2008a), where $\eta_{\rm AD}$ is the AD resistivity (Pinto et al. 2008; Pinto & Galli 2008) and $l_{\rm B}$ is the coherence length of the magnetic field, which is approximated by the minimum of the Alfvén damping scale and the integral scale.

The evolution of the magnetic field energy $E_{\rm B} = B^2/8\pi$ is then calculated

2

as

$$\frac{\mathrm{d}E_{\mathrm{B}}}{\mathrm{d}t} = \begin{cases} 2\alpha E_{\mathrm{B}} \ln \dot{\rho_{\mathrm{b}}} - L_{\mathrm{AD}} & z \geqslant z_{\mathrm{vir}}, \\ 2E_{\mathrm{B}} \left(\ln \dot{B}\right)_{\mathrm{dynamo}} & z_{\mathrm{vir}} > z. \end{cases} \tag{2.19}$$

## D THERMAL EVOLUTION

The evolution of the temperature $T$ is given by the equation (see, e.g., Peebles 1993; Sethi et al. 2008)

$$\frac{\mathrm{d}T}{\mathrm{d}t} = \frac{\gamma - 1}{\rho_{\mathrm{b}}} \left[ T \frac{d\rho_{\mathrm{b}}}{dt} + \frac{\mu m_{\mathrm{H}}}{k_{\mathrm{B}}} \left(L_{\mathrm{h}} - L_{\mathrm{c}}\right) \right] + k_{\mathrm{C}} x_{\mathrm{e}} \left(T_{\mathrm{CMB}} - T\right). \tag{2.20}$$

The first term is the adiabatic heating/cooling rate due to collapse/expansion, the second term incorporates various other heating (magnetic energy dissipation by AD and turbulent dissipation, discussed below) and cooling (through $H_2$ and H) volume rates, and the third term represents Compton heating/cooling.

Part of the turbulent energy will go to driving the small-scale dynamo, and part of it will be transferred from large scales to smaller ones in a cascade process, until it is dissipated by viscosity at small enough scales. The rate (per unit mass) at which energy is injected into the system is $\epsilon_{\mathrm{in}} = E_{\mathrm{in}}/m/t_{\mathrm{ed}}(\lambda_{\mathrm{J}}) = v_{\mathrm{in}}^3/\left(2\lambda_{\mathrm{J}}\right)$(Shu 1992). The volume heating rate from the accretion-driven turbulence (ADT) can then be estimated as $L_{\mathrm{ADT}} = f_{\mathrm{t}}\rho_{\mathrm{b}}\epsilon_{\mathrm{in}}$, where $f_{\mathrm{t}}$ is the fraction of the injected energy that is dissipated. Its fiducial value is taken to be 0.1; changing this by a factor of a few does not significantly affect the results.

The model is initialized at a redshift of 800. The dissipation of magnetic fields into the IGM after recombination can significantly influence its temperature and ionization, so this early start provides the proper initial conditions. We follow the evolution of a $10^9\,\mathrm{M}_{\odot}$ halo that virializes at $z = 10$. The radiation background is switched on at turnaround ($z \approx 15$); an earlier or slightly later turn-on is also possible, but does not change the results significantly. The integration is stopped when $n_{\mathrm{b}} \approx 10^7\,\mathrm{cm}^{-3}$. The model is not suitable for higher densities; this would require the inclusion of additional physical processes, e.g., three-body interactions which increase the $H_2$ formation rate.

### 2.B.3 RESULTS

After the radiation background is turned on, $H_2$ is destroyed rapidly. It cannot self-shield as the density is too low; this happens even for a low intensity of $J_{21} = 1$. The cooling is dominated by H, so the temperature is high and the ionized fraction becomes elevated, which in turn aids the formation of $H_2$. For intensities that are low enough, $H_2$ succeeds in reforming, and becomes the dominant coolant. The density at which this occurs depends on the radiation intensity, and is higher for higher intensities. However, at a certain intensity $H_2$ cannot reform fast enough to become an important coolant. If turbulence is unimportant and the initial magnetic field is $\leqslant 0.01$ nG (virtually identical to the zero-field case), this critical intensity is found to be $10 < J_{21} \leqslant 10^2$ for a T4 background and $10^4 < J_{21} \leqslant 10^5$ for a T5 background. Similar results are found for $B_0 = 1$ nG, always assuming $\alpha = 2/3$. However, if the magnetic field is increased to 2 nG, $H_2$ never becomes an important coolant for a T4 background with $J_{21} = 10$, while for $J_{21} = 1$ an instability occurs at high densities. Here, the AD heating becomes too strong to be compensated by the $H_2$ cooling and the temperature suddenly increases, because much of the $H_2$ is destroyed by collisional dissociation and H cooling is still very strongly suppressed below $\sim 8 \times 10^3$ K. The gas stays hot afterwards, with cooling dominated by H. For a T5 background with the same $B_0$, a similar instability occurs for $J_{21} = 10^4$, while with $B_0 = 3$ nG the $H_2$ fraction is never large enough for significant cooling at this intensity. Thus, when turbulence is not important, a larger initial magnetic field decreases the critical intensity required to keep the gas in the halo hot, as illustrated in Figure 2.B.1.

During collapse, turbulence quickly brings the magnetic field strength to $B_{max}$ through the small-scale dynamo, either by amplifying smaller fields or by draining energy from larger fields. Since $B_{max}$ only increases as $\propto n_b^{1/2}$, heating from turbulent dissipation always dominates over AD heating, and because $H_2$ cooling grows more steeply with density than turbulent heating does, the gas always cools through $H_2$ when there is no radiation background present. Since heating is dominated by turbulence, halos with different $B_0$ converge to approximately the same evolutionary track; the turbulence has a moderating effect.

The effects of different radiation intensities of a T4 and T5 background on the temperature, electron fraction, and $H_2$ fraction are shown in Figure 2.B.2 for a turbulent halo with $B_0 = 1\,\mathrm{nG}$. However, the critical intensity does not depend on $B_0$ when turbulent effects are important. The critical intensity is found to be $1 < J_{21} \leqslant 10$ for a T4 background and $10^3 < J_{21} \leqslant 10^4$ for a T5 background.

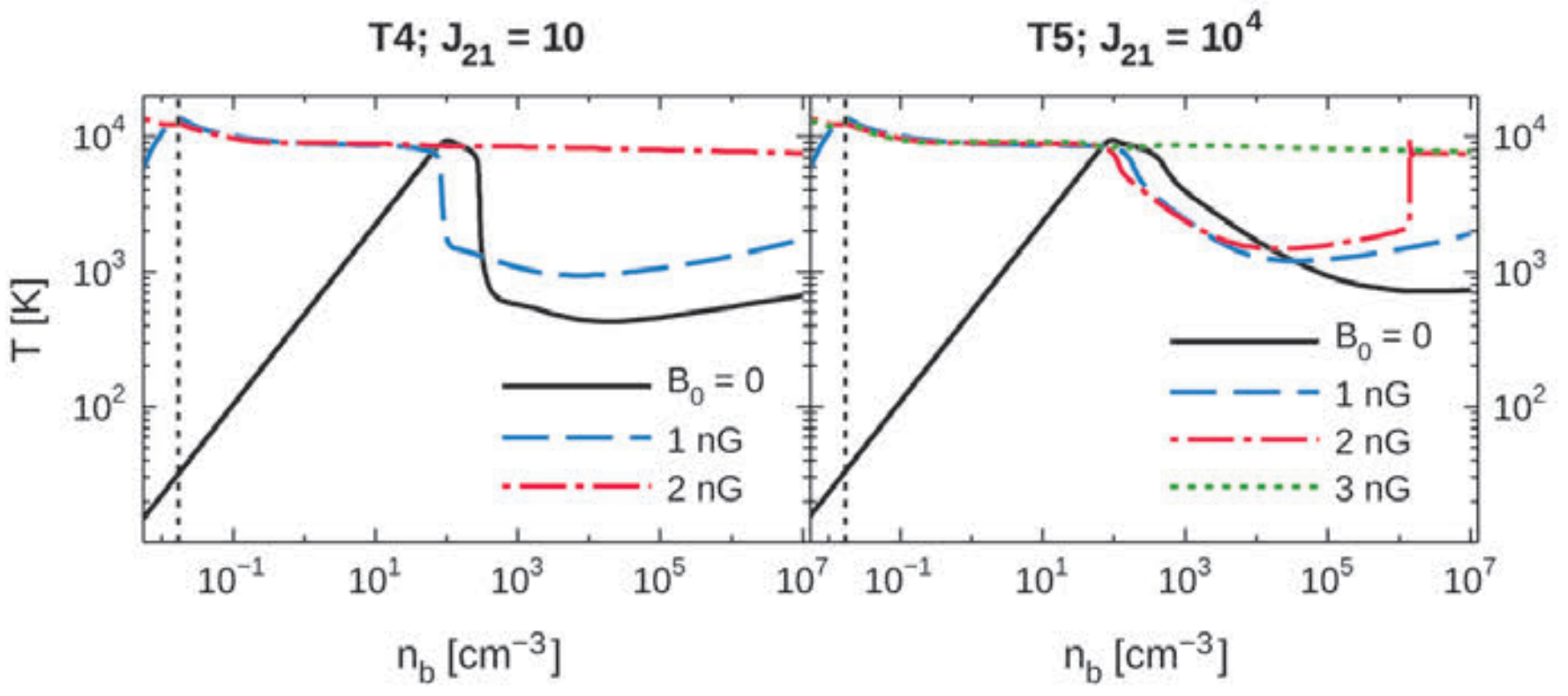

**Figure 2.B.1** – Gas temperature as a function of density for a T4 background with $J_{21} = 10$ (left) and a T5 background with $J_{21} = 10^4$ (right), for several different initial magnetic field strengths (no turbulence). The dotted vertical line indicates virialization.

### 2.B.4 DISCUSSION

The effects of magnetic fields and turbulence on the critical intensity of a UV radiation background were examined. Note that these results only hold in a zero or very low metallicity environment, because metals and dust are much more efficient coolants than $H_2$. For a halo not significantly influenced by turbulence or magnetic fields, the critical intensity was found to be $10 < J_{21}^{\mathrm{crit}} \leqslant 10^2$ for a T4 background, and $10^4 < J_{21}^{\mathrm{crit}} \leqslant 10^5$ for a T5 background. These limits are consistent with those found by Shang et al. (2010) and lower by a factor ~10 than previous estimates by e.g., Omukai (2001) and Bromm & Loeb (2003a), most likely as a result of the different $H_2$ dissociation rates used. For $B_0 = 1\,\mathrm{nG}$, these limits do not change; however, when the field is increased to ~2 nG, the critical intensity is lowered by a factor ~10, and the stronger the field, the lower $J_{21}^{\mathrm{crit}}$. Such magnetic fields alone do not give rise to sufficient AD heating to overcome $H_2$ cooling, but in combination with a radiation background they do influence the ability of the gas to stay hot. Note that the amount of AD heating depends on the scaling of $B$ with $\rho$; it is therefore important to obtain a correct model for this relationship. The current upper limit on the primordial magnetic field is ~1 nG comoving (Schleicher & Miniati 2011; Trivedi et al. 2012), so a 2 nG field would be reached by the ~$2\sigma$ upward fluctuations.

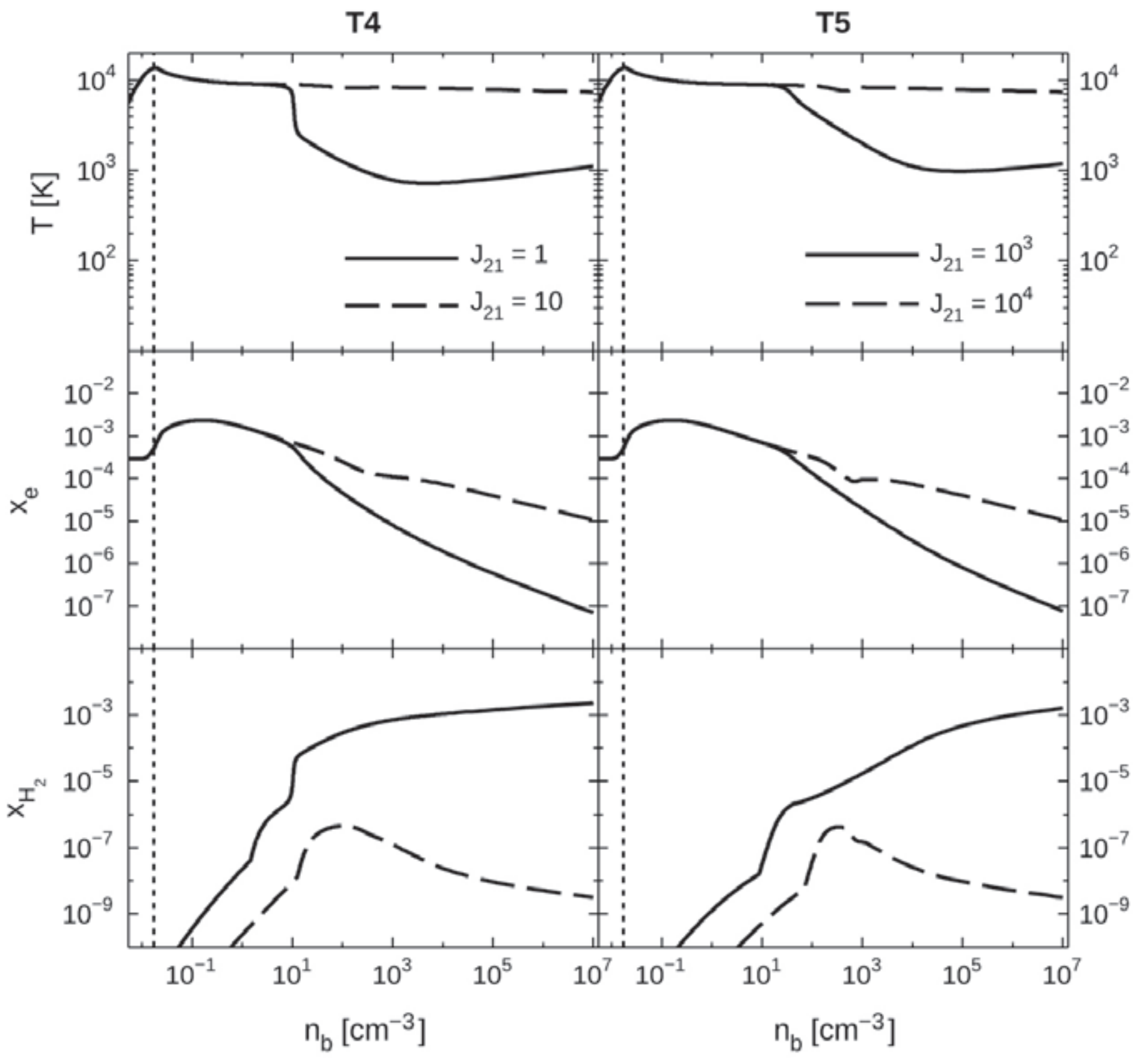

**Figure 2.B.2** – Gas temperature, electron fraction, and $H_2$ fraction as a function of density for a turbulent halo with $B_0 = 1\,\mathrm{nG}$, a T4 (left) and a T5 (right) background, and different $J_{21}$ as indicated in the top panels. The dotted vertical line indicates virialization.

In a $10^9\,\mathrm{M}_\odot$ turbulent halo, the critical intensity was found to be $1 < J_{21}^{\rm crit} \leqslant 10$ for a T4 background, and $10^3 < J_{21}^{\rm crit} \leqslant 10^4$ for a T5 background;

these are a factor ~10 lower than for a halo not affected by turbulence or magnetic fields. Since this is due to the turbulent heating in such halos, larger halos and/or halos with stronger turbulent heating will have an even lower $J_{21}^{\rm crit}$. Note that the results for the non-turbulent halos are independent of halo mass. Interestingly, in turbulent halos with $M \gtrsim 10^{11}\,{\rm M}_{\odot}$ (depending on the strength of the turbulence), the turbulent heating alone is able to keep the gas hot without any UV background.

The fact that the values of $J_{21}^{\rm crit}$ that have been found here are smaller than previous estimates is quite important. The mean cosmic UV background is expected to be around $J_{21}^{\rm bg} \sim 40$, which is smaller than the critical intensities, especially in the case of a T5 background. However, the background will fluctuate spatially, and thus a fraction of all halos will be irradiated by a supercritical intensity. Then, the lower the critical intensity, the larger the fraction of halos which are suitable candidates for direct SMBH formation; e.g., according to the distribution proposed by Dijkstra et al. (2008), a decrease in $J_{21}^{\rm crit}$ from $10^4$ to $10^3$ means an increase in the fraction of irradiated $T_{\rm vir} \approx 10^4$ K halos from negligibly small ($\lesssim 10^{-8}$) to $\sim 10^{-6}$. For $J_{21}^{\rm crit} \sim 10^2$, the halo fraction even increases to $\sim 10^{-3}$. With a sufficiently low $J_{21}^{\rm crit}$, one could argue that this mechanism provided many, if not all, seeds for the SMBHs observed in galaxies today.

2

## ACKNOWLEDGMENTS

C. Van Borm acknowledges research funding by the German Science Foundation (DFG) under grant SFB 963/1 (project A12).

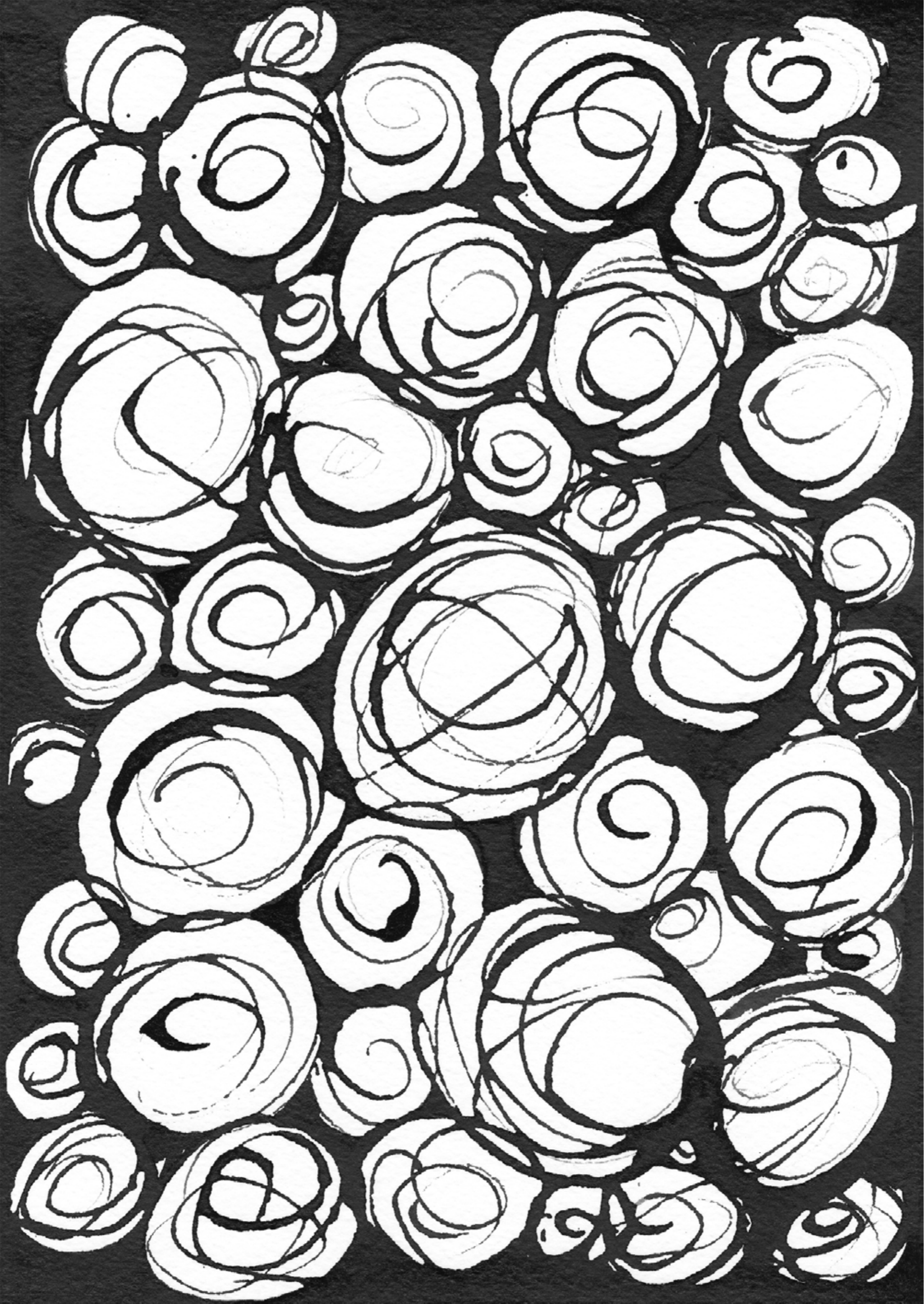

# Part II

# ISM Thermodynamics and Fragmentation

*THE SECRET TO CREATIVITY IS KNOWING HOW TO HIDE YOUR SOURCES.*

— ADAPTED FROM C. E. M. JOAD (1926)

# CHAPTER

# 3

# THE PDR-Zz CODE

## ABSTRACT

In this work, we describe our photodissociation region (PDR) code, PDR-Zz in detail. This code is based on the PDR code by Meijerink & Spaans (2005), which we updated and significantly improved, and also extended for use in a wider range of physical conditions (e.g. much higher densities, metallicities from essentially 0 to solar, redshifts >0). The input parameters to the code are the total hydrogen density, the metallicity, the incident radiation field strength, the cosmic ray ionization rate, and the redshift; the combination of which defines the coupled chemical and thermal solution for the simulated region. This computational code can be used to study traditional PDRs, and can additionally be applied to many other situations where one is interested in the equilibrium state of gas at different optical depths, under various conditions; from diffuse and molecular clouds in the Milky Way, to clouds near AGN and material in galaxies in the high-redshift Universe.

## 3.1 INTRODUCTION

The life of a star is strongly linked to the interstellar medium (ISM) of the galaxy in which it resides. Stars, both at the present time and in the early Universe, are born from this material, and change it during their lifetime by ejecting newly formed elements and dust into it, as well as through the radiation they emit. The interstellar medium has a complex thermal and ionization structure, and contains several phases characterized by different temperatures, densities, and ionization fractions.

The original 'two-phase' ISM model was described by Field et al. (1969), who identified two stable phases, warm and cold, by considering the stability of clouds in local thermal pressure equilibrium. Later work, notably by McKee & Ostriker (1977), expanded upon this model by adding a third phase of hot, low-density medium, in order to consider the effect of supernova remnant blast waves. Currently, the ISM is thought to contain at least five main phases (e.g. Tielens 2005; Draine 2011). The coldest, densest phase is the cold molecular phase, comprising molecular clouds of mostly $H_2$, at temperatures of ~10-20 K and densities $\gtrsim 200\,\mathrm{cm}^{-3}$. Most molecular clouds are gravitationally bound, with the densest cores likely unstable to collapse and thus viable sites for star formation. In between the molecular clouds we find more diffuse, mainly atomic gas, in the cold neutral medium (CNM) and warm neutral medium (WNM) phases, which coexist in approximate thermal and pressure equilibrium. The CNM comprises cold neutral hydrogen (HI) gas distributed in sheets and filaments, with characteristic temperatures of ~50-100 K and densities of ~50 $cm^{-3}$. The WNM (sometimes also called warm intercloud medium) is found mainly in diffuse photon-dominated regions (see below), on the boundaries between ionized hydrogen (HII) regions and molecular clouds, with typical temperatures of ~8000 K and densities of ~0.5 $cm^{-3}$. Some of the warm gas is nearly fully ionized, which is then called the warm ionized medium (WIM), with characteristic temperatures of ~8000 K and densities of ~0.1 $cm^{-3}$. The source of ionization is not entirely clear; possibilities include photons from O and B type stars and shocks or collisional ionization. Finally, there exists also a pervasive, hot, tenuous, intercloud phase called the hot ionized medium (HIM, sometimes also 'corona'), with typical temperatures of ~3-10 × $10^5$ K and densities of ~$10^{-3}$ $cm^{-3}$. This gas is heated and ionized through shocks driven by stellar winds from early-type stars and by supernova explosions, hence reflecting mainly the macroscopic processes in the ISM, as opposed to the other phases which better reflect the microphysics of the medium. The HIM fills most of the volume of the halo of the galaxy.

Often, a sixth component is included as well, consisting of the highly ionized HII regions, with temperatures ~$10^4$ K and a wide range of densities, ~1-$10^5$ $cm^{-3}$ (though its filling fraction is rather low). HII regions are formed around young, massive stars (O & B type), which emit a large amount of photons beyond the Lyman limit at 13.6 eV, and ionize and heat the surrounding molecular clouds from which they were born.

Also far-ultraviolet (FUV) radiation, with photon energies in the range 6-13.6 eV, will affect the chemistry and heating of neutral atomic gas and much of the molecular gas in galaxies. Predominantly neutral regions of the interstellar medium in which far-UV photons significantly influence the gas chemistry and thermal balance are called photon-dominated regions (PDRs, sometimes also called photodissociation regions). Such regions can be found for example near massive stars, in the neutral envelopes of planetary nebulae, in diffuse clouds, in the outer regions of molecular clouds, in protoplanetary disks, as well as near active galactic nuclei (AGN). In terms of physical and chemical processes, essentially most of the ISM is in PDRs.

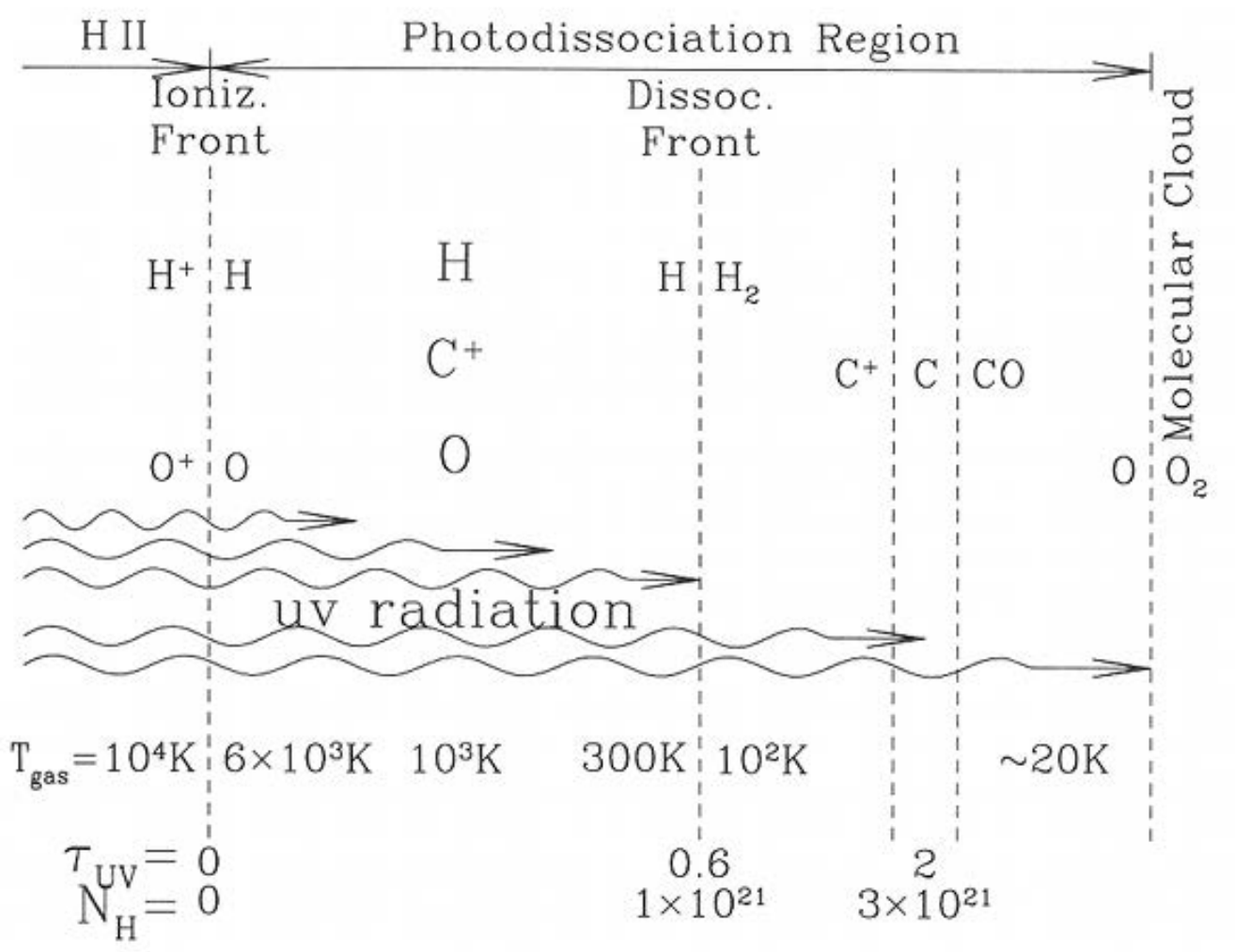

**Figure 3.1** – Structure of a typical PDR, at the interface between an HII region and a dense molecular cloud (Figure 31.2 from Draine 2011). $N_{\rm H}$ is given in units of $\rm cm^{-2}$.

The physical properties of a PDR change with distance (optical depth/ column density), as the photons are gradually absorbed while traveling into the region (e.g. Spitzer 1978; Draine 2011). A schematic overview of the structure of a typical PDR can be seen in Figure 3.1 (adopted from Draine (2011)), showing the interface between an HII region and a dense molecular cloud, illuminated from the left. As the UV radiation becomes attenuated, ionized species first give way to neutral atomic species, and deeper into the cloud molecules such as $H_2$, CO, $O_2$, and PAHs (polycyclic aromatic hydrocarbons) become stable. The boundary between the regions dominated by ionized hydrogen and atomic hydrogen is called the ionization front, while the boundary between the regions dominated by atomic hydrogen and molecular hydrogen is called the dissociation front. The transition from H to $H_2$ occurs when the Lyman and Werner bands in the FUV which lead to dissociation of molecular hydrogen become optically thick, and $H_2$ becomes shielded. Somewhat deeper in the region, the carbon-ionizing flux has dropped sufficiently and ionized carbon recombines and forms carbon monoxide, resulting in a $C^+$/C/CO transition. Even deeper into the region, oxygen becomes predominantly molecular as well. The increase in optical depth is also associated with a gradual drop in temperature. A thorough overview of PDRs, including physical and chemical processes and observations, can be found in Hollenbach & Tielens (1999).

The first theoretical PDR models date back to the 1970's, focusing on equilibrium modeling of the transition from H to $H_2$ and $C^+$ to CO (e.g. Hollenbach & Salpeter 1971; Jura 1974; Glassgold & Langer 1975; Black & Dalgarno 1977). In the years after that, models considering the chemical and thermal structure of clouds irradiated by an FUV radiation field were developed (e.g. de Jong et al. 1980; Tielens & Hollenbach 1985; van Dishoeck & Black 1988; Sternberg & Dalgarno 1989; Hollenbach et al. 1991; Le Bourlot et al. 1993; Störzer et al. 1996). The main input parameters of these early models were the total hydrogen density and the intensity of the radiation field. Following these base models, a number of models were created that focused on specific aspects of PDR chemistry and physics, e.g. using time-dependent chemistry, or modeling either clumped or turbulent media (e.g. Hill & Hollenbach 1978; Wagenblast & Hartquist 1988; de Boisanger et al. 1992; Bertoldi & Draine 1996; Lee et al. 1996; Hegmann & Kegel 1996; Spaans 1996; Nejad & Wagenblast 1999; Röllig et al. 2002). A more detailed review of the history of PDR modeling can be found in Hollenbach & Tielens (1999).

More recently, a comparison and benchmarking study has been performed by Röllig et al. (2007)[1], in which 11 different PDR codes participated. These are, amongst others, CLOUDY (Ferland et al. 2013), the HTBKW PDR code (Kaufman et al. 1999), the Meudon PDR code (Le Petit et al. 2006), KOSMA-$\tau$ (Röllig et al. 2006), UCL_PDR (Bell et al. 2005), as well as the code by Meijerink & Spaans (2005), which forms the basis for the code described here. As all of these codes were developed with a specific goal in mind, they differ in many aspects, for example, geometry (finite or semi-finite plane-parallel, spherical, disk), chemistry (steady-state or time-dependent, chemical network and species, reaction rates), radiative transfer, shielding, treatment of dust and PAHs, treatment of cooling and heating processes, numerical methods, and input parameters (which quantities, their range).

In this chapter, we will describe a computational code based on the existing PDR model by Meijerink & Spaans (2005), but extended for use in a wider range of physical conditions and more robust than the original code. This model can be used to study PDRs, and additionally it can be applied to many other situations where one is interested in the equilibrium state of gas at different optical depths, under various conditions; from clouds in the Milky Way to AGN and the high-redshift Universe.

[1] Summary of results available online at `http://www.astro.uni-koeln.de/site/pdr-comparison/`

## 3.2 THE CODE

Our code, PDR-ZZ, is based on the PDR code described in Meijerink & Spaans (2005) (hereafter the MSPDR code), but improved and extended. The model solved by the computational code considers a plane-parallel, semi-infinite slab illuminated from one side by a radiation field. The modeled region is divided into 'depth' zones, perpendicular to the surface of the slab. The depth parameter can be represented by either the physical distance from the surface, the total hydrogen column density $N_{\rm H}$, or the optical extinction $A_{\rm V}$. The distance would be a good choice if one wants to investigate clouds of a certain size, while the optical extinction is useful to understand the full range of impact of the photoprocesses. However, since extinction depends on metallicity through the dust abundance, high extinctions combined with low metallicities correspond to unrealistically large columns, so in that case the column density will be more representative. By default, the optical extinction range is set to $10^{-3} < A_{\rm V} < 20\,{\rm mag}$, or the column density range to $10^{18} \leqslant N_{\rm H} \leqslant 5 \times 10^{22}\,{\rm cm}^{-2}$, with the number of zones set to 150, including a surface zone. The size of the zones is calculated in such a way that the depth parameter is sufficiently resolved in the desired range. At each depth, the thermal and chemical balance have to be solved simultaneously to determine the temperature and species abundances. It is computationally much more feasible to solve both balances separately, though because of their interdependency this must be done iteratively.

Several parameters can be set in this model, the combination of which fully defines the chemo-thermal solution for the simulated region.

- The total hydrogen density $n_{\rm H}$ (in units of ${\rm cm}^{-3}$), which affects both the chemical reaction rates and the cooling and heating rates.

- The metallicity $Z$ (in units of $Z_\odot$), which sets the amount of metals and dust available and thus the chemical structure, affecting the cooling and heating rates.
- The redshift $z$, which sets the cosmic microwave background (CMB) temperature, and therefore affects the thermal balance and the radiation field.
- The incident radiation field (both its intensity and shape), which is regulated through the metallicity (dust abundance), the redshift (CMB radiation), and through a newly defined UV scaling parameter $W_0$ (see further; $W_0$ is analogous in function to, e.g., $G_0$, the UV radiation scaling parameter in units of the Habing (1968) radiation field).
- The cosmic ray (CR) ionization rate $\zeta$ (in units of $s^{-1}$), which has an impact on the electron abundance and contributes to the heating rate.

### 3.2.1 RADIATION

The basic shape of the radiation field in the model is that of the solar neighborhood interstellar radiation field (ISRF), as given by Mezger et al. (1982); Mathis et al. (1983), and updated by Black (1994) using results from the FIRAS experiment on COBE (Wright et al. 1991). The interstellar radiation field for wavelengths above the Lyman limit of 912 Å can be represented by the sum of five components (ignoring radio emission):

1. a UV component, due to emission from massive stars,
2. an optical-near infrared (V-NIR) component, due to emission from A-type stars (optical) and late type stars (far-red to NIR),
3. a mid-infrared (MIR) component, due to emission from small non-thermally heated grains,
4. a far-infrared (FIR) component, due to diffuse dust emission, and
5. a microwave component, due to the cosmic microwave background.

Fits to these five components are shown in Figure 3.1.

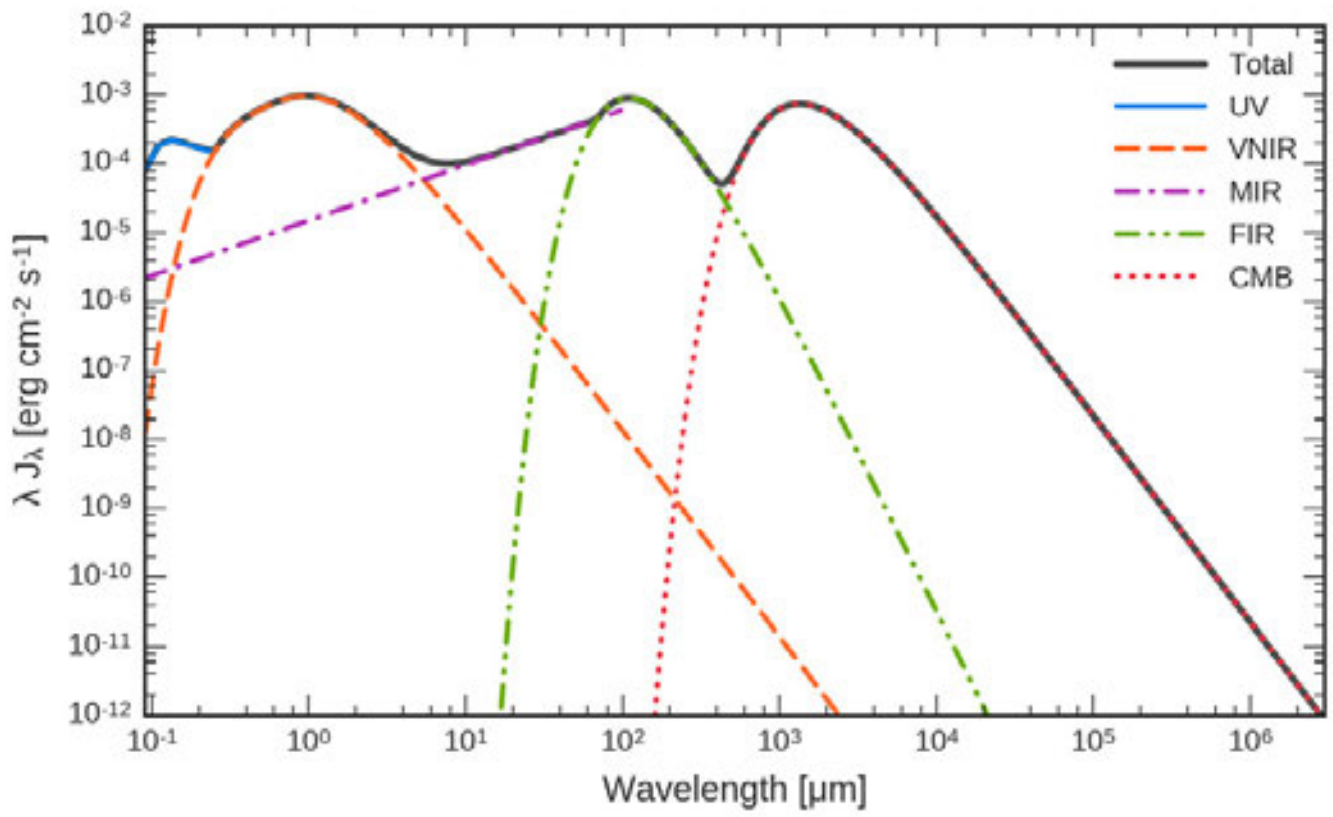

**Figure 3.1** – Interstellar radiation field, expressed as $\lambda J_\lambda$ $(= \nu J_\nu)$ in units of $\mathrm{erg\,cm^{-2}\,s^{-1}}$ as a function of wavelength, for $W_0 = 1$ and $z = 0$. A plot of mean intensity $J_\nu$ as a function of frequency $\nu$ is identical to Figure 1 of Black (1994).

We use the prescription of Mezger et al. (1982) for the UV component:

$$4\pi J_\lambda^{\mathrm{UV}} = \begin{cases} 38.57 \times 10^4\, \lambda_{\mu\mathrm{m}}^{3.4172} & \text{if } 912\,\text{Å} \leqslant \lambda < 0.110\,\mu\mathrm{m}, \\ 2.045 \times 10^2 & \text{if } 0.110\,\mu\mathrm{m} \leqslant \lambda < 0.134\,\mu\mathrm{m}, \\ 7.115\, \lambda_{\mu\mathrm{m}}^{-1.6678} & \text{if } 0.134\,\mu\mathrm{m} \leqslant \lambda < 0.246\,\mu\mathrm{m}, \end{cases} \tag{3.1}$$

with $\lambda_{\mu\mathrm{m}}$ the wavelength in μm; and

$$J_\nu = J_\lambda \frac{\lambda^2}{c} \tag{3.2}$$

to obtain the mean intensity as function of frequency $\nu$, $J_\nu^{\mathrm{UV}}$, in units of $\mathrm{erg\,cm^{-2}\,s^{-1}\,Hz^{-1}}$.

For the other four components, we use the parametrization given by Zucconi et al. (2001), with the values of the parameters listed in Table 3.1. The V-NIR, FIR, and CMB components can be represented by a modified blackbody (or a sum thereof) with temperature $T_{\mathrm{i}}$, of the form

$$J_\nu^{(\mathrm{X})} = \frac{2h\nu^3}{c^2} \left(\frac{\lambda_{\mathrm{p}}}{\lambda}\right)^p \sum_i \frac{W_i}{e^{h\nu/(k_{\mathrm{B}} T_i)} - 1}, \tag{3.3}$$

where $\lambda_\mathrm{p}$ is the peak wavelength of $J_\lambda$, and $W_i$ is a dilution factor. As usual, $h$ is the Planck constant, $k_\mathrm{B}$ is the Boltzmann constant, and $c$ is the speed of light. The MIR component can be approximated by a power law of the form

$$J_\nu^\mathrm{MIR} = W_i \frac{2h(c/\lambda_\mathrm{p})^3}{c^2}\left(\frac{\lambda_\mathrm{p}}{\lambda}\right)^p, \tag{3.4}$$

for $\lambda < \lambda_\mathrm{p}$.

| | $\lambda_\mathrm{p}$ | $p$ | $W_i$ | $T_i$ (K) |
|---|---|---|---|---|
| V-NIR | 0.4 µm | 0 | $1\times10^{-14}$ | 7500 |
| | 0.75 µm | 0 | $1\times10^{-13}$ | 4000 |
| | 1.0 µm | 0 | $4\times10^{-13}$ | 3000 |
| MIR | 100 µm | -1.8 | $5\times10^{-7}$ | |
| FIR | 140 µm | 1.65 | $2\times10^{-4}$ | 23.3 |
| CMB | 1.06 mm | 0 | 1 | 2.725(1+z) |

**Table 3.1** – Parameters for the interstellar radiation field, from Zucconi et al. (2001)

The redshift dependence of the radiation field is enclosed in the CMB temperature, as $T_\mathrm{CMB} = T_\mathrm{CMB,0}(1+z)$, with $T_\mathrm{CMB,0} = 2.725\,\mathrm{K}$. To handle metallicity dependence, the MIR and FIR components are multiplied by the metallicity, as their intensity depends on the amount of dust present. Finally, all components except for the CMB component are multiplied by a UV scaling factor $W_0$ (see below for a definition).

### INTERMEZZO: UV RADIATION FIELD ESTIMATES

There are two ways in which UV radiation field comparisons are made: either by comparing the field strength at a certain wavelength, e.g. 1000 Å, or by comparing the integrated energy density over a certain range of wavelengths. The former is useful when calculating, e.g., the photodissociation rate of $H_2$ or CO, which requires photons with energies above a certain threshold (>10 eV), while the latter is more appropriate when calculating e.g. the photoelectric heating rate, which is sensitive to a wider range of photon energies.

There are quite a few different UV scaling parameters to be found in the literature, resulting from e.g. different estimates of the UV radiation field, and a different integration range for calculating the integrated energy density. We will briefly discuss the most common ones, and motivate the choice we have made. Plots of some different estimates of the UV field are shown in Figure 3.2.

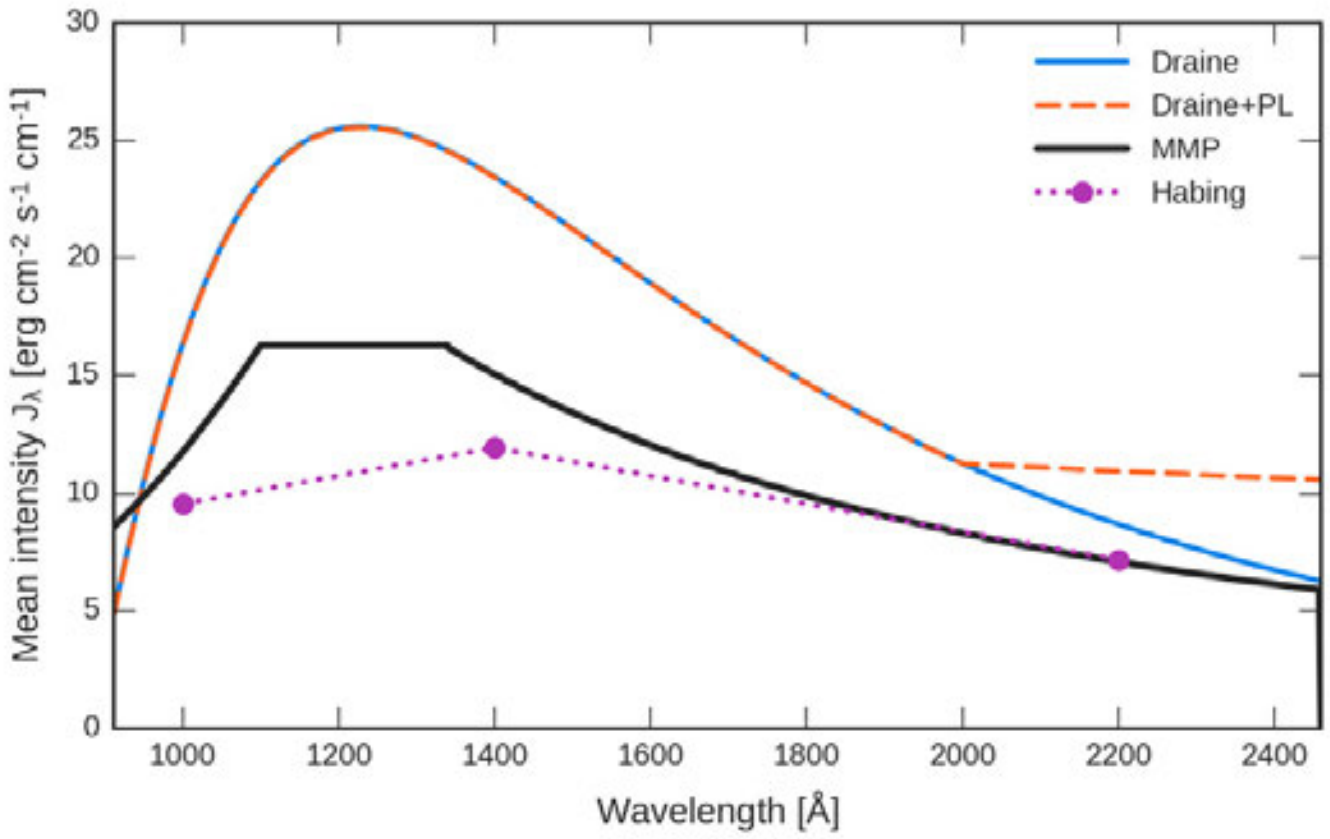

**Figure 3.2** – Different estimates of the ultraviolet radiation field in the solar neighborhood, shown as $J_\lambda$ in units of $\mathrm{erg\,cm^{-2}\,s^{-1}\,cm^{-1}}$, as a function of wavelength; from Habing (1968),Draine (1978) (including a version with power-law (PL) component), and Mathis et al. (1983, MMP).

One of the earliest estimates of the UV radiation field was performed by Habing (1968), who suggested the following energy density:

$$u_\lambda^{\mathrm{Hab}} = \begin{cases} 4.0\times10^{-9} & \text{at } \lambda = 1000\,\text{Å}, (= 12.4\,\mathrm{eV}), \\ 5.0\times10^{-9} & \text{at } \lambda = 1400\,\text{Å}, (= 8.9\,\mathrm{eV}), \\ 3.0\times10^{-9} & \text{at } \lambda = 2200\,\text{Å}, (= 5.6\,\mathrm{eV}), \end{cases} \tag{3.5}$$

in units of $\mathrm{erg\,cm^{-3}\,cm^{-1}}$. The energy density relates to the mean intensity by $u_\lambda = \frac{4\pi}{c} J_\lambda$, and equivalently $u_\nu = \frac{4\pi}{c} J_\nu$. Two commonly used dimensionless parameters are based on this representation. One is $\chi$, defined as

$$\chi \equiv \frac{u_\lambda(1000\,\text{Å})}{u_\lambda^{\mathrm{Hab}}(1000\,\text{Å})} = \frac{u_\lambda(1000\,\text{Å})}{4.0\times10^{-9}\,\mathrm{erg\,cm^{-3}\,cm^{-1}}}, \tag{3.6}$$

which can be used to compare the radiation field strength at 1000 Å. The other is $G_0$, which is the integrated energy density normalized with respect to the integral of the Habing field:

$$G_0 = \frac{U_{\rm UV}(\lambda_{\rm c})}{U_{\rm UV}^{\rm Hab}(\lambda_{\rm c})}, \quad (3.7)$$

where the integrated UV energy density $U_{\rm UV}$ is given by

$$U_{\rm UV}(\lambda_{\rm c}) = \int_{912\,\text{Å}}^{\lambda_{\rm c}} \frac{4\pi}{c} J_\lambda \, {\rm d}\lambda, \quad (3.8)$$

in units of erg cm$^{-3}$. The value of the integrated energy density depends on the integration boundaries used. However, there seems to be some variation in the literature on the values of these boundaries, in particular on the value of the upper cutoff $\lambda_{\rm c}$ (corresponding to the lower cutoff energy; usually in the range 5-6 eV), and thus on the value of the integral. When we choose $\lambda_{\rm c}$ = 2460 Å, corresponding to ~5 eV, the total Habing energy density is $U_{\rm UV}^{\rm Hab} = 6.13 \times 10^{-14}$ erg cm$^{-3}$, after linear interpolation of the three points given for $u_\lambda^{\rm Hab}$.

Another widely used estimate for the UV radiation field comes from Draine (1978). Some different versions of this field exist in the literature, probably due to rewriting the expression, though not all of these give the same result. Additionally, sometimes a power-law component is added for $\lambda > 2000$ Å (e.g. Le Petit et al. 2006), though the difference for the integrated energy density is small. According to the original paper, the photon flux is given by:

$$F^{\rm Dra} = 1.658 \times 10^6\, E - 2.152 \times 10^5\, E^2 + 6.919 \times 10^3\, E^3, \quad (3.9)$$

in units of photons cm$^{-2}$ s$^{-1}$ sr$^{-1}$ eV$^{-1}$, with $E = h_{\rm eV} \frac{c}{\lambda}$ the photon energy in eV. To convert the photon flux to mean intensity,

$$J_\nu^{\rm Dra} = F^{\rm Dra} h_{\rm eV} E_{\rm erg} \quad (3.10)$$

in units of erg cm$^{-2}$ s$^{-1}$ Hz$^{-1}$ (if the radiation field is isotropic). The integrated energy density of this field, using our chosen $\lambda_{\rm c}$, is $U_{\rm UV}^{\rm Dra} = 1.03 \times 10^{-13}$ erg cm$^{-3}$. With an additional power-law component as in Le Petit et al. (2006), this becomes $U_{\rm UV}^{\rm Dra+PL} = 1.07 \times 10^{-13}$ erg cm$^{-3}$. When compared to the Habing field, the Draine field has $\chi = 1.71$ and $G_0 = 1.68$ ($G_0 = 1.75$ with power-law) for our chosen $\lambda_{\rm c}$.

As mentioned, here we use yet another estimate for the UV radiation field, namely the one from Mezger et al. (1982) (MMP), which should be a better match with available observations (Draine 2011). Its integrated energy density is $U_{\rm UV}^{\rm MMP} = 7.13 \times 10^{-14}\,{\rm erg\,cm^{-3}}$, resulting in $\chi = 1.23$ and $G_0 = 1.16$ when compared to the Habing field.

As we use the MMP field, it makes little sense to use $G_0$ to scale the field to higher or lower intensities. Therefore, we introduce a new scaling parameter $W_0$, defined relative to the integrated energy density of the MMP field:

$$W_0 = \frac{U_{\rm UV}(\lambda_{\rm c})}{U_{\rm UV}^{\rm MMP}(\lambda_{\rm c})}, \tag{3.11}$$

again using $\lambda_{\rm c} = 2460\,$Å. With this definition, $W_0 = 0.86$ for the Habing field, and 1.44 for the Draine field (1.50 with power law).

### 3.2.2 CHEMISTRY

The majority of the included chemical reactions and their rates are taken from the fifth release of the UMIST database, RATE12[2] (McElroy et al. 2013). The rate coefficients for two-body reactions in the UMIST database are parametrized using the typical Arrhenius-type formula, with fitting parameters $\alpha$, $\beta$, and $\gamma$:

$$k = \alpha \left(\frac{T}{300\,{\rm K}}\right)^{\beta} \exp\left(-\frac{\gamma}{T}\right)\,{\rm cm^3\,s^{-1}}, \tag{3.12}$$

with T the gas temperature. We use all 13 elements present in this database (H, He, C, N, O, F, Na, Mg, Si, P, S, Cl, and Fe), but only include species composed of 6 atoms or less, as we do not expect the larger molecules to have a significant impact. We also included eight additional species, resulting in a total of 346 species. These are five deuterium species (D, $D^+$, $D^-$, HD, $HD^+$), and polycyclic aromatic hydrocarbons and its ions (PAH, $PAH^+$, $PAH^-$). For solar metallicity, we adopted the element abundances from Table 1 in Asplund et al. (2009), but converted to bulk abundances following their recommendations (see also Draine 2011). The gas-phase abundances are typically lower, reflecting that some amount of the elements is trapped in dust grains. To account for this element depletion, the bulk abundances were multiplied by gas-phase factors as computed from the fits by Jenkins (2009, Table 4) for an intermediately depleted region ($F^* = 0.5$). The exceptions are F and Na, which were not included in the study by Jenkins (2009). For F, we adopt a gas-phase abundance of $1.8 \times 10^{-8}$ relative to H, as suggested by Neufeld et al. (2005). The amount of depletion of Na is not known, though circumstantial evidence suggests that it does not strongly deplete (Weingartner & Draine 2001b; Draine 2011), and therefore we adopt a somewhat arbitrary gas-phase factor of 0.9. The gas-phase factors and both the total and gas-phase abundances are listed in Table 3.2. Furthermore, we adopted a deuterium abundance of $2.0 \times 10^{-5}$ (Prodanović et al. 2010) and a PAH abundance of $1.0 \times 10^{-7}$ (Tielens 2008). For a metallicity different from solar, all metal abundances and the PAH fraction are scaled by this metallicity $Z$.

[2]Publicly available at http://udfa.ajmarkwick.net/

## A THREE-BODY RATES

The UMIST database also provides an additional network containing three-body reactions, which we included. However, when using these rates, care must be taken to handle situations when the temperature solution falls outside of the temperature range specified for each rate, as some of these three-body rates will diverge unphysically. To deal with this, when out of range we set the rate to the value at the closest point for which it is defined. We have verified that this does not lead to undesired behavior.

Additionally, we added two more three-body reactions which are missing from UMIST, namely $H + H + H \longrightarrow H_2 + H$, adopting the rate from Forrey (2013), and $H_2 + H + H \longrightarrow H_2 + H_2$, adopting the rate from the former, but divided by 8 (Palla et al. 1983).

## B DEUTERIUM NETWORK

Our deuterium network consists of 25 reactions, involving five deuterium species: D, $D^+$, $D^-$, HD, $HD^+$; see Table 3.3 for an overview of the deuterium network and rate coefficient references. We included all 18 reactions from the network compiled by Nakamura & Umemura (2002), two additional reactions (D8, D11), HD photodissociation (D13), two three-body reactions analogous to the hydrogen three-body reactions (D22-23), and two collisional dissociation reactions (D24-25). For the reaction rate coefficients, we adopted in part the rates recommended by Glover & Abel (2008), and in part the rates from the corresponding hydrogen reactions already in our network. The HD photodissociation rate was taken from Wolcott-Green & Haiman (2011).

| | $(n_X/n_H)_{tot}$ [a] | $f_{gas}$ [b] | $(n_X/n_H)_{gas}$ [c] |
|---|---|---|---|
| H | 1.0 | | 1.0 |
| He | $9.55 \times 10^{-2}$ | | $9.55 \times 10^{-2}$ |
| C | $2.95 \times 10^{-4}$ | 0.688 | $2.03 \times 10^{-4}$ |
| N | $7.41 \times 10^{-5}$ | 0.778 | $5.76 \times 10^{-5}$ |
| O | $5.37 \times 10^{-4}$ | 0.753 | $4.04 \times 10^{-4}$ |
| F | $2.88 \times 10^{-8}$ | | $1.80 \times 10^{-8}$ [d] |
| Na | $2.04 \times 10^{-6}$ | ~0.9 | $1.84 \times 10^{-6}$ |
| Mg | $4.37 \times 10^{-5}$ | 0.170 | $7.43 \times 10^{-6}$ |
| Si | $3.55 \times 10^{-5}$ | 0.162 | $5.75 \times 10^{-6}$ |
| P | $2.82 \times 10^{-7}$ | 0.665 | $1.88 \times 10^{-7}$ |
| S | $1.45 \times 10^{-5}$ | 0.530 | $7.69 \times 10^{-6}$ |
| Cl | $1.86 \times 10^{-7}$ | 0.663 | $1.23 \times 10^{-7}$ |
| Fe | $3.47 \times 10^{-5}$ | 0.0255 | $8.85 \times 10^{-7}$ |
| D | | | $2.00 \times 10^{-5}$ [d] |
| PAH | | | $1.00 \times 10^{-7}$ [d] |

**Table 3.2** – Element abundances and depletion.

[a] Total element abundances (number fractions) relative to hydrogen, adapted from Asplund et al. (2009).
[b] Fraction of total abundance in the gas phase for intermediate depletion ($F^* = 0.5$), computed from Jenkins (2009) with the exception of Na.
[c] Gas-phase abundance (number fractions) relative to hydrogen.
[d] See text for references.

| Nr. | Reaction | Ref. |
|---|---|---|
| D1 | $D + H^+ \longrightarrow HD^+ + \gamma$ | 1 |
| D2 | $D^+ + H \longrightarrow HD^+ + \gamma$ | 1 |
| D3 | $D + e^- \longrightarrow D^- + \gamma$ | 2 |
| D4 | $D^+ + e^- \longrightarrow D + \gamma$ | 2 |
| D5 | $HD^+ + H \longrightarrow HD + H^+$ | 2 |

**Table 3.3** – Continued from previous page

| Nr. | Reaction | Ref. |
|---|---|---|
| D6 | $HD^+ + e^- \longrightarrow H + D$ | 3 |
| D7 | $H^+ + D^- \longrightarrow H + D$ | 2 |
| D8 | $D^+ + H^- \longrightarrow H + D$ | 2 |
| D9 | $D^+ + D^- \longrightarrow D + D$ | 2 |
| D10 | $D^- + H \longrightarrow HD + e^-$ | 4 |
| D11 | $H^- + D \longrightarrow HD + e^-$ | 4 |
| D12 | $D + H \longrightarrow HD + \gamma$ | 5 |
| D13 | $HD + \gamma \longrightarrow D + H$ | 6 |
| D14 | $D + H^+ \longrightarrow D^+ + H$ | 7 |
| D15 | $D^+ + H \longrightarrow D + H^+$ | 7 |
| D16 | $D + H_2 \longrightarrow HD + H$ | 8 |
| D17 | $HD + H \longrightarrow D + H_2$ | 9 |
| D18 | $D^+ + H_2 \longrightarrow HD + H^+$ | 10 |
| D19 | $HD + H^+ \longrightarrow D^+ + H_2$ | 10 |
| D20 | $H^- + D \longrightarrow D^- + H$ | 11 |
| D21 | $D^- + H \longrightarrow H^- + D$ | 11 |
| D22 | $D + H + H \longrightarrow HD + H$ | 12 |
| D23 | $H_2 + D + H \longrightarrow HD + H_2$ | 12 |
| D24 | $H + HD \longrightarrow D + H + H$ | 2 |
| D25 | $H_2 + HD \longrightarrow H_2 + D + H$ | 2 |

**Table 3.3** – Deuterium reaction network.
References: [1] Glover & Abel (2008); [2] same as corresponding H reaction in RATE12; [3] Strömholm et al. (1995); [4] 0.5 times the corresponding H reaction in RATE12; [5] Dickinson (2005); [6] Wolcott-Green & Haiman (2011); [7] Savin (2002); [8] fit by Glover & Abel (2008) to data from Mielke et al. (2003); [9] Shavitt (1959); [10] Gerlich (1982); [11] Dalgarno & McDowell (1956), scaled by D reduced mass; [12] same as corresponding H reaction.

## C SHIELDING AGAINST RADIATION

UV radiation gets attenuated with increasing column density, meaning that processes such as photodissociation become gradually less effective. To quantify this, the rate coefficients of important photoreactions are multiplied by (self-)shielding factors. Regarding photodissociation and photoexcitation of $H_2$, both the columns in atomic hydrogen HI and in $H_2$ will shield the molecules from Lyman-Werner radiation, as HI also has absorption lines in this range. We use the shielding factors from Wolcott-Green et al. (2011), given by

$$f_{\rm sh}^{\rm H_2,H_2} = \frac{0.965}{\left(1+x_{\rm H_2}/\delta v_5\right)^{1.1}} + \frac{0.035}{\left(1+x_{\rm H_2}\right)^{0.5}} \exp\left(-8.5\times10^{-4}\left(1+x_{\rm H_2}\right)^{0.5}\right), \quad (3.13)$$

for self-shielding by $H_2$, also accounting for line overlap, and

$$f_{\rm sh}^{\rm H_2,HI} = \frac{1}{(1+x_{\rm HI})^{1.62}} \exp\left(-1.49\times10^{-1} x_{\rm HI}\right), \quad (3.14)$$

for shielding by HI. Here, $x_{\rm H_2} = N_{\rm H_2}/\left(5\times10^{14}\,{\rm cm^{-2}}\right)$ is the scaled $H_2$ column density, $x_{\rm HI} = N_{\rm HI}/\left(2.85\times10^{23}\,{\rm cm^{-2}}\right)$ is the scaled HI column density, and $\delta v_5 = \delta v/\left(10^5\,{\rm cm\,s^{-1}}\right)$ is the scaled Doppler broadening parameter, with $\delta v = \left(2k_{\rm B}T/m + v_{\rm turb}^2\right)^{1/2}$. The total shielding factor is then the product of the two factors above, and it is multiplied with the relevant rate coefficients.

We also include HD shielding in much the same fashion, except that shielding now occurs by columns of both atomic and molecular hydrogen, together with HD self-shielding. We use the shielding factors from Wolcott-Green & Haiman (2011), given by

$$f_{\rm sh}^{\rm HD,HD} = \frac{0.965}{(1+x_{\rm HD}/b_5)^{1.1}} + \frac{0.035}{(1+x_{\rm HD})^{0.5}} \exp\left(-8.5\times10^{-4}(1+x_{\rm HD})^{0.5}\right), \quad (3.15)$$

for self-shielding by HD,

$$f_{\rm sh}^{\rm HD,H_2} = \frac{1}{\left(1+x_{\rm H_2}\right)^{0.238}} \exp\left(-5.20\times10^{-3} x_{\rm H_2}\right), \quad (3.16)$$

for shielding by $H_2$, and

$$f_{\rm sh}^{\rm HD,HI} = f_{\rm sh}^{\rm H_2,HI}, \quad (3.17)$$

for shielding by HI. Here, $x_{\rm HD} = N_{\rm HD}/\left(5\times10^{14}\,{\rm cm^{-2}}\right)$ is the scaled HD column density, $x_{\rm H_2} = N_{\rm H_2}/\left(2.34\times10^{19}\,{\rm cm^{-2}}\right)$ is the scaled $H_2$ column density (with a different scaling factor than in $f_{\rm sh}^{\rm H_2,H_2}$), $x_{\rm HI}$ is again the scaled HI column density (same as above), and the Doppler parameter now depends on the mass of HD. The total shielding factor is then the product of the three factors above.

Photodissociation by UV radiation is an important destruction mechanism for carbon monoxide, though CO molecules can be shielded against photodissociation by large enough columns of CO and $H_2$. To determine the shielding factor for CO as a function of $N_{CO}$ and $N_{H_2}$, we interpolate the tables of Visser et al. (2009)[3]. In order to handle situations where $N_{CO}$ and/or $N_{H_2}$ are outside the tabulated range, we set the factor to the closest tabulated value.

## D PAH REACTIONS

Polycyclic aromatic hydrocarbons (PAHs) are large planar molecules (sometimes also considered very small dust grains) containing 20 – 100 carbon atoms. They can play an important role in the ionization balance of the interstellar medium, in particular through the neutralization of atomic cations with PAH anions, and through charge exchange between atomic cations and neutral PAHs:

$$X^+ + PAH^- \longrightarrow X + PAH, \tag{3.18}$$

$$X^+ + PAH^0 \longrightarrow X + PAH^+. \tag{3.19}$$

PAH ions are created and destroyed through the following reactions:

$$PAH + \gamma \longrightarrow PAH^+ + e^-, \tag{3.20}$$

$$PAH^- + \gamma \longrightarrow PAH + e^-, \tag{3.21}$$

$$PAH^+ + e^- \longrightarrow PAH, \tag{3.22}$$

$$PAH + e^- \longrightarrow PAH^-. \tag{3.23}$$

Rate coefficients for these reactions were adopted from the MSPDR code. More information on the chosen PAH model can be found in Meijerink & Spaans (2005) and Wolfire et al. (2003). For a more in-depth discussion of the PAH charge balance, we refer the interested reader to Cox & Spaans (2006).

[3] Available online at http://home.strw.leidenuniv.nl/~ewine/photo/index.php?file=co.php

## E $H_2$ AND HD FORMATION ON DUST

Molecular hydrogen is the most abundant molecule in the Universe. However, it has been known for many years that molecular hydrogen does not form efficiently in the gas phase Gould & Salpeter (1963). In the seventies, the first quantum mechanical models were developed that described the formation of molecular hydrogen by recombination of hydrogen atoms on grain surfaces (Hollenbach & Salpeter 1970, 1971; Goodman 1978). It was found that this process can be quite efficient and account for the observed molecular hydrogen abundance. On surfaces, hydrogen atoms can be physisorbed and chemisorbed, and they can move between absorption sites by quantum mechanical tunneling or thermal hopping. Cazaux & Tielens (2002, 2004) developed a prescription for astrophysically relevant conditions, which takes into account all of the aforementioned processes, and which has been verified against results from laboratory experiments. It was also applied to study $H_2$ formation on dust at high redshifts Cazaux & Spaans (2004), and later updated using realistic grain surface barriers and including HD formation (Cazaux & Spaans 2009). The formation rate of $H_2$ and HD on grains can be expressed as

3

$$R_{\rm d}({\rm H_2}) = \frac{1}{2} n_{\rm H} I v_{\rm H} S(T, T_{\rm d}) \left[ (n_{\rm gr}\sigma\epsilon_{\rm H_2})^{\rm carbon} + (n_{\rm gr}\sigma\epsilon_{\rm H_2})^{\rm silicate} \right], \tag{3.24}$$

$$R_{\rm d}({\rm HD}) = n_{\rm D} v_{\rm D} S(T, T_{\rm d}) \left[ (n_{\rm gr}\sigma\epsilon_{\rm HD})^{\rm carbon} + (n_{\rm gr}\sigma\epsilon_{\rm HD})^{\rm silicate} \right], \tag{3.25}$$

where $v_{\rm H}$ and $v_{\rm D}$ are the thermal velocities of H and D atoms, $\epsilon_{\rm H_2}$ and $\epsilon_{\rm HD}$ are the formation efficiencies of $H_2$ and HD, respectively, $S(T, T_{\rm d})$ is the sticking coefficient, and $n_{\rm gr}\sigma$ is the grain density times their cross section for either carbon or silicate grains. The sticking coefficient is given by (Hollenbach & McKee 1979; Burke & Hollenbach 1983)

$$S(T, T_{\rm d}) = \left[ 1 + 0.4 \left( \frac{T + T_{\rm d}}{100} \right)^{0.5} + 0.2 \left( \frac{T}{100} \right) + 0.08 \left( \frac{T}{100} \right)^2 \right]^{-1}, \tag{3.26}$$

where $T$ is the gas temperature and $T_{\rm d}$ is the dust temperature. For solar metallicity, we adopt a mean cross section for collisions between grains and atoms per H atom, $n_{\rm gr}\sigma / n_{\rm H}$, of $1.6 \times 10^{-21}$ cm$^{-2}$ for PAHs (carbon, size $<$100 Å) and $1.0 \times 10^{-21}$ cm$^{-2}$ for silicate grains, using the Milky Way grain size distribution of Weingartner & Draine (2001a). For metallicities other than solar, we assume that the dust abundance scales linearly with the overall metallicity. The grain size distribution at high redshift is unknown, but models suggest there may be more small grains, which could increase the total cross section. Therefore, adopting a linear scale likely gives a lower limit to the formation rates. The rates can now be written as (Cazaux & Spaans 2009)

$$R_{\rm d}({\rm H_2}) = 7.25\times10^{-15} n_{\rm H} I \sqrt{\frac{T}{100}} S(T,T_{\rm d}) \left[1.75\epsilon_{\rm H_2}^{\rm carbon} + 1.1\epsilon_{\rm H_2}^{\rm silicate}\right] n_{\rm H} Z, \tag{3.27}$$

$$R_{\rm d}({\rm HD}) = 1.1\times10^{-14} n_{\rm D} \sqrt{\frac{T}{100}} S(T,T_{\rm d}) \left[1.75\epsilon_{\rm HD}^{\rm carbon} + 1.1\epsilon_{\rm HD}^{\rm silicate}\right] n_{\rm H} Z, \tag{3.28}$$

and the formation efficiencies are given by

$$\epsilon_{\rm H_2}^{\rm carbon} = \epsilon_{\rm HD}^{\rm carbon} = \frac{1-T_{\rm H}}{1+\frac{1}{4}\left(1+\sqrt{\frac{E_{\rm chem}-E_{\rm S}}{E_{\rm phys}-E_{\rm S}}}\right)^2 \exp\left(-\frac{E_{\rm S}}{T_{\rm d}}\right)} \tag{3.29}$$

for carbonaceous grains, and

$$\epsilon_{\rm X}^{\rm silicate} = \frac{1}{1+\frac{16T_{\rm d}}{E_{\rm chem}-E_{\rm S}}\exp\left(-\frac{E_{\rm phys}}{T_{\rm d}}\right)\exp\left(f_{\rm X} a_{\rm pc}\sqrt{E_{\rm phys}-E_{\rm S}}\right)} \tag{3.30}$$

$$+2\frac{\exp\left(-\frac{E_{\rm phys}-E_{\rm S}}{E_{\rm phys}+T}\right)}{\left(1+\sqrt{\frac{E_{\rm chem}-E_{\rm S}}{E_{\rm phys}-E_{\rm S}}}\right)^2} \tag{3.31}$$

for silicate grains, with $f_{\rm H_2} = 4\times10^9$ and $f_{\rm HD} = 5.6\times10^9$. The grain surface parameters $E_{\rm phys}$, $E_{\rm chem}$, $E_{\rm S}$ and $a_{\rm pc}$ can be found in Table 3.4.

| | $E_{\rm phys}$ [K] | $E_{\rm chem}$ [K] | $E_{\rm S}$ [K] | $a_{\rm pc}$ [Å] |
|---|---|---|---|---|
| PAHs (carbon) | 800 | 7000 | -2300 | 1.5 |
| Silicates | 700 | 15000 | -1000 | 1.7 |

**Table 3.4** – Grain surface characteristics.

## F CHEMISTRY SOLVER

We found that there are a few bugs in the chemistry solver of the MSPDR code, and that convergence behavior is erratic, especially outside the 'easier' region of parameter space. We therefore decided to implement a new solver algorithm. The system of rate equations can be represented by $N$ (number of species) reactions of the type

$$\frac{\partial x_i}{\partial t} = F_i(\mathbf{x}), \tag{3.32}$$

where we have elected to express these in terms of number fractions instead of number densities, and $x_i = n_i/n_{\mathrm{H}}$, with $n_{\mathrm{H}}$ the total hydrogen number density. Here, $F_i$ is a function containing all formation and destruction terms for species $i$, and thus depends on the number fractions of many species. For a few species ($e^-$, H, He, ...), $F_i$ is replaced by an explicit conservation equation for the corresponding element (as well as for the charge and the PAH abundance).

The steady-state approximation is used, meaning that we consider the state of the system on a timescale longer than the timescale of any of the processes involved. This is one of the main limitations of the model, meaning that results cannot be used to make reliable predictions for rapidly evolving systems. Setting the time derivatives to zero, the system to solve now contains $N$ equations of the type

$$F_i(\mathbf{x}) = 0, \tag{3.33}$$

which can be solved using a root-finding method.

The Newton-Raphson method, which is derived from the Taylor series expansion of a function in the neighborhood of a point, has been chosen to solve this non-linear system of equations (Press et al. 2002). In the neighborhood of $\mathbf{x}$, each of the functions $F_i$ can be expanded in a Taylor series:

$$F_i(\mathbf{x}+\delta\mathbf{x}) = F_i(\mathbf{x}) + \sum_{j=1}^{N} \frac{\partial F_i}{\partial x_j}\delta x_j + \mathcal{O}(\delta\mathbf{x}^2), \tag{3.34}$$

where $\frac{\partial F_i}{\partial x_j} = J_{ij}$ is an element of the Jacobian matrix $\mathbf{J}$. By neglecting terms of order $\delta\mathbf{x}^2$ and higher and by setting $F_i(\mathbf{x}+\delta\mathbf{x}) = 0$, we obtain a set of linear equations for the corrections $\delta\mathbf{x}$ that move each function $F_i$ closer to zero simultaneously, namely

$$\mathbf{J}\cdot\delta\mathbf{x} = -\mathbf{F}. \tag{3.35}$$

The elements of vector **F** and the elements of the derivative matrix **J** can be easily calculated, and then the linear system of equations can be solved for $\delta\mathbf{x}$ using LU factorization. The LAPACK library[4] was used for the factorization step (Anderson et al. 1999), which greatly improves the computation time compared to the old routine. The calculated corrections are then added to the abundance vector,

$$\mathbf{x}_{\text{new}} = \mathbf{x}_{\text{old}} + f_{\text{step}}\delta\mathbf{x} \tag{3.36}$$

and the process is iterated until convergence is reached. The factor $f_{\text{step}}$ controls whether a full Newton step is taken, or a fraction thereof. Initially, the full step is tried ($f_{\text{step}} = 1$) as this will generally lead to convergence most quickly, but if convergence is not reached then $f_{\text{step}}$ is decreased until a solution is found, or until the maximum number of tries is exceeded.

Convergence is declared when all $F_i$ have gotten sufficiently small in absolute terms (typically $<10^{-16}$), and all $\delta x_i$ have gotten sufficiently small relative to $x_i$ (typically $<10^{-5}$). Before the solution is accepted, it is checked that all elemental abundances as well as the charge are conserved (typically within $<10^{-8}$).

### 3.2.3 COOLING & HEATING

#### A ATOMIC FINESTRUCTURE & METASTABLE LINES

The cooling rate due to a transition $i \to j$ of a species $X$ can be calculated as (e.g. Tielens 2005)

$$\Lambda_{X,ij} = n_{X,i} A_{X,ij} h\nu_{ij} \beta(\tau_{ij}) \frac{S_{X,ij} - J_{\text{bg}}(\nu_{ij})}{S_{X,ij}}, \tag{3.37}$$

where $\nu_{ij}$ is the frequency of the line, $n_{X,i}$ is the population density of species $X$ in the upper level $i$, $A_{X,ij}$ is the spontaneous transition probability, $\beta(\tau_{ij})$ is the probability that a photon created at an optical depth $\tau_{ij}$ can escape,

[4] Available online at http://www.netlib.org/lapack

3

$S_{X,ij}$ is the source function, and $J_{\rm bg}(\nu_{ij})$ is the intensity of the background radiation field at the line frequency. The total cooling rate $\Lambda_X$ is the sum over all levels of $\Lambda_{X,ij}$. The factor containing $J_{\rm bg}$ takes into account that levels can also become populated by warm background radiation through absorption, which, if followed by collisional de-excitation, results in net heating, rather than cooling.

To calculate the mean background intensity, we have used our new implementation of the radiation field; see Section 3.2.1. The source function is given by

$$S_{X,ij} = \frac{2h\nu_{ij}^3}{c^2}\left(\frac{g_{X,i}n_{X,j}}{g_{X,j}n_{X,i}} - 1\right)^{-1}, \tag{3.38}$$

where $g_i$ and $g_j$ are the statistical weights of the upper and lower level, respectively. The level populations are calculated by assuming statistical equilibrium, and the equation for level $i$ is given by

$$n_{X,i}\sum_{j\neq i}^{k} R_{X,ij} = \sum_{j\neq i}^{k} n_{X,j}R_{X,ji}, \tag{3.39}$$

3

with $k$ the total number of levels that is taken into account, and

$$R_{X,ij} = \begin{cases} A_{X,ij}\beta(\tau_{ij})\left(1 + \dfrac{c^2}{2h\nu_{ij}^3}J_{\rm bg}(\nu_{ij})\right) + C_{X,ij} & \text{if } i > j, \\ \dfrac{g_{X,i}}{g_{X,j}}A_{X,ij}\beta(\tau_{ij})\dfrac{c^2}{2h\nu_{ij}^3}J_{\rm bg}(\nu_{ij}) + C_{X,ij} & \text{if } i < j, \end{cases} \tag{3.40}$$

where $C_{X,ij}$ is the collisional transition rate from level $i$ to $j$. One of the equations for $n_{X,i}$ must be replaced by the conservation equation $n_X = \sum_{j=0}^{k} n_{X,j}$.

The collisional de-excitation rate coefficients $C_{X,ij}$ are taken from the literature, and depend on the collision partners. Collisional excitation rate coefficients are related as follows:

$$C_{X,ji} = \frac{g_{X,i}}{g_{X,j}}C_{X,ij}\exp\left(-\frac{E_{X,ij}}{k_{\rm B}T}\right), \tag{3.41}$$

with $E_{X,ij}$ the energy difference between levels $i$ and $j$.

For a semi-infinite slab, the line-averaged optical depth is given by

$$\tau_{ij} = \frac{A_{X,ij}c^3}{8\pi\nu_{ij}^3}\int_0^L n_{X,i}(l)\left(\frac{g_{X,i}n_{X,j}(l)}{g_{X,j}n_{X,i}(l)} - 1\right)\frac{dl}{\delta v}, \tag{3.42}$$

where $\delta v$ is again the effective Doppler broadening parameter (depending on both thermal and turbulent motions), and the integration should be done from the edge of the slab to the depth $L$ where the transition occurs. For a plane-parallel slab, the escape probability can be approximated by (de Jong et al. 1980)

$$\beta(\tau_{ij}) = \begin{cases} \dfrac{1-\exp(-2.34\tau_{ij})}{4.68\tau_{ij}} & \text{if } \tau_{ij} < 7, \\ \dfrac{1}{4\tau_{ij}\left(\ln\left(\frac{\tau_{ij}}{\sqrt{\pi}}\right)\right)^{1/2}} & \text{if } \tau_{ij} \geqslant 7. \end{cases} \tag{3.43}$$

For small optical depths, the escape probability goes to 1/2, as photons can escape only through the edge of the semi-infinite slab. For large optical depths, $\beta$ scales with $\tau_{ij}^{-1}$.

**FINESTRUCTURE LINES** In regions where most of the gas is atomic, atomic finestructure emission lines will be important coolants. We adopt the approach from the MSPDR code and take into account finestructure lines from three transitions in C, O, S, Si, and $Fe^+$, and a single transition in $C^+$ and $Si^+$. Of particular importance are the $C^+$ line at 158 μm, and the O lines at 63 μm and 146 μm. Collisions with electrons, H, ortho- and para-$H_2$, as well as with $H^+$ are taken into account. An overview of the included transitions and their wavelengths can be found in Appendix 3.A, and the corresponding rate coefficients can be found in the references therein, or summarized in Table 5.1 in Meijerink (2006). The spontaneous transition rate coefficients are adopted from Hollenbach & McKee (1989), which is a compilation of data from Aller (1984); Garstang (1958, 1962, 1964, 1968); Grevesse et al. (1971); Wiese et al. (1966, 1969), and the collisional de-excitation rate coefficients are taken from Hollenbach & McKee (1989) (calculated from equations given by Bahcall & Wolf (1968)), Aannestad (1973); Chambaud et al. (1980); Dufton & Kingston (1994); Flower & Launay (1977); Jaquet et al. (1992); Johnson et al. (1987); Launay & Roueff (1977a,b); Mendoza (1983); Roueff & Le Bourlot (1990); Sampson et al. (1994); Schroder et al. (1991); Sternberg & Dalgarno (1995).

**METASTABLE LINES** In atoms, the conservation of electronic spin can lead to metastable excited levels, as transitions are typically those forbidden by electric dipole selection rules. Transition rates are therefore relatively low. Nevertheless, metastable transitions of the most abundant metal species can dominate the atomic contribution to the cooling between ~5000 K and $10^4$ K. We adopt the approach from the MSPDR code and include the metastable cooling lines of C, $C^+$, O, $O^+$, S, $S^+$, Si, $Si^+$, Fe, and $Fe^+$, taking into account collisions with electrons and atomic hydrogen. An overview of the included transitions and their wavelengths can be found in Appendix 3.B. The spontaneous transition rate coefficients are adopted from Hollenbach & McKee (1989), which is a compilation of data from Aller (1984); Garstang (1958, 1962, 1964, 1968); Grevesse et al. (1971); Nussbaumer & Storey (1980); Raymond (1979); Wiese et al. (1966, 1969), and the collisional de-excitation rate coefficients are taken from Hollenbach & McKee (1989), which is a compilation from Aller (1984); Bahcall & Wolf (1968); Czyzak et al. (1968); Henry et al. (1969); Osterbrock (1974); Raymond (1979), with the exception of the coefficients for $C^+$ (Sampson et al. 1994), $O^+$ (McLaughlin & Bell 1993), and $Si^+$ (Dufton & Kingston 1994). The rate coefficients for H collisions are estimated to be $10^{-12}\,\mathrm{cm^3\,s^{-1}}$ (Federman & Shipsey 1983).

3

In the MSPDR code, interaction with the background radiation field as well as optical depth effects were only taken into account for the finestructure lines, but not for the metastable lines. We have expanded the methodology described above to the metastable lines, now accounting for possible stimulated emission and absorption in these lines as well.

## B CO, $H_2O$ AND OH COOLING

To treat rotational and vibrational cooling from CO, $H_2O$, and OH, we follow Neufeld & Kaufman (1993) and express the cooling rate per unit volume for species X in the form

$$\Lambda_X = L_X n_{\mathrm{H_2}} n_X \,\mathrm{erg\,cm^{-3}\,s^{-1}}, \tag{3.44}$$

where the cooling rate coefficient $L_X$ has the units of $\mathrm{erg\,cm^6\,s^{-1}}$, and is a function of the temperature, the $H_2$ density, and an effective optical depth parameter $\tilde{N}_X$ in units of $\mathrm{cm^{-2}}$ per $\mathrm{km\,s^{-1}}$. For a static plane-parallel slab, this parameter is given by $N_X/\delta v$, with $N_X$ the column density of species $X$ and $\delta v$ the line broadening parameter, see Section c.

**ROTATIONAL COOLING** For rotational cooling from CO, $H_2O$, and OH, the rate coefficient $L_X$ is expressed as a four-parameter fit of the form

$$\frac{1}{L_X} = \frac{1}{L_0} + \frac{n_{H_2}}{L_{LTE}} + \frac{1}{L_0}\left(\frac{n_{H_2}}{n_{1/2}}\right)^{\alpha}\left(1 - \frac{L_0}{L_{LTE}} n_{1/2}\right), \tag{3.45}$$

where $L_0$ is the cooling rate in the low-density limit and depends only on the temperature, $L_{LTE}$ is the cooling rate per molecule when the level populations are in local thermodynamical equilibrium (LTE), and $n_{1/2}$ is the $H_2$ number density where $L_X = L_0/2$. Values for the fitting parameters $L_0$, $L_{LTE}$, $n_{1/2}$, and $\alpha$ are found by interpolating tables in temperature and effective optical depth (physical parameters). For rotational cooling from $H_2O$, we adopt the tabulated values from Neufeld & Kaufman (1993) for the high-temperature regime, and from Neufeld et al. (1995) for the low-temperature regime. Rotational cooling from CO was recalculated by Omukai et al. (2010) using updated level energies and transitional coefficients, and we therefore adopt their tabulated values, as well as the ones they provide for OH cooling.

3

For CO rotational cooling, the physical parameters are tabulated in the range $14.0 \leqslant \log(\tilde{N}_{CO}) \leqslant 19.0$ for $10\,K \leqslant T \leqslant 2000\,K$, and for OH cooling they are tabulated in the range $10.0 \leqslant \log(\tilde{N}_{CO}) \leqslant 18.0$ for $30\,K \leqslant T \leqslant 600\,K$. For $H_2O$ rotational cooling, the physical parameters are tabulated in the range $10.0 \leqslant \log(\tilde{N}_{H_2O}) \leqslant 19.0$ for $10\,K \leqslant T \leqslant 4000\,K$, with separate rates for ortho- and para-$H_2O$ listed for $10\,K \leqslant T \leqslant 100\,K$. We assume a fixed ortho:para ratio of 3:1 for $H_2O$.

To extend the rotational cooling rates beyond these limits, we follow the approach of Glover et al. (2010), except at low temperatures. Below 10 K, instead of switching the cooling off completely, we multiply it by a rapidly decreasing factor, $f_T = \exp[1 - (10/T)^5]$. Doing this enables the code to find a temperature solution and continue running also in the cases where rotational cooling is the dominant cooling channel at 10 K. When this occurs, the resulting temperatures (below 10 K) should be treated with consideration. At high temperatures, >2000 K for CO and >4000 K for $H_2O$, we adopt fitting parameters equal to the fitting parameters at the largest tabulated temperature. As we expect the CO and $H_2O$ abundances to be small at these temperatures, the effect on the total cooling rate should be small. The same approach is used for the OH cooling. For gas with $\tilde{N}_X$ below the tabulated range, we adopt fitting parameters equal to the fitting parameters at the smallest tabulated effective optical depth, which already corresponds to approximately optically thin gas. To handle the case where $\tilde{N}_X$ exceeds the upper tabulated value, we again use fitting parameters equal to the fitting parameters at the largest tabulated $\tilde{N}_X$. This may result in overestimating the cooling rate somewhat in very dense, highly shielded gas.

The treatment by Neufeld & Kaufman (1993) and Neufeld et al. (1995) assumes that only collisions with $H_2$ are important, which is only appropriate for fully molecular gas. In other situations, collisions with atomic hydrogen and electrons may be important as well. Following the method in the MSPDR code, based on Yan (1997), we replace $n_{H_2}$ by an effective number density. For CO, this is given by

$$n^{\rm rot}_{\rm eff,CO} = n_{\rm H_2} + \frac{\sqrt{2}\sigma_{\rm H}}{\sigma_{\rm H_2}} n_{\rm HI} + \frac{1.3\times10^{-8}\,{\rm cm^3\,s^{-1}}}{\sigma_{\rm H_2} v_{\rm e}} n_{\rm e}, \tag{3.46}$$

with

$$\sigma_{\rm H} = 2.3\times10^{-15}\,{\rm cm}^2, \tag{3.47}$$

$$\sigma_{\rm H_2} = 3.3\times10^{-16}\left(\frac{T}{10^3\,{\rm K}}\right)^{-1/4}{\rm cm}^2, \tag{3.48}$$

$$v_{\rm e} = 1.03\times10^4\,T^{1/2}\,{\rm cm\,s^{-1}}, \tag{3.49}$$

while for $H_2O$ this is given by

$$n^{\rm rot}_{\rm eff,H_2O} = n_{\rm H_2} + 10 n_{\rm HI} + \frac{k_{\rm e}}{k_{\rm H_2}} n_{\rm e}, \tag{3.50}$$

with

$$k_{\rm H_2} = 7.4\times10^{-12}\,T^{1/2}\,{\rm cm^3\,s^{-1}}, \tag{3.51}$$

$$k_{\rm e} = {\rm dex}\left(-8.020 + \frac{15.749}{T^{1/6}} - \frac{47.137}{T^{1/3}} + \frac{76.648}{T^{1/2}} - \frac{60.191}{T^{2/3}}\right). \tag{3.52}$$

As suggested by Glover et al. (2010), we have replaced $k_{\rm e}$ from the MSPDR code by a new expression from Faure et al. (2004), as the original one diverges at low temperatures. For OH, we assume an effective density for the collision partners of $n_{\rm H_2} + n_{\rm HI}$, which likely gives a lower limit to the cooling rate.

**VIBRATIONAL COOLING** For vibrational cooling from CO and $H_2O$, the rate coefficient $L_X$ is expressed as a two-parameter fit of the form

$$\frac{1}{L_X} = \frac{1}{L_0} + \frac{n_{H_2}}{L_{LTE}}. \tag{3.53}$$

Vibrational cooling from OH is not included. Values for $L_0$ can be calculated from the analytical fits given in Neufeld & Kaufman (1993); Neufeld et al. (1995), and values for $L_{LTE}$ can be found by interpolating in their tables, which are tabulated in the ranges $100\,K \leqslant T \leqslant 4000\,K$ and $13.0 \leqslant \log(\tilde{N}_X) \leqslant 20.0$. The limits are extended in the same way as described above for rotational cooling, and $n_{H_2}$ is again replaced by an effective number density. For CO, this is given by

$$n^{vib}_{eff,CO} = n_{H_2} + 50 n_{HI} + \frac{k_{CO,e}}{k_{CO,0}} n_e, \tag{3.54}$$

with

$$k_{CO,e} = 1.03 \times 10^{-10} \left(\frac{T}{300\,K}\right)^{0.938} \exp\left(-\frac{3080}{T}\right), \tag{3.55}$$

$$k_{CO,0} = 1.14 \times 10^{-14}\, T \exp\left(-\frac{68.0}{T^{1/3}}\right) \exp\left(-\frac{3080}{T}\right), \tag{3.56}$$

while for $H_2O$ this is given by

$$n^{vib}_{eff,H_2O} = n_{H_2} + 10 n_{HI} + \frac{k_{H_2O,e}}{k_{H_2O,0}} n_e, \tag{3.57}$$

with

$$k_{H_2O,e} = 2.6 \times 10^{-6}\, T^{-0.5} \exp\left(-\frac{2325}{T}\right), \tag{3.58}$$

$$k_{H_2O,0} = 0.64 \times 10^{-14}\, T \exp\left(-\frac{47.5}{T^{1/3}}\right) \exp\left(-\frac{2325}{T}\right). \tag{3.59}$$

### C $H_2$ COOLING

To treat rovibrational cooling from $H_2$, we adopt the prescription from Glover & Abel (2008), with modifications as suggested in Glover (2015). This takes into account not only collisions with atomic hydrogen, but also collisions with $H_2$, He, $e^-$, and $H^+$. They computed cooling for these collision partners separately for ortho-$H_2$ and para-$H_2$, in the low-density limit where the density is much smaller than the critical density ($\sim 10^4$ cm$^{-3}$).

To calculate the $H_2$ ortho-para ratio as a function of temperature, we use a simple prescription, taking into account the first 6 rotational levels ($J = 0-5$) and assuming the level populations are given by the Boltzmann distribution.

Cooling due to collisions with H is based on the rate coefficients from Wrathmall & Flower (2007). Cooling due to collisions between two $H_2$ molecules is based on Flower et al. (2000), using rates for the excitation of ortho- and para-$H_2$ by ground-state para-$H_2$ derived from Flower & Roueff (1998), and rates for the excitation of ortho-$H_2$ and para-$H_2$ by ground-state ortho-$H_2$ derived from Flower & Roueff (1999). Cooling due to collisions with He is derived from the calculations of Balakrishnan et al. (1999) for temperatures below 100 K, and from Flower & Roueff (1998) for temperatures in the range $100\,\mathrm{K} < T < 6000\,\mathrm{K}$. For cooling due to collisions with $H^+$, the rotational excitation rates of Gerlich (1990) and the vibrational cross-sections of Krstić (2002) were used. Finally, for cooling due to collisions with $e^-$, a rate based on data from the compilation by Yoon et al. (2008) was used. Most of these rates have been fit in a form

$$\log L_{\mathrm{H_2},X} = \sum_{i=0}^{N} a^i \log(T_3)^i, \tag{3.60}$$

where $X$ is the collision partner, $T_3 = T/10^3$ K, and the fitting coefficients $a_\mathrm{i}$ can be found the tables in Glover & Abel (2008), and in the appendix of Glover (2015) for collisions with electrons. In the low-density limit, the total cooling rate per $H_2$ molecule is given by the sum of the rates due to the different collision partners,

$$L_{\mathrm{H_2},n\to 0} = \sum_X L_{\mathrm{H_2},X}\, n_X, \tag{3.61}$$

in units of erg s$^{-1}$.

At sufficiently high densities, the $H_2$ level populations are in LTE, thus the $H_2$ cooling rate per molecule becomes independent of the chemical composition of the gas, and is instead largely determined by the magnitude of the transition probabilities. Since these are small, the LTE cooling rate is also small. We have used the analytical fit for the LTE cooling from Hollenbach & McKee (1979), using $T_3 = T/1\times10^3$ K,

$$L^{\rm rot}_{\rm H_2,LTE} = \left(\frac{9.5\times10^{-22}\,T_3^{3.76}}{1.0+0.12\,T_3^{2.1}}\right)\exp\left[-\left(\frac{0.13}{T_3}\right)^{3.0}\right] + 3.0\times10^{-24}\exp\left(-\frac{0.51}{T_3}\right), \tag{3.62}$$

$$L^{\rm vib}_{\rm H_2,LTE} = 6.7\times10^{-19}\exp\left(-\frac{5.86}{T_3}\right) + 1.6\times10^{-18}\exp\left(-\frac{11.7}{T_3}\right), \tag{3.63}$$

$$L_{\rm H_2,LTE} = L^{\rm rot}_{\rm H_2,LTE} + L^{\rm vib}_{\rm H_2,LTE}, \tag{3.64}$$

where $L^{\rm rot}_{\rm H_2,LTE}$ and $L^{\rm vib}_{\rm H_2,LTE}$ are the cooling rate coefficients for rotational LTE in $v=0$ and vibrational LTE in $v=0,1,2$. The total optically thin cooling rate is then given by

$$\Lambda_{\rm H_2,thin} = \frac{L_{\rm H_2,LTE}}{1+L_{\rm H_2,LTE}/L_{{\rm H_2},n\to0}}\,n_{\rm H_2}, \tag{3.65}$$

in erg cm$^{-3}$ s$^{-1}$. To account for optical depth effects, we use the opacity table as function of temperature and $H_2$ column density that comes with the KROME package[5] (Grassi et al. 2014), provided under the option 'Omukai'. This factor is simply multiplied with the optically thin rate, to arrive at the final $H_2$ cooling rate.

### D HD COOLING

To treat cooling by HD molecules, we use the cooling function of Lipovka et al. (2005), with a few modifications as described in Glover & Abel (2008). This rate only takes into account collisions with atomic hydrogen, which should be an acceptable approximation in the case of HD, as opposed to the $H_2$ case. The parametrization of the cooling rate should be valid only in the range $100\,{\rm K} \leqslant T \leqslant 2\times10^4$ K. However, Glover & Abel (2008) have found that the rate remains reasonably accurate down to lower temperatures as well, with at most a factor ∼2 error at 30 K.

To prevent HD cooling below the CMB temperature at the relevant redshift, we follow Glover & Abel (2008) and use a modified cooling rate, given by

$$L'_{\rm HD}(T) = L_{\rm HD}(T) - L_{\rm HD}(T_{\rm CMB}) \tag{3.66}$$

[5] Available online at http://www.kromepackage.org/

where $L_{\rm HD}(T)$ and $L_{\rm HD}(T_{\rm CMB})$ are the unmodified HD cooling rates per molecule at the gas temperature and the CMB temperature, respectively.

The parametrization of the cooling rate should be valid for densities in the range $1\,{\rm cm}^{-3} \leqslant n_{\rm H} \leqslant 10^8\,{\rm cm}^{-3}$. Following Glover & Abel (2008), we extend the function down to smaller densities by assuming that in that case the cooling rate per molecule is directly proportional to $n_{\rm H}$, and thus

$$L_{\rm HD}(n'_{\rm H}) = n'_{\rm H} L_{\rm HD}(n_{\rm H} = 1\,{\rm cm}^{-3}), \tag{3.67}$$

for $n'_{\rm H} < 1\,{\rm cm}^{-3}$. To extend the function to higher densities, we can assume HD to be in LTE and thus the cooling rate per molecule has become independent of density,

$$L_{\rm HD}(n''_{\rm H}) = L_{\rm HD}(n_{\rm H} = 10^8\,{\rm cm}^{-3}), \tag{3.68}$$

for $n''_{\rm H} > 10^8\,{\rm cm}^{-3}$. This should be justified as the HD critical density lies well inside the density range of the parametrization.

## E CIE COOLING

When a collision takes place between an $H_2$ molecule and either another $H_2$ molecule, a He atom, or a H atom, the interacting pair briefly acts as a 'supermolecule' with a non-zero electric dipole, and there is a high probability of a photon being emitted. This process is called collision-induced emission (CIE), and may become important at high densities ($n_{\rm H} \gtrsim 10^{14}\,{\rm cm}^{-3}$), depending on the gas temperature. Our CIE cooling rate function is based on data from Ripamonti & Abel (2004), but re-fit and extended by Grassi et al. (2014). As it is uncertain whether the fit is still valid for extremely dissociated media, a check was added to ensure the cooling is 0 when $f_{\rm H_2} < 0.1$, and does not become important before $f_{\rm H_2} \sim 0.5$, where $f_{\rm H_2}$ is the $H_2$ mass fraction relative to the total hydrogen mass fraction.

## F ATOMIC HYDROGEN COOLING

An important cooling channel at temperatures $\gtrsim 5000\,{\rm K}$ is cooling by atomic hydrogen. We use the well-known cooling rate (Spitzer 1978)

$$\Lambda_{\rm H} = 7.3 \times 10^{-19} \exp\left(-\frac{118\,400}{T}\right) n_{\rm e} n_{\rm HI}, \tag{3.69}$$

where $n_{\mathrm{HI}}$ is the number density of atomic hydrogen. This rate only accounts for the contribution of collisional excitation by electrons, but in practice this process will be the dominant one.

### G PAH / GRAIN RECOMBINATION COOLING

Recombination of electrons with small grains and PAHs can also become an important cooling process in gas with temperatures $\gtrsim 5000\,\mathrm{K}$. The cooling rate increases with increasing $W_0 T^{1/2}/n_{\mathrm{e}}$, due to an increase in grain charge, resulting in stronger Coulomb interaction. Additionally, the cooling increases with increasing gas temperature, as electrons with higher energies become bound to the grains with each recombination reaction. Bakes & Tielens (1994) fitted an analytical expression to their numerical results, which have been computed for a wide range of physical conditions. We use a slightly modified version of their fit, to account for the attenuation of UV radiation and for different metallicities:

$$\Lambda_{\mathrm{rec}} = 3.49\times10^{-30}\,T^{0.944}\left(1.16 W_0 e^{(-k_{\mathrm{UV}}A_{\mathrm{V}})}\,T^{1/2} n_{\mathrm{e}}^{-1}\right)^{\frac{0.735}{T^{0.068}}} Z n_{\mathrm{H}} n_{\mathrm{e}}, \tag{3.70}$$

where $W_0$ is the UV radiation scaling parameter, $Z$ is the metallicity, $A_{\mathrm{V}}$ is the extinction at visible wavelengths, and $k_{\mathrm{UV}}$ is a parameter used to take into account the increased extinction at ultraviolet wavelengths, here set to 1.8 (Tielens & Hollenbach 1985).

### H GAS-GRAIN COLLISIONAL COOLING / HEATING

Gas and dust are often not in thermodynamic equilibrium, and gas particles colliding with dust grains can transfer energy. In the case that the dust is cooler than the gas, it will act as a coolant, while in the case of warm dust the gas will be heated up. The rate for this process is given by (Hollenbach & McKee 1979, 1989)

$$\begin{aligned}\Lambda_{\mathrm{gg}} = 1.2\times10^{-31}&\left(\frac{T}{10^3\,\mathrm{K}}\right)^{1/2}\left(\frac{100\,\mathrm{\AA}}{a_{\mathrm{min}}}\right)^{1/2}\\ &\times\left(1-0.8\exp\left(-\frac{75}{T}\right)\right)(T_{\mathrm{d}}-T)\,Z n_{\mathrm{H}}^2,\end{aligned} \tag{3.71}$$

where $a_{\rm min}$ is the minimum grain size, which we set to 10 Å, and $T_{\rm d}$ is the dust temperature. Like above, this rate is scaled by metallicity $Z$ to account for the change in dust abundance.

**DUST TEMPERATURE** We created a new prescription for the dust temperature, based on the method used by Zucconi et al. (2001), but re-calculated and re-fitted, and including redshift dependence. We have made separate fits for all five components of the background radiation field, UV, V-NIR, MIR, FIR, and CMB, for easy scaling with radiation intensity, metallicity, and redshift.

The dust temperature at a given position is found from the classical equilibrium between the cooling and heating of a spherical grain at a position $r$ in the cloud:

$$\int_0^\infty Q_{\nu,\rm e} B_\nu(T_{\rm d}(r)) {\rm d}\nu = \int_0^\infty Q_{\nu,\rm a} J_\nu(r) {\rm d}\nu \tag{3.72}$$

Here, the left hand side gives the cooling rate for a grain with a temperature $T_{\rm d}$, which depends on the depth $r$, with emission efficiency $Q_{\nu,\rm e}$ and blackbody spectrum given by the Planck function $B_\nu(T)$. The right hand side gives the heating rate of the grain due to an incident radiation field with mean intensity $J_\nu(r)$, where $Q_{\nu,\rm a}$ is the grain absorption efficiency.

We use the dust absorption efficiency from Zucconi et al. (2001), which is a piecewise power-law fit to the data from Ossenkopf & Henning (1994). They assume that in a sufficiently large range around the peak wavelength $\lambda_{\rm a}$ of each component, the absorption efficiency $Q_{\nu,\rm a}$ can be approximated by a power-law of the form

$$Q_{\nu,\rm a} = Q_{\rm a} \left(\frac{\lambda_{\rm a}}{\lambda}\right)^{\alpha_{\rm a}}. \tag{3.73}$$

The values of the parameters $Q_{\rm a}$, $\lambda_{\rm a}$, and $\alpha_{\rm a}$ can be found in Table 3.5. For dust emission, we assume values for $Q_{\rm e}$, $\lambda_{\rm e}$, and $\alpha_{\rm e}$ equal to the values for absorption in the range 10-400 µm.

As discussed in Section 3.2.1, the radiation field can be described by the sum of five components, three of which can be represented by modified blackbodies (V-NIR, FIR, CMB), one by a power-law (MIR), and one by a piecewise power-law (UV). With the assumptions about the absorption, we can express the radiation field at depth $r$ as

$$J_\nu(r) = J_\nu \exp\left(-\tau_\nu(r)\right) = J_\nu \exp\left[-\tau_a(r)\left(\frac{\nu}{\nu_a}\right)^{\alpha_a}\right], \tag{3.74}$$

where $\tau_a(r)$ is the optical depth at depth $r$ and at frequency $\nu_a$ (which corresponds to wavelength $\lambda_a$).

| Range | $\lambda_a$ | $Q_a$ | $\alpha_a$ |
|---|---|---|---|
| 0.1-10 µm | 1 µm | $3.9 \times 10^{-22}$ | 1.4 |
| 10-400 µm | 140 µm | $1.5 \times 10^{-24}$ | 1.6 |
| 400 µm-10 mm | 1.06 mm | $3.3 \times 10^{-26}$ | 2.0 |

**Table 3.5** – Dust absorption efficiency parameters, from Zucconi et al. (2001). $Q_a$ is $Q_{\nu,a}$ at $\lambda_a$, in units of $\mathrm{cm^2\,Hz^{-1}}$.

Using the above approximations together with the equilibrium equation 3.72, we can write the following:

$$\begin{aligned}\int_0^\infty Q_e\left(\frac{\nu}{\nu_e}\right)^{\alpha_e}\frac{2h\nu^3}{c^2}\frac{1}{e^{h\nu/(k_B T_d)}-1}\mathrm{d}\nu \\ = \int_0^\infty Q_a\left(\frac{\nu}{\nu_a}\right)^{\alpha_a} J_\nu \exp\left[-\tau_a(r)\left(\frac{\nu}{\nu_a}\right)^{\alpha_a}\right]\mathrm{d}\nu.\end{aligned} \tag{3.75}$$

We can simplify this expression by introducing a few parameters:

$$\beta_d \equiv \frac{k_B T_d}{h\nu_e} \tag{3.76}$$

and

$$\mathscr{A} \equiv \int_0^\infty \left(\frac{\nu}{\nu_e}\right)^{\alpha_e+3}\frac{1}{e^{(\nu/\nu_e)\beta_d^{-1}}-1}\mathrm{d}\left(\frac{\nu}{\nu_e}\right), \tag{3.77}$$

$$\mathscr{B} \equiv \int_0^\infty \left(\frac{\nu}{\nu_a}\right)^{\alpha_a} J_\nu \exp\left[-\tau_a(r)\left(\frac{\nu}{\nu_a}\right)^{\alpha_a}\right]\mathrm{d}\nu, \tag{3.78}$$

so that the above equation becomes

$$\mathscr{A} Q_e \nu_e^4 \frac{2h}{c^2} = Q_a \mathscr{B}. \tag{3.79}$$

The integral in $\mathscr{A}$ can be rewritten in the form $\int_0^\infty \frac{u^{x-1}}{e^u-1}\mathrm{d}u$, which has a known solution of the form $\Gamma(x)\zeta(x)$, with $\Gamma(x)$ the gamma function and $\zeta(x)$ the Riemann zeta function, so that

$$\mathscr{A} = \Gamma(\alpha_\mathrm{e}+4)\zeta(\alpha_\mathrm{e}+4)\beta_\mathrm{d}^{\alpha_\mathrm{e}+4}. \tag{3.80}$$

The expression for $T_\mathrm{d}$ then becomes

$$T_\mathrm{d}(r) = \frac{h\nu_\mathrm{e}}{k_\mathrm{B}}\left[\frac{c^2}{2h}\frac{Q_\mathrm{a}}{Q_\mathrm{e}}\frac{1}{\nu_\mathrm{e}^4}\frac{\mathscr{B}}{\Gamma(\alpha_\mathrm{e}+4)\zeta(\alpha_\mathrm{e}+4)}\right]^{\frac{1}{\alpha_\mathrm{e}+4}} \tag{3.81}$$

The integral in $\mathscr{B}$ is somewhat more tricky, as it depends on the components of the radiation field, which have different frequency dependencies. For the UV, V-NIR, FIR and CMB components, we integrate numerically over the respective frequency ranges of each component. For the MIR component, which is given as a simple power-law in the range $[\nu_\mathrm{l}, \nu_\mathrm{u}]$, it is possible to derive an analytical expression:

$$\mathscr{B} = \int_{\nu_\mathrm{l}}^{\nu_\mathrm{u}} \left(\frac{\nu}{\nu_\mathrm{a}}\right)^{\alpha_\mathrm{a}} W_i \frac{2h\nu_\mathrm{p}^3}{c^2}\left(\frac{\nu}{\nu_\mathrm{p}}\right)^p \exp\left[-\tau_\mathrm{a}(r)\left(\frac{\nu}{\nu_\mathrm{a}}\right)^{\alpha_\mathrm{a}}\right]\mathrm{d}\nu \tag{3.82}$$

$$= \frac{2hW_i}{c^2}\nu_\mathrm{p}^{3-p}\nu_\mathrm{a}^p \int_{\nu_\mathrm{l}}^{\nu_\mathrm{u}} \left(\frac{\nu}{\nu_\mathrm{a}}\right)^{\alpha_\mathrm{a}+p} \exp\left[-\tau_\mathrm{a}(r)\left(\frac{\nu}{\nu_\mathrm{a}}\right)^{\alpha_\mathrm{a}}\right]\mathrm{d}\nu \tag{3.83}$$

Substituting $t = \tau_\mathrm{a}(r)\left(\frac{\nu}{\nu_\mathrm{a}}\right)^{\alpha_\mathrm{a}}$, and $\frac{\mathrm{d}t}{t} = \alpha_\mathrm{a}\frac{\mathrm{d}\nu}{\nu}$,

$$\mathscr{B} = \frac{2hW_i}{c^2}\nu_\mathrm{p}^{3-p}\nu_\mathrm{a}^p \frac{\nu_\mathrm{a}}{\alpha_\mathrm{a}} \int_{t_\mathrm{l}}^{t_\mathrm{u}} \left(\frac{t}{\tau_\mathrm{a}(r)}\right)^{\frac{\alpha_\mathrm{a}+p+1}{\alpha_\mathrm{a}}} \frac{e^{-t}}{t}\mathrm{d}t. \tag{3.84}$$

Defining a parameter $s$ as $s \equiv \frac{\alpha_\mathrm{a}+p+1}{\alpha_\mathrm{a}}$, we get

$$\mathscr{B} = \frac{2hW_i}{c^2}\nu_\mathrm{p}^{3-p}\frac{\nu_\mathrm{a}^{p+1}}{\alpha_\mathrm{a}\tau_\mathrm{a}^s(r)} \int_{t_\mathrm{l}}^{t_\mathrm{u}} t^{s-1}e^{-t}\mathrm{d}t. \tag{3.85}$$

Finally, the integral can now be expressed as the difference between two lower incomplete gamma functions,

$$\int_{t_\mathrm{l}}^{t_\mathrm{u}} t^{s-1}e^{-t}\mathrm{d}t = \gamma(s, t_\mathrm{u}) - \gamma(s, t_\mathrm{l}) \tag{3.86}$$

Filling this into the expression for the dust temperature, we get

$$T_{\rm d}(r) = \frac{h\nu_{\rm e}}{k_{\rm B}}\left[W_i\frac{Q_{\rm a}}{Q_{\rm e}}\left(\frac{\nu_{\rm a}}{\nu_{\rm p}}\right)^4\frac{\nu_{\rm a}\nu_{\rm p}^3}{\nu_{\rm e}^4}\frac{1}{\alpha_{\rm a}\tau_{\rm a}^s(r)}\frac{\gamma(s,t_{\rm u})-\gamma(s,t_{\rm l})}{\Gamma(\alpha_{\rm e}+4)\zeta(\alpha_{\rm e}+4)}\right]^{\frac{1}{\alpha_{\rm e}+4}} \tag{3.87}$$

The depth dependence of $T_{\rm d}$ is contained in $\tau_{\rm a}$. For our computational model, it is more convenient to express the dust temperature in optical extinction $A_{\rm V}$. To make the conversion, we use (Zucconi et al. 2001)

$$A_{\rm V} = 1.086\tau_{\rm V} = 1.086\frac{Q_{\rm V}}{Q_{\rm a}}\tau_{\rm a} = 1.086\left(\frac{\lambda_{\rm a}}{\lambda_{\rm V}}\right)^{\alpha_{\rm a}}\tau_{\rm a}, \tag{3.88}$$

with $\lambda_{\rm V} = 551\,{\rm nm}$ the midpoint of the visual wavelength band.

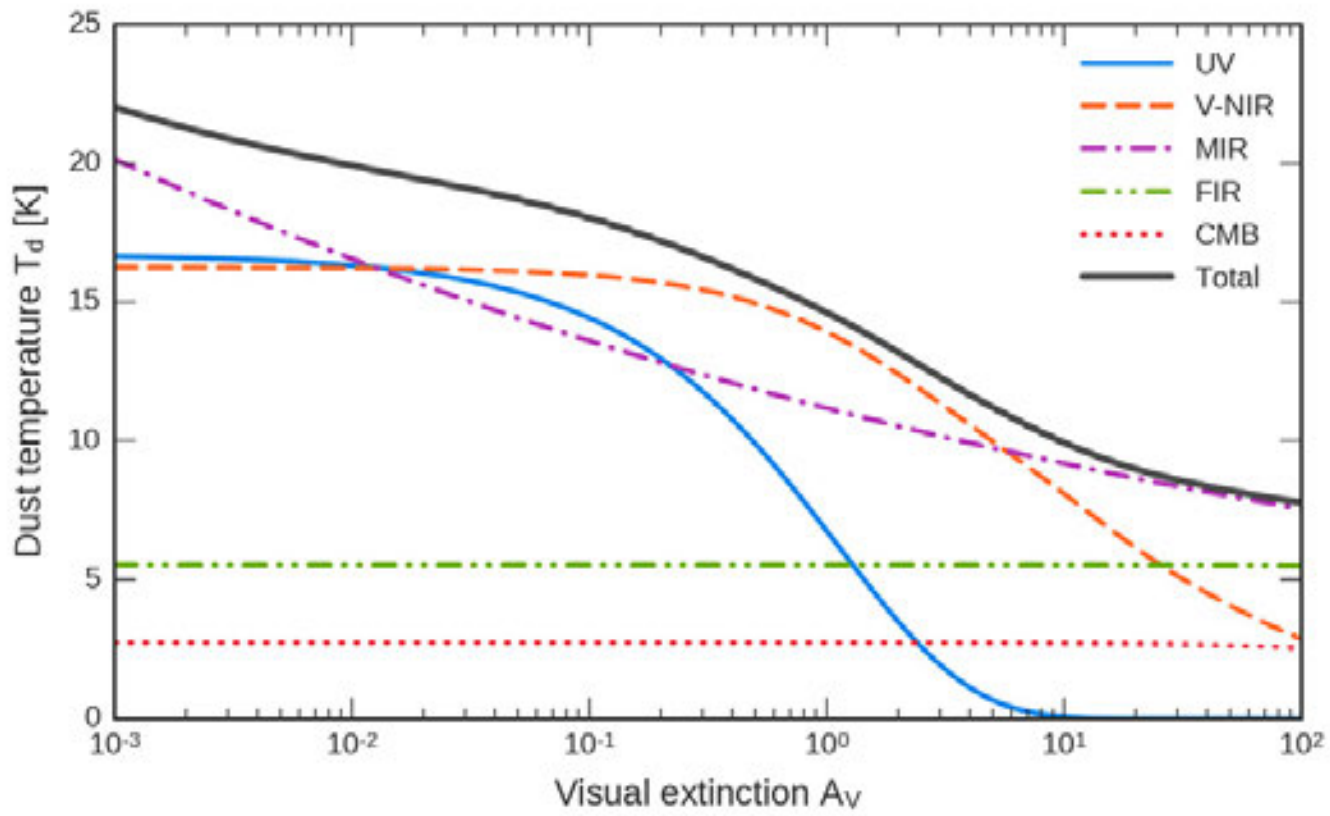

**Figure 3.3** – Fits to the different components of the dust temperature due to different parts of the radiation spectrum, as function of extinction in the V band. Here, $W_0 = 1$, $Z = 1$, $z = 0$.

For faster computation, we have made accurate fits based on our numerical calculations of the dust temperature for the separate components of the radiation field.

$$T_{\rm d}^{\rm UV} = 16.712020 \exp\left(-0.907349 A_{\rm V}^{0.789715}\right), \tag{3.89}$$
$$T_{\rm d}^{\rm V-NIR} = 1.960721 + 5.600740\left(0.904133^{A_{\rm V}}\right) \tag{3.90}$$
$$+ 3.872140\left(0.557763^{A_{\rm V}}\right) + 4.814906\left(0.983259^{A_{\rm V}}\right), \tag{3.91}$$
$$T_{\rm d}^{\rm MIR} = 11.173358 \exp\left(-0.085324 \log\left(A_{\rm V}\right)\right), \tag{3.92}$$
$$T_{\rm d}^{\rm FIR} = 5.534396 - 2.994759 \times 10^{-4} A_{\rm V}, \tag{3.93}$$
$$T_{\rm d}^{\rm CMB} = T_{\rm CMB,0}(1+z) + \left(-1.948150 \times 10^{-3} + 4.462962 \times 10^{-4} z\right. \tag{3.94}$$
$$- 4.149061 \times 10^{-5} z^2 + 2.126257 \times 10^{-7} z^3 \tag{3.95}$$
$$\left.-3.474845 \times 10^{-9} z^4\right) A_{\rm V}, \tag{3.96}$$

with $T_{\rm CMB,0}$ the CMB temperature at $z = 0$. These can be combined into an expression for the total dust temperature in the following way:

$$T_{\rm d} = \left[\left(\left(T_{\rm d}^{\rm UV}\right)^q + \left(T_{\rm d}^{\rm V-NIR}\right)^q + \left(T_{\rm d}^{\rm MIR}\right)^q Z + \left(T_{\rm d}^{\rm FIR}\right)^q Z\right) W_0 + \left(T_{\rm d}^{\rm CMB}\right)^q\right]^{1/q}, \tag{3.97}$$

with $q = \alpha_{\rm e} + 4$, equal to 5.6 in our case. Here we assume that all components except for the CMB scale roughly with the intensity of the UV radiation, expressed through $W_0$. For the sources that we consider, we do not expect dust emission to be optically thick in the MIR and FIR ranges, so we can scale these components linearly with metallicity $Z$.

3

## I GAS-GRAIN VISCOUS HEATING

Another way in which dust can heat the gas is through viscous heating. Radiation pressure will accelerate dust grains relative to the gas, which results in heating due to gas drag. Tielens & Hollenbach (1985) find that the grain acceleration timescale is small compared to other relevant timescales, so that the grains can be considered as moving at their local drift velocity $v_{\rm d}$. Their calculations show that all momentum gained by the grains is transferred to the gas, mainly through Coulomb interaction, and that steady-state drift velocities are typically small, $<10^3$ cm s$^{-1}$. For the resulting heating rate, they give the following equation:

$$\Gamma_{\rm gg,v} = 8\pi q^4 Z n_{\rm d} Z_{\rm d}^2 (k_{\rm B} T)^{-1} \ln(L)\, v_{\rm d} \left[ n_{\rm C^+} G\left(\frac{v_{\rm d}}{v_{\rm th,C^+}}\right) + n_{\rm e} G\left(\frac{v_{\rm d}}{v_{\rm th,e^-}}\right)\right], \tag{3.98}$$

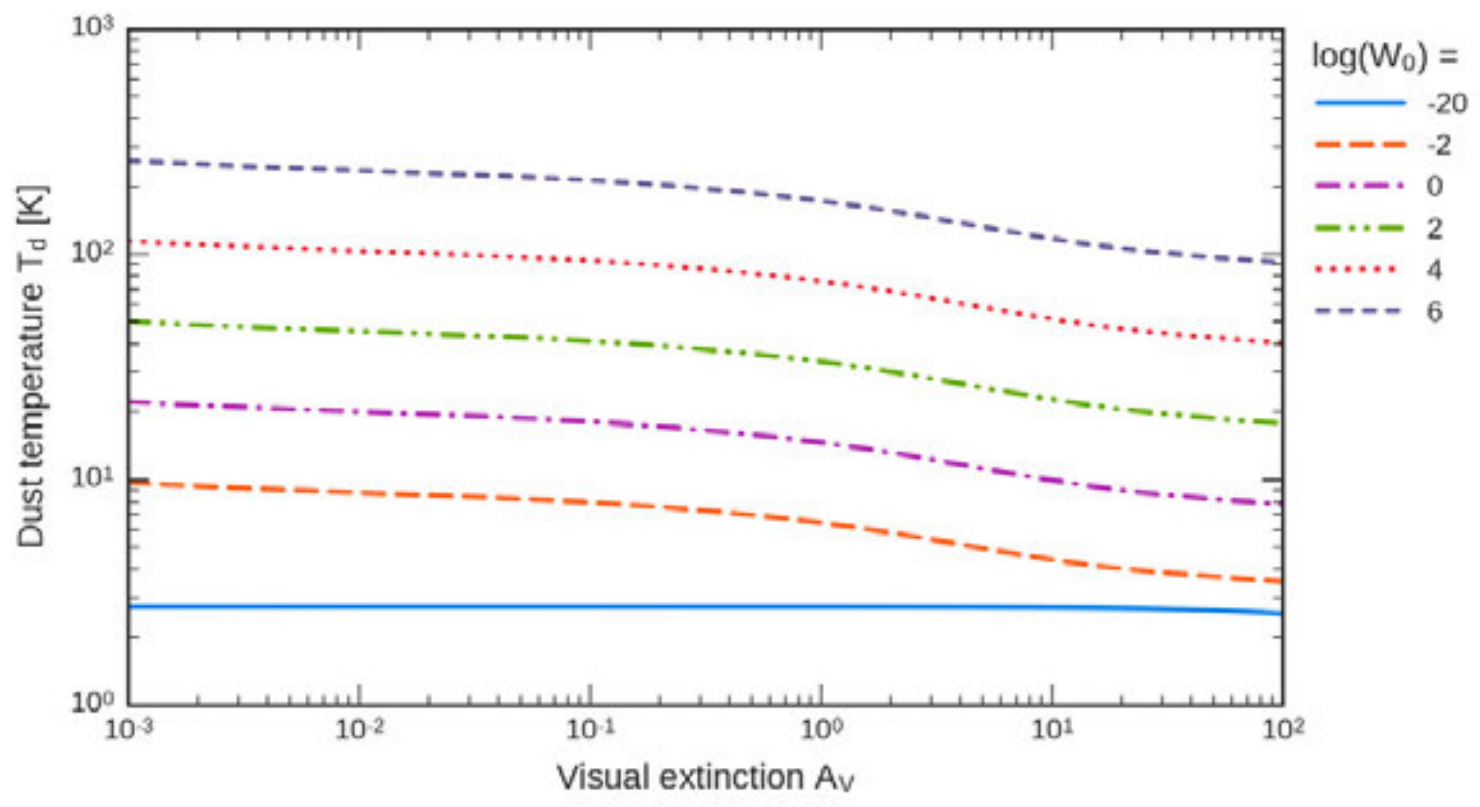

**Figure 3.4** – Dust temperature fits for different values of the UV background scaling parameter $W_0$. Here, $Z = 1$, $z = 0$.

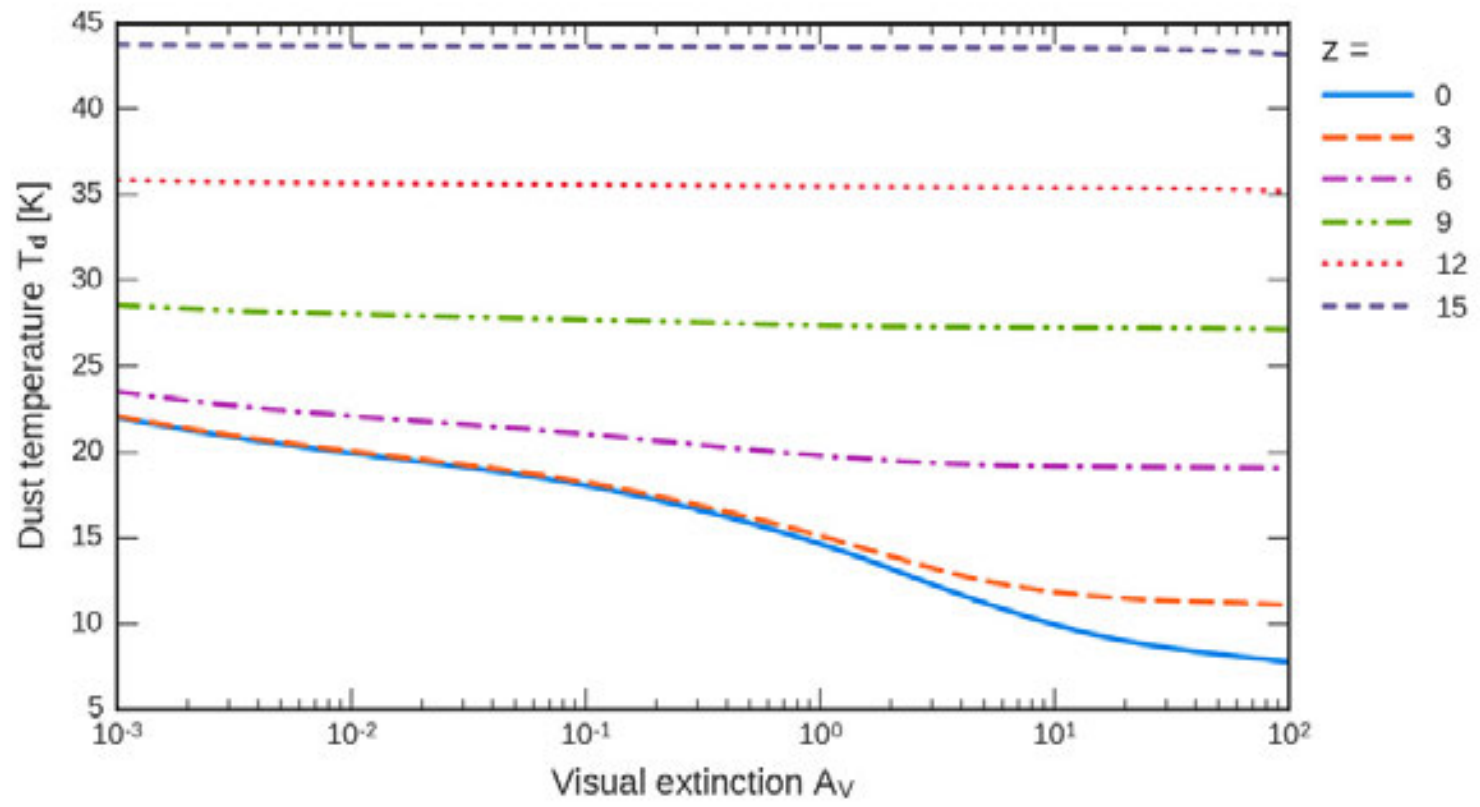

**Figure 3.5** – Dust temperature fits for different values of the redshift $z$. Here, $W_0 = 1$, $Z = 1$. At redshifts $\gtrsim 6$ and for a standard radiation field, the CMB component dominates the dust temperature.

where $q$ is the elemental charge in statcoulomb ($\mathrm{cm^{3/2}\,g^{1/2}\,s^{-1}}$), $n_\mathrm{d}$ is the grain number density, $Z_\mathrm{d}$ is the grain charge, $v_\mathrm{th}$ is the thermal velocity of the indicated ion or electron, and the functions $L$ and $G(y)$ are given by

$$L = 1.5 Z_\mathrm{d}^{-1} q^{-3} (k_\mathrm{B} T)^{1.5} (\pi n_\mathrm{e})^{-0.5}, \tag{3.99}$$

$$G(y) = \frac{1}{2y^2}\left[\mathrm{erf}(y) - \frac{2}{\sqrt{\pi}} y \exp\left(-y^2\right)\right]. \tag{3.100}$$

Here, $\mathrm{erf}(y)$ is the error function, which is defined as

$$\mathrm{erf}(y) = \frac{2}{\sqrt{\pi}} \int_0^y \exp(-t^2)\,\mathrm{d}t. \tag{3.101}$$

Following the implementation in the MSPDR code, we assume a small drift velocity $v_\mathrm{d} = 10^2\,\mathrm{cm\,s^{-1}}$ and a dust density $n_\mathrm{d} = 1.9 \times 10^{-8} n_\mathrm{H}$. The grain charge is given by Tielens & Hollenbach (1985)

$$Z_\mathrm{d} = (x - x_\mathrm{d})(h\nu_\mathrm{H})_\mathrm{erg}\, a q^{-2}, \tag{3.102}$$

where $a = 100\,\text{Å}$ is the grain radius, $h\nu_\mathrm{H} = 13.6\,\mathrm{eV}$ is the energy corresponding to the Lyman limit (in erg), and the grain charge parameter $x$ can be found by solving (de Jong 1977)

$$x^3 + (x_\mathrm{k} - x_\mathrm{d} + \gamma_\mathrm{d})x^2 - \gamma_\mathrm{d} = 0, \tag{3.103}$$

which results from the fact that the charge of a dust grain can be found by equating the rate of photoejection of electrons from a dust grain, and the rate of recombination of electrons with a grain. The parameters in this cubic equation are $x = \nu_0/\nu_\mathrm{H}$, $x_\mathrm{k} = k_\mathrm{B} T / h\nu_\mathrm{H}$ and $x_\mathrm{d} = \nu_\mathrm{d}/\nu_\mathrm{H}$, where $\nu_\mathrm{d}$ is the photoelectric threshold for the neutral dust material and $\nu_0$ corresponds to the energy barrier that electrons have to overcome in order to leave the grain, including the grain-charge effect. The dimensionless parameter $\gamma_\mathrm{d}$ is given by

$$\gamma_\mathrm{d} = 2.9 \times 10^{-4}\, Y\, (1.16 W_0)\, e^{-k_\mathrm{UV} A_\mathrm{V}}\, T^{1/2} n_\mathrm{e}^{-1}, \tag{3.104}$$

with $Y$ the photoelectric yield. Following de Jong et al. (1980), we assume $Y = 0.1$ and $h\nu_\mathrm{d} = 6\,\mathrm{eV}$. A simple, approximate solution to the above cubic equation can be found by assuming that $x_\mathrm{k}$ can be neglected with respect to $x_\mathrm{d}$, so that

$$x = 1 - \frac{1 - x_\mathrm{d}}{2\gamma_\mathrm{d} + 1}, \tag{3.105}$$

which is exact for $\gamma_\mathrm{d} \gg x_\mathrm{d}/2$ and $\gamma_\mathrm{d} \ll x_\mathrm{d}/2$, and accurate to $\sim$10% for $\gamma_\mathrm{d} \approx x_\mathrm{d}/2$.

### J PHOTOELECTRIC HEATING

Radiation impinging on PAHs and dust grains can free electrons, which deposit their kinetic energy in the gas through elastic collisions, and thus contribute to the heating rate. In traditional PDRs, this process is the most important source of heating. The majority of the energy originates from PAHs and small grains. The contribution from grains larger than 100 Å is negligible, in part because small grains dominate the FUV absorption extinction and in part because electrons can escape more easily from PAHs and small grains than from large grains. The heating rate is given by (Bakes & Tielens 1994):

$$\Gamma_{\rm pe} = 10^{-24}\epsilon(1.16W_0)e^{-k_{\rm UV}A_{\rm V}}Zn_{\rm H}, \tag{3.106}$$

where $\epsilon$ is the photoelectric heating efficiency, given by

$$\begin{aligned}\epsilon = & \frac{4.87\times10^{-2}}{1+4\times10^{-3}\left(1.16W_0e^{-k_{\rm UV}A_{\rm V}}T^{1/2}/n_{\rm e}\right)^{0.73}} \\ & + \frac{3.65\times10^{-2}\left(T/10^4\right)^{0.7}}{1+2\times10^{-4}\left(1.16W_0e^{-k_{\rm UV}A_{\rm V}}T^{1/2}/n_{\rm e}\right)}.\end{aligned} \tag{3.107}$$

The heating efficiency depends on the charge of the grain or PAH, and thus on the ratio of the photoionization rate over the recombination rate, which can be represented by $W_0e^{-k_{\rm UV}A_{\rm V}}T^{1/2}/n_{\rm e}$. When this quantity is small, grains and PAHs are mostly neutral, and the efficiency is highest.

### K HEATING FROM $H_2$ PHOTOREACTIONS

FUV photons with energies just below the Lyman limit (~11.2-13.6 eV) can be absorbed by molecular hydrogen, exciting it from the electronic ground state to a higher electronic state. Radiative decay occurs rapidly, and in about 10% of the cases this results in dissociation of the molecule, a phenomenon also known as the two-step Solomon process (suggested by Solomon in 1965, see Field et al. 1966; Stecher & Williams 1967). The remaining 90% of decays lead to population of excited rotational-vibrational levels of the electronic ground state (Black & Dalgarno 1976). Both of these processes also heat the gas.

**$H_2$ PHOTODISSOCIATION HEATING** The energy that is released at every dissociation goes to kinetic energy of the hydrogen atoms, resulting in a heating rate of the following form:

$$\Gamma_{H_2,\mathrm{diss}} = 0.1 k_{\mathrm{ex}} E_{\mathrm{diss}}, \tag{3.108}$$

where $k_{\mathrm{ex}}$ is the excitation rate of $H_2$, and $E_{\mathrm{diss}}$ is the mean kinetic energy of the resulting hydrogen atoms, which is taken to be 0.4 eV (Stephens & Dalgarno 1973). The excitation rate is

$$k_{\mathrm{ex}} = 3.40 \times 10^{-10} \chi e^{-2.5 A_V} f_{\mathrm{sh}}^{H_2}, \tag{3.109}$$

where $\chi$ is a parameter used to compare radiation field strengths just below the Lyman limit (see Equation (3.6) for a definition) and is equal to $1.23 W_0$ in our case, using the MMP field. Self-shielding of molecular hydrogen is taken into account through the shielding factor parameter $f_{\mathrm{sh}}^{H_2}$, and is described in detail in Section c.

**$H_2$ COLLISIONAL DE-EXCITATION HEATING** After becoming excited by absorbing a photon in the Lyman-Werner bands, molecular hydrogen will in 90% of the cases decay to rovibrational levels in the electronic ground state. Vibrational heating will be important when the FUV radiation field provides a significant pumping to higher vibrational states. To approximate the complicated cascade process, (London 1978; Tielens & Hollenbach 1985) proposed a model in which a single pseudo-level $v^* = 6$ represents the vibrational levels of ground state $H_2$. This vibrationally excited $H_2$ is treated as a separate species in the MSPDR code, which requires extending the chemical network with many reactions. However, Röllig et al. (2006) found that the heating rate resulting from this model does not fit the heating from the full system very well. The main problem is that this approximation does not properly account for cooling via rapid excitation to $v = 1$, since this is prevented by the large energy gap between the ground state and the excited pseudo-level. It also fails to account for heating via FUV pumping and collisional de-excitation from all 15 vibrational levels. Therefore, we adopt the prescription by Röllig et al. (2006) to compute the net heating rate. They introduce an updated version of the two-level approximation, providing the same heating rate as the full system of 15 vibrational levels in the electronic ground state (neglecting the rotational structure).

Using this approximation, the heating rate can then be expressed as

$$\Gamma_{H_2,\mathrm{coll}} = \frac{\chi' P_{\mathrm{tot}} \Delta E_{\mathrm{eff}}}{1 + (A_{\mathrm{eff}} + \chi' D_{\mathrm{eff}})/(C_{\mathrm{eff}} n_{\mathrm{H}})} n_{H_2}, \tag{3.110}$$

where $\chi'$ represents the attenuated UV radiation field strength in units of the Draine field,

$$\chi' = 0.69 W_0 e^{-2.5 A_V} f_{\mathrm{sh}}^{H_2}, \tag{3.111}$$

with $f_{\mathrm{sh}}^{H_2}$ the shielding factor. The effective coefficients have been obtained by considering different asymptotic values of the density and the radiation field, yielding

$$\begin{aligned} &P_{\mathrm{tot}}\,\Delta E_{\mathrm{eff}} = 9.4 \times 10^{-22}\,\mathrm{erg\,s^{-1}}, \quad A_{\mathrm{eff}} = 1.9 \times 10^{-6}\,\mathrm{s^{-1}}, \\ &C_{\mathrm{eff}} = C_{1,0}, \quad D_{\mathrm{eff}} = 4.7 \times 10^{-10}\,\mathrm{s^{-1}}, \end{aligned} \tag{3.112}$$

where $C_{1,0}$ is the collisional rate coefficient for the $v = 1 \rightarrow 0$ transition, see further.

The net heating rate is reduced by vibrational cooling at high gas temperatures, which is most effective at low $\chi$, where there is a significant $H_2$ abundance. Assuming most of the cooling results from collisional excitation to $v = 1$ followed by either radiative decay or photodissociation, the cooling rate is given by

$$\Lambda_{H_2,\mathrm{coll}} = \Delta E_{1,0} C_{1,0} \exp\left(-\frac{\Delta E_{1,0}}{k_{\mathrm{B}} T}\right) \frac{A_{1,0} + \chi D_1}{C_{1,0} n + A_{1,0} + \chi D_1} n_{\mathrm{H}} n_{H_2}. \tag{3.113}$$

The molecular constants for the lowest vibrational transition are

$$\begin{aligned} &\Delta E_{1,0} = 6592\,\mathrm{K}, \quad A_{1,0} = 8.6 \times 10^{-7}\,\mathrm{s^{-1}}, \\ &C_{1,0} = 5.4 \times 10^{-13} \sqrt{T}\,\mathrm{cm^{-3}\,s^{-1}}, \quad D_1 = 2.6 \times 10^{-11}\,\mathrm{s^{-1}}. \end{aligned} \tag{3.114}$$

where $\Delta E_{1,0}$ has been increased by ~10% to obtain a better fit. The net heating rate resulting from $H_2$ collisional de-excitation is then

$$\Gamma_{H_2,\mathrm{coll}}^{\mathrm{net}} = \Gamma_{H_2,\mathrm{coll}} - \Lambda_{H_2,\mathrm{coll}}. \tag{3.115}$$

## L CARBON IONIZATION HEATING

Ionization of carbon atoms releases $E_C = 1.06\,\text{eV}$ per ionization event, resulting in a heating rate of the form

$$\Gamma_C = k_{C^+} n_C E_C, \tag{3.116}$$

where $k_{C^+}$ is the carbon ionization rate, given by (Tielens & Hollenbach 1985) as

$$k_{C^+} = 1.6 \times 10^{-10} (1.16 W_0) \exp\left(-2.4 A_V - \tau_C - \frac{\tau_{H_2} b}{\pi v_1^2}\right)\left(1 + \frac{\tau_{H_2} b}{\pi v_1^2}\right)^{-1}. \tag{3.117}$$

Here, the exponential factor with $\tau_C$ describes the attenuation by carbon self-absorption, with $\tau_C = 10^{-17} N_C$ (Werner 1970), and the factor containing $\tau_{H_2}$ describes the attenuation by $H_2$, with $\tau_{H_2} = 1.2 \times 10^{-14} N_{H_2}/\delta v$, and with parameters $b = 9.2 \times 10^{-3}/\delta v$ and $v_1 = 5 \times 10^2/\delta v$ (de Jong et al. 1980), where $N_C$ and $N_{H_2}$ are the C and $H_2$ column densities, respectively, and $\delta v$ is the line broadening parameter.

3

## M COSMIC RAY HEATING

As cosmic rays (CRs) are much less attenuated than UV radiation, they can penetrate much deeper into the region, and become an important heating source at significant optical depths where most of the UV radiation has already been absorbed. The heating rate resulting from ionization events by cosmic rays can be expressed as

$$\Gamma_{CR} = \zeta_n n_H Q, \tag{3.118}$$

with $\zeta_n$ the total cosmic ray ionization rate, and $Q$ the heating energy per produced ion pair. We take into account all cosmic ray ionization reactions that are included in RATE12, so that

$$\zeta_n n_H = \sum_X \zeta_X n_X, \tag{3.119}$$

for all species $X$ that have a rate specified. Here, $\zeta_X = \alpha_X \zeta / 1.36 \times 10^{-17}\,\text{s}^{-1}$, where $\alpha_X$ is the rate coefficient normalized to a total cosmic ray rate of $1.36 \times 10^{-17}\,\mathrm{s}^{-1}$, and $\zeta$ is a global parameter used to scale the total CR rate. To calculate the heating energy per ion pair $Q$, we follow the prescription of Glassgold et al. (2012). They state that the heating energy can be computed as a sum of the effects from elastic collisions plus rotational excitation ($Q_{\mathrm{el/rot}}$), excitation of $H_2$ vibrational levels ($Q_{\mathrm{vib}}$), dissociation of $H_2$ ($Q_{\mathrm{diss}}$), and chemical heating ($Q_{\mathrm{chem}}$), all of which are given below.

The heating from elastic scattering and rotational excitation is given by

$$Q_{\mathrm{el/rot}} = Q_{\mathrm{el}}(\mathrm{H,e}) + Q_{\mathrm{el/rot}}(\mathrm{H_2,e}) \tag{3.120}$$

$$= \frac{x_{\mathrm{H}}\eta(\mathrm{H,e}) + x_{\mathrm{H_2}}\eta(\mathrm{H_2,e})}{x_{\mathrm{H}} + x_{\mathrm{H_2}}} W(\mathrm{H_2,e}), \tag{3.121}$$

where $x_{\mathrm{H}}$ and $x_{\mathrm{H_2}}$ are the H and $H_2$ number fractions, respectively, $W(\mathrm{H_2,e})$ is the average energy to make an ion pair, ~37 eV, and the heating efficiency $\eta$ is given in the form

$$\eta = 1 - \frac{1-\eta_0}{1+Cx_{\mathrm{e}}^{\alpha}}. \tag{3.122}$$

The fitting parameters $\eta_0$, $\alpha$ and $C$ are given in Table 7 of Dalgarno et al. (1999) for different electron energies. We have adopted the values appropriate for $E = 1\,\mathrm{keV}$:

$$\eta_0(\mathrm{H,e}) = 0.117, \quad \alpha = 0.678, \quad C = 7.95, \tag{3.123}$$

$$\eta_0(\mathrm{H_2,e}) = 0.055, \quad \alpha = 0.366, \quad C = 2.17. \tag{3.124}$$

The heating from dissociation of $H_2$ is given by (Dalgarno et al. 1999, Table 5)

$$Q_{\mathrm{diss}} = \frac{2.14\,\mathrm{eV}}{1+22.0x_{\mathrm{e}}^{0.574}} \frac{x_{\mathrm{H_2}}}{x_{\mathrm{H}} + x_{\mathrm{H_2}}} \tag{3.125}$$

The heating from vibrational excitation is given by

$$Q_{\mathrm{vib}} = f_{\mathrm{vib}}\left(Q_{\mathrm{dir/vib}} + Q_{\mathrm{BC/vib}}\right) \frac{x_{\mathrm{H_2}}}{x_{\mathrm{H}} + x_{\mathrm{H_2}}}, \tag{3.126}$$

where $f_{\mathrm{vib}} = \min(1, n/n_{\mathrm{cr}})$ is a factor which takes into account that excitation energy goes into heating only if the densities are high enough for the levels to be collisionally de-excited, with $n_{\mathrm{cr}}$ the critical density for the de-excitation of the vibrational transitions of the $H_2$ ground level. The heating from de-excitation following direct collisional excitation is given by

$$Q_{\rm dir/vib} = 19.0\,{\rm eV}\left(\frac{1}{\epsilon_1} + \frac{2}{\epsilon_2}\right), \tag{3.127}$$

and the heating from de-excitation following population of vibrational levels by decay from excited B and C states is given by

$$Q_{\rm BC/vib} = 147.0\,{\rm eV}\left(\frac{1}{\epsilon_{\rm B}} + \frac{1}{\epsilon_{\rm C}}\right), \tag{3.128}$$

with (Dalgarno et al. 1999, Table 5)

$$\epsilon_1 = 7.81\,{\rm eV}\left(1 + 23500.0 x_{\rm e}^{0.955}\right), \tag{3.129}$$
$$\epsilon_2 = 109.0\,{\rm eV}\left(1 + 10700.0 x_{\rm e}^{0.907}\right), \tag{3.130}$$
$$\epsilon_{\rm B} = 117.0\,{\rm eV}\left(1 + 7.09 x_{\rm e}^{0.779}\right), \tag{3.131}$$
$$\epsilon_{\rm C} = 132.0\,{\rm eV}\left(1 + 6.88 x_{\rm e}^{0.802}\right). \tag{3.132}$$

3

Chemical heating mainly originates from reactions initiated by the primary cosmic ray ions, $H_2^+$, $H^+$, and $He^+$, with neutral species (here focusing on the potentially most abundant neutrals: CO, $H_2O$, and O) and electrons,

$$Q_{\rm chem} = \left(Q_{\rm chem}({\rm H_2}^+) + Q_{\rm chem}({\rm H}^+) + Q_{\rm chem}({\rm He}^+)\right)\frac{x_{\rm H_2}}{x_{\rm H} + x_{\rm H_2}}. \tag{3.133}$$

Heating by $H_2^+$ is given by

$$Q_{\rm chem}({\rm H_2}^+) = F_{\rm H_2^+} P({\rm H_2}^+, {\rm H_3}^+)\left[\sum_i B_i({\rm H_2}^+)\bar{q}_i({\rm H_2}^+) + B_{\rm e}\bar{q}_{\rm e}({\rm H_2}^+)\right], \tag{3.134}$$

where $F_{\rm H_2^+} = 0.88$ is a weighting factor, $P({\rm H_2}^+, {\rm H_3}^+)$ is the probability for $H_3^+$ production,

$$P({\rm H_2}^+, {\rm H_3}^+) = \frac{x_{\rm H_2}}{{\rm H_2}^+ + 100 T^{-0.5} x_{\rm e}}, \tag{3.135}$$

the heating energies are $\bar{q}_i({\rm H_2}^+) = [11.1, /, 8.4, /, 6.4]$, and $\bar{q}_{\rm e}({\rm H_2}^+) = 7.6\,{\rm eV}$. The branching ratios $B_i$ and $B_{\rm e}$ are given by

$$B_i = \frac{k_i x_i}{\sum_{i=1}^{3} k_i x_i + \beta' x_{\rm e}}, \quad B_{\rm e} = \frac{\beta' x_{\rm e}}{\sum_{i=1}^{3} k_i x_i + \beta' x_{\rm e}}, \tag{3.136}$$

with

$$\beta' = 4.5 \times 10^{-6} T^{-0.65}, \tag{3.137}$$

and

$$k_1 = 1.6 \times 10^{-9}, \quad k_2 = 5.3 \times 10^{-9}, \quad k_3 = 0.8 \times 10^{-9}, \tag{3.138}$$

in units of $cm^3 s^{-1}$, with $i = 1, 2, 3$ for CO, $H_2O$, and O, respectively. Heating by $H^+$ is given by

$$Q_{chem}(H^+) = F_{H^+} B_2(H^+) q_2(H^+), \tag{3.139}$$

with weighting factor $F_{H^+} = 0.04$, branching ratio $B_2(H^+) \approx 1.0$, and heating energy $q_2(H^+) = q_2(H^+) - 1.83\,\text{eV}$. Finally, heating by $He^+$ is given by

$$Q_{chem}(He^+) = F_{He^+} B_1(He^+) q_1(He^+), \tag{3.140}$$

with weighting factor $F_{He^+} = 0.08$, branching ratio $B_1(He^+) = 1.0$, and heating energy $q_1(H^+) = 15.6\,\text{eV}$.

## N THERMAL BALANCE SOLVER

The equilibrium temperature in a depth zone is a solution to the equation

$$\Gamma((x), T, A_V) - \Lambda((x), T, A_V) = 0, \tag{3.141}$$

with $\Gamma((x), T)$ the heating function and $\Lambda((x), T)$ the cooling function, where both functions depend on the species abundances $\mathbf{x}$, the temperature $T$, and the optical extinction $A_V$. This is a one-dimensional root-finding problem in $T$, provided that the abundances are given by solving the chemical balance (see Section f for information on the chemistry solver). As the abundances also depend on the temperature, a first guess for the temperature is needed to calculate them, before computing the temperature from the above equation. Therefore, we always solve first the chemical balance, then the thermal balance, and repeat this process until convergence has been reached for both the abundances and the temperature. Equation (3.141) can have multiple solutions, though generally only one of these is a stable solution. Let us define

$$F(T) = \Gamma((x), T, A_V) - \Lambda((x), T, A_V), \tag{3.142}$$

for a given zone and abundances. A stable solution will then have $\frac{dF}{dT} > 0$, so that a slight increase in $T$ leads to an increase in the cooling, returning the temperature to its equilibrium value; and vice versa.

In the MSPDR code, the root finding was done using a simple bisection method, which is rather slow, and it was not checked whether the obtained solution was stable or unstable. Additionally, the initial temperature guess could not be too far off in order for a solution to be found. We have improved upon these points in the following ways, drawing inspiration from the convergence algorithm employed by the CLOUDY code (Ferland et al. 2013).

At the start of a run, for the first zone by the edge of the region, a rough search for a temperature interval which brackets the solution is performed. This is done by starting from an initial guess for $T$, and then defining a new temperature given by

$$T_{i+1} = T_i - \frac{F_i}{\mathrm{d}F_i/\mathrm{d}T}, \tag{3.143}$$

where the derivative is approximated as

$$\frac{\mathrm{d}F_i}{\mathrm{d}T} = \frac{F_i - F_{i-1}}{T_i - T_{i-1}}, \tag{3.144}$$

with the subscript $i$ denoting the values from the previous iteration, and $i-1$ the values from the iteration before that. To prevent too large steps, the new temperature is restricted to a value in the interval $[T_i/f, fT_i]$, where for most situations a factor of $f = 1.5$ suffices, though in case convergence fails, a smaller factor (1.1) is also tried. The iteration stops when the sign of $F$ changes. Should convergence not be reached within a set number of iterations (200 by default), the program will try again using either a smaller factor, a lower initial guess for the temperature, or a higher initial guess, consecutively.

For all other zones, the equilibrium temperature of the previous zone is used as starting guess. We will use the Brent method for root finding, as described below, and this method requires an initial interval that brackets the solution in order to start from. As we expect our current $T$ guess to be not too far away from the solution, a finer bracket search than the one described above for the first zone can be performed. A new temperature is now calculated as

$$T_{i+1} = T_i + |\Delta T_i|\,\mathrm{sgn}(-F_i), \tag{3.145}$$

where $\mathrm{sgn}(x)$ is the sign function, which returns the sign of a real number. The step $\Delta T_i$ is set to $f_1 \Delta T_{i-1}$, with $f_1 = \sqrt{2}$, provided this value is smaller than the maximum allowed step value $f_2 T_i$, with $f_2$ set to 0.2. Again, the iteration is stopped when the sign of $F$ changes, as this means a bracket has been found.

While the bisection method for root-finding is a very reliable algorithm, it converges only linearly. Superlinear convergence may be achieved by using the Wijngaarden-Dekker-Brent method, hereafter *Brent's method* for brevity (Brent 1973; Press et al. 2002). It combines the bisection method and inverse quadratic interpolation, giving it the reliability of bisection, but the potential to be as fast as some of the less-reliable methods. When possible, the algorithm tries to use the potentially fast-converging inverse quadratic interpolation, but it falls back to the more robust bisection method when necessary. Brent's algorithm is the method of choice to find a bracketed root of a general one-dimensional function, when you cannot easily compute its derivative. For inverse quadratic interpolation, three point pairs are needed: the bracket endpoints $[T_{\rm a}, F_{\rm a}]$ and $[T_{\rm c}, F_{\rm c}]$, and the current best guess for the root inside this interval $[T_{\rm b}, F_{\rm b}]$. The next root estimate can then be written as

$$T_{\rm x} = T_{\rm b} + \frac{P}{Q}, \tag{3.146}$$

where

$$P = F_{\rm b}\left[F_{\rm a}\left(F_{\rm b} - F_{\rm a}\right)\left(T_{\rm c} - T_{\rm b}\right) - F_{\rm c}\left(F_{\rm c} - F_{\rm b}\right)\left(T_{\rm b} - T_{\rm a}\right)\right], \tag{3.147}$$

$$Q = \left(F_{\rm a} - F_{\rm c}\right)\left(F_{\rm b} - F_{\rm c}\right)\left(F_{\rm b} - F_{\rm a}\right), \tag{3.148}$$

and $P/Q$ should be a small correction, though this may not always be the case, depending on the function's behavior. When the new root estimate falls outside of the interval, or when the bounds are not collapsing rapidly enough, a bisection step is taken, so the next root estimate is

$$T_{\rm x} = \frac{|T_{\rm b} - T_{\rm a/c}|}{2}, \tag{3.149}$$

where $T_{\rm a/c}$ is $T_{\rm a}$ or $T_{\rm c}$, depending on whether the sign change in $F$ occurs between $T_{\rm b}$ and $T_{\rm a}$, or between $T_{\rm b}$ and $T_{\rm c}$, respectively. Convergence is reached when both the cooling function and the change in temperature have become sufficiently small, which is checked by

$$\frac{|F_{\rm b}|}{\max\left(\Lambda_{\rm b}, \Gamma_{\rm b}\right)} < \epsilon_{\rm F}, \quad \frac{|T_{\rm c} - T_{\rm a}|}{T_{\rm b}} < \frac{\epsilon_{\rm F}}{3.0}, \tag{3.150}$$

where in practice $\epsilon_{\rm F} = 0.05$ proves to be adequate.

3

Due to the way the convergence algorithm is set up, usually only stable solutions will be found. In the exceptional case where the solution turns out to be unstable, a warning flag is added to the output from that zone.

# APPENDICES

## 3.A FINESTRUCTURE LINES

| Species | Upper | Lower | $\lambda$ (µm) | Partner | Ref. |
|---|---|---|---|---|---|
| $C^+$ | $^2P_{3/2}$ | $^2P_{1/2}$ | 157.7 | $e^-$ | 1 |
| | | | | H | 2 |
| | | | | $H_2$ | 3 |
| $Si^+$ | $^2P_{3/2}$ | $^2P_{1/2}$ | 34.8 | $e^-$ | 4 |
| | | | | H,$H_2$ | 5 |
| C | $^3P_1$ | $^3P_0$ | 609.7 | $e^-$ | 6 |
| | | | | H | 7 |
| | | | | o-$H_2$, p-$H_2$ | 8 |
| | | | | $H^+$ | 9 |
| C | $^3P_2$ | $^3P_1$ | 370.4 | $e^-$ | 6 |
| | | | | H | 7 |
| | | | | o-$H_2$, p-$H_2$ | 8 |
| | | | | $H^+$ | 9 |
| C | $^3P_2$ | $^3P_0$ | 230.4 | $e^-$ | 6 |
| | | | | H | 7 |
| | | | | o-$H_2$, p-$H_2$ | 8 |
| | | | | $H^+$ | 9 |
| O | $^3P_1$ | $^3P_2$ | 63.1 | $e^-$ | 10 |
| | | | | H | 7 |
| | | | | o-$H_2$, p-$H_2$ | 11 |
| | | | | $H^+$ | 12 |
| O | $^3P_0$ | $^3P_1$ | 145.5 | $e^-$ | 10 |

**Table 3.A.1** – Continued from previous page

| Species | Upper | Lower | $\lambda$ (µm) | Partner | Ref. |
|---|---|---|---|---|---|
| | | | | H | 7 |
| | | | | o-$H_2$, p-$H_2$ | 11 |
| | | | | $H^+$ | 12 |
| O | $^3P_0$ | $^3P_2$ | 44.1 | $e^-$ | 10 |
| | | | | H | 7 |
| | | | | o-$H_2$, p-$H_2$ | 11 |
| | | | | $H^+$ | 12 |
| S | $^3P_1$ | $^3P_2$ | 25.2 | H, $H_2$ | 13 |
| | | | | $H^+$ | 5 |
| S | $^3P_0$ | $^3P_1$ | 56.3 | H, $H_2$ | 13 |
| | | | | $H^+$ | 5 |
| S | $^3P_0$ | $^3P_2$ | 17.4 | H, $H_2$ | 13 |
| | | | | $H^+$ | 5 |
| Si | $^3P_1$ | $^3P_0$ | 129.7 | H, $H_2$ | 5 |
| | | | | $H^+$ | 5 |
| Si | $^3P_2$ | $^3P_1$ | 68.5 | H, $H_2$ | 5 |
| | | | | $H^+$ | 5 |
| Si | $^3P_2$ | $^3P_0$ | 44.8 | H, $H_2$ | 5 |
| | | | | $H^+$ | 5 |
| $Fe^+$ | $^6D_{7/2}$ | $^6D_{9/2}$ | 25.2 | $e^-$ | 14 |
| | | | | H, $H_2$ | 14 |
| $Fe^+$ | $^6D_{5/2}$ | $^6D_{7/2}$ | 35.3 | $e^-$ | 14 |
| | | | | H, $H_2$ | 14 |
| $Fe^+$ | $^6D_{5/2}$ | $^6D_{9/2}$ | 15.0 | $e^-$ | 14 |
| | | | | H, $H_2$ | 14 |

**Table 3.A.1** – Atomic finestructure line data; energy level information is from the NIST Atomic Spectra Database (Kramida et al. 2015).
*References.* Spontaneous transition rate coefficients: compilation by Hollenbach & McKee (1989), from Aller (1984); Garstang (1958, 1962, 1964, 1968); Grevesse et al. (1971); Wiese et al. (1966, 1969).
Collisional rate coefficients: [1] Sampson et al. (1994), [2] Launay & Roueff (1977a), [3] Flower & Launay (1977), [4] Dufton & Kingston (1994), [5] Hollenbach & McKee (1989) (calculated from formulae given by Bahcall & Wolf (1968)), [6] Johnson et al. (1987), [7] Launay & Roueff (1977b), [8] Schroder et al. (1991), [9] Roueff & Le Bourlot (1990), [10] Mendoza (1983), [11] Jaquet et al. (1992), [12] Chambaud et al. (1980), [13] Sternberg & Dalgarno (1995), [14] Aannestad (1973).

## 3.B METASTABLE LINES

| Species | Upper | Lower | $\lambda$ (µm) | Ref. |
|---|---|---|---|---|
| C | $^1D$ | $^3P$ | 0.9823, 0.9850 | 1 |
| | $^1S$ | $^1D$ | 0.8727 | 1 |
| | $^1S$ | $^3P$ | 0.4622 | 1 |
| $C^+$ | $^4P$ | $^2P$ | 0.2326 | 2 |
| O | $^1D$ | $^3P$ | 0.6300, 0.6363 | 1 |
| | $^1S$ | $^1D$ | 0.5577 | 1 |
| | $^1S$ | $^3P$ | 0.2972 | 1 |
| $O^+$ | $^2D_{5/2}$ | $^4S$ | 0.3729 | 3 |
| | $^2D_{3/2}$ | $^2D_{5/2}$ | 508 | 3 |
| | $^2D_{3/2}$ | $^4S$ | 0.3726 | 3 |
| S | $^1D$ | $^3P$ | 1.08, 1.13 | 1 |
| | $^1S$ | $^1D$ | 0.7725 | 1 |
| | $^1S$ | $^3P$ | 0.4589 | 1 |
| $S^+$ | $^2D_{3/2}$ | $^4S$ | 0.6731 | 1 |
| | $^2D_{5/2}$ | $^2D_{3/2}$ | 318 | 1 |
| | $^2D_{5/2}$ | $^4S$ | 0.6717 | 1 |
| Si | $^1D$ | $^3P$ | 1.62 | 1 |
| | $^1S$ | $^1D$ | 1.1 | 1 |
| | $^1S$ | $^3P$ | 0.6527 | 1 |
| $Si^+$ | $^4P$ | $^2P$ | 0.2240 | 4 |
| Fe | $^5F_5$ | $^5D_4$ | 1.44 | 1 |
| | $^5F_4$ | $^5F_5$ | 22.3 | 1 |
| | $^5F_4$ | $^5D_4$ | 1.36 | 1 |

**Table 3.B.1** – Continued from previous page

| Species | Upper | Lower | $\lambda$ (µm) | Ref. |
|---|---|---|---|---|
| $Fe^+$ | $^4F_{9/2}$ | $^6D_{9/2}$ | 5.34 | 1 |
| | $^4D_{7/2}$ | $^4F_{9/2}$ | 1.64 | 1 |
| | $^4D_{7/2}$ | $^6D_{9/2}$ | 1.26 | 1 |

**Table 3.B.1** – Atomic metastable line data.
*References.* Spontaneous transition rate coefficients: compilation by Hollenbach & McKee (1989), from Aller (1984); Garstang (1958, 1962, 1964, 1968); Grevesse et al. (1971); Nussbaumer & Storey (1980); Raymond (1979); Wiese et al. (1966, 1969).
Collisional rate coefficients: [1] Hollenbach & McKee (1989) (a compilation from Aller (1984); Bahcall & Wolf (1968); Czyzak et al. (1968); Henry et al. (1969); Osterbrock (1974); Raymond (1979)), [2] Sampson et al. (1994), [3] McLaughlin & Bell (1993), [4] Dufton & Kingston (1994).

*WHEN I WAS YOUNG, I ADMIRED CLEVER PEOPLE. NOW THAT I AM OLD, I ADMIRE KIND PEOPLE.*

— ABRAHAM JOSHUA HESCHEL

CHAPTER

4

# ISM THERMODYNAMICS THROUGH THE AGES: A SYSTEMATIC STUDY

## ABSTRACT

**Context** Feedback processes from stars and black holes shape the interstellar medium (ISM) out of which new generations of luminous objects form. To understand the properties of these objects, e.g. the stellar initial mass function, it is vital to have a firm grasp of the chemical and thermodynamical properties of the feedback-regulated ISM.
**Aims** Since the conditions in the ISM may vary greatly, we here aim to explore in a systematic way the overall impact of various feedback effects, both radiative and chemical, on the chemical and thermal balance of the gas in different regimes.
**Methods** A grid of models is run using our numerical code PDR-Zz, which provides 1D models using a steady-state approach. The grid covers a sizable range in total hydrogen density, metallicity, redshift, UV radiation field scaling parameter, and cosmic ray ionization rate.
**Results** We provide insight in the most important processes and dominant chemical species in different regimes. We show the effects of radiative and chemical feedback on the effective gas temperature both in the outer layers and in the bulk of the cloud, and describe the differences. Finally, we examine the equations of state in our parameter space, identify several regions of interest, and relate these to the IMF.
**Conclusion** We find a transition in EOS behavior around $10^{-1}\,Z_\odot$, with an overall similar EOS for the bulk gas for $Z \leq 10^{-1}\,Z_\odot$ due to the main heating and cooling channels scaling roughly with metallicity in gas optically thin to the UV continuum. For low metallicities, we find a relatively soft EOS, though coupled with the higher gas temperatures this may still lead to massive stars. Additionally, starburst-like conditions give rise to both a stiffer EOS and higher temperatures, resulting in a more top-heavy IMF.

4

## 4.1 INTRODUCTION

At present, the complex physics associated with galaxy formation and evolution is still not well understood. Major ingredients to the recipe for galaxy assembly are the feedback processes from stars and black holes, which shape the interstellar medium (ISM) inside galaxies, or their smaller progenitor systems, out of which new generations of luminous objects will form.

The first stars fundamentally transformed the early Universe through their radiative, chemical, and mechanical feedback (for a review, see Ciardi & Ferrara 2005). They were the first sources of hydrogen-ionizing photons, beginning the process of reionizing the Universe, and of photons capable of dissociating molecular hydrogen (for a review, see Meiksin 2009; Zaroubi 2013). They produced the first metals, or chemical elements that are heavier than hydrogen and helium (aside from the trace amounts of lithium and deuterium resulting from the Big Bang) and also the first dust grains (for a review, see Karlsson et al. 2013). Large amounts of energy were injected into the environment through stellar winds or when ending their lives as supernovae (SNe), creating shocks and turbulence, and possibly expelling gas out of the host halo (e.g. Wise & Abel 2008). Additionally, the first stars may have been the sites where dynamically significant magnetic fields were created for the first time (e.g. Tan & Blackman 2004; Schleicher et al. 2010a; Sur et al. 2010; Widrow et al. 2012). And of course, following the epoch of the first stars, further generations of stars continue to shape galaxies through their feedback processes. In addition to stars, also black holes can be responsible for feedback effects, including radiative feedback, in particular their X-ray emission, and mechanical feedback in the form of winds and jets (e.g. Haiman et al. 2000; Aykutalp et al. 2014).

### 4.1.1 RADIATIVE FEEDBACK

When metals and dust are present, far-ultraviolet (FUV) radiation, with photon energies in the range 6-13.6 eV, profoundly impacts the chemistry and thermal balance of the material, through many different processes (e.g. photodissociation of molecules, photoelectric heating on grains, heating of the dust).

In (nearly) metal-free gas, the main effect of UV radiation is that a sufficiently intense photon flux in the Lyman-Werner bands (11.2-13.6 eV) will photodissociate $H_2$, while lower energy FUV photons will photodetach $H^-$, which is an important intermediary species for the formation of $H_2$. If low-metallicity gas is irradiated by ultraviolet (UV) radiation that is efficient at dissociating $H_2$ molecules, the temperature will be higher, which may result in higher accretion rates, since $\dot{M} \propto T^{3/2}$, and a larger final mass for a star formed out of such gas (e.g. Hirano et al. 2015).

On the other hand, radiative feedback may also act to increase the ionization fraction in low-metallicity gas. The larger abundance of free electrons boosts $H_2$ formation and cooling, possibly lowering the temperature to the point where HD cooling can become effective, which may even enable the gas to cool down to the temperature floor set by the CMB (e.g. Nakamura & Umemura 2002; Tan & McKee 2008). The lower temperatures and change in thermodynamical evolution may then result in a lower final mass compared to a typical Pop III star.

Several other processes could also lead to an enhanced ionization fraction and potentially allow the gas to reach the HD-dominated regime. Collisional ionization due to the strong shocks occurs when more massive halos undergo virialization (Johnson et al. 2008; Greif et al. 2008), though it may not be trivial to prevent the gas in these halos from having already been enriched with metals.

Another possibility is that second generation stars form in relic HII regions, which have previously been ionized by Pop III stars but have recombined, typically resulting in a boosted abundance of $H_2$ and HD (Yoshida et al. 2007). This may be possible as the volume that becomes ionized by a single massive Pop III star is much larger than the volume that becomes enriched with metals after the star dies as a supernova (Greif et al. 2010). A hurdle for this scenario may be the low density in such environments, as a consequence of outflows driven by the photoionization heating (Alvarez et al. 2006).

Finally, an enhanced ionization fraction could also be the result of irradiation by X-rays or cosmic rays (CRs; energetic (~GeV) protons). While the presence of an X-ray background appeared to be a promising pathway (Haiman et al. 2000), more recent simulations show that when employing realistic models for the X-ray background that also account for the simultaneous growth of the soft UV background, star formation is relatively insensitive to the X-rays (Glover & Brand 2003; Machacek et al. 2003; Hummel et al. 2015; Latif et al. 2015).

More important may thus be the presence of cosmic rays (especially low-energy CRs, ~100 MeV), which is known to significantly influence the interstellar medium in local galaxies, by heating the ISM and ionizing deeply embedded gas clouds, as the absorption cross sections for CRs are small and they can propagate far (e.g. Glassgold & Langer 1973; Caselli et al. 1998; Glassgold et al. 2012; Tielens 2005). Another effect is that primary electrons produced by cosmic ray ionization can still have enough energy to excite $H_2$, which will decay radiatively and produce FUV photons. As a result, there can still be FUV photons present deep into a molecular cloud, driving photoprocesses in a region where the FUV radiation from external sources has become strongly attenuated (e.g. Prasad & Tarafdar 1983; Gredel et al. 1989). Cosmic ray effects are therefore particularly important in regions where UV radiation does not penetrate, e.g., inside dense molecular clouds, and in regions close to bright sources of energetic photons or particles, e.g., near massive stars, supernovae, and active galactic nuclei (AGN). However, the magnitude of the cosmic ray ionization rate in the early Universe is still largely unknown (Jasche et al. 2007; Stacy & Bromm 2007; Inayoshi & Omukai 2011; Nakauchi et al. 2014).

### 4.1.2 CHEMICAL FEEDBACK

As soon as the first metals and dust have been created and dispersed by energetic Pop III supernovae, the physics and chemistry of subsequent star formation will be fundamentally changed. Efficient cooling through atomic lines and dust grains becomes possible, and conditions will start to resemble the present-day interstellar medium in the Milky Way. It has been argued that there may be a threshold level of enrichment, a 'critical metallicity', that characterizes the transition from Pop III star formation mode to the more normal Pop II mode (Omukai 2000; Bromm et al. 2001a).

Currently, there are two main scenarios. One of them assumes that the main cooling channels are CII and OI finestructure lines, setting critical abundances on carbon and oxygen of $[C/H]_{\rm crit} \sim -3.5$ and $[O/H]_{\rm crit} \sim -3.0$ (Bromm & Loeb 2003b; Santoro & Shull 2006; Safranek-Shrader et al. 2014). The other assumes dust is the main coolant, and predicts critical abundances that are typically smaller by a factor of 10-100 (Schneider et al. 2003, 2012a; Omukai et al. 2005; Omukai 2012; Clark et al. 2008; Dopcke et al. 2011). On the other hand, it has also been suggested that there may not be a critical metallicity or dust fraction below which fragmentation does not occur at all (Jappsen et al. 2009a,b; Dopcke et al. 2013). To make matters even more complex, there may also be other factors at work, like the redshift, which sets the CMB temperature floor, or magnetic fields.

### 4.1.3 MECHANICAL FEEDBACK

Supernovae from massive stars, stellar outflows and jets deposit a large amount of energy into the interstellar medium through shocks and turbulence. The effects on the ISM of heating due to mechanical feedback have been explored by e.g. Loenen et al. (2008); Kazandjian et al. (2012, 2015). Energetic supernova and multi-supernova events can also remove gas from the galaxy in which they occur, depending on its mass and dark matter content. However, mechanical feedback appears to be less efficient at completely blowing out material than expected from simple energetic arguments, as off-center SN explosions also drive shocks inward that tend to collect and compress gas in the central regions of a galaxy (Mori et al. 2002; Bromm et al. 2003). SN explosions may also promote star formation, by sweeping up and compressing the interstellar material into dense regions (Vishniac 1983; Norman & Ferrara 1996); thus, mechanical feedback plays an important role in regulating star formation (e.g. Schaye et al. 2010; Kimm et al. 2015).

### 4.1.4 INITIAL MASS FUNCTION

Intimately linked to stellar feedback is the initial mass function (IMF), as the conditions and properties of the medium out of which the stars form determine their masses, while in turn the mass of a star affects the types and strength of its feedback which, to a large extent, determine said conditions and properties of the ISM.

Observations have provided us with a decent idea of what the initial mass function of present-day star formation in the Milky Way looks like. The IMF above $\sim 1\,\mathrm{M}_\odot$ has a power-law slope, $\mathrm{d}N_*/\mathrm{d}M_* \propto M_*^{\alpha}$, with $\alpha = 2.3(5)$, originally identified by Salpeter (1955). The IMF peaks at a few tenths of a solar mass, which sets the typical stellar mass. Below a solar mass, the distribution flattens (Miller & Scalo 1979; Kroupa 2001; Chabrier 2003), and there appears to be an upper mass limit of $\sim 150\,M_\odot$ (Elmegreen 2000; Figer 2005).

Current evidence suggests that the IMF is near-universal over a wide range of star-forming environments, especially throughout the Milky Way and local galaxies. However, systematic variations of the IMF have been reported for extreme starburst environments and massive elliptical galaxies, potentially due to their unique formation environment (e.g. Spiniello et al. 2014). Trends with metallicity have also been observed, where the IMF may become increasingly bottom-heavy with increasing metallicity (e.g. Martín-Navarro et al. 2015). For a comprehensive review of the observational evidence regarding the universality of the IMF and the potential variations in extreme environments, see Kroupa et al. (2013); Offner et al. (2014).

Stars form out of cold, molecular gas, embedded in regions of more tenuous and predominantly neutral, atomic gas, with a gradual transition in between. Dense condensations, often containing embedded protostars, have been observed within molecular clouds; these substructures are called 'cores'. The distribution of core masses (CMF) is remarkably similar to the IMF, suggesting that these cores serve as gas reservoirs from which stars accrete most of their mass (e.g. Caselli et al. 2002; André et al. 2010; Könyves et al. 2010). The presence of supersonic turbulence plays an essential role in defining the CMF. Turbulence is ubiquitous in the interstellar medium, and has a dual effect: it can both locally compress the gas, and provide support against gravity on larger scales (e.g. Larson 1981; Padoan & Nordlund 2002; Mac Low & Klessen 2004; McKee & Ostriker 2007; Federrath & Klessen 2012). The most robust features of theoretical models describing the turbulent origins of the CMF are the Salpeter-like slope (similar to the IMF slope above a solar mass), and the log-normal-like turnover towards lower masses. The location of the turnover may be closely tied to the scale below which thermal (or magnetic) support dominates over turbulence. As a result, the location of the peak of the CMF, and consequently also of the IMF, will depend strongly on the gas thermodynamics (Scalo et al. 1998; Spaans & Silk 2000; Larson 2005; Jappsen et al. 2005; Bonnell et al. 2006; Klessen et al. 2007). In any case, star formation regions vary so widely in their physical properties, including their turbulence and magnetic fields, that these do not really provide a compelling explanation for a typical mass scale, which seems universal at least in the Milky Way. It is therefore conceivable that the thermal state of the gas will be involved in some way.

The equilibrium state of gas is described by its equation of state (EOS). Assuming a polytropic EOS, $P \propto \rho^{\gamma_p}$, the polytropic exponent $\gamma_p$ then encompasses most of the chemical and thermodynamical properties of the gas (Vazquez-Semadeni et al. 1996; Scalo et al. 1998; Passot & Vázquez-Semadeni 1998; Spaans & Silk 2000; Scalo & Biswas 2002), and is a measure for the compressibility. It is likely that the stiffness of the EOS is related to the density probability density function (pdf) of gas in the turbulent ISM. Numerical simulations have also shown that the amount of fragmentation is very sensitive to the exact temperature-density relation in collapsing clouds (Li et al. 2003; Jappsen et al. 2005; Bonnell et al. 2006; Klessen et al. 2007; Hocuk & Spaans 2010; Federrath & Banerjee 2015; Hocuk et al. 2016). The thermodynamical properties, and thus the polytropic exponent, depend strongly on e.g. metallicity and the background radiation field, which is related to the redshift through the CMB. Therefore, in order to determine the IMF of the first, as well as later generations of stars, it is vital to understand the chemical and thermodynamical properties of the gas under the relevant conditions.

### 4.1.5 THIS WORK

Since the conditions in the ISM may vary greatly, we here aim to explore in a systematic way the overall impact of various feedback effects, both radiative and chemical, on the thermal balance of the gas in different regimes. A grid of models is run with our numerical code PDR-Zz, which has been described in Chapter 3 and provides 1D models using a steady-state approach. In this chapter, we describe the results of this grid. First, we provide insight in the most important processes and dominant chemical species in different regimes. Then, we show the effects of radiative and chemical feedback on the effective gas temperature both at the surface and in the bulk of the cloud. And finally, we examine the equation of state under different conditions, and relate it to the IMF.

## 4.2 GRID SETUP

To compute the grid, we use the default code setup as previously described. The adopted element abundances can be found in Table 3.2 in Chapter 3, and for the turbulent velocity $v_{\rm turb}$ a typical value of $2.7\,{\rm km\,s^{-1}}$ was used. The actual abundance ratios and amount of turbulence in the medium at high redshift are currently still uncertain, so we choose to use local values to provide reference models. At very high redshift, when not many supernova explosions have occurred yet, it is expected that there was less turbulence in the interstellar medium as compared to today, and our adopted turbulent velocity may be somewhat overestimated. However, it is straightforward to edit these parameters in the code and experiment with different values.

The input parameters that have been varied to obtain the grid are

- the total hydrogen density $n_{\rm H}$ (in units of ${\rm cm^{-3}}$), which affects both the chemical reaction rates and the cooling and heating rates;
- the metallicity $Z$ (in units of $Z_\odot$), which sets the amount of metals and dust available and thus the chemical structure, affecting the cooling and heating rates;
- the redshift $z$, which sets the cosmic microwave background (CMB) temperature, and therefore affects the thermal balance and the radiation field;
- the UV scaling parameter $W_0$ (in units of the MMP radiation field (Mezger et al. 1982); $W_0$ is analogous in function to, e.g., $G_0$, the UV scaling parameter in units of the Habing (1968) radiation field), which, together with the metallicity (through the dust abundance) and the redshift (through the CMB radiation), sets the incident radiation field (see Section 3.2.1 for more information on how the radiation field is defined); and
- the cosmic ray (CR) ionization rate $\zeta$ (in units of ${\rm s^{-1}}$), which has an impact on the electron abundance and contributes to the heating rate.

The parameter ranges and step sizes that have been used to calculate the grid are listed in Table 4.1.

4

| | $\log(n_H/\mathrm{cm}^{-3})$[a] | $\log(Z/Z_\odot)$ | $\log(W_0)$ | $\log(\zeta/\mathrm{s}^{-1})$ | $z$ |
|---|---|---|---|---|---|
| Range | −2 to 21 | −6 to 0 | −20; −2 to 5 | −20; −17; −15 | 0 to 30 |
| Step | 0.5 | 1 | 1 | – | 5 |
| Number | 47 | 7 | 9 | 3 | 7 |

**Table 4.1** – Parameter ranges used to calculate the grid.

[a] Maximum density may not always be reached for all models.

The grid has been run by using ‘density-chains’. This means that all parameters except for the density are set to a certain value, and a first model is calculated starting from the lowest density, for which we chose $10^{-2}$ cm$^{-3}$, using generic initial abundances and a guess for the initial temperature. Once this model is complete, a new model is calculated with the next, higher density, using the converged species abundances and temperature from the first zone in the previous density model as initial values. This decreases the calculation time for the first zone, as well as making it easier to compute models that have difficulty converging to a suitable starting point. In theory, we would like to calculate each of these chains up to a density of $10^{21}$ cm$^{-3}$, though in practice not all of them reach this value, due to convergence issues which arise at a certain point.

## 4.3 RESULTS

### 4.3.1 MAJOR CHEMICAL AND THERMAL PROCESSES

To better understand the results from the grid, we have selected a few models to examine in detail, which will illustrate the most important processes. The adopted fiducial model represents a diffuse cloud with a density of 100 cm$^{-3}$ in the Milky Way at the present time, with solar metallicity, and exposed to an interstellar radiation field with UV scaling factor $W_0 = 10$ and a cosmic ray ionization rate of $10^{-17}$ s$^{-1}$. Then, for most parameters we have chosen a lower and a higher value to show which physical and chemical processes are affected by this quantity. The exceptions are the metallicity, where we have chosen two lower values, and the redshift, where we chose only a higher value. The high-density case ($n_H = 10^5$ cm$^{-3}$) corresponds to a cloud that is expected to be self-gravitating. All parameters are shown in Table 4.1. Runs are performed with four of the parameters set to their fiducial values, and the fifth parameter set to one of the alternate values. We chose to use the optical extinction as depth parameter for these model comparisons, in the range $10^{-3} < A_V < 20$ mag, as it shows the full range of impact of the photoprocesses. However, one must keep in mind that high extinctions combined with low metallicities correspond to unrealistically large columns.

| | $n_H$ (cm$^{-3}$) | $Z$ ($Z_\odot$) | $W_0$ | $\zeta$ (s$^{-1}$) | $z$ |
|---|---|---|---|---|---|
| Fiducial | $10^2$ | $10^0$ | $10^1$ | $10^{-17}$ | 0 |
| Low[a] | $10^0$ | $10^{-6}$ | $10^{-2}$ | $10^{-20}$ | |
| High[b] | $10^5$ | $10^{-3}$ | $10^5$ | $10^{-15}$ | 20 |

**Table 4.1** – Model parameters for the fiducial model and parameter values for the alternate models.

[a] No lower redshift value.

[b] Metallicity value is higher than the low value, but actually lower than the fiducial value.

## A FIDUCIAL MODEL

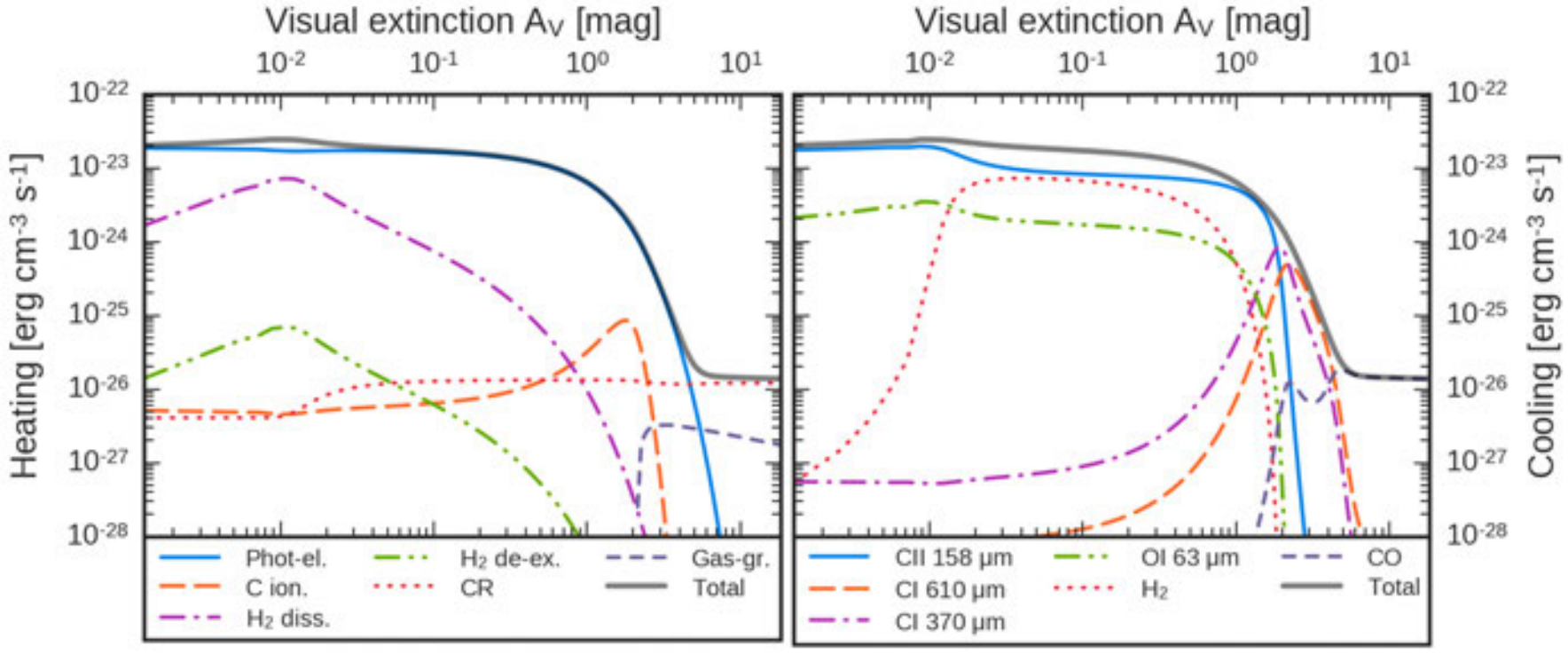

**Figure 4.1** – Heating and cooling rates as a function of visual extinction for the fiducial model.

In our fiducial model, the gas temperature is around 150 K in the outer layers of the cloud, and drops to ~9 K at an extinction of approximately

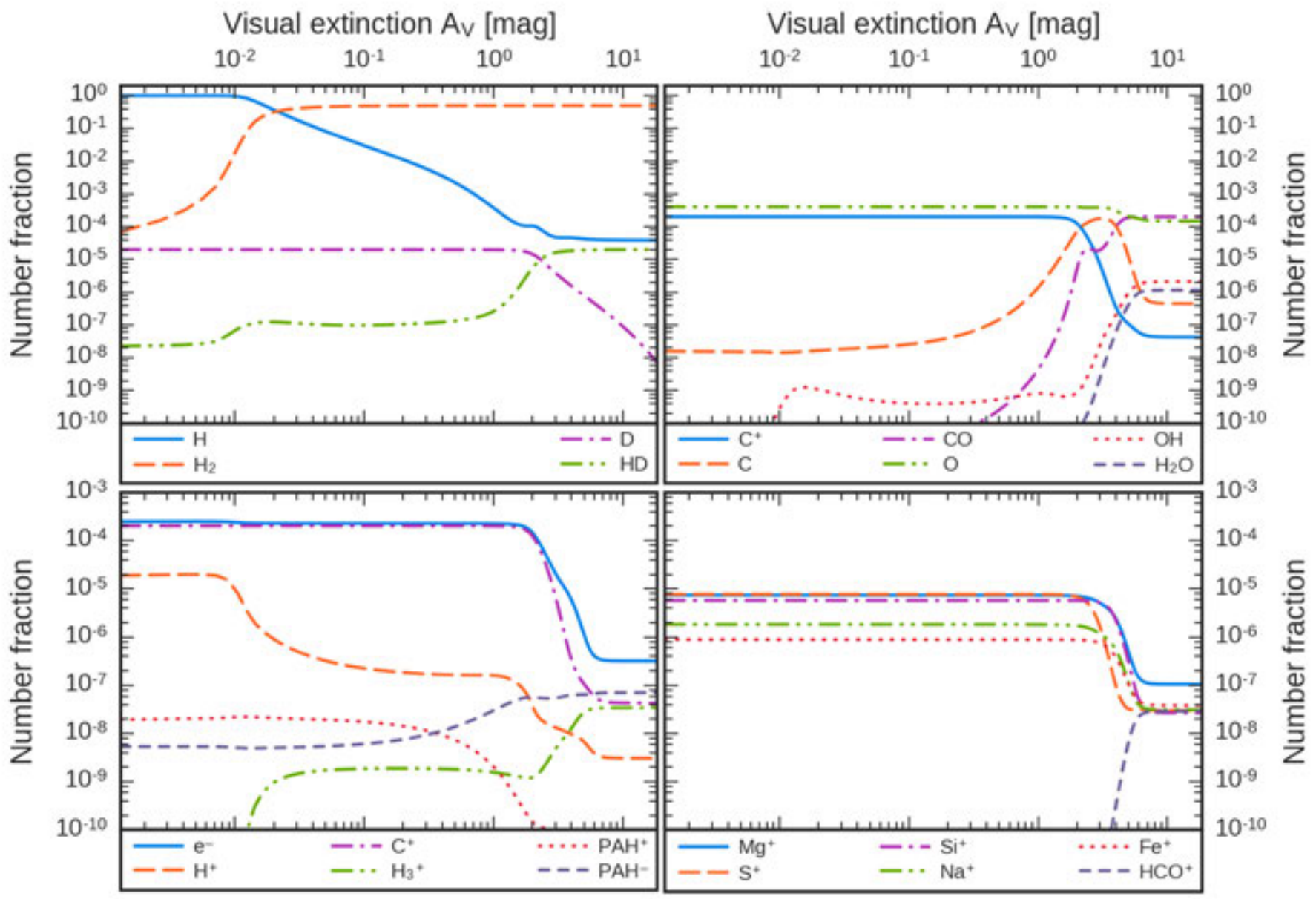

**Figure 4.2** – Abundances of some major species, in number fractions relative to the total hydrogen number density, as a function of visual extinction for the fiducial model.

2 magnitudes, as seen in the top left panel of Figure 4.3. The dust temperature slowly decreases from ~30 K at the edge to ~14 K at our model's maximum extinction. Heating and cooling rates for the most important processes are shown in Figure 4.1. The heating rate is dominated by photoelectric heating until ~5 mag, where cosmic rays become the main heating source. The drop in the photoelectric heating rate is due in part to the FUV radiation getting more and more absorbed, and in part to the significant decrease in electron fraction due to recombination. The small bump in temperature at ~0.01 mag is due to $H_2$ dissociation heating briefly contributing noticeably to the total heating rate, though the rate related to this process drops again at higher extinctions, as molecular hydrogen becomes shielded from the dissociating UV radiation. Until an extinction of about 2 mag, the cooling rate is dominated by emission of the CII finestructure line at 158 μm, with a significant contribution from $H_2$ rovibrational line emission between ~0.02-0.6 mag. At ~2 mag, emission of the CI 370 μm finestructure line peaks and briefly becomes dominant. Between ~2-5 mag, another carbon finestructure line, namely CI 610 µm, is the main coolant, and at even higher extinctions emission from CO rotational lines dominates. Some of the more important species abundances are shown in Figure 4.2. The hydrogen in the gas transitions from mostly atomic to mostly molecular around an extinction of 0.02 mag, while the $C^+$/C/CO transition and D/HD transition occur around 2-3 mag. While the gas is warm, the most abundant ion and main electron donor is $C^+$, with an abundance of $\sim 2 \times 10^{-4}$ relative to hydrogen. When the gas gets cold, the electron fraction drops to $\sim 3 \times 10^{-7}$, and while they are still the most abundant negatively charged species, $PAH^-$ gets close with an abundance of $\sim 7 \times 10^{-8}$. In this cold gas, the main positive ion is $Mg^+$, though quite a few other ions are present at comparable abundances; that is, in order of decreasing abundance: $C^+$, $Fe^+$, $H_3{}^+$, $Na^+$, $S^+$, $HCO^+$, and $Si^+$.

## B PARAMETER STUDY

Figure 4.3 shows the resulting temperature as a function of visual extinction for the different models. The fiducial model is shown in the first panel, and the other panels show the effect of varying the parameters on both the gas and the dust temperature. Below we will discuss each of these models in more detail, using plots to illustrate the most important heating and cooling processes and major chemical species, both neutral and ions.

**DENSITY ($n_H$)** The effect of changing the number density on the heating and cooling rates is shown in Figure 4.4. A decrease in density to 1 cm$^{-3}$ results in a significantly higher gas temperature, while an increase in density to $10^5$ cm$^{-3}$ results in a somewhat lower gas temperature, closer to the dust temperature. This is related to the fact that heating rates tend to scale like $\sim n$, while cooling rates scale like $\sim n^2$. In the low-density case, the H/$H_2$ and D/HD transitions are pushed deeper into the cloud, to extinctions of 1 and 15 mag, respectively. The $C^+$/C transition occurs around 4 mag, while a possible transition to CO falls outside our simulated extinction range. Where the gas is hot, at extinctions below ~1 mag in the low-density model, the most abundant ion is $H^+$, as opposed to $C^+$. When the temperature drops to ~200 K, at an extinction of about 1 mag, $C^+$ does become the main electron donor, with contributions from mainly $He^+$, but also $Mg^+$ and $Si^+$, when the gas gets even colder. In the high-density case, the majority of hydrogen is in molecular form. The D/HD and $C^+$/C/CO transitions move to lower extinctions, with CO becoming the dominant carbon species from ~0.1 mag. $C^+$ is the most abundant ion until ~0.2 mag, where $Mg^+$ and $Si^+$ become the main electron donors. Most of these abundances, including the electron fraction, continue to decrease steeply, and from an extinction of about 3 mag, the most abundant ions are $PAH^-$ and $SiOH^+$.

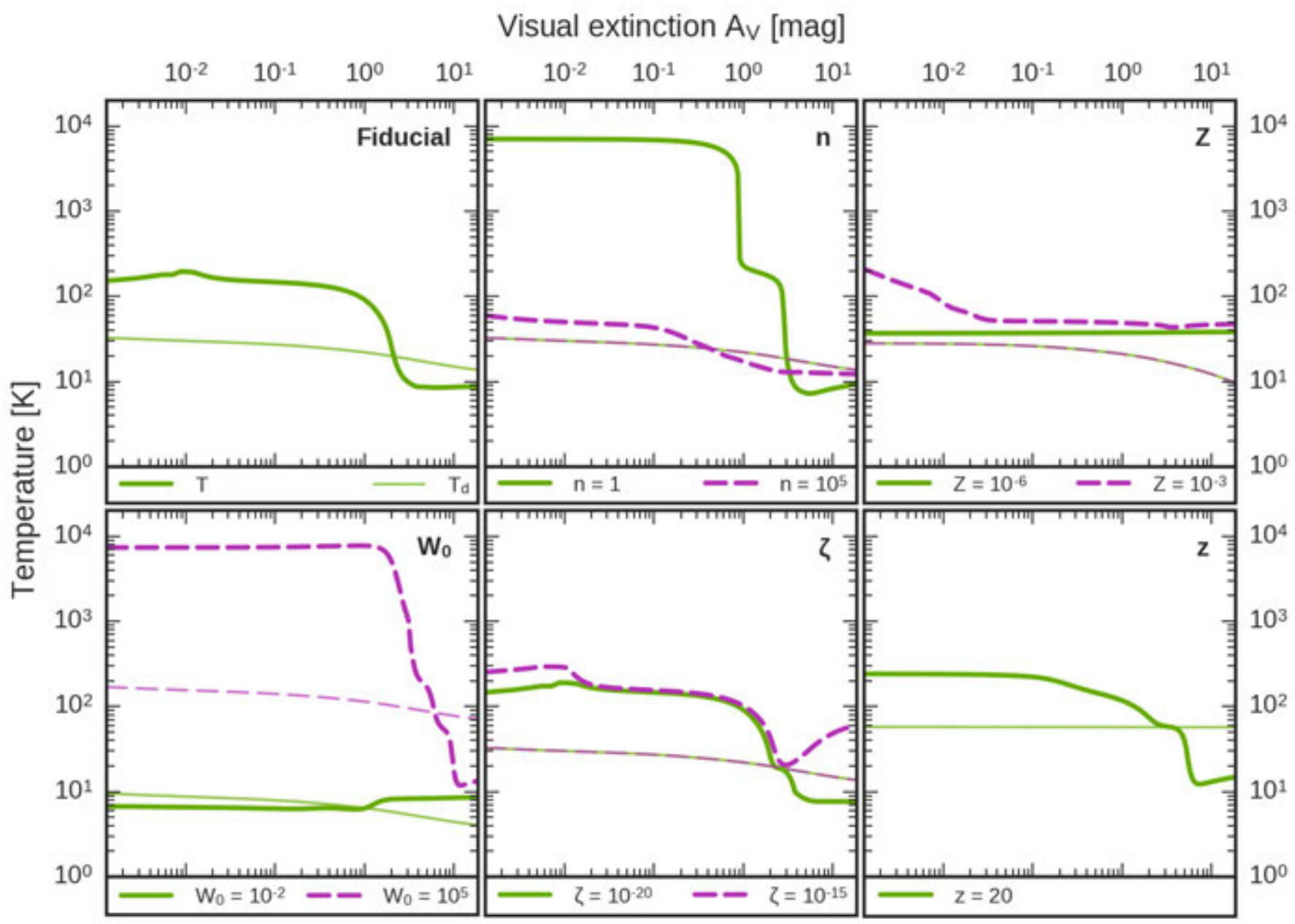

**Figure 4.3** – Temperature as a function of visual extinction for the fiducial model (top left), varying density ($n_H$ in units of cm$^{-3}$; top center), metallicity ($Z$ in units of $Z_\odot$; top right), UV radiation strength ($W_0$ in units of the MMP field; bottom left), cosmic ray ionization rate ($\zeta$ in units of s$^{-1}$; bottom center), and higher redshift ($z$; bottom right). Thick lines represent the gas temperature and thin lines represent the dust temperature. With the exception of the first and last panel, full (green) lines represent the lower parameter value and dashed (purple) lines represent the higher value (note that for metallicity both alternate values are lower than the fiducial one).

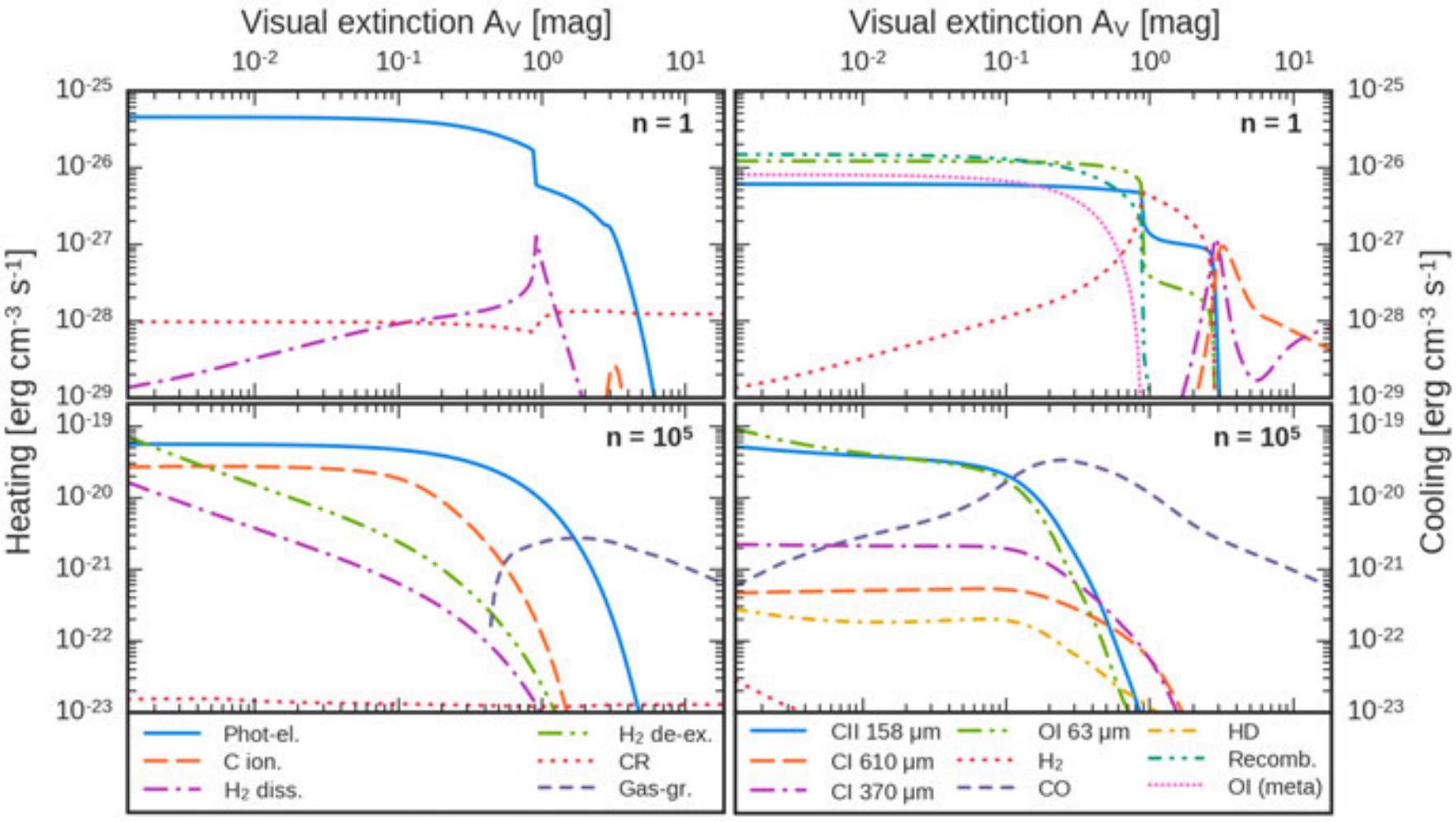

**Figure 4.4** – Heating (left) and cooling (right) rates as a function of visual extinction for a lower number density $n_\mathrm{H}$ of 1 cm$^{-3}$ (top) and a higher number density of $10^5$ cm$^{-3}$ (bottom), as compared to the fiducial value of $10^2$ cm$^{-3}$.

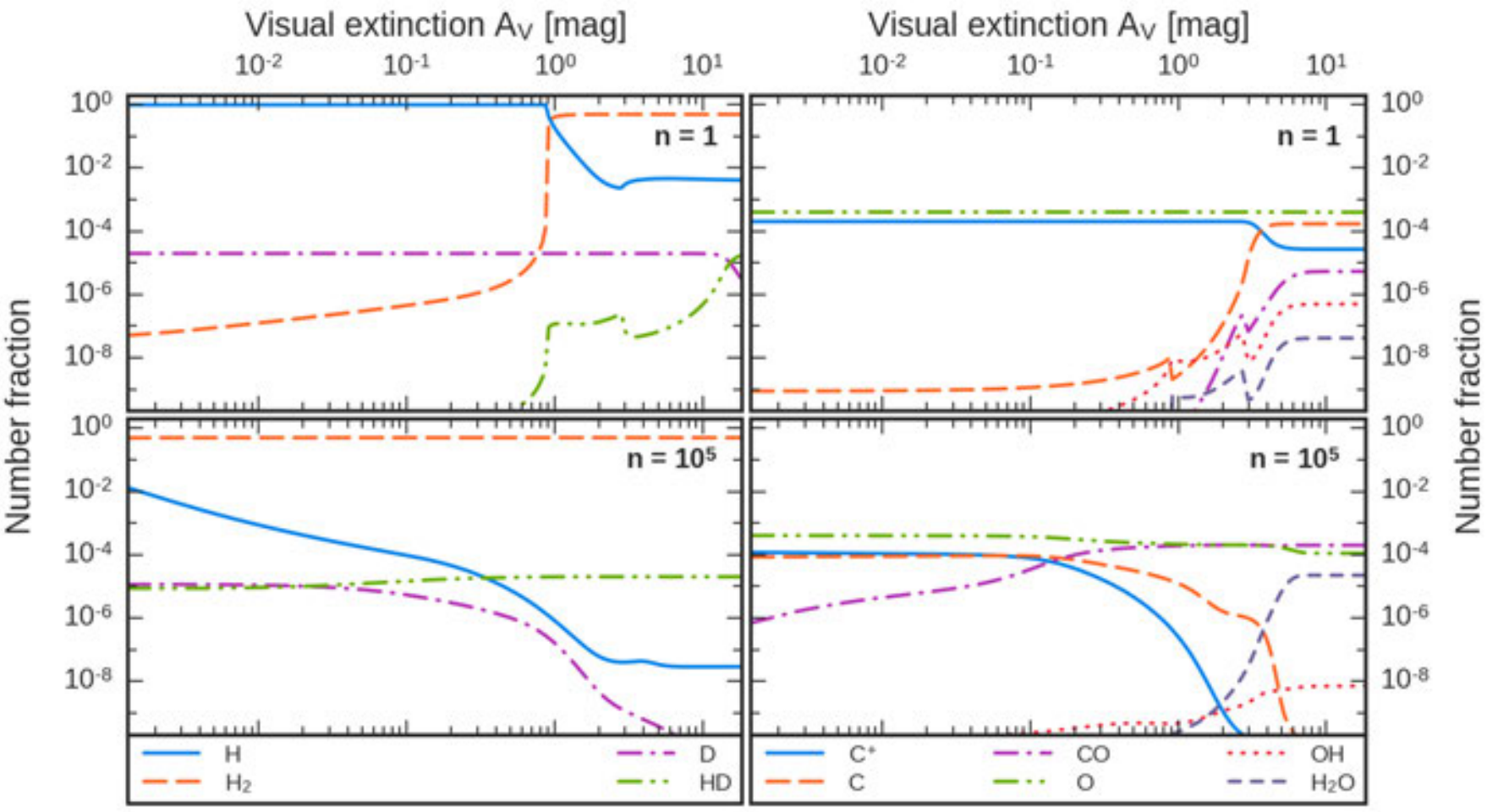

**Figure 4.5** – Abundances of several major species, in number fractions relative to the total hydrogen number density, as a function of visual extinction for a lower number density $n_\mathrm{H}$ of 1 cm$^{-3}$ (top) and a higher number density of $10^5$ cm$^{-3}$ (bottom), as compared to the fiducial value of $10^2$ cm$^{-3}$.

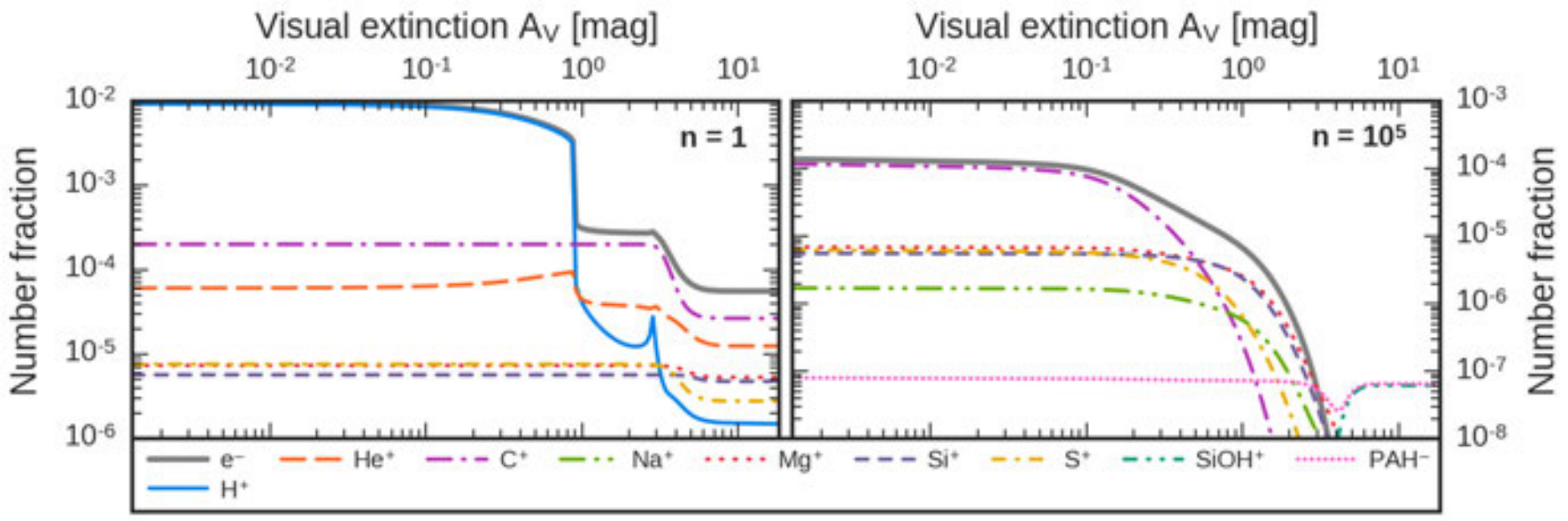

**Figure 4.6** – Ion abundances, in number fractions relative to the total hydrogen number density, as a function of visual extinction for a lower number density $n_H$ of 1 cm$^{-3}$ (left) and a higher number density of $10^5$ cm$^{-3}$ (right), as compared to the fiducial value of $10^2$ cm$^{-3}$.

**METALLICITY ($Z$)** The effect of changing the metallicity on the heating and cooling rates is shown in Figure 4.7. At the lowest metallicity, $10^{-6}$ in units of $Z_\odot$, the temperature is ~40 K and very flat, which is lower than in the fiducial case. The very low abundance of metals and dust means that the heating is now dominated by cosmic rays, the cooling by HD, and the main electron donor is $H^+$. Hydrogen is mostly atomic, though most of the deuterium is in the form of HD. This reflects the fact that molecular hydrogen forms most efficiently on grains, and thus its formation rate is tied to the metallicity, while HD can form efficiently in the gas phase as well. The case with a slightly higher metallicity of $10^{-3}$ $Z_\odot$ resembles the fiducial model somewhat more at the edge, due to an increased $H_2$ formation rate, but deeper into the cloud the temperature is similar to the lowest metallicity model, due to the heating and cooling rates again being dominated by cosmic rays and HD, respectively. The contributions from photoelectric heating, and from $H_2$ lines, CII 158 μm, and CI 370 μm finestructure lines to the cooling rate are now more significant than in the $10^{-6}\,Z_\odot$ model, however. Like in the very low metallicity case, $H^+$ is still the main electron donor, however, most of the hydrogen does become molecular, at an extinction of ~0.03 mag. In both of these models, the dust temperature is also slightly lower, due to the metallicity dependence of the MIR and FIR components.

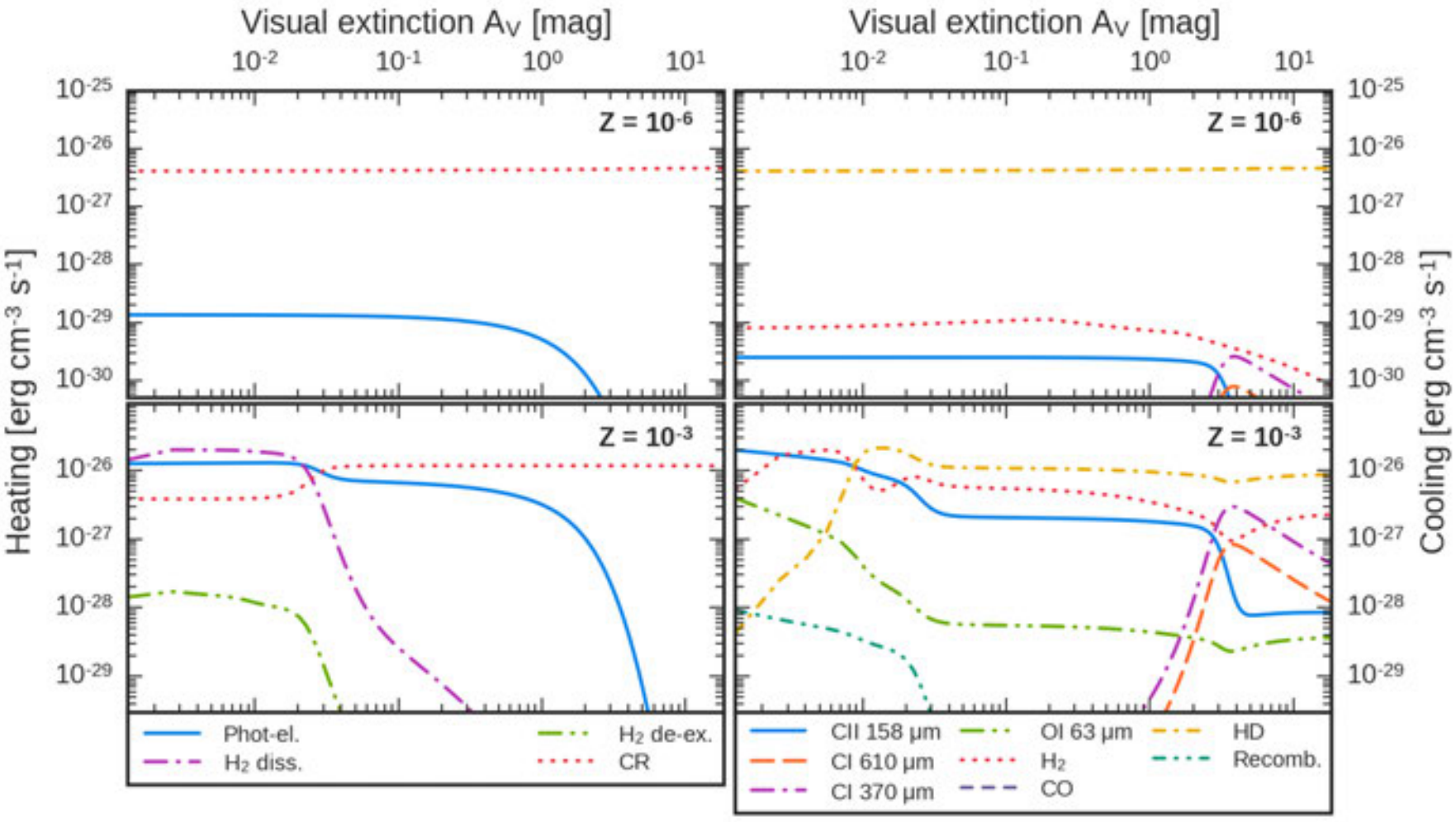

**Figure 4.7** – Heating (left) and cooling (right) rates as a function of visual extinction for a very low metallicity $Z$ of $10^{-6}$ $Z_{\odot}$ (top) and a low metallicity of $10^{-3}$ $Z_{\odot}$ (bottom).

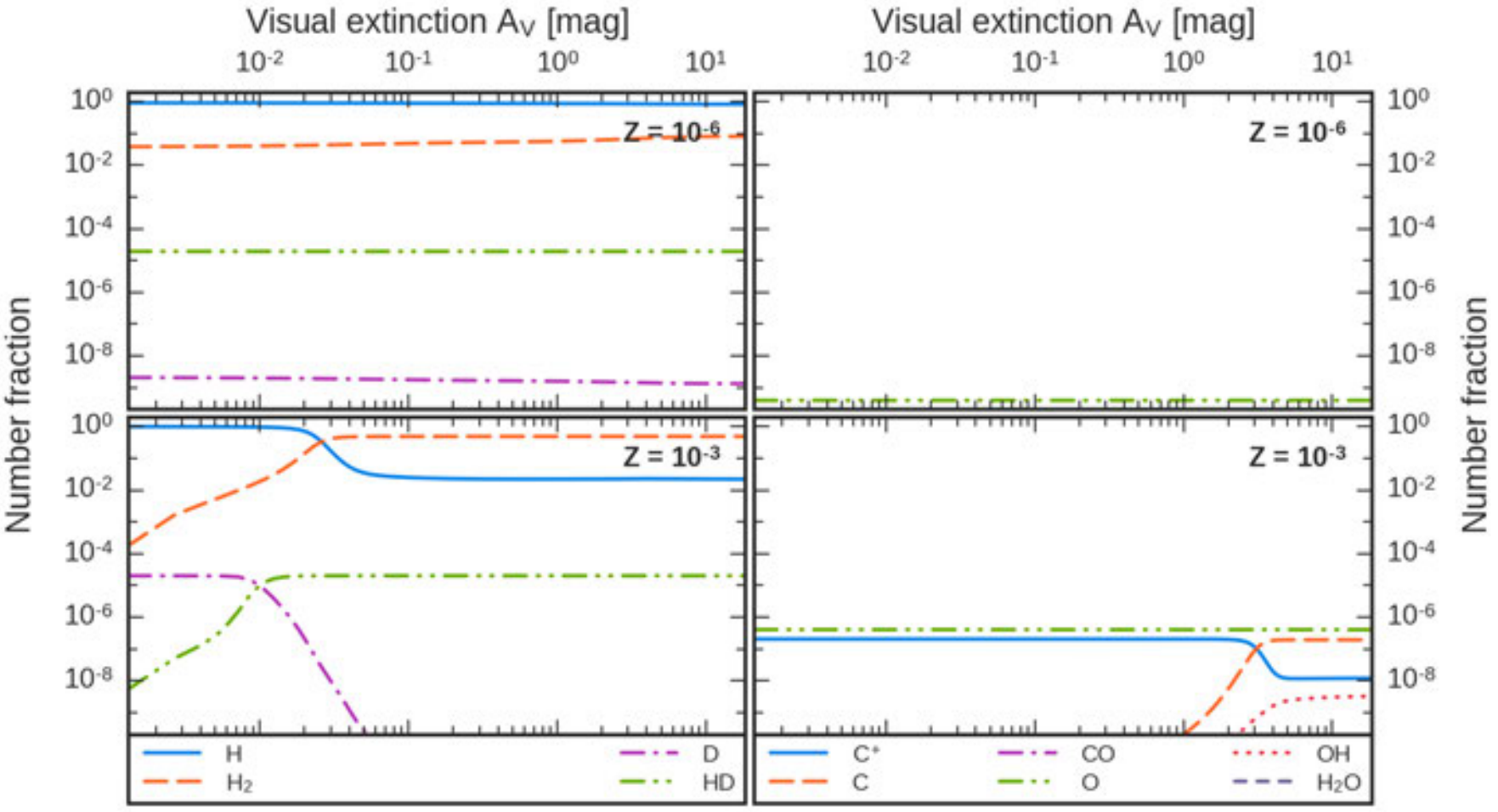

**Figure 4.8** – Abundances of several major species, in number fractions relative to the total hydrogen number density, as a function of visual extinction for a very low metallicity $Z$ of $10^{-6}$ $Z_{\odot}$ (top) and a low metallicity of $10^{-3}$ $Z_{\odot}$ (bottom).

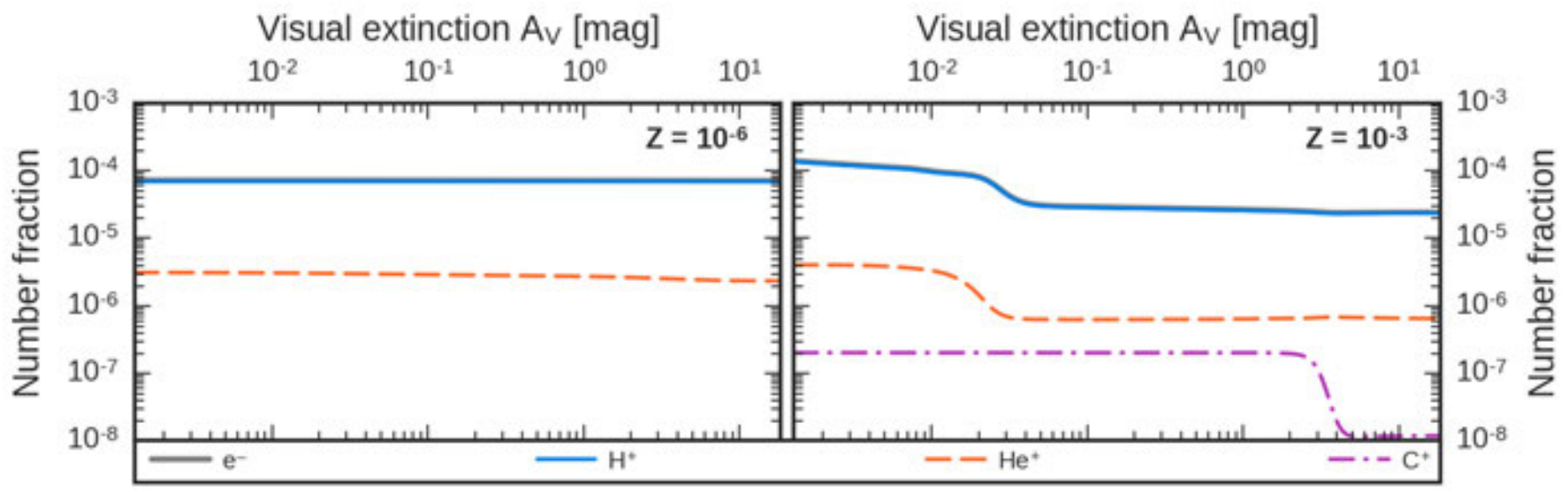

**Figure 4.9** – Ion abundances, in number fractions relative to the total hydrogen number density, as a function of visual extinction for a very low metallicity $Z$ of $10^{-6}\,Z_\odot$ (left) and a low metallicity of $10^{-3}\,Z_\odot$ (right). Note that in both models the electron fraction is equal to the $H^+$ fraction everywhere.

**RADIATION STRENGTH ($W_0$)** Decreasing or increasing the intensity of the background radiation field has a profound effect on both the gas and dust temperatures, as also apparent from the heating and cooling rates in Figure 4.10. In the weak radiation field case with $W_0 = 10^{-2}$, the temperatures are lower mainly due to a lower photoelectric heating rate in the case of the gas, and due to the $W_0$ dependence of the temperature in the case of the dust. Hydrogen is fully molecular in this model. Until an extinction of ~0.1 mag, the C and $C^+$ abundances are almost equal, resulting in C contributing more significantly to the thermal processes than in the fiducial model. The carbon ionization heating rate is comparable to the photoelectric heating rate until ~0.1 mag, and where in the fiducial model cooling is dominated by $C^+$ 158 μm finestructure line emission, here the C 610 μm finestructure line is most important. In the strong radiation field case with $W_0 = 10^5$, both gas and dust temperatures are much higher, for the same reasons as why they are lower in the weak field case. Photoelectric heating dominates until 8 mag, and at higher extinction the gas is heated by absorption in the finestructure lines, mainly by CI 370 µm and OI 63 µm, and above an extinction of 10 mag also by gas-grain collisions. Until ~2 mag, the main cooling channel is recombination of electrons with small grains and PAHs. At higher extinction, when the temperature drops rapidly, several other cooling channels become dominant in succession: the OI 63 µm finestructure line, $H_2$ rovibrational lines, the CI 370 µm and 610 µm finestructure lines, and CO rotational lines. Both the H/$H_2$ and $C^+$/C/CO transitions are pushed deeper into the cloud, and deuterium remains in atomic form. Where the gas is hot, at extinctions lower than ~3 mag, the most abundant ion is $H^+$.

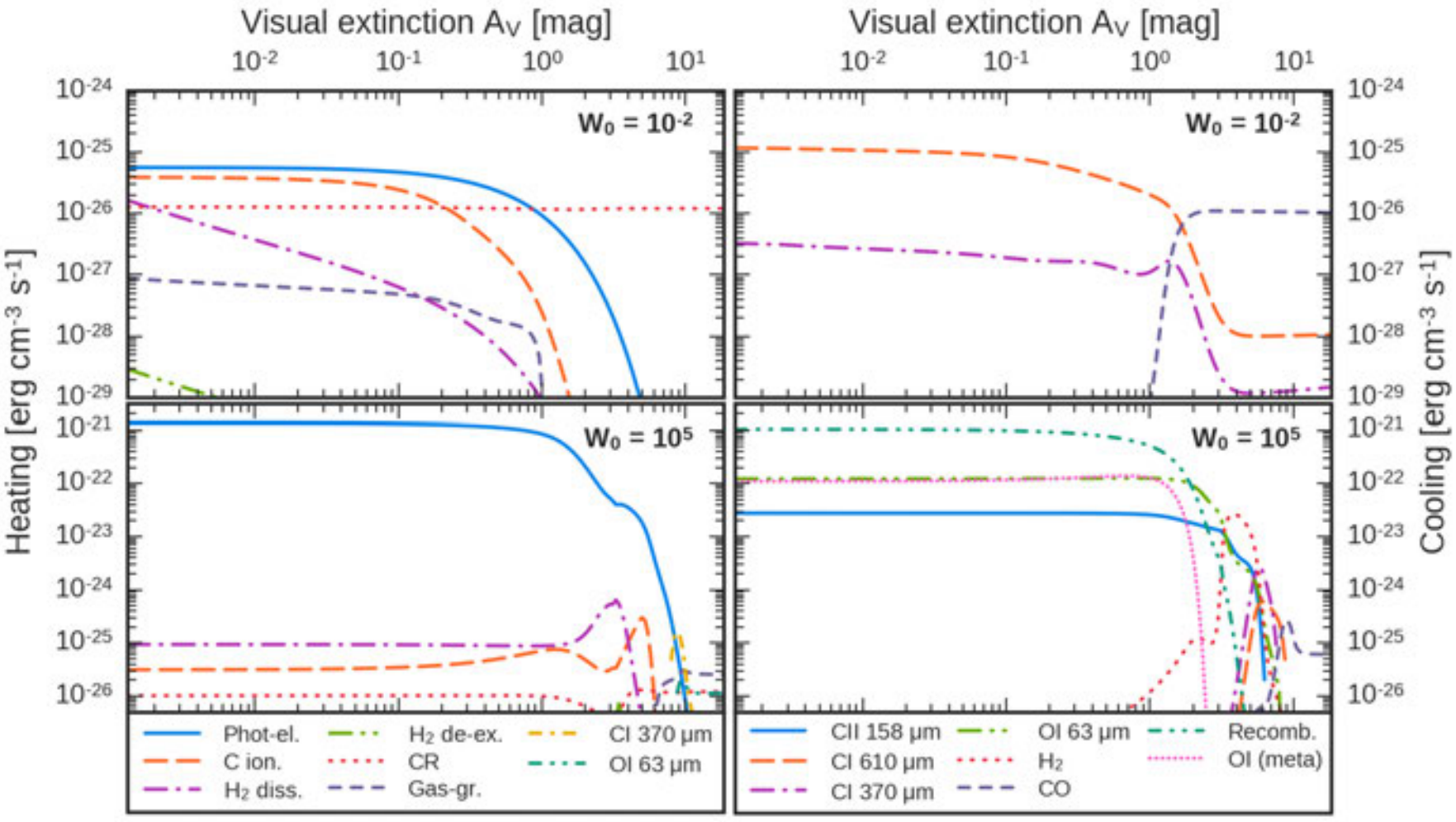

**Figure 4.10** – Heating (left) and cooling (right) rates as a function of visual extinction for a lower UV radiation field strength $W_0$ of $10^{-2}$ (top) and a higher radiation strength of $10^5$ (bottom), as compared to the fiducial value of 10.

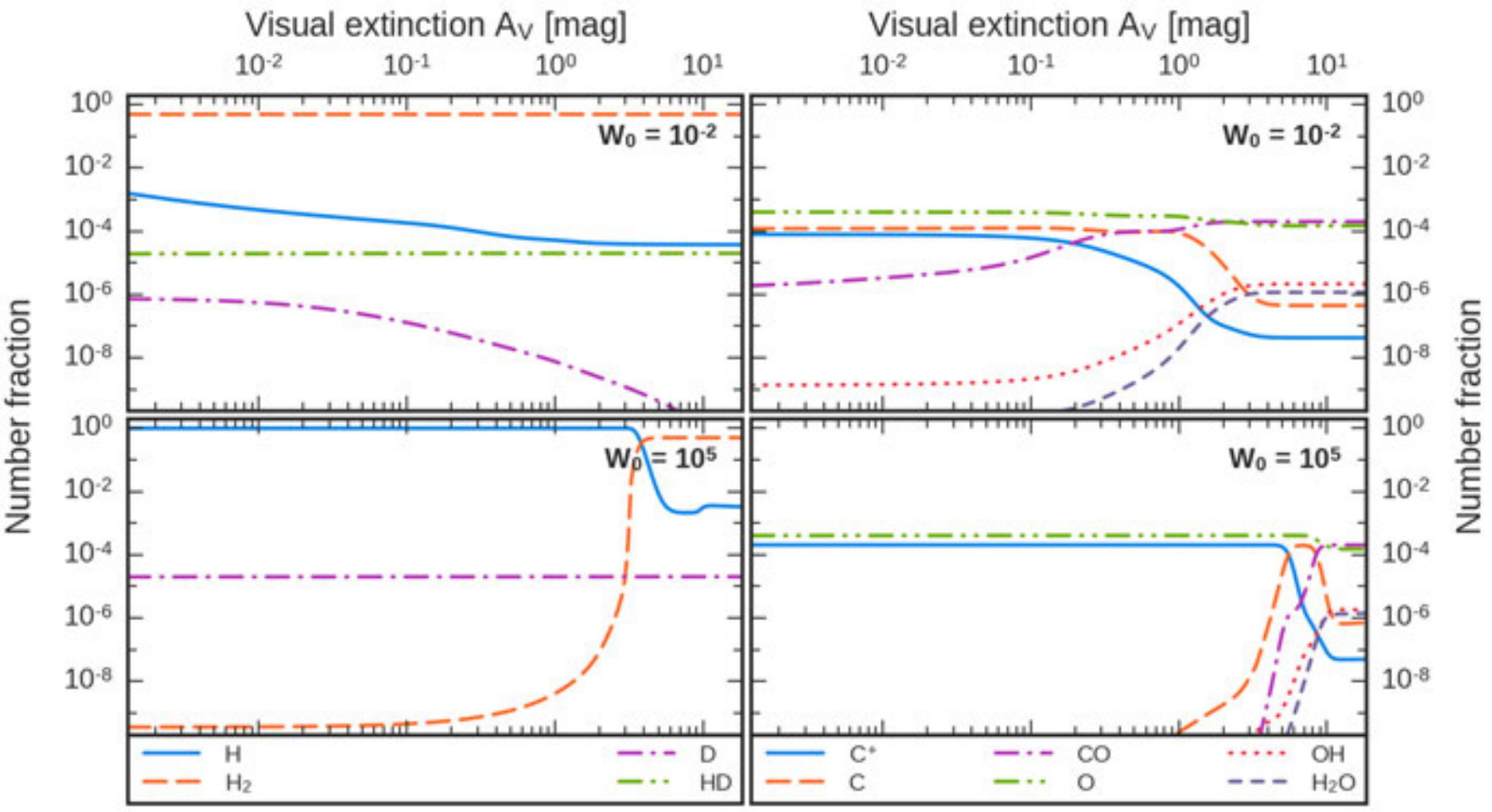

**Figure 4.11** – Abundances of several major species, in number fractions relative to the total hydrogen number density, as a function of visual extinction for a lower UV radiation field strength $W_0$ of $10^{-2}$ (top) and a higher radiation strength of $10^5$ (bottom), as compared to the fiducial value of 10.

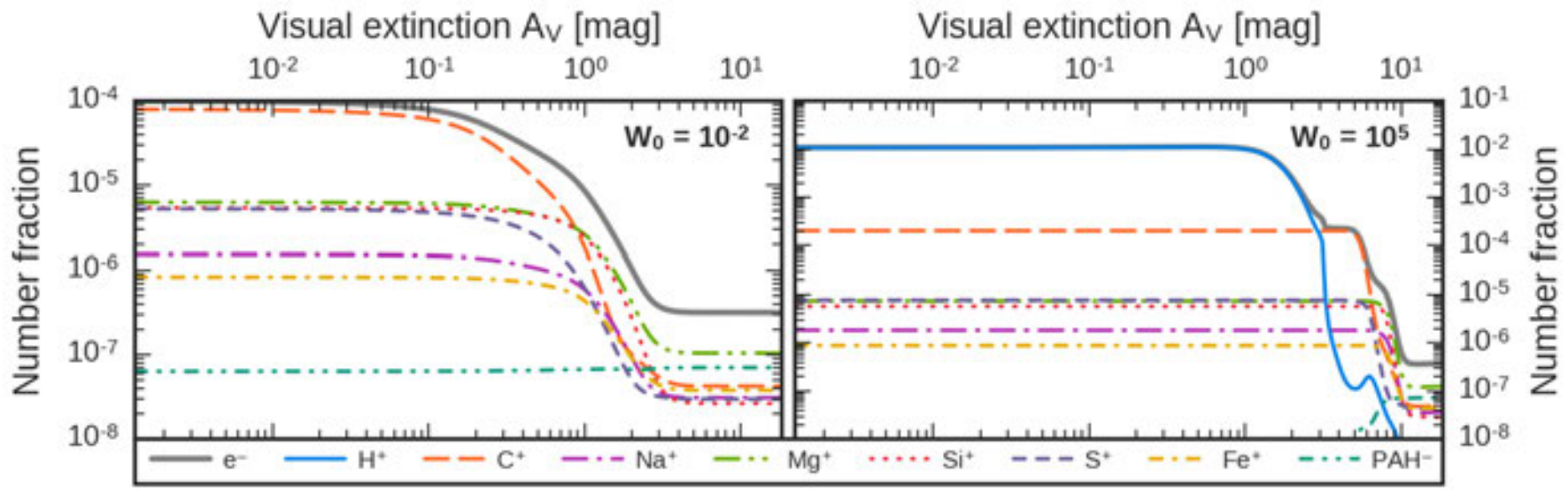

**Figure 4.12** – Ion abundances, in number fractions relative to the total hydrogen number density, as a function of visual extinction for a lower UV radiation field strength $W_0$ of $10^{-2}$ (left) and a higher radiation strength of $10^5$ (right), as compared to the fiducial value of 10.

**COSMIC RAY RATE ($\zeta$)** The effect of a different cosmic ray ionization rate is more subtle, and most noticeable at low and high extinctions, as also apparent from the heating and cooling rates in Figure 4.13. With a low CR rate of $10^{-20}$ s$^{-1}$, the main effect is the change of heating source at extinctions above ~5 mag from cosmic ray heating to gas-grain collisional heating, which has a small effect on the gas temperature. Abundances of most ions are also considerably lower. With a higher CR rate, $10^{-15}$ s$^{-1}$, the temperature increase at the edge and at extinctions above 3 mag is due to a higher cosmic ray heating rate. The higher temperature far inside the cloud prevents sufficient CO from forming, which means the cooling in that region is no longer dominated by CO rotational lines, but instead by CI 370 µm finestructure emission, with contributions from the CI 610 µm and CII 158 µm lines. Ion abundances are now significantly higher, with $H^+$ the main electron donor until ~0.01 mag.

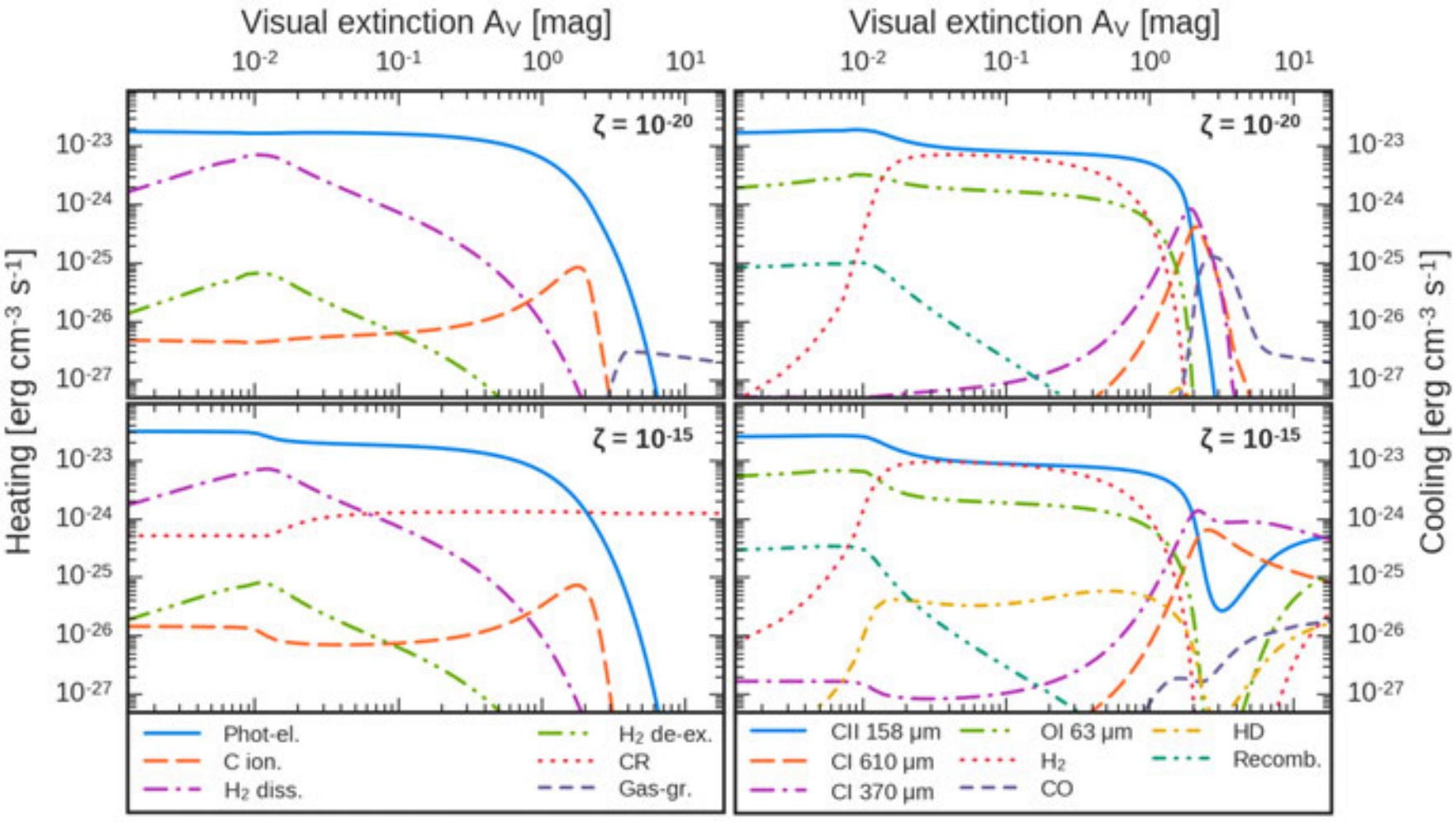

**Figure 4.13** – Heating (left) and cooling (right) rates as a function of visual extinction for a lower cosmic ray ionization rate $\zeta$ of $10^{-20}\,\mathrm{s}^{-1}$ (top) and a higher CR rate of $10^{-15}\,\mathrm{s}^{-1}$ (bottom), as compared to the fiducial value of $10^{-17}\,\mathrm{s}^{-1}$.

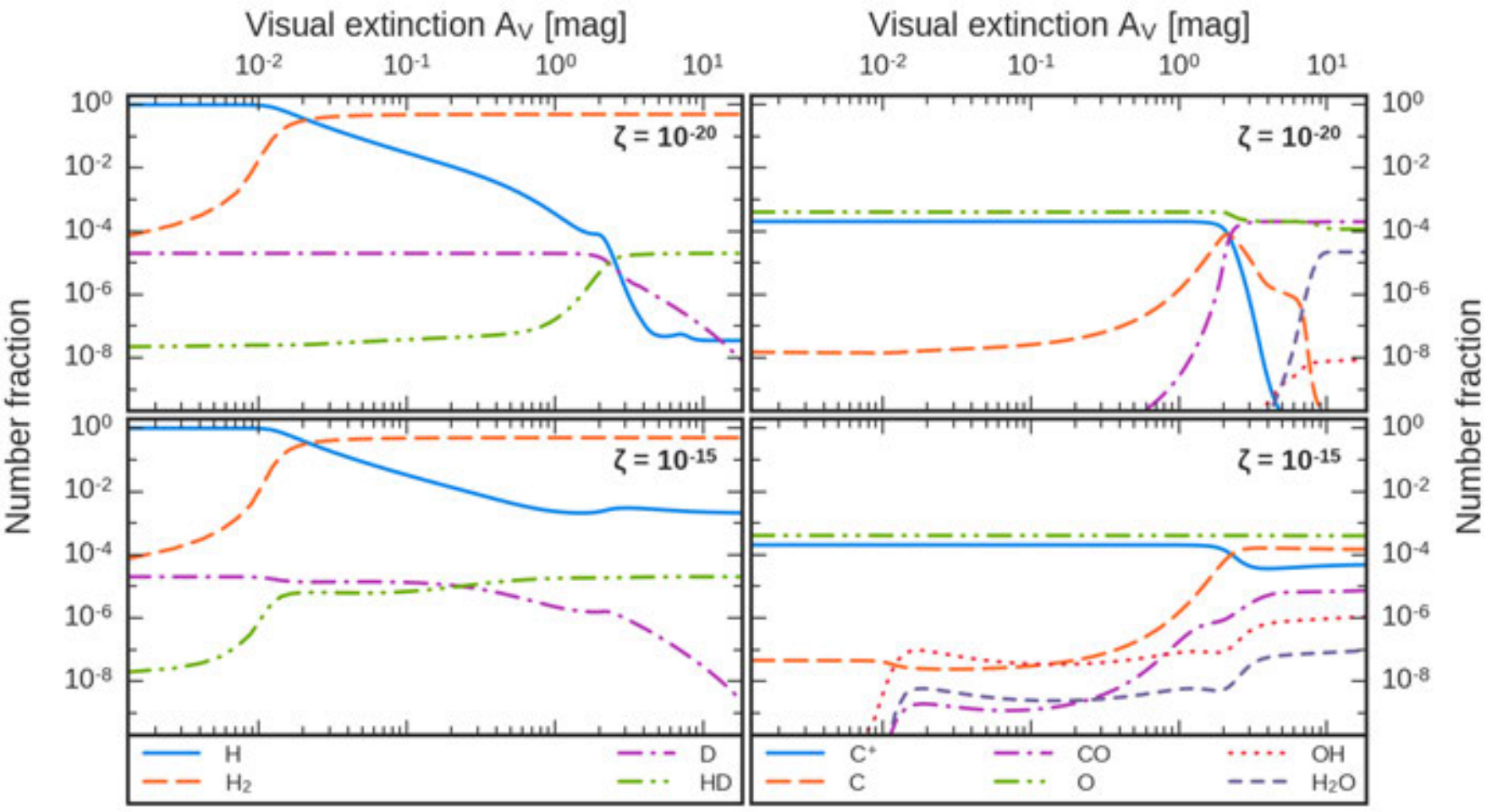

**Figure 4.14** – Abundances of several major species, in number fractions relative to the total hydrogen number density, as a function of visual extinction for a lower cosmic ray ionization rate $\zeta$ of $10^{-20}\,\mathrm{s}^{-1}$ (top) and a higher CR rate of $10^{-15}\,\mathrm{s}^{-1}$ (bottom), as compared to the fiducial value of $10^{-17}\,\mathrm{s}^{-1}$.

4

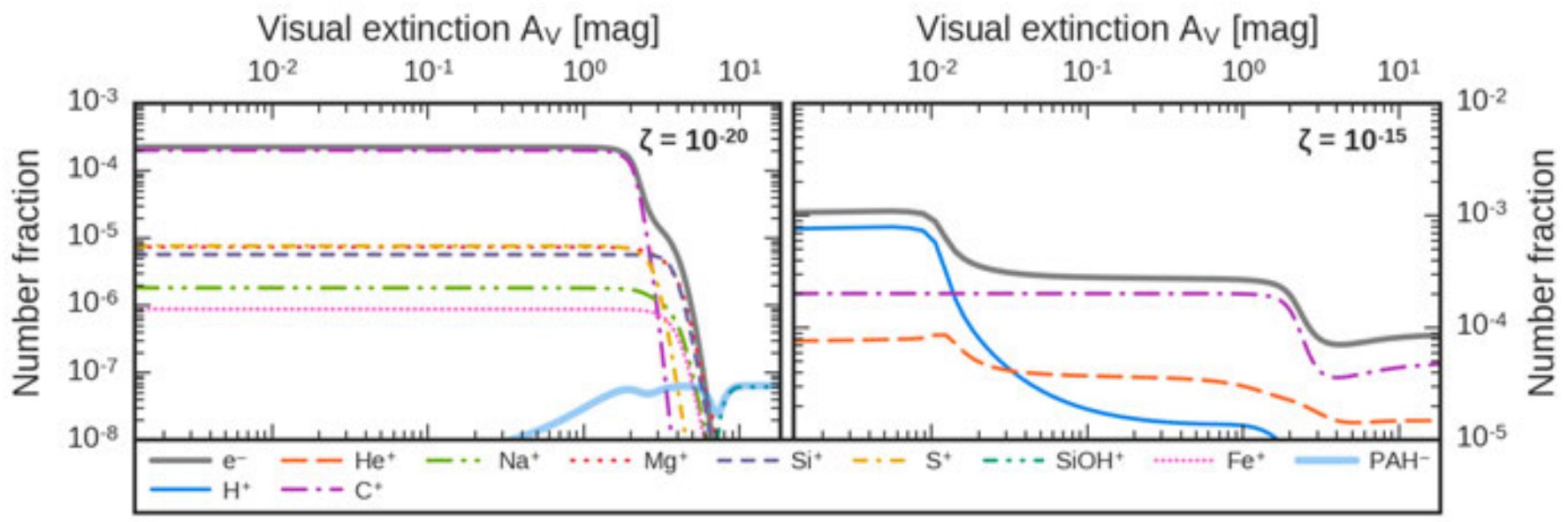

**Figure 4.15** – Ion abundances, in number fractions relative to the total hydrogen number density, as a function of visual extinction for a lower cosmic ray ionization rate $\zeta$ of $10^{-20}\,\mathrm{s}^{-1}$ (left) and a higher CR rate of $10^{-15}\,\mathrm{s}^{-1}$ (right), as compared to the fiducial value of $10^{-17}\,\mathrm{s}^{-1}$.

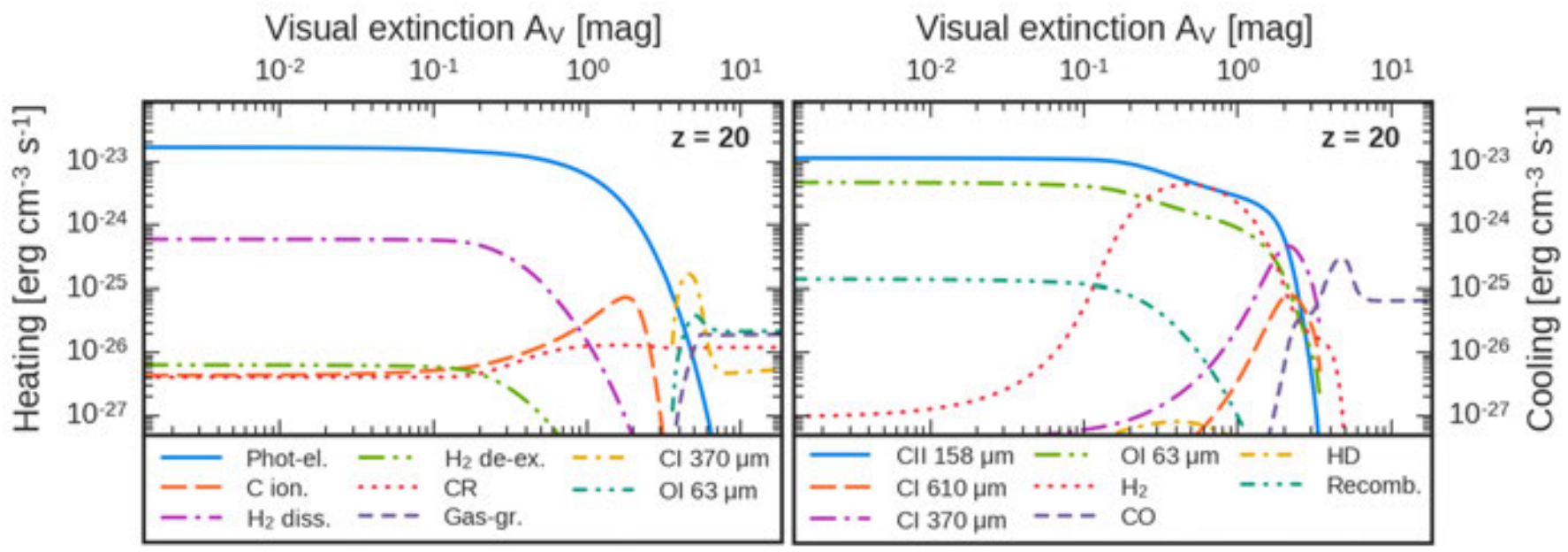

**Figure 4.16** – Heating (left) and cooling (right) rates as a function of visual extinction for a higher redshift $z$ of 20, as compared to the fiducial value of $z = 0$.

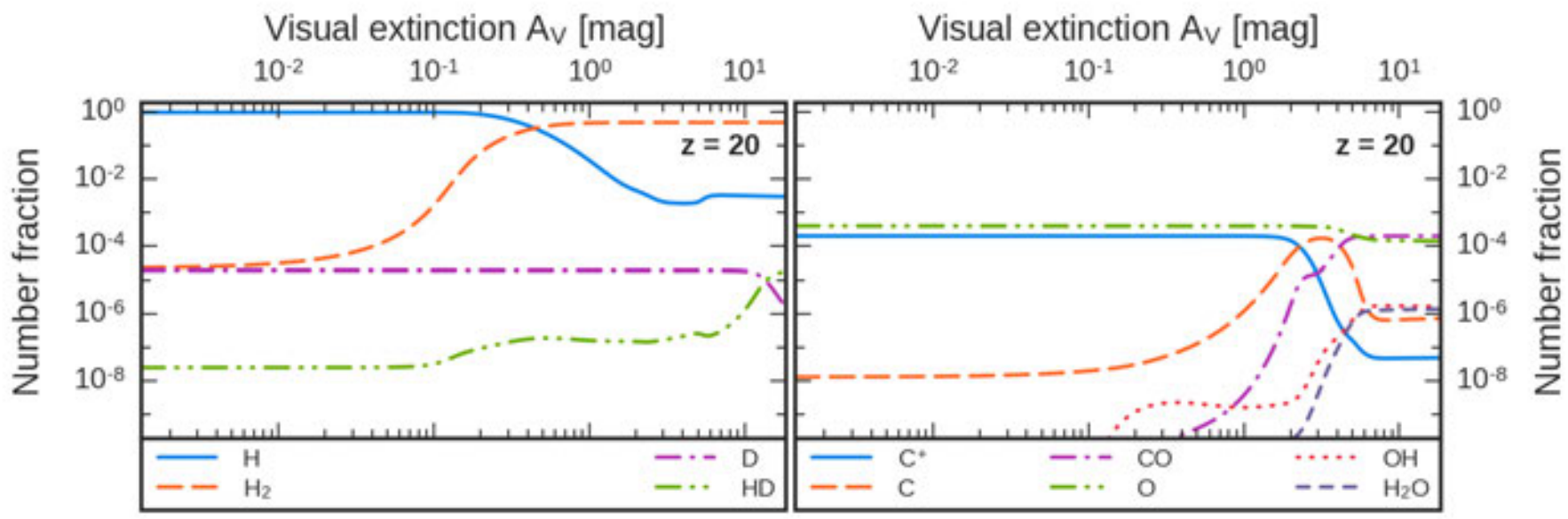

**Figure 4.17** – Abundances of several major species, in number fractions relative to the total hydrogen number density, as a function of visual extinction for a higher redshift $z$ of 20, as compared to the fiducial value of $z = 0$.

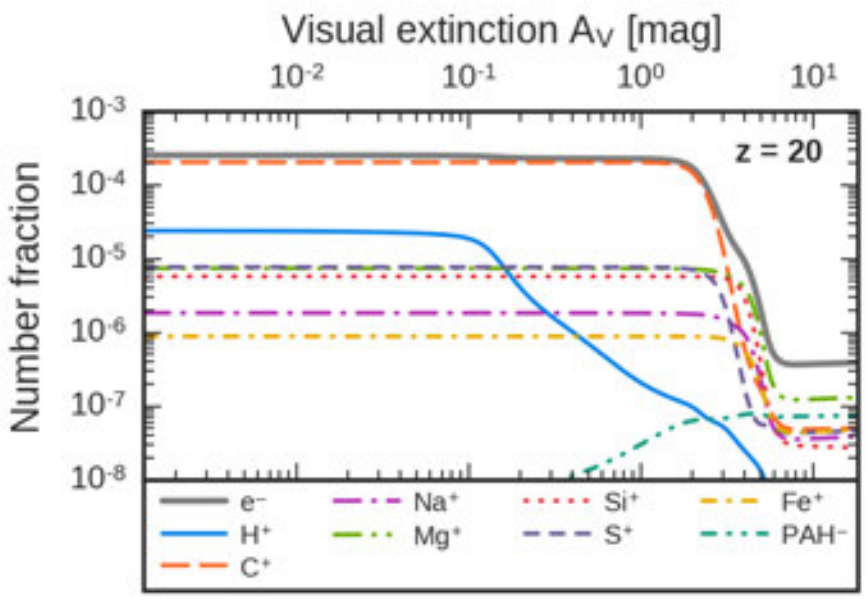

**Figure 4.18** – Ion abundances, in number fractions relative to the total hydrogen number density, as a function of visual extinction for a higher redshift $z$ of 20, as compared to the fiducial value of $z = 0$.

**REDSHIFT ($z$)** Finally, changing only the redshift results in a slight increase in gas temperature up to 1 mag, and a more pronounced increase at higher extinctions. The dust temperature is higher overall, reflecting the fact that it is now dominated by the contribution from the CMB. A result of the higher dust temperature is that the formation efficiencies of $H_2$ and HD molecules on grains drop, which becomes apparent by the positions of the H/$H_2$ and D/HD transitions, now located deeper into the cloud. The heating and cooling rates for this model are shown in Figure 4.16. The bump in gas temperature between ~3-6 mag is due to CI 370 μm absorption. At higher extinctions, cosmic rays contribute again to the heating rate, but more important heating sources are OI 63 μm absorption and gas-grain collisions. Like in the fiducial model, at first cooling is dominated by CII 158 μm finestructure emission, though now with more contribution from the OI 63 μm line and less from $H_2$ rovibrational lines, due to the lower $H_2$ abundance. Between 2-3 mag, CI 370 μm emission takes over, followed by CO rotational lines at higher extinctions.

### 4.3.2 TEMPERATURE TRENDS

4

We will now switch from an extinction-based approach to a column density based approach, as this will be more representative of realistic clouds, considering that high extinctions combined with low metallicities correspond to unrealistically large columns. The conversion between column density and visual extinction is calculated as follows (Bohlin et al. 1978):

$$A_V = \frac{N_H}{1.87 \times 10^{21}\,\mathrm{cm}^{-2}}\,\mathrm{mag}, \quad (4.1)$$

with $N_H$ the total hydrogen column density, and assuming $A_V = 3.1E(B-V)$, with $E(B-V)$ the color excess. The range of total hydrogen column densities considered is $10^{18} \leqslant N_H \leqslant 5 \times 10^{22}\,\mathrm{cm}^{-2}$, including a surface zone where $N_H = 0$. For solar metallicity, this corresponds to visual extinctions in the range $5 \times 10^{-4} \lesssim A_V \lesssim 27\,\mathrm{mag}$.

We will look at the behavior of the temperature in three different areas of the simulated clouds: the surface, which is a single zone at the illuminated face of the cloud, the bulk, which is the gas present at a column density higher than a certain value, and the outer layers, which is the gas sitting at a lower column density than our chosen value, but does not include the surface. We choose a column density of $10^{22}$ cm$^{-2}$ to separate the outer layers from the bulk. In solar metallicity gas at redshift zero, this corresponds to an optical extinction of ~5, so the cloud has become optically thick, and in most cases also molecular (see e.g. the discussion of the fiducial model in Section a). Mean temperatures in both parts are then calculated as averages of the temperatures in all simulated zones in each part, weighted by column density (which corresponds to mass-weighting).

The surface temperature can be useful because it is usually the hottest part of the cloud, and therefore related to the observability. Some other studies have also looked at surface temperatures, and thus a comparison can be made. However, not just the surface, but also the rest of the optically thin gas (what we call the outer layers) contributes to the molecular and atomic line emissions, often used as diagnostic tools. The bulk temperature, on the other hand, may be more representative for the state of the majority of the gas.

## A RADIATION FIELD STRENGTH

Figures 4.19, 4.20 and 4.21 show the surface temperature, average temperature in the outer layers, and average bulk temperature, respectively, as a function of the total hydrogen number density and the radiation field strength, for redshift $z = 0$ and a cosmic ray ionization rate of $10^{-17}$ s$^{-1}$. The y-axis has been split, to show both the range in $W_0$ from $10^{-2}$ to $10^5$, and the case where the radiation field is practically zero, $W_0 = 10^{-20}$.

**SURFACE** The surface temperature (see Figure 4.19) shows the influence of different metallicities on the heating/cooling balance, without the effects of dust extinction. In the solar metallicity case, the hottest gas is the very low density gas irradiated by fields with $W_0$ in the range $10^{-1}$-$10^2$. Perhaps surprising is that at these low densities, the temperature decreases somewhat for increasing $W_0$. Under these conditions, the temperature is set by a balance between photoelectric heating and recombination cooling. The grains become very positively charged, resulting in a reduced heating efficiency as it is more difficult to eject electrons from the positively charged grains. Addi-tionally, the efficiency of electron-grain recombination increases, resulting in a higher recombination cooling rate. This effect was also seen by Kaufman et al. (1999). They note that the results in this corner of parameter space may not be realistic, as neither model takes the drifting of grains through the gas due to radiation pressure into account.

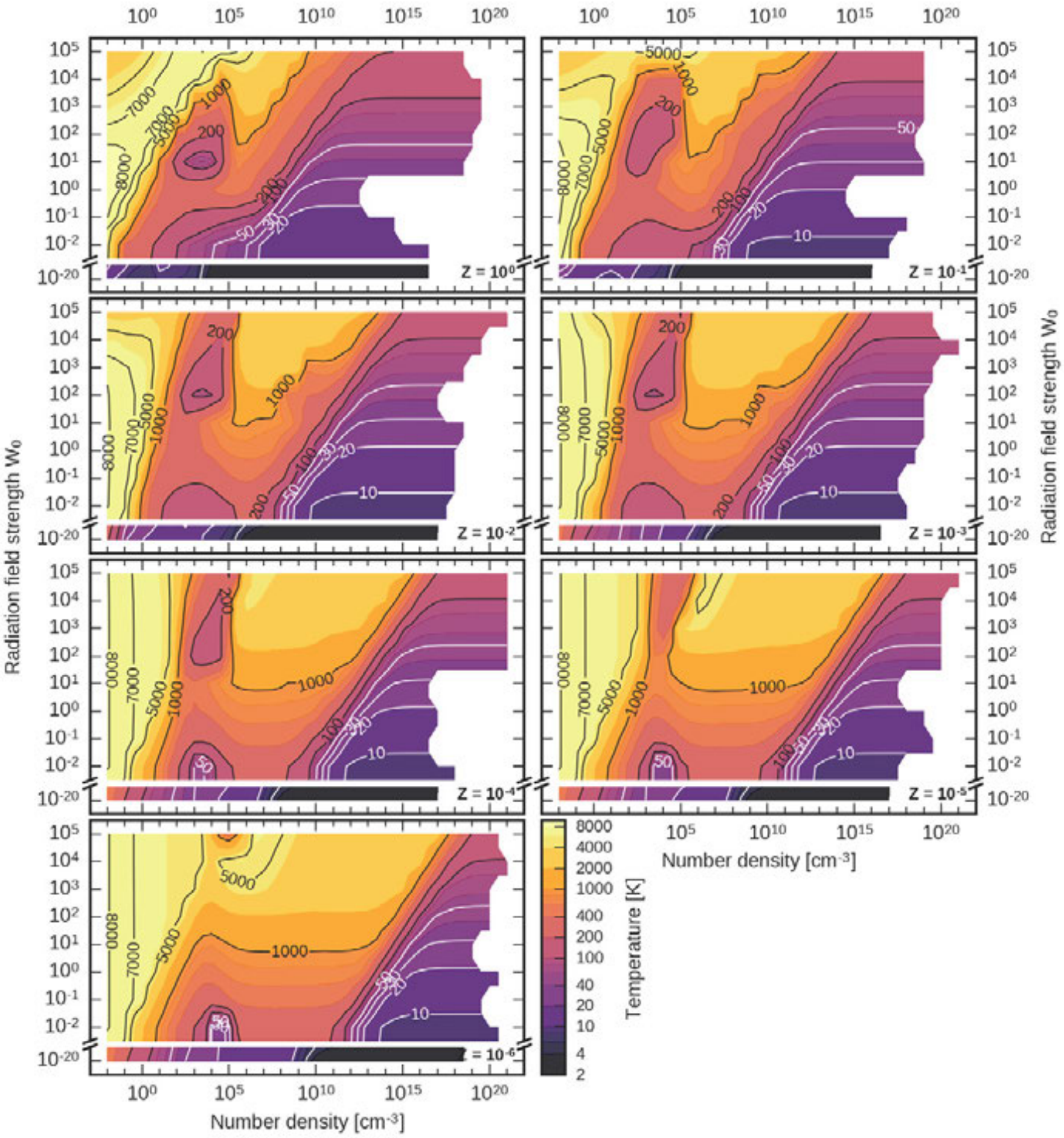

**Figure 4.19** – Gas temperature at the illuminated surface of the region, as a function of number density and radiation field strength, at redshift $z = 0$ and cosmic ray ionization rate $\zeta = 10^{-17}\,\mathrm{s}^{-1}$. The subplots show the temperature for different metallicities, starting with solar metallicity in the top left, and ending with $Z = 10^{-6}\,\mathrm{Z}_{\odot}$ at the bottom.

At high densities of $\sim 10^7\,\mathrm{cm}^{-3}$ and above, the surface temperature increases more or less in a straightforward fashion with increasing radiation field strength for a fixed density. When looking along a fixed $W_0$, there is always a certain density where the gas and dust become so strongly coupled that the temperature of the gas becomes equal to the dust temperature, and no longer changes with a further increase in density.

For radiation field strengths below $W_0 = 1$, the gas becomes gradually cooler moving from low to high density. For stronger radiation fields however, something interesting happens for gas in the density range of $10 \sim 10^7\,\mathrm{cm}^{-3}$. For $W_0 = 1$ and starting from $10\,\mathrm{cm}^{-3}$, the gas stays around 300 K for several orders of magnitude in density, then slightly increases in temperature, and finally gradually decreases once more until hitting the dust temperature. For $W_0 = 10$, instead of loitering at the same temperature, the gas cools down significantly, until almost 50 K at $n_\mathrm{H} = 10^{3.5}\,\mathrm{cm}^{-3}$. After this point, the gas temperature again first increases and finally decreases with increasing density, flattening off where the gas-dust coupling becomes strong. For higher radiation fields, there is a similar trend, though with increasingly less dramatic drops in temperature, until $W_0 = 10^5$, where temperature again decreases steadily with density, without a significant dip. The decrease in temperature with increasing density (the 'cold spot') is due to efficient cooling by the CII 158 µm and OI 63 µm finestructure lines, as before these lines reach LTE, the cooling rate per unit volume increases as $n^2$, while the density dependence of the photoelectric heating rate (the main heating process) is less steep. When the density is higher than the critical density of these coolants ($n_{\mathrm{cr,CII}} \sim 3 \times 10^3\,\mathrm{cm}^{-3}$, $n_{\mathrm{cr,OI}} \sim 5 \times 10^5\,\mathrm{cm}^{-3}$), the heating rate increases faster than the cooling and the temperature increases again. This 'warm valley', where the surface temperature of the gas is warmer than for lower density gas, is mainly a result of efficient $H_2$ collisional de-excitation heating in that part of parameter space.

With decreasing metallicity, a few trends are seen. The slightly cooler corner at very low densities and high radiation field strengths disappears, as both photoelectric heating and recombination cooling are metallicity-dependent, and are replaced by cosmic ray heating and HI cooling as main heating and cooling channels, respectively. The 'cool spot' situated around a density of $10^3$-$10^4$ cm$^{-3}$ moves upwards to higher radiation strengths, and becomes somewhat warmer. Note that in all cases, the gas in the cool spot is never colder than the dust temperature. The density at which the dust-gas coupling brings the temperature down to the dust temperature moves up about a factor ten in density for each factor ten decrease in metallicity, due to the lowered dust-to-gas ratio. The result is that the 'warm valley', which occurs at densities just higher than those where the cool spot is located, extends down to lower radiation field strengths, and also becomes warmer compared to a higher metallicity case. At a metallicity of $Z = 10^{-2}\,Z_\odot$, it has extended down to a radiation field of $W_0 = 10^{-2}$, separating the cool gas at densities $\sim 10^2$-$10^5$ cm$^{-3}$ from the cooler gas at $\sim 10^8$ cm$^{-3}$ and above. This effect never extends down to our lowest radiation field strength of $W_0 = 10^{-20}$, however. There, the temperature always decreases steadily with increasing density. Such a weak field is not able to heat the gas much above the dust temperature in the solar metallicity case. The difference between the temperature at low densities and at high densities increases with decreasing metallicity, as it becomes increasingly warmer at low densities. However, at higher redshifts this gradient will become less pronounced, as also the dust temperature will increase due to the warmer CMB.

Using the original implementation of the PDR code, Meijerink et al. (2007) also ran a grid of PDR surface temperatures, spanning a range in densities of $10^2$-$10^6$ cm$^{-3}$ and in radiation strength $G_0$ of 3-$10^5$, though with a significant amount of missing data points at low $n_\mathrm{H}$, high $G_0$ and at high $n_\mathrm{H}$, low $G_0$. Their results were obtained for solar metallicity at redshift zero, and with a cosmic ray ionization rate of $5 \times 10^{17}$ s$^{-1}$, which is slightly higher than what is shown in our Figure 4.19. Our surface temperatures in the relevant part of parameter space roughly agree, with the exception that we find higher temperatures for the higher end of their density range ($10^5$-$10^6$ cm$^{-3}$) for intermediate radiation fields. This is likely due to different prescriptions for the $H_2$ collisional de-excitation heating, and a difference in net $H_2$ formation rate. Similar high temperatures in this $n_\mathrm{H}$-$W_0$ range were also found by Xue & Huang (2009), who used the CLOUDY code to calculate comparable PDR models.

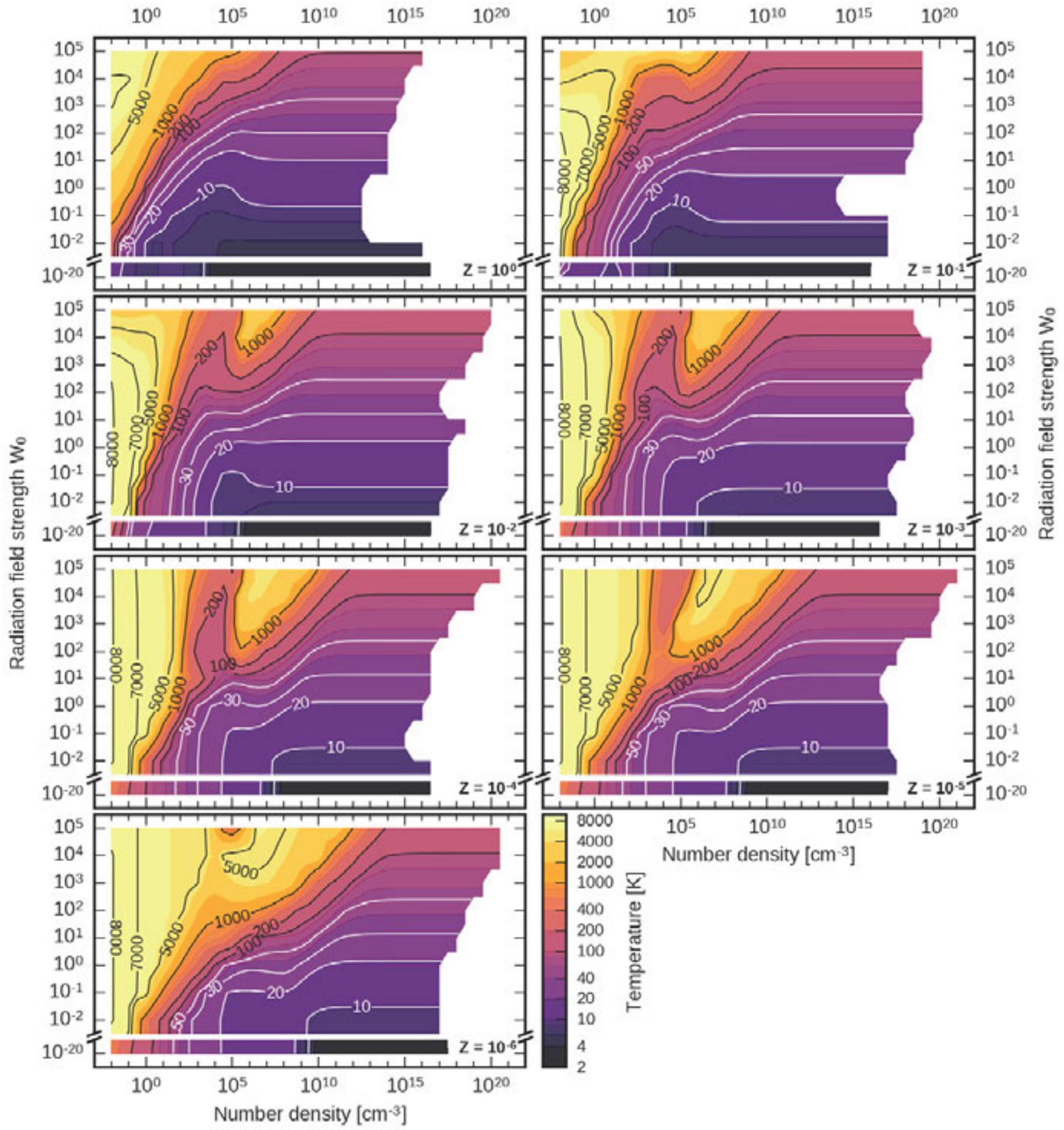

**Figure 4.20** – Mass-weighted gas temperature in the outer layers of the cloud ($N_\mathrm{H} < 10^{22}\,\mathrm{cm}^{-2}$) as a function of number density and radiation field strength, at redshift $z = 0$ and cosmic ray ionization rate $\zeta = 10^{-17}\,\mathrm{s}^{-1}$. The subplots show the temperature for different metallicities, starting with solar metallicity in the top left, and ending with $Z = 10^{-6}\,\mathrm{Z}_\odot$ at the bottom.

**OUTER LAYERS** When the incident FUV field heats up and pressurizes the surface layers of a gas cloud, it can cause some of the gas to disperse away, a process known as photoevaporation (e.g. Bertoldi 1989; Gorti & Hollenbach 2002). Therefore, it may be useful to consider not just the temperature at the very surface of the cloud, but also the average temperature of the low column density gas, which is optically thin to the UV continuum.

In Figure 4.20, the average temperature of the gas at column densities lower than $10^{22}$ cm$^{-3}$ (the outer layers) is shown. For solar metallicity, the warm valley (and, consequently, also the cold spot) has now disappeared, indicating that the interplay causing this effect is only of importance for gas sitting at very low extinctions. For decreasing metallicity, the cold spot/warm valley feature becomes more pronounced again, though never extending down to such weak radiation fields as when considering only the surface temperatures. This also points to the fact that it is an effect occurring only at low extinction, since at constant column density, a factor ten drop in metallicity corresponds to a factor ten drop in extinction. With the exception of the cold spot, which moves towards stronger radiation fields for lower metallicities, the overall trend is that the point at which the gas temperature converges to the dust temperature occurs for increasingly higher densities for increasing radiation field strengths. This point also moves to higher densities for lower metallicities, as was the case for the surface temperature. However, it no longer shifts by almost exactly a factor ten, and the transition region now broadens with each downwards step in metallicity.

4

**BULK** In Figure 4.21, we show the average temperature of the gas at column densities higher than $10^{22}$ cm$^{-3}$ (the bulk). At solar metallicity, almost all the of the bulk gas sits at a temperature below 100 K; with the exception of the most intense radiation field cases, the entire bulk even has a temperature of 50 K or lower. For radiation fields with $W_0 = 10$ and above, there is no longer a trend of decreasing temperature from low to high density. On the contrary, there is a significant density range, from $\sim 10^{-1-1}$ cm$^{-3}$ to $\sim 10^{4-6}$ cm$^{-3}$ in which the gas is actually cooler than the dust. This effect can also be seen in the fiducial model, discussed in Section a, and is due to insufficient gas-dust coupling and efficient CO rotational cooling.

The change of temperature structure from solar metallicity to $Z = 10^{-1}\,\mathrm{Z}_\odot$ is much more dramatic than the change for further metallicity steps. This is likely related to the shift from looking at mostly optically thick gas for solar metallicity, to partially optically thin gas for $Z = 10^{-1}\,Z_{\odot}$. For this metallicity and below, a part of the low-density gas is still heated to temperatures of 1000 K and higher. While for $Z = 10^{-1}\,Z_{\odot}$ the transition region from warm to cold gas is quite narrow, for $Z = 10^{-6}\,Z_{\odot}$ there is a significant amount of warm gas present at densities above $10\,\mathrm{cm}^{-3}$.

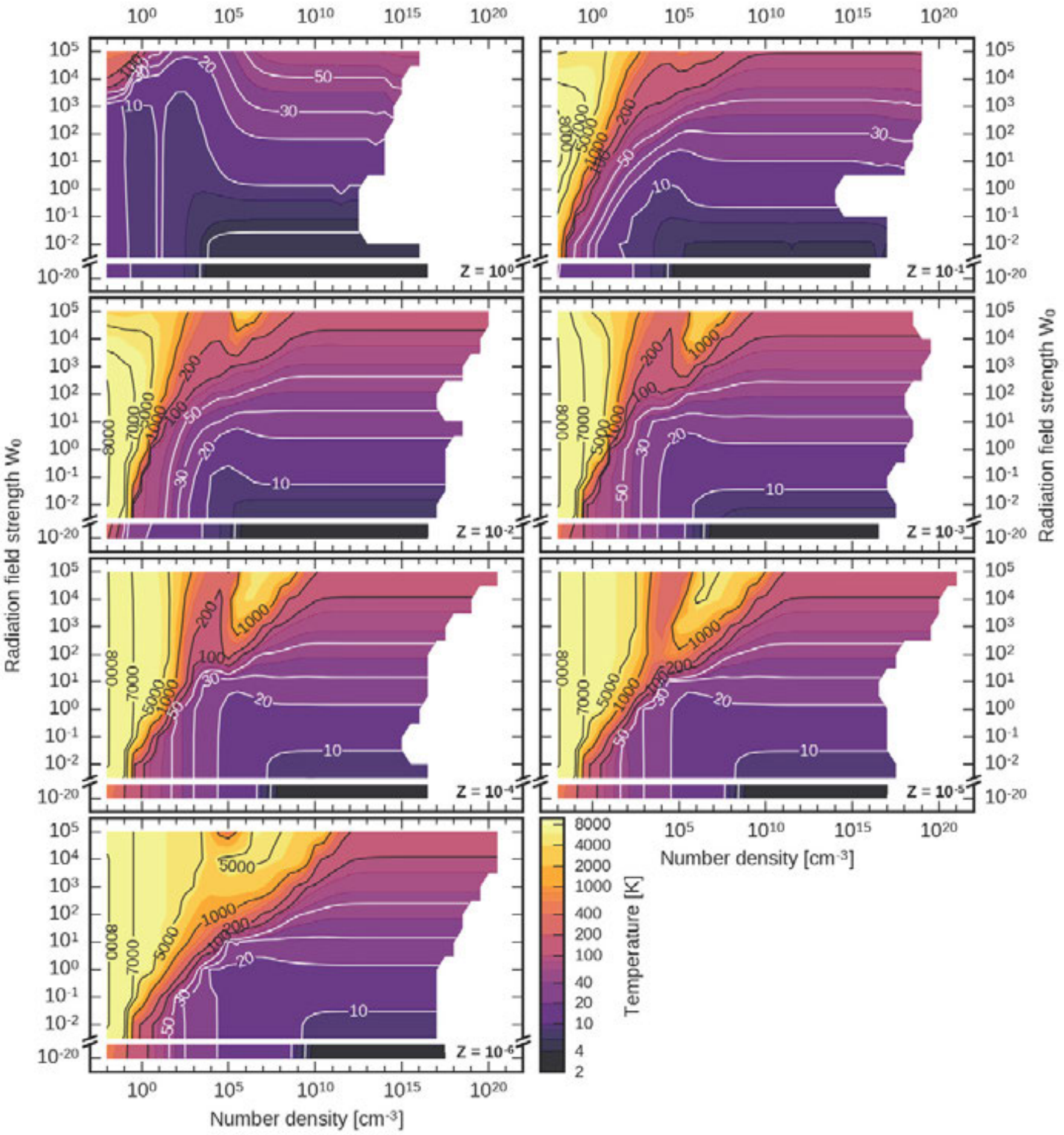

**Figure 4.21** – Mass-weighted gas temperature in the bulk of the cloud ($N_\mathrm{H} \geqslant 10^{22}\,\mathrm{cm}^{-2}$) as a function of number density and radiation field strength, at redshift $z = 0$ and cosmic ray ionization rate $\zeta = 10^{-17}\,\mathrm{s}^{-1}$. The subplots show the temperature for different metallicities, starting with solar metallicity in the top left, and ending with $Z = 10^{-6}\,\mathrm{Z}_\odot$ at the bottom.

At high densities, the gas temperature becomes more and more similar between the edge and bulk when going to lower metallicities. This is due to the dust temperature, to which the gas is coupled, becoming less dependent on the column density for decreasing metallicity.

## B REDSHIFT

**OUTER LAYERS** In Figure 4.22 we show the temperature in the outer layers of the cloud as a function of total hydrogen number density and redshift, for different metallicities, at a radiation field strength of $W_0 = 1$ and a cosmic ray ionization rate of $10^{-17}\,\mathrm{s}^{-1}$. At high densities, the gas temperature in both the outer layers and the bulk increases with redshift, due to the warmer CMB which sets the dust temperature. At low densities, the gas is consistently warm at all redshifts, with a larger warm zone for a lower metallicity. At intermediate densities, there are 'cold peaks' visible, inside of which the gas is colder than the dust. At solar metallicity, the peak occurs around $\sim 10^4\,\mathrm{cm}^{-3}$, and moves approximately a factor ten up in density for a factor ten decrease in metallicity. However, it also decreases in redshift from $Z = 1\,Z_{\odot}$ to $Z = 10^{-2}\,Z_{\odot}$, practically disappears at $Z = 10^{-3}\,Z_{\odot}$, and then increases in redshift again between $Z = 10^{-4} - 10^{-6}\,Z_{\odot}$. The location of the peak in density space is mainly related to the column density chosen to split the cloud into outer layers and bulk. On the other hand, both the location of the peak in metallicity-space and the height of the peak (so the magnitude of the difference between the coldest temperature inside the peak, and the dust temperature) are related to a change in dominant cooling channels above a certain column density, plus the range in column densities for which this coolant is most effective. For metallicities above $Z = 10^{-2}\,Z_{\odot}$, the main cooling channel is CO rotational cooling. The abundance of CO drops with metallicity, and at $Z = 10^{-3}\,Z_{\odot}$ it has become so low that the temperature is set by gas-grain collisions. As the dust abundance also decreases with metallicity, for even lower metallicities, the gas-dust coupling becomes weak enough that gas can again cool below the dust temperature, by means of OH and $H_2O$ rotational cooling.

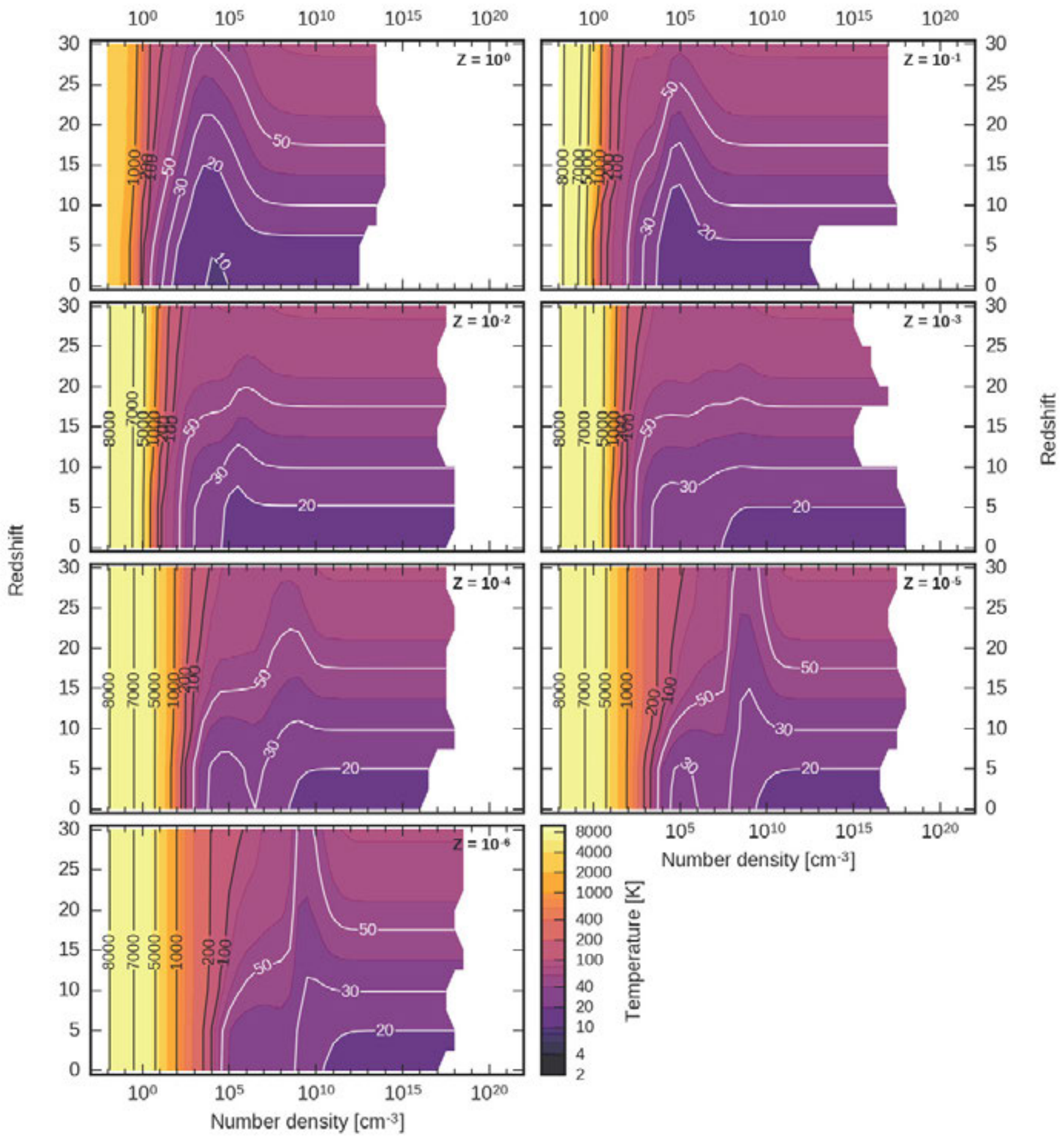

**Figure 4.22** – Mass-weighted gas temperature in the outer layers of the cloud ($N_{\rm H} < 10^{22}\,\rm cm^{-2}$) as a function of number density and redshift, at radiation field strength $W_0 = 1$ and cosmic ray ionization rate $\zeta = 10^{-17}\,\rm s^{-1}$. The subplots show the temperature for different metallicities, starting with solar metallicity in the top left, and ending with $Z = 10^{-6}\,\rm Z_\odot$ at the bottom.

One must keep in mind that only the situation for $W_0 = 1$ is shown here, and that for stronger or weaker radiation fields the temperature trends will look somewhat different.

**BULK** The difference between the outer layers and the bulk, the latter of which is shown in Figure 4.23, is not very large at the different redshifts. At solar metallicity, the warm edge at low densities disappears and the bulk gas has a temperature equal to or lower than the dust. At lower metallicities, the warm edge is only slightly narrower in density space than for the outer layers. At densities behind the warm edge, the bulk gas cools down to the dust floor (or below) in a significantly narrower density range than in the outer layers.

## C COSMIC RAY IONIZATION RATE

In Figure 4.24, we show the average temperature in the outer layers of the cloud as a function of number density and cosmic ray ionization rate, at a radiation field strength of $W_0 = 1$ and redshift $z = 0$. We ran models for just three different ionization rates, $10^{-20}$, $10^{-17}$, and $10^{-15}\,\mathrm{s}^{-1}$, so only the overall trends in these temperature plots should be taken into consideration. The largest effect of the cosmic rays is seen at low densities, where the temperature increases significantly with a higher ionization rate, due to additional heating. At higher densities, the difference between CR ionization rates of $10^{-20}$ and $10^{-17}\,\mathrm{s}^{-1}$ is rather insignificant. For a cosmic ray ionization rate of $10^{-15}\,\mathrm{s}^{-1}$, the temperatures are slightly higher than for a lower rate, though the difference disappears at high densities where the dust temperature is reached (or slightly warmer than the dust, in the case of solar metallicity).

In Figure 4.25, we show the average bulk temperature of the gas, for the same parameters ($W_0 = 1$, $z = 0$). At solar metallicity and $\zeta = 10^{-20}\,\mathrm{s}^{-1}$, the bulk gas at all densities is cold, $<10\,\mathrm{K}$, while at low densities and $\zeta = 10^{-15}\,\mathrm{s}^{-1}$, some of the bulk gets heated. In contrast to the outer layers, there is a decreasing trend in bulk temperature from an cosmic ray ionization rate of $10^{-17}$ to $10^{-20}\,\mathrm{s}^{-1}$.

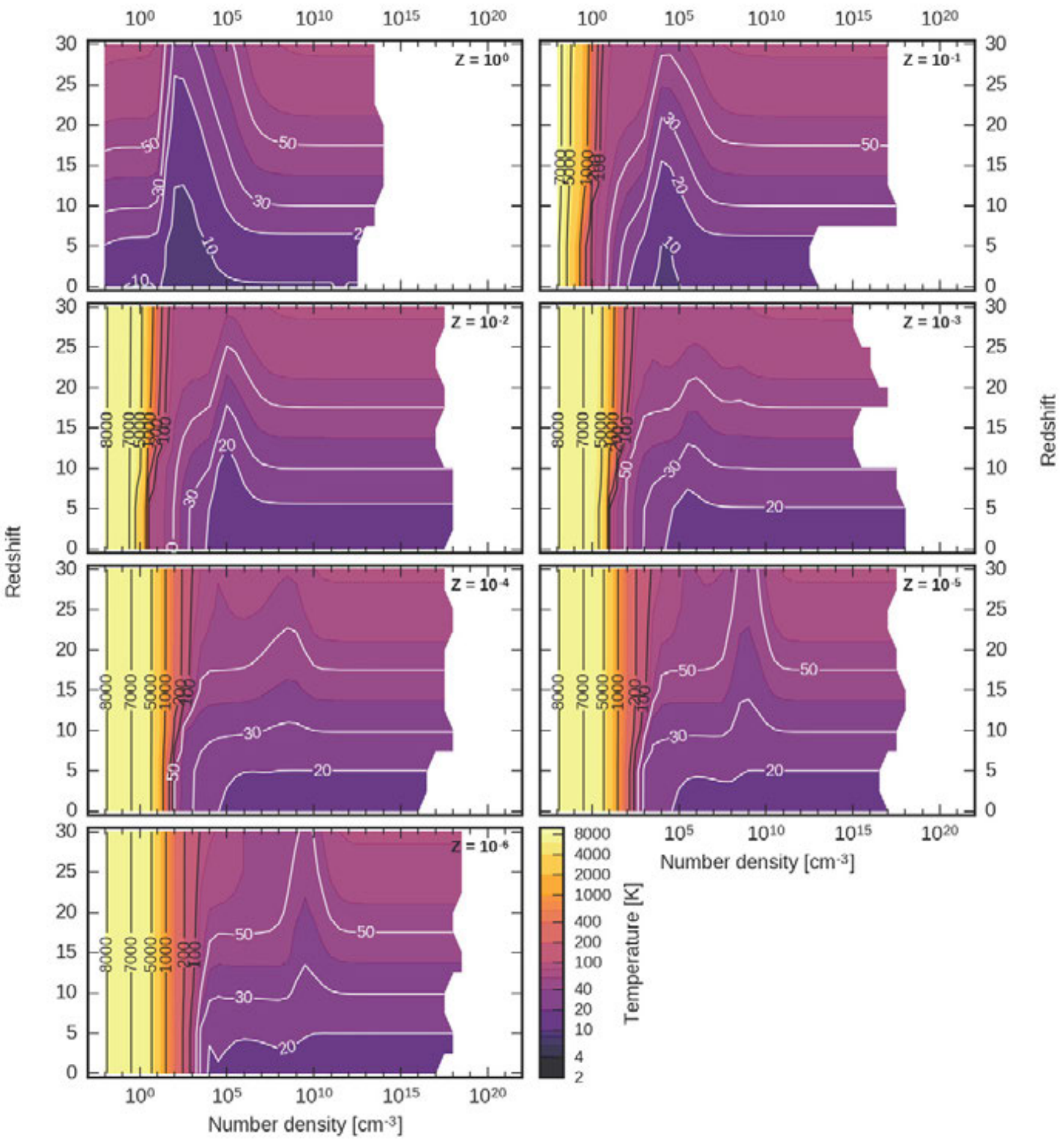

**Figure 4.23** – Mass-weighted gas temperature in the bulk of the cloud ($N_\mathrm{H} \geqslant 10^{22}\,\mathrm{cm}^{-2}$) as a function of number density and redshift, at radiation field strength $W_0 = 1$ and cosmic ray ionization rate $\zeta = 10^{-17}\,\mathrm{s}^{-1}$. The subplots show the temperature for different metallicities, starting with solar metallicity in the top left, and ending with $Z = 10^{-6}\,\mathrm{Z}_\odot$ at the bottom.

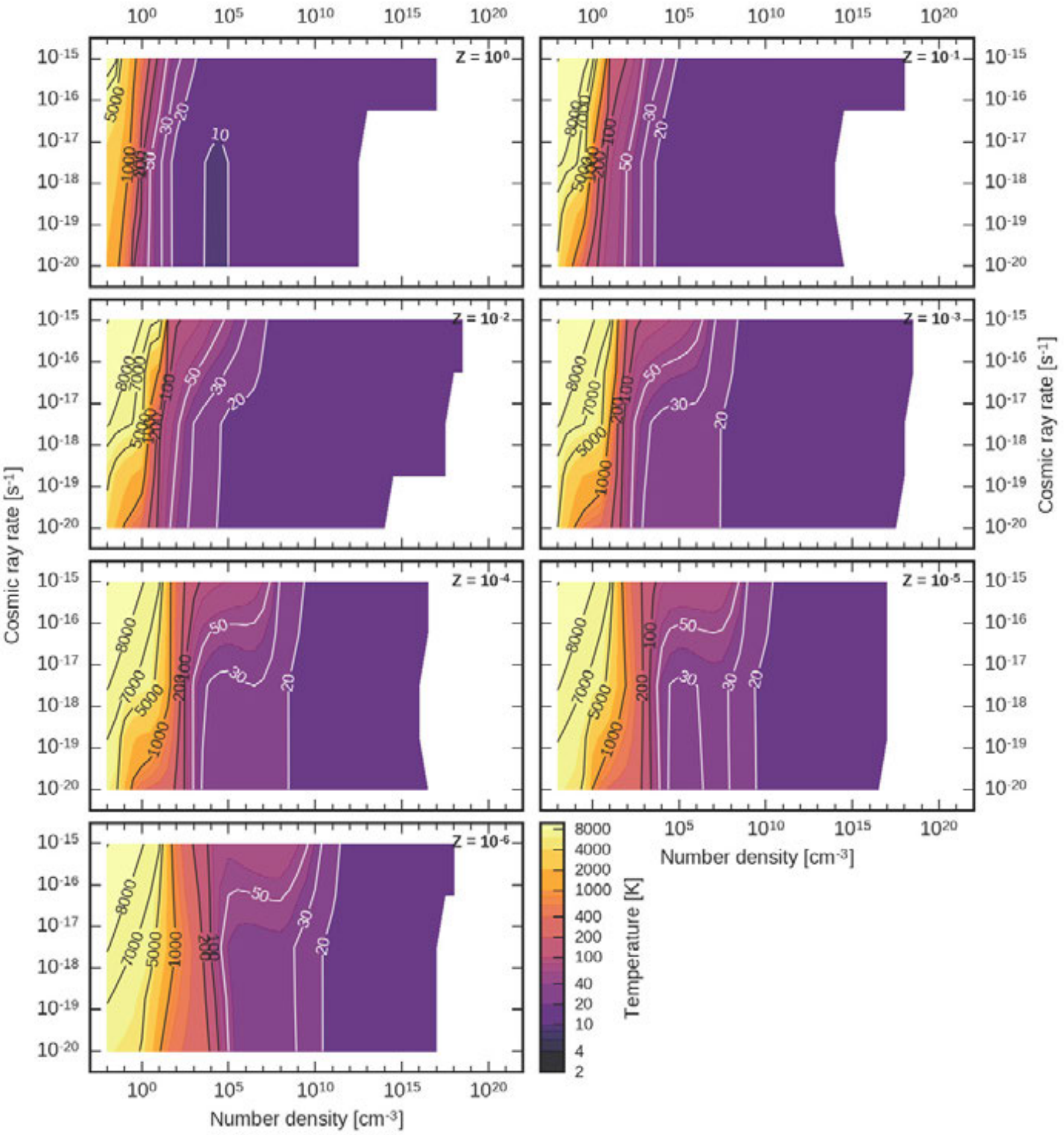

**Figure 4.24** – Mass-weighted gas temperature in the outer layers of the cloud ($N_{\mathrm{H}} < 10^{22}\,\mathrm{cm}^{-2}$) as a function of number density and cosmic ray ionization rate, at radiation field strength $W_0 = 1$ and redshift $z = 0$. The subplots show the temperature for different metallicities, starting with solar metallicity in the top left, and ending with $Z = 10^{-6}\,\mathrm{Z}_{\odot}$ at the bottom.

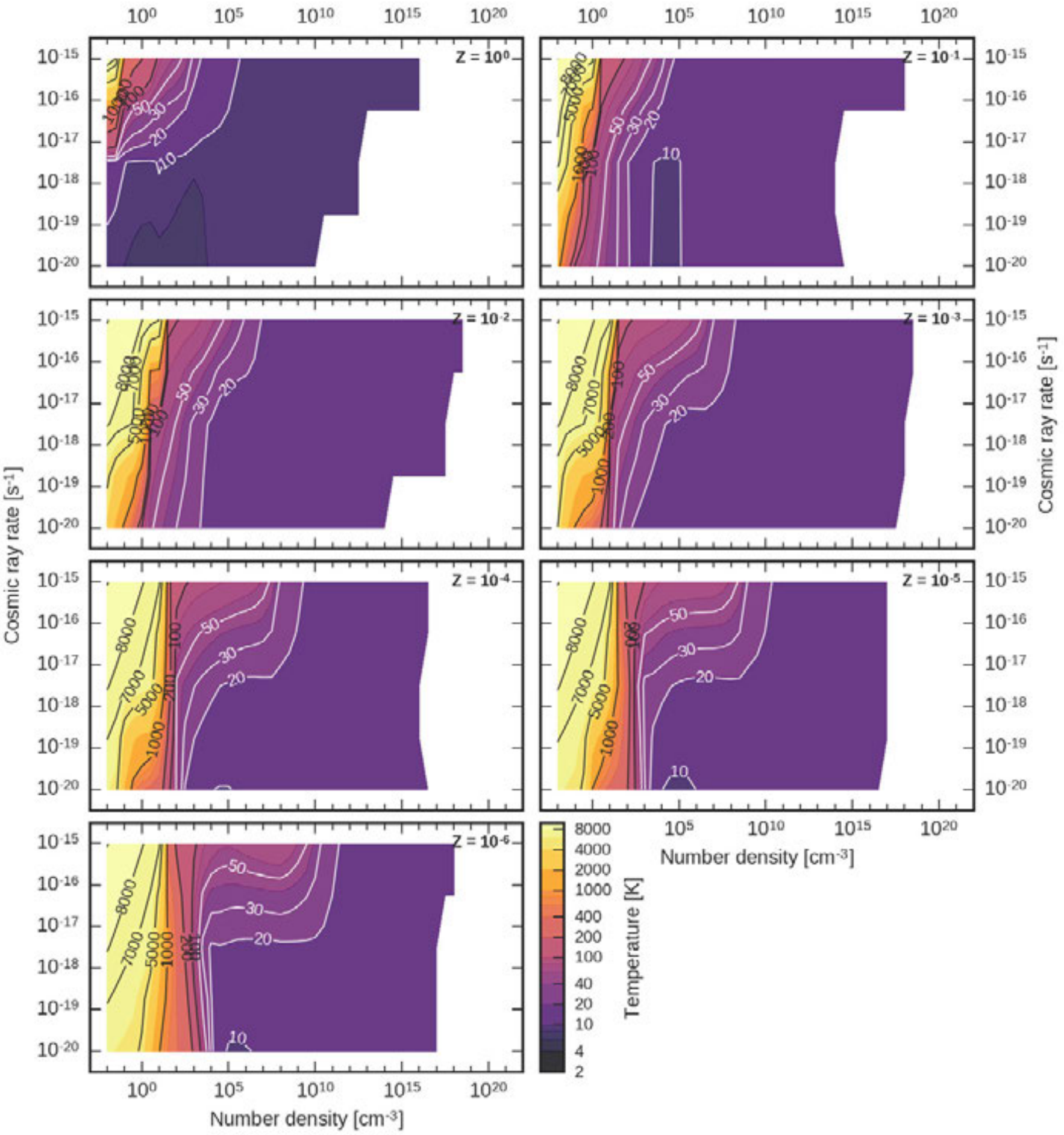

**Figure 4.25** – Mass-weighted gas temperature in the bulk of the cloud ($N_H \geqslant 10^{22}\,\mathrm{cm}^{-2}$) as a function of number density and cosmic ray ionization rate, at radiation field strength $W_0 = 1$ and redshift $z = 0$. The subplots show the temperature for different metallicities, starting with solar metallicity in the top left, and ending with $Z = 10^{-6}\,Z_\odot$ at the bottom.

For higher radiation fields (not shown), the cold spot discussed in Section 4.3.2 (a) is significantly larger and colder for a low CR ionization rate of $10^{-20}\,\mathrm{s}^{-1}$, and much smaller or even absent for low metallicities and a high ionization rate of $10^{-15}\,\mathrm{s}^{-1}$. This is related to the change in electron fraction: a higher cosmic ray ionization rate means a higher electron fraction, which results in a higher electron-grain recombination rate, and thus a higher photoelectric heating efficiency and a higher temperature.

### 4.3.3 EQUATION OF STATE

The heating caused by FUV radiation in the outer layers of the cloud can create strong photoevaporative mass flows and drive shocks into the bulk material, compressing it to high densities (e.g. Bertoldi 1989; Gorti & Hollenbach 2002). Other processes which can perturb the bulk and trigger collapse include cloud collisions, supernova blast waves, and (weak) AGN jets. We will now investigate the fragmentation properties of the bulk gas. It has been shown that the equation of state provides an indication for the amount of fragmentation expected in interstellar gas clouds (e.g. Spaans & Silk 2000; Li et al. 2003). We will assume that the physical state of the gas is described by a polytropic equation of state (EOS), of the form

$$P = K\rho^{\gamma_\mathrm{p}}, \tag{4.2}$$

where $P$ is the gas pressure, $K$ is a proportionality constant, $\rho$ is the mass density of the gas, and $\gamma_\mathrm{p}$ is the polytropic exponent. A unique temperature (and thus pressure) can be associated with a density if the gas is in thermal equilibrium, meaning that the heating and cooling processes will balance each other out on a timescale shorter than the dynamical timescale of the gas motions $t_\mathrm{cool} < t_\mathrm{dyn}$ (Vazquez-Semadeni et al. 1996; Scalo & Biswas 2002).

This equation implies that many of the factors affecting the state of the gas are encompassed by the polytropic exponent; in the case of this work, these are the UV radiation intensity, the metallicity or chemical composition, the cosmic ray ionization rate, and the redshift (affecting both the radiation field shape and the dust temperature). The velocity field is constant in all our models, and we assume that neither turbulent nor magnetic pressure dominates the total pressure of the cloud. Assuming an ideal gas, its equation of state is given by $P = nk_\mathrm{B}T$, with $n$ the total gas density (note that this differs from the previously defined $n_{\rm H}$, which refers to the total hydrogen density; $n \propto n_{\rm H}/\mu$ with $\mu$ the mean molecular weight). The polytropic exponent can then be expressed in terms of density and gas temperature, after taking the logarithm and differentiating:

$$\gamma_{\rm p} = 1 + \frac{{\rm d}\log T}{{\rm d}\log\rho} - \frac{{\rm d}\log\mu}{{\rm d}\log\rho}. \tag{4.3}$$

In previous studies, the change in molecular weight is nearly always neglected. We have checked this assumption and found it does not lead to an appreciable change in $\gamma_{\rm p}$, meaning it is justified to neglect $\mu$ in the expression for the polytropic exponent:

$$\gamma_{\rm p} = 1 + \frac{{\rm d}\log T}{{\rm d}\log\rho}. \tag{4.4}$$

For the sake of completeness, we will nonetheless include the mean molecular weight in our calculations.

To understand how fragmentation behavior is affected by changes in the polytropic exponent, we can express the Jeans mass in terms of $\gamma_{\rm p}$:

$$M_{\rm J} = \left(\frac{5k_{\rm B}T}{G\mu m_{\rm H}}\right)^{3/2}\left(\frac{3}{4\pi\rho}\right)^{1/2}, \tag{4.5}$$

$$\propto \left(\frac{T^3}{\mu^3\rho}\right)^{1/2}, \tag{4.6}$$

$$\propto \rho^{\frac{3}{2}(\gamma_{\rm p}-\frac{4}{3})}. \tag{4.7}$$

When $\gamma_{\rm p} < 1$, and thus temperature decreases with increasing density, the Jeans mass decreases, favoring the occurrence of fragmentation. When $\gamma_{\rm p} = 1$ the gas is isothermal, and when $1 < \gamma_{\rm p} < 4/3$ the temperature increases with density; under these circumstances the Jeans mass is still decreasing, though much more slowly, indicating fragmentation is much less likely. Finally, if $\gamma_{\rm p} > 4/3$, fragmentation is expected to cease entirely. Such behavior, where the degree of fragmentation decreases with increasing polytropic exponent, has also been confirmed in numerical simulations of turbulent gas (Li et al. 2003; Jappsen et al. 2005; Bonnell et al. 2006; Klessen et al. 2007; Hocuk & Spaans 2010; Federrath & Banerjee 2015; Hocuk et al. 2016). Consistently low values of $\gamma_{\rm p}$ are expected to lead to the formation of clusters of low-mass stars, while $\gamma_p > 1$ probably results in the formation of isolated and massive stars.

In Figures 4.26 to 4.29, we show plots of the polytropic exponent, similar to the temperature maps in the previous section. In the white-to-green areas, $\gamma_p < 1$ and fragmentation is likely, while in the pink-to-black areas, $\gamma_p > 1$ and fragmentation is suppressed. The blue, dark green and dark purple areas are the transition regions where $0.95 \leqslant \gamma_p \leqslant 1.05$, so there the gas is isothermal or very close to, and the exactly isothermal points, $\gamma_p = 1$, are connected by a white contour line. The (near-)white areas indicate regions where the polytropic exponent is small and fragmentation is vigorous, while the (near-)black areas indicate regions of large polytropic exponent, $\gamma_p > 4/3$, and here fragmentation is halted. Of particular interest are the points where $\gamma_p$ changes from smaller than 1 to nearly 1 or larger, as here the EOS changes from 'soft' to 'stiff', and the typical fragment scale of the system may be set.

## A RADIATION FIELD STRENGTH

**LOCAL UNIVERSE** In Figure 4.26, we show the mass-weighted polytropic exponent in the bulk of the cloud, at column densities exceeding $10^{22}$ cm$^{-2}$, as a function of total hydrogen number density $n_H$ (note that $\rho$ is directly proportional to $n_H$) and radiation field strength, at redshift $z = 0$ and cosmic ray ionization rate $\zeta = 10^{-17}$ s$^{-1}$.

For solar metallicity, there is a significant part of parameter space where the EOS is stiff, around a density of $10^4$-$10^6$ cm$^{-3}$, indicating that fragmentation is likely limited at such densities, and gas may tend to linger in that phase. These are precisely the kind of densities that are observed for molecular cloud cores. There is a clear difference between the $W_0 \leqslant 10^{-1}$ cases and the $W_0 > 10^{-1}$ cases. If irradiated by a weak UV field, the EOS stiffens briefly around a density of 1-10 cm$^{-3}$, and then softens again until $10^5$ cm$^{-3}$, where the gas becomes very close to isothermal due to gas-dust coupling ($T \simeq T_{dust}$ is here indicated by the blue area). For radiation fields of $W_0 = 1 - 10^3$, again the EOS stiffens briefly around 1-10 cm$^{-3}$, softens again, and finally begins to stiffen once more at $\sim 10^2$-$10^3$ cm$^{-3}$. For these UV fields, contrary to weaker fields, the EOS stiffens considerably before gas-dust coupling sets in. For the very strongest radiation fields, $W_0 > 10^3$, the EOS stiffens for the first time at $\sim 10^2$-$10^3$ cm$^{-3}$, while at lower densities the EOS is rather soft and significant fragmentation can be expected there. While the densities where we obtain a stiff EOS does not differ much between a local mean interstellar radiation field and a very strong field with $W_0 = 10^5$, such as found e.g. in starburst regions, the temperature is higher for intenser radiation and thus a more top-heavy IMF could be expected.

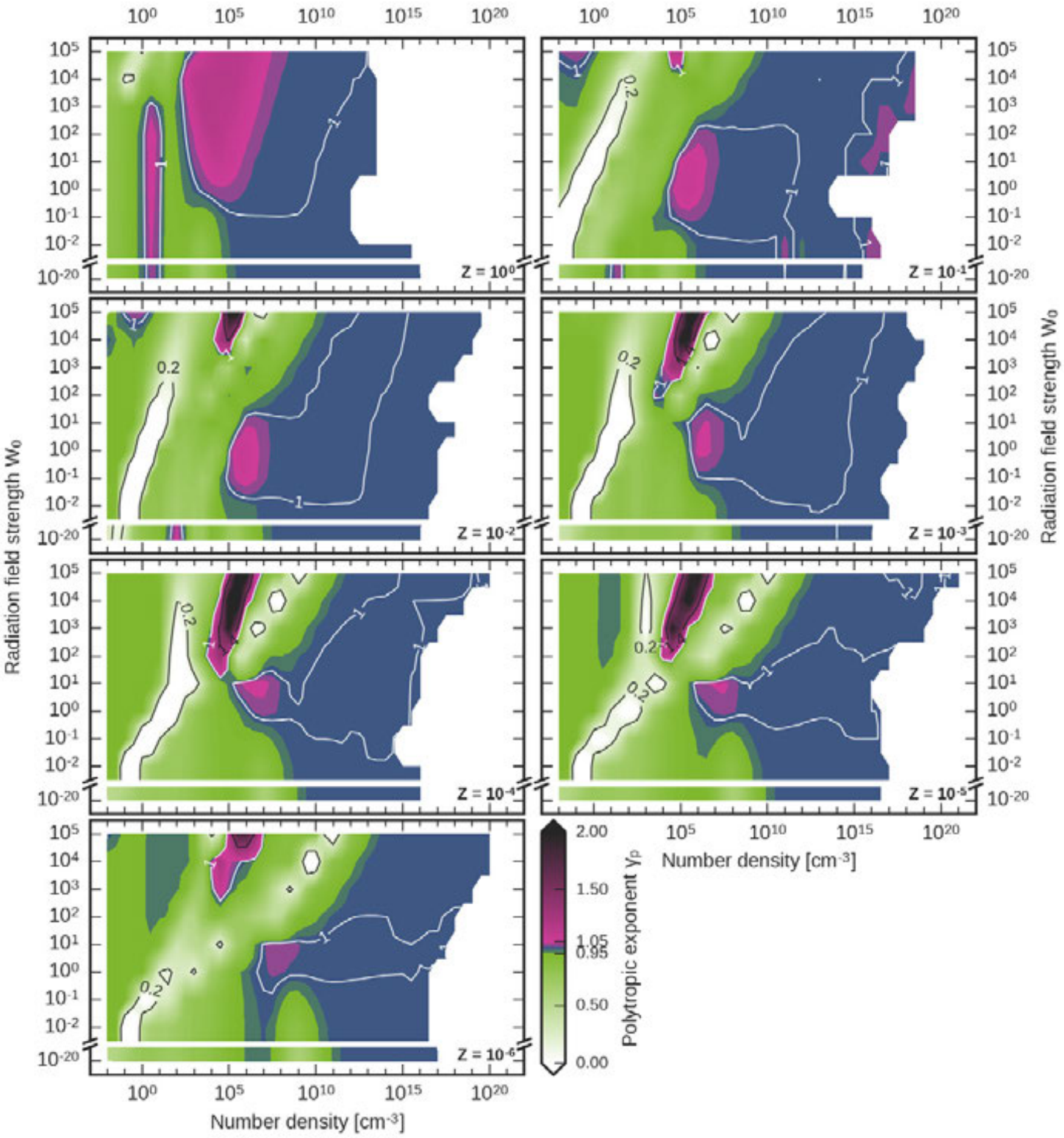

**Figure 4.26** – Mass-weighted polytropic exponent in the bulk of the cloud ($N_H \geqslant 10^{22}\,\mathrm{cm}^{-2}$) as a function of number density and radiation field strength, at redshift $z = 0$ and cosmic ray ionization rate $\zeta = 10^{-17}\,\mathrm{s}^{-1}$. The subplots show the temperature for different metallicities, starting with solar metallicity in the top left, and ending with $Z = 10^{-6}\,\mathrm{Z}_\odot$ at the bottom.

For metallicities below solar, the EOS for the bulk is quite different from the solar case, with more extreme values for the polytropic exponent. There is a region of very soft EOS at low densities, around $\sim 1\,\mathrm{cm}^{-3}$, disappearing only for the strongest radiation fields. On the other hand, there is also a region of very stiff EOS at somewhat higher densities, around $\sim 10^5\,\mathrm{cm}^{-3}$, occurring only for the stronger radiation fields. However, for a local mean radiation field, $W_0 \sim 1$, and metallicities $Z \gtrsim 10^{-3}\,\mathrm{Z}_\odot$, we still find a stiff EOS around the same densities as for the solar case, $10^4$-$10^6\,\mathrm{Z}_\odot$.

The highest density at which the EOS changes from soft to stiff varies quite a bit for different radiation intensities and metallicities. For $W_0 \gtrsim 100$, this density moves up roughly an order of magnitude for each decade in both radiation field (increasing), and metallicity (decreasing). For $W_0 \leqslant 10^{-1}$, there is also a trend of increasing density with decreasing metallicity, though the trend with radiation field is less clear. For intermediate radiation intensities, the highest density where the EOS stiffens stays approximately the same, $\sim 10^5\,\mathrm{cm}^{-3}$, for the different metallicities.

4

**EARLY UNIVERSE** Figure 4.27 is similar to Figure 4.26, though showing the polytropic exponent for redshift $z = 20$ instead of $z = 0$. In the solar metallicity case, the EOS is now stiff around $10^4$-$10^6\,\mathrm{Z}_\odot$ for all UV field intensities. The disappearance of the radiation field dependence is due to the warm dust at this redshift, setting a higher temperature floor. Because of the increased temperature, a somewhat more top-heavy IMF can be expected also for the weakest radiation fields, provided it is possible for gas to be enriched up to solar metallicity already at this redshift.

For metallicities below solar, the impact of a strong UV background dominates over the effects of a warmer CMB, and the EOS, as well as the absolute temperatures, are similar to those at $z = 0$. At intermediate radiation fields, the stiff EOS region now also reaches all the way down to the weakest intensities, and moves approximately a factor ten up in density for each order-of-magnitude decrease in metallicity, though somewhat less for $Z \leqslant 10^{-3}$ and somewhat more for $Z \geqslant 10^{-4}$.

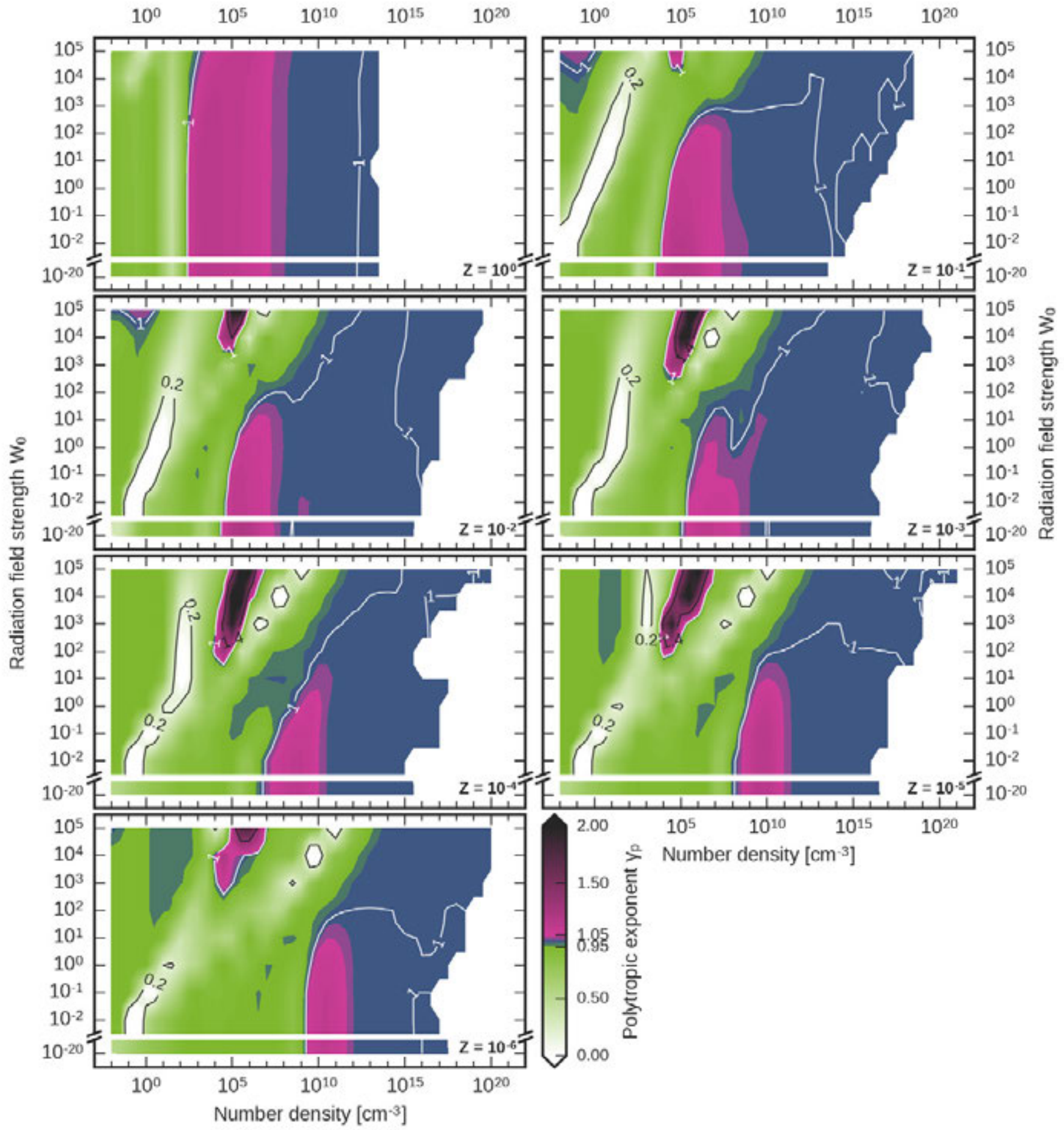

**Figure 4.27** – Mass-weighted polytropic exponent in the bulk of the cloud ($N_{\rm H} \geqslant 10^{22}\,{\rm cm}^{-2}$) as a function of number density and radiation field strength, at redshift $z = 20$ and cosmic ray ionization rate $\zeta = 10^{-17}\,{\rm s}^{-1}$. The subplots show the temperature for different metallicities, starting with solar metallicity in the top left, and ending with $Z = 10^{-6}\,{\rm Z}_{\odot}$ at the bottom.

## B REDSHIFT

Figure 4.28 shows the mass-weighted polytropic exponent in the bulk of the cloud, at column densities exceeding $10^{22}\,\mathrm{cm}^{-2}$, as a function of total hydrogen number density $n_\mathrm{H}$ and redshift, for a local mean radiation field, $W_0 = 1$ and cosmic ray ionization rate $\zeta = 10^{-17}\,\mathrm{s}^{-1}$. For metallicities $Z \gtrsim 10^{-3}\,\mathrm{Z}_\odot$, the density at which the EOS stiffens does not change much over the redshift range considered, and the stiff region stays around $10^5\,\mathrm{cm}^{-3}$, though it does become narrower in density range. However, for metallicities $Z \lesssim 10^{-4}\,\mathrm{Z}_\odot$ and $z < 5$, the EOS goes from soft to near-isothermal, without stiffening significantly.

## C COSMIC RAY IONIZATION RATE

**LOCAL UNIVERSE** Figure 4.29 shows the mass-weighted polytropic exponent in the bulk of the cloud, at column densities exceeding $10^{22}\,\mathrm{cm}^{-2}$, as a function of total hydrogen number density $n_\mathrm{H}$ and cosmic ray ionization rate, for a local mean radiation field, $W_0 = 1$, and redshift $z = 0$. Note that models were computed for just three ionization rates, $10^{-20}$, $10^{-17}$, and $10^{-15}\,\mathrm{s}^{-1}$, so only the overall trends should be taken into consideration. For all metallicities, there seems to be a trend towards a softer EOS for higher CR ionization rate. For metallicities $1 > Z \gtrsim 10^{-3}\,\mathrm{Z}_\odot$, the differences between $\zeta = 10^{-20}$ and $10^{-17}\,\mathrm{s}^{-1}$ are small, while in the case of $\zeta = 10^{-15}\,\mathrm{s}^{-1}$ the EOS gradually moves from soft to (nearly) isothermal, without a phase where $\gamma_\mathrm{p} > 1$, and reaching the isothermal state at a higher density than where the stiffening of the EOS occurs for lower CR ionization rates. For metallicities $Z \lesssim 10^{-4}\,\mathrm{Z}_\odot$ and high CR ionization rates, a small isothermal region appears at lower densities, still followed by a soft region before gas-dust coupling is reached. This results in a significant difference between the highest density with a soft EOS for a low and high cosmic ray ionization rate; e.g. for $Z = 10^{-6}\,\mathrm{Z}_\odot$, this density is $\sim 10^5\,\mathrm{cm}^{-3}$ for $\zeta = 10^{-20}\,\mathrm{s}^{-1}$, and $\sim 10^{12}\,\mathrm{cm}^{-3}$ for $\zeta = 10^{-15}\,\mathrm{s}^{-1}$.

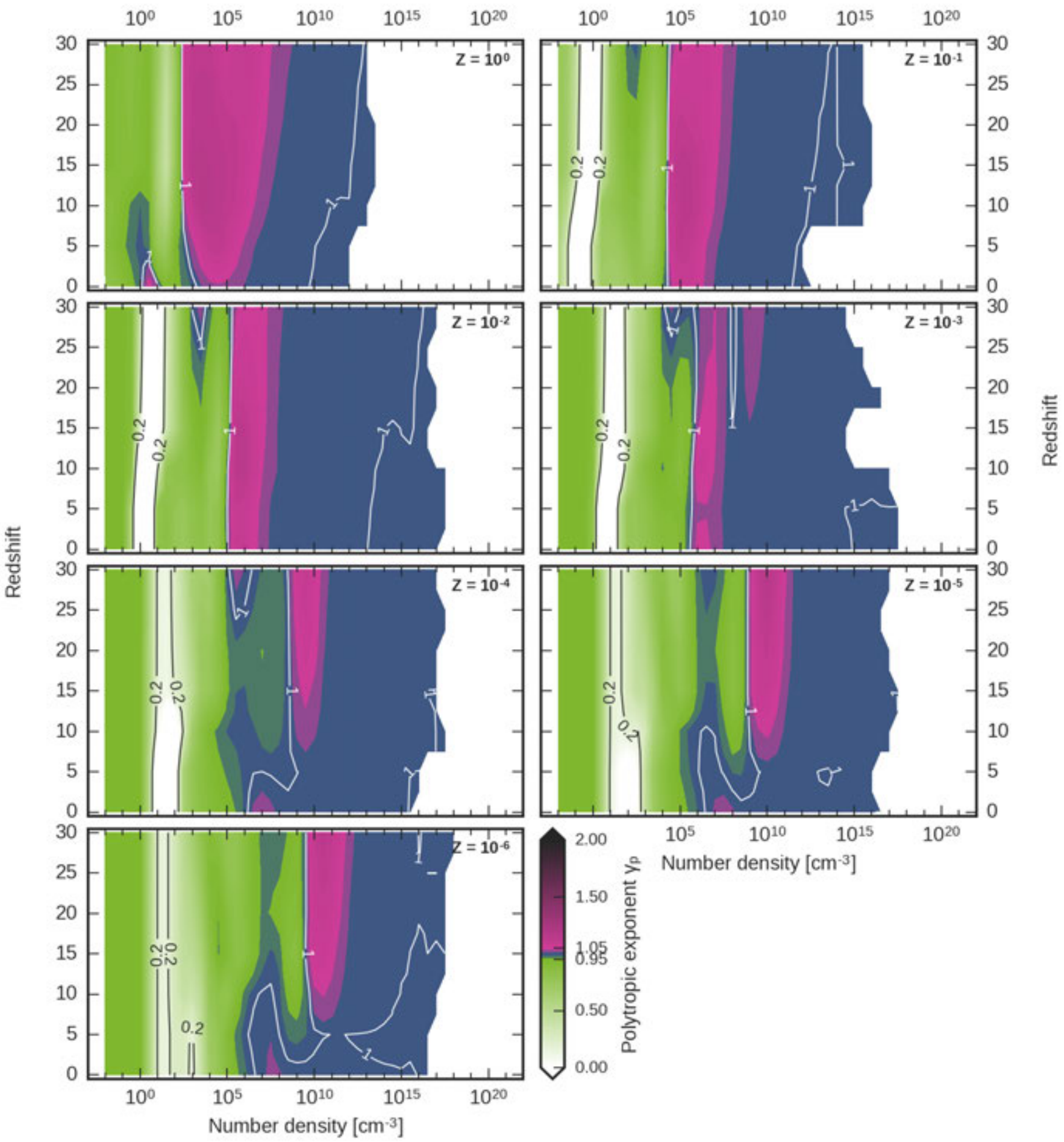

**Figure 4.28** – Mass-weighted polytropic exponent in the bulk of the cloud ($N_\mathrm{H} \geqslant 10^{22}\,\mathrm{cm}^{-2}$) as a function of number density and redshift, at radiation field strength $W_0 = 1$ and cosmic ray ionization rate $\zeta = 10^{-17}\,\mathrm{s}^{-1}$. The subplots show the temperature for different metallicities, starting with solar metallicity in the top left, and ending with $Z = 10^{-6}\,\mathrm{Z}_\odot$ at the bottom.

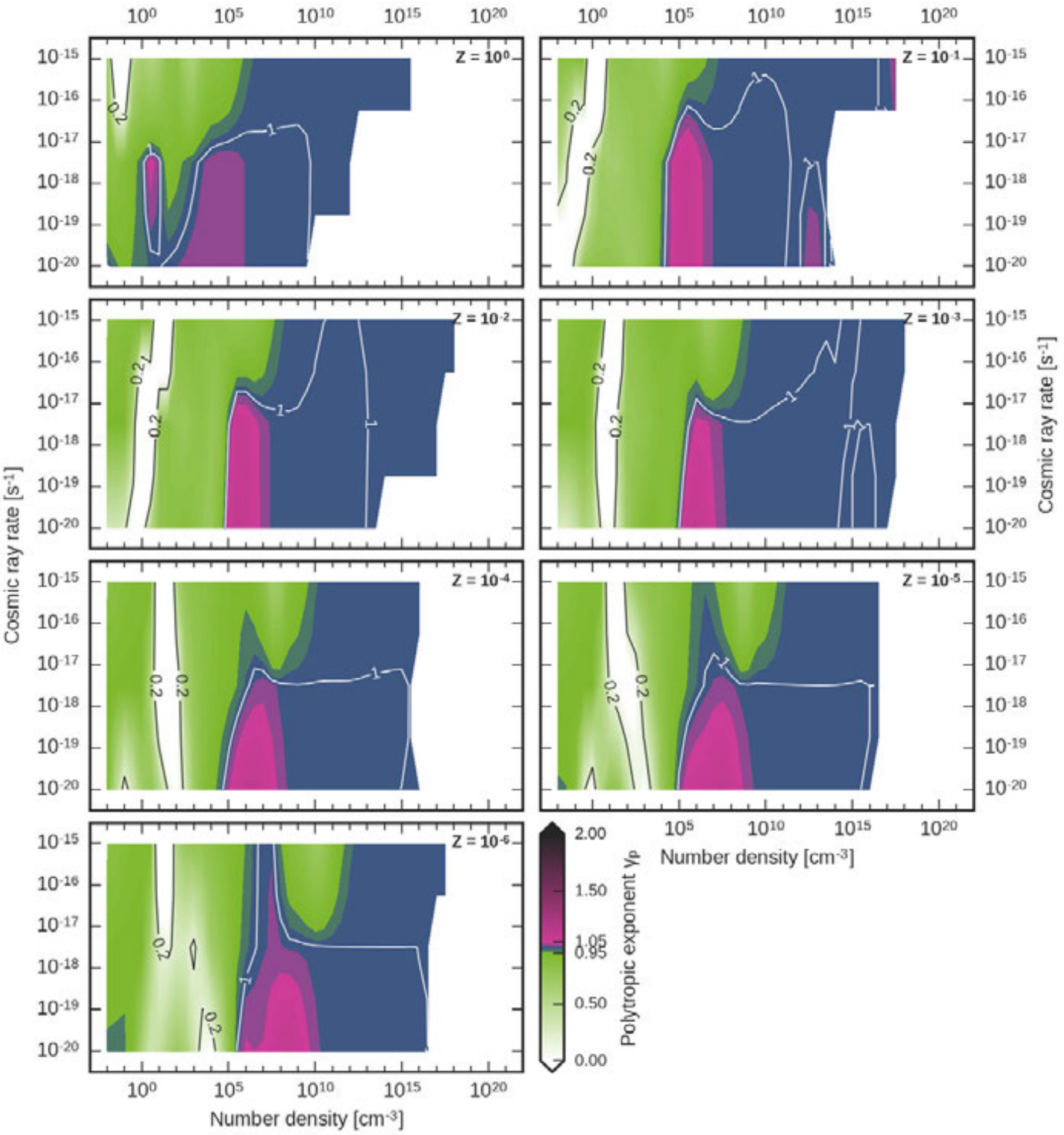

**Figure 4.29** – Mass-weighted polytropic exponent in the bulk of the cloud ($N_H \geqslant 10^{22}\,cm^{-2}$) as a function of number density and cosmic ray ionization rate, at radiation field strength $W_0 = 1$ and redshift $z = 0$. The subplots show the temperature for different metallicities, starting with solar metallicity in the top left, and ending with $Z = 10^{-6}\,Z_\odot$ at the bottom.

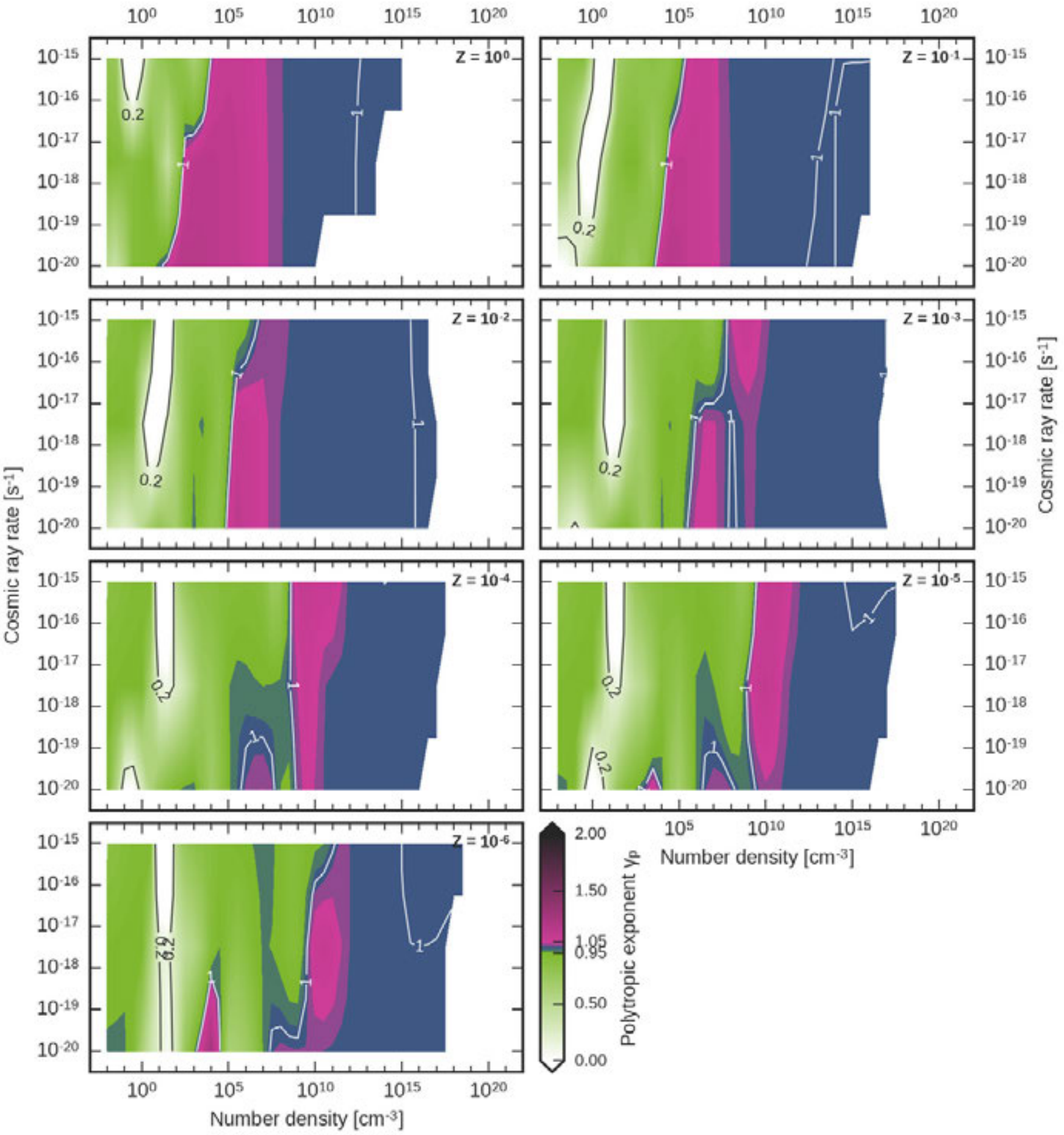

**Figure 4.30** – Mass-weighted polytropic exponent in the bulk of the cloud ($N_H \geqslant 10^{22}\,\mathrm{cm}^{-2}$) as a function of number density and cosmic ray ionization rate, at radiation field strength $W_0 = 1$ and redshift $z = 20$. The subplots show the temperature for different metallicities, starting with solar metallicity in the top left, and ending with $Z = 10^{-6}\,Z_\odot$ at the bottom.

**EARLY UNIVERSE** Figure 4.30 is similar to Figure 4.29, though showing the polytropic exponent for redshift $z = 20$ instead of $z = 0$. At such high redshifts, the effect of the cosmic ray ionization rate on the stiffness of the EOS is much milder than at $z = 0$ due to the mediating effect of the warm dust, and the density where the EOS stiffens changes at most by a factor of 100.

## 4.4 DISCUSSION

One thing to note is that the overall change of temperature structure in the bulk, and thus the EOS, from solar metallicity to $Z = 10^{-1}\,Z_\odot$ is much more dramatic than the change between further metallicity steps. This occurs because at solar metallicity, the bulk gas has become optically thick to the impinging FUV radiation, while at $Z = 10^{-1}\,Z_\odot$ and lower metallicity, the bulk gas is still affected significantly by the FUV photons. This is consistent with what was found by Norman & Spaans (1997).

A defining feature we find in molecular gas typical for the local Universe, so at $z = 0$, $\zeta = 10^{-17}$, with metallicities of $Z \geq 10^{-3}\,Z_\odot$, and irradiated by a UV field with $W_0 \sim 1$, is a stiffening of the EOS around a density of $10^4$-$10^6\,\mathrm{cm}^{-3}$. This means that gas at these densities is not very compressible and less likely to fragment. It is interesting to note that these are the same densities observed for dense cores inside molecular clouds in the local Universe. Similar results were found from time-dependent (not steady-state) models of collapsing clouds (e.g. Omukai et al. 2005). For solar metallicity, gas with a density of $10^5\,\mathrm{cm}^{-3}$ has a temperature of ~8.5 K and a mean molecular weight of 2.3. Using Equation (4.5), this results in a Jeans mass of $\sim 1.5\,M_\odot$, which roughly corresponds to the observed typical mass of the dense structures inside molecular clouds. For $Z = 10^{-3}\,Z_\odot$, the temperature is about a factor two higher at the relevant densities, meaning the Jeans mass will be a factor ~3 larger. For starburst-like conditions, $Z \sim Z_\odot$, $W_0 = 10^5$, and $\zeta = 10^{-15}\,\mathrm{s}^{-1}$, for example in NGC 253 (e.g. Rieke et al. 1980) and Arp 220 (e.g. Downes & Solomon 1998), the temperature is even ~4 times higher, meaning the Jeans mass will be 8 times larger. The final mass of the newborn star critically depends on the accretion rate and on how long accretion can continue, which may to a large extent be regulated by radiative feedback from the protostar itself.

We find that the warmer CMB towards higher redshifts leads to a trend of somewhat stiffer equations of state. Conversely, a higher cosmic ray ionization rate leads to a significantly softer EOS, though the differences become smaller at higher redshift. Since the value of the CR ionization rate at high redshift is still largely unknown, it may be helpful to know that its precise value is actually relatively unimportant for the thermal evolution of the gas.

If the clouds are not static or loitering, but in the process of collapsing, an additional heating source will enter the thermal balance, due to adiabatic compression of the gas. Adiabatic heating can be estimated as $nk_B T/t_{ff}$, with $t_{ff} = (3\pi/(32G\rho))^{0.5}$ the collapse timescale for pressure-free collapse (pure free-fall). Figure 4.1 shows the ratio of adiabatic heating to total heating at $z = 0$ for $\zeta = 10^{-17}\,s^{-1}$; green indicates the regime where adiabatic heating will not be important and our results are robust. For metallicities $\gtrsim 10^{-3}\,Z_\odot$, photoelectric heating and gas-grain heating will dominate over adiabatic heating until well into the gas-dust coupling regime, at which point we expect the major fragmentation episodes to have occurred already. For lower metallicities, adiabatic heating will become important already at lower densities, and will likely affect the EOS. A warmer CMB will not significantly change this picture, as the additional heat input is modest compared to adiabatic heating. However, the heating resulting from a cosmic ray ionization rate of $10^{-15}\,s^{-1}$ will dominate over adiabatic heating up to e.g. $\sim 10^7\,cm^{-3}$ for $Z = 10^{-6}\,Z_\odot$, which can be inferred from Figure 4.2; so our results again become more robust. For all redshifts, such a high CR ionization rate leads to an overall softer EOS; see the appendix, Figures 4.A.1 and 4.A.2.

Note that we did not include opacity in the dust cooling function. The bulk is expected to become opaque to its cooling radiation and evolve adiabatically above densities of $\sim 10^{10}\,cm^{-3}$ for solar metallicity and $\sim 10^{16}\,cm^{-3}$ for $Z = 10^{-6}\,Z_\odot$, far above the densities where we expect the fragmentation scale to be set.

We chose not to include X-rays in our models, as we do not expect them to be of importance in the majority of the ISM volume. X-rays are mostly produced by accreting black holes, and the X-ray flux drops off like $\propto 1/r^2$; this means that for a $\sim 10^8\,M_\odot$ black hole accreting at a significant fraction of the Eddington rate, the heating and ionization impact becomes negligible beyond ~300 pc. Additionally, even though X-rays heat more efficiently than cosmic rays, their ionization effect is roughly mimicked by the CRs. Detailed models and 3D simulations using realistic models for the X-ray background, that also account for the simultaneous growth of the soft UV background, show that star formation is relatively insensitive to the X-rays (Glover & Brand 2003; Machacek et al. 2003; Hummel et al. 2015; Latif et al. 2015).

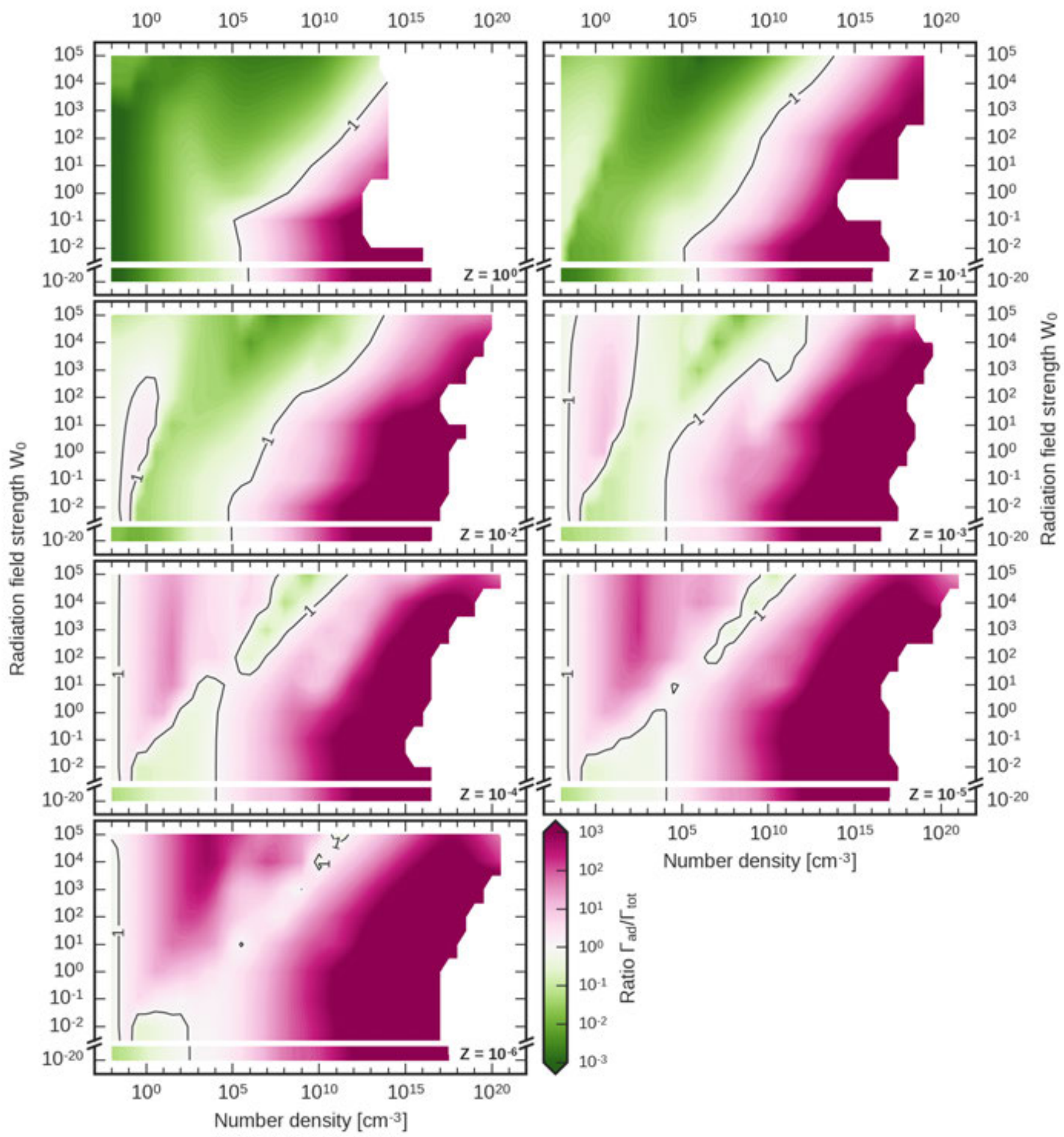

**Figure 4.1** – Ratio of adiabatic heating to total heating in the bulk of the cloud ($N_{\rm H} \geqslant 10^{22}\,{\rm cm}^{-2}$) as a function of number density and radiation field strength, at redshift $z = 0$ and cosmic ray ionization rate $\zeta = 10^{-17}\,{\rm s}^{-1}$. The subplots show the temperature for different metallicities, starting with solar metallicity in the top left, and ending with $Z = 10^{-6}\,{\rm Z}_{\odot}$ at the bottom.

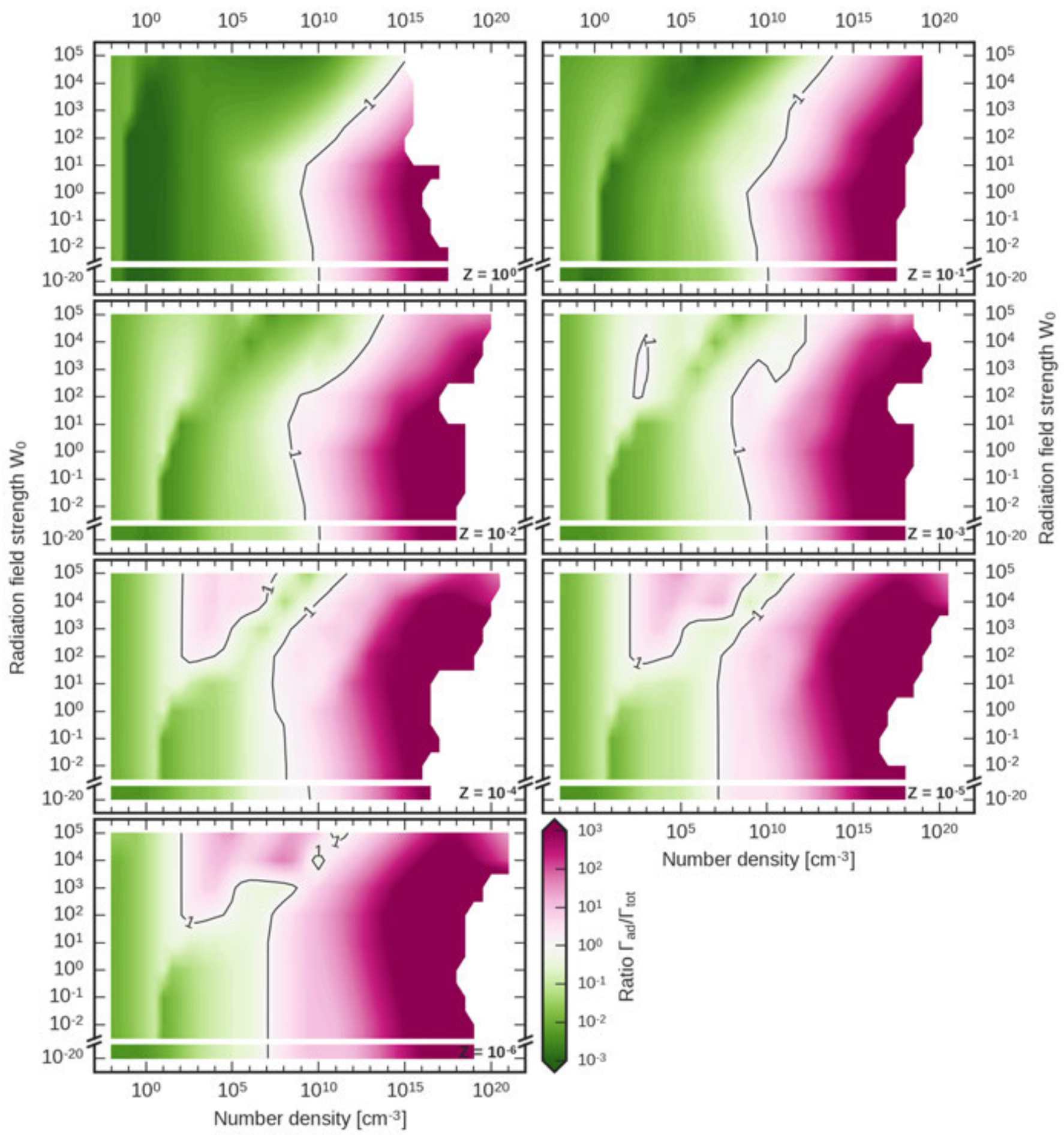

**Figure 4.2** – Ratio of adiabatic heating to total heating in the bulk of the cloud ($N_\mathrm{H} \geqslant 10^{22}\,\mathrm{cm}^{-2}$) as a function of number density and radiation field strength, at redshift $z = 0$ and cosmic ray ionization rate $\zeta = 10^{-15}\,\mathrm{s}^{-1}$. The subplots show the temperature for different metallicities, starting with solar metallicity in the top left, and ending with $Z = 10^{-6}\,\mathrm{Z}_\odot$ at the bottom.

One of the limitations of the model is the use of the steady-state approximation (often also termed an ‘equilibrium’ model), meaning that we consider the state of the system on a timescale longer than the timescale of the processes involved in setting the chemical and thermal equilibrium.

Thus, results cannot be used to make reliable predictions for rapidly evolving systems, for example, in regions where a hot O or B star has just switched on and an ionization front travels into the cloud, or when the radiation field varies on short timescales (see e.g. Bertoldi & Draine 1996; Störzer & Hollenbach 1998).

## 4.5 CONCLUSIONS

We have run a grid of 1D models with our numerical code PDR-Zz, using a steady-state approach. We have provided insight in the most important processes and dominant chemical species in different regimes. We have show the effects of radiative and chemical feedback on the effective gas temperature both at the surface and in the bulk of the cloud. And finally, we have examined the equation of state in different parts of the grid, and related it to the IMF.

We find that the main physical processes that set the thermodynamics of the ISM and affect the behavior of the EOS are the following:

- both the photoelectric heating, which is the main heating channel in many (but not all) cases for gas that is optically thin to the UV continuum and has not yet reached the gas-dust coupling regime, and the net cooling rate scale roughly with metallicity, resulting in an overall similar EOS for $Z \leq 10^{-1}\,Z_{\odot}$;
- gas-dust coupling always takes over at some density if there is any dust present, though the scaling of this density with radiation intensity is non-trivial; and
- the CMB sets a temperature floor for the dust grains, though the gas can cool somewhat below this floor at densities where gas-dust coupling has not yet been reached.

We also note that a correct prescription of $H_2$ collisional de-excitation heating and the $H_2$ formation and destruction rates are important to accurately describe the gas temperature in regions where the optical depth to the UV continuum is low, e.g. the surface of a solar metallicity PDR or the bulk of a lower metallicity PDR.

We have identified several regions in EOS parameter space that should prove interesting to explore further using detailed hydrodynamical simulations of the ISM. These points of interest are

1. the transition in EOS behavior around $10^{-1}\,Z_{\odot}$, since such metallicities are relevant for the bulk of star-forming galaxies around the peak in the star formation history of the Universe;
2. the starburst-like conditions, with solar or super-solar metallicities, and $W_0 \geq 10^5$, which give rise to a stiffer EOS and higher temperatures, resulting in a more top-heavy IMF; and
3. the relatively soft EOS for low metallicities, which implies moderate to strong fragmentation, but which is also coupled with higher gas temperatures, possibly resulting in higher accretion rates and a larger final mass for the star.

We envision that through a combination of detailed models with a simple geometry as the ones described here, and 3D models that can account for dynamical effects, an understanding of the complex physics giving rise to the IMF can be obtained.

# APPENDICES

## 4.A EQUATION OF STATE FOR HIGH $\zeta$

4

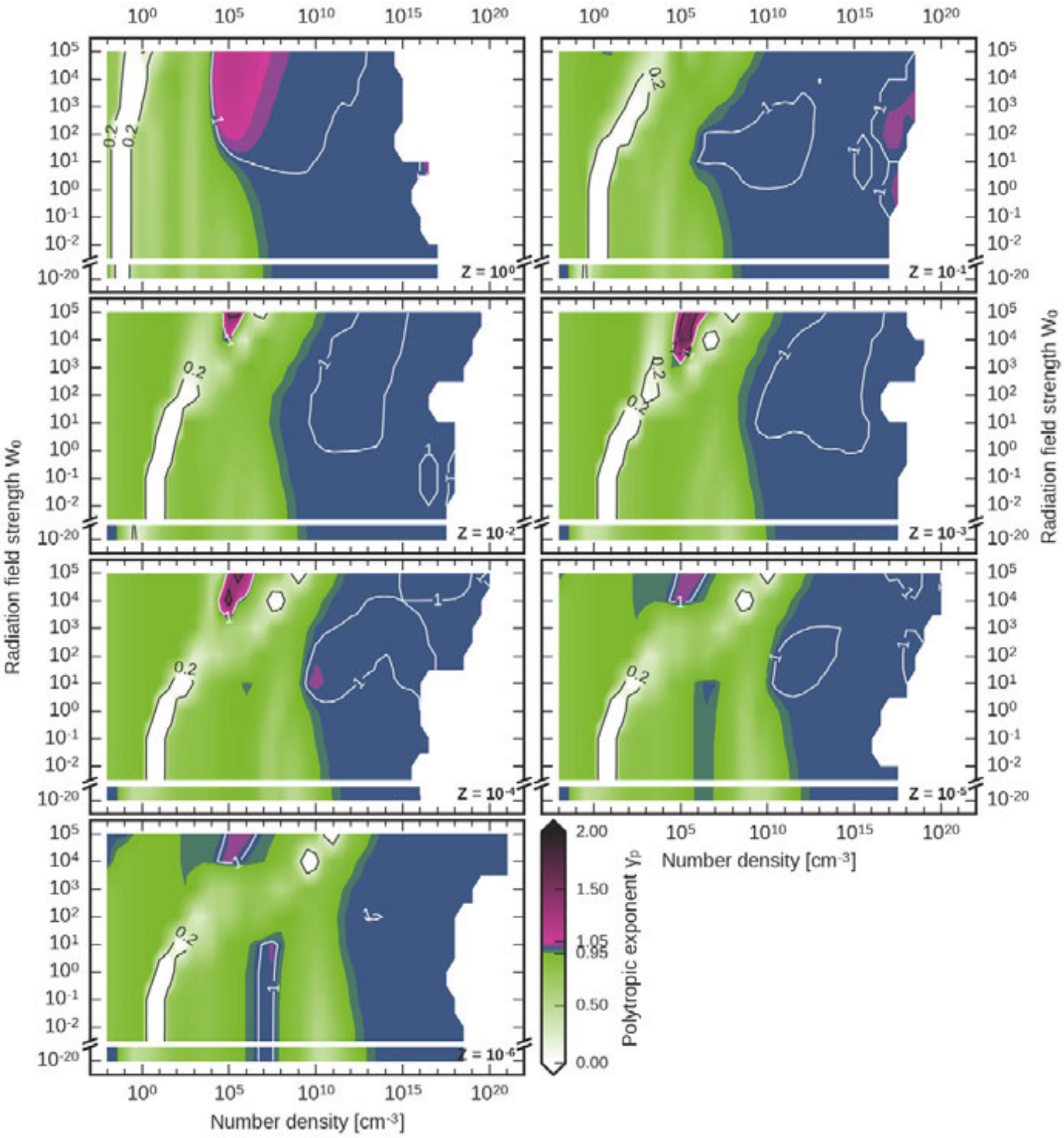

**Figure 4.A.1** – Mass-weighted polytropic exponent in the bulk of the cloud ($N_H \geqslant 10^{22}$ cm$^{-2}$) as a function of number density and radiation field strength, at redshift $z = 0$ and cosmic ray ionization rate $\zeta = 10^{-15}$ s$^{-1}$. The subplots show the temperature for different metallicities, starting with solar metallicity in the top left, and ending with $Z = 10^{-6}$ Z$_\odot$ at the bottom.

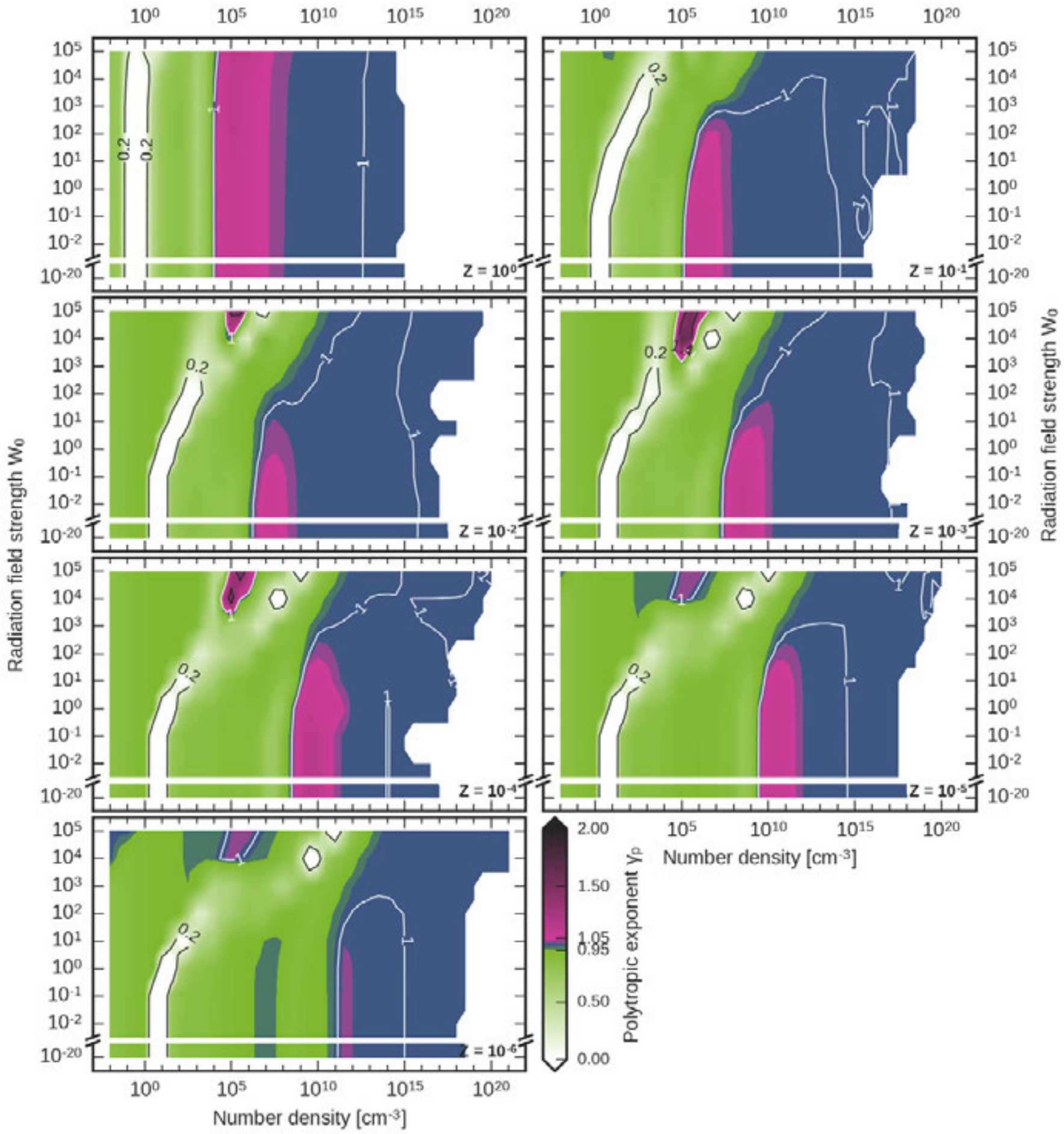

**Figure 4.A.2** – Mass-weighted polytropic exponent in the bulk of the cloud ($N_\mathrm{H} \geqslant 10^{22}\,\mathrm{cm}^{-2}$) as a function of number density and radiation field strength, at redshift $z = 20$ and cosmic ray ionization rate $\zeta = 10^{-15}\,\mathrm{s}^{-1}$. The subplots show the temperature for different metallicities, starting with solar metallicity in the top left, and ending with $Z = 10^{-6}\,\mathrm{Z}_\odot$ at the bottom.

*WE WILL NEVER BE THE SAME AGAIN. BUT HERE'S A LITTLE SECRET FOR YOU: NO ONE IS EVER THE SAME THING AGAIN AFTER ANYTHING. YOU ARE NEVER THE SAME TWICE, AND MUCH OF YOUR UNHAPPINESS COMES FROM TRYING TO PRETEND THAT YOU ARE. ACCEPT THAT YOU ARE DIFFERENT EACH DAY, AND DO SO JOYFULLY, RECOGNIZING IT FOR THE GIFT IT IS. WORK WITHIN THE DESIRES AND GOALS OF THE PERSON YOU ARE CURRENTLY, UNTIL YOU AREN'T THAT PERSON ANYMORE, AND EVERYTHING CHANGES ONCE AGAIN.*

— WELCOME TO NIGHT VALE
(EPISODE 75 - THROUGH THE
NARROW PLACE)

CHAPTER

# 5

# OUTLOOK

## 5.1 THEORY AND SIMULATIONS

There are still several topics that could be explored using our PDR-Zz code. One could vary the turbulent velocity, which would affect the shielding of e.g. $H_2$ and CO, and the optical depth for the cooling lines, and investigate the impact on the thermal balance and the EOS. Also interesting would be to evaluate the impact of abundance ratios that differ from the solar values and different dust-to-gas mass fractions, as the element production in the first supernovae may be quite different from local values (e.g. Cayrel et al. 2004; Spaans & Silk 2005). Additionally, switching on a simple mechanical heating term (already present in the code) will allow for estimating the impact of mechanical energy input through shocks as a result of supernova events or gas outflows (e.g. Loenen et al. 2008; Kazandjian et al. 2012, 2015).

Some improvements to the code are also still possible. For example, the addition of continuum opacity to the gas-dust collisional cooling process, and the addition of adiabatic heating which would result from cloud collapse, as discussed in Section 4.4. Of course, then an estimate must be made for the timescale on which the collapse is occurring. The addition of X-ray processes, most importantly X-ray heating, would make it possible to also consider matter in the vicinity of an accreting (seed) black hole (e.g. Meijerink & Spaans 2005; Meijerink et al. 2007; Aykutalp et al. 2014).

The grid presented in Chapter 4 allows one to identify thermodynamic trends in the behavior of the IMF. Of course, to truly understand the origins of the IMF, follow-up 3D (magneto)hydrodynamical simulations in specific parts of the grid's parameter space (metallicity, redshift, radiation field and cosmic ray ionization rate) are required to assess the full impact of both thermodynamics and collapse dynamics. The regions of interest for the EOS that we have identified can provide a starting point for these detailed simulations.

Simulating the interstellar medium is computationally challenging, as many complex processes have to be taken into account. Aside from chemical reactions and cooling and heating processes, also turbulence, magnetic fields, and feedback from stars and black holes, including radiative transfer, are important elements. Since the relevant spatial scales and timescales both span a huge range, the cost of calculating all involved processes on all relevant scales becomes prohibitively expensive, and some sacrifices must be made.

Simulations of the ISM with supersonic turbulence, thought to be important in defining the IMF, often assume isothermal gas. This can be significantly improved by using a more realistic EOS like the ones we have presented, as it has been found from previous simulations that the EOS strongly affects the density probability density function, and thus the IMF (Klessen et al. 2007; Hocuk & Spaans 2010; Federrath & Banerjee 2015; Hocuk et al. 2016).

A somewhat better approach would be to extract cooling and heating tables from our model and couple these to detailed 3D simulations, as interpolating from a table is much faster than evaluating the rates at runtime. In this way, heating and cooling resulting from dynamical processes, like the dissipation of turbulence and ambipolar diffusion, can be taken into account as well, though it is still assumed that chemical equilibrium can be reached on a simulation time step. If one wants to do simulations exploring the growth of seed black holes, of course it will be important to also include X-ray feedback.

Moving away from an equilibrium approach and calculating everything in a time-dependent manner at runtime would be preferable, though at the moment this is only feasible for much smaller reaction networks than what we have used in our code, and assumptions about e.g. the initial conditions will need to be made (e.g. Bovino et al. 2016; Richings & Schaye 2016).

## 5.2 OBSERVATIONS

Although a lot of progress has been made from a theoretical perspective, more observational evidence of the first stars and black holes will be required to properly constrain the models. Observations of CR7, the brightest Ly$\alpha$ emitter at $z = 6.6$ known to date, suggest that it might host a combination of Pop III stars and a more normal stellar population (Sobral et al. 2015), which may be the first indisputable evidence for Pop III stars in a distant galaxy. Alternatively, it may instead be powered by accretion on to a direct collapse black hole (Pallottini et al. 2015). In any case, the observed properties of CR7 indicate that this galaxy is most likely powered by sources formed from pristine gas.

The X-ray radiation from massive ($>10^5\,M_\odot$), accreting black holes excites strong emission lines of $H_2$ and CO, which may be observable with ALMA (Atacama Large Millimeter Array) at redshifts $z = 5-20$, provided they radiate close to the Eddington limit (Spaans & Meijerink 2008). Such detections would provide information about the conditions in the ambient gas.

Dijkstra et al. (2016) suggest that the direct collapse scenario for the formation of a black hole seed provides optimal conditions for 3-cm fine structure maser action. The signal of the redshifted 3-cm emission line may be detectable with ultra-deep surveys being planned with SKA1-MID, and could provide direct evidence for the direct collapse scenario.

Pacucci et al. (2016) propose that a seed black hole can be characterized by a steep spectrum in the infrared (1.6-4.5 μm), and found two candidate objects in the CANDELS/GOODS-S survey with $z > 6$. Upcoming space-based observatories like JWST may make it possible to find many more.

Also the observation of gravitational waves from a directly collapsing black hole has been proposed, though it may be difficult to separate the expected weak signal from the background and foreground noise (Pacucci et al. 2015). Mergers of the compact remnants of Pop III stars may also be observable through their gravitational wave signal, which may help to constrain their IMF (Stacy et al. 2010; Kinugawa et al. 2014; Hartwig et al. 2016).

It will probably not be possible to probe the complete functional form of the Pop III IMF any time soon. Concerning the high-mass end, stars with a mass in the range ~140-260 $M_\odot$ are predicted to end their lives as hypernovae or very energetic pair-instability supernovae. Those events would be bright enough to be observed by the JWST (James Webb Space Telescope) out to high redshift (Hummel et al. 2012; Pan et al. 2012). If the mass and rotation are right, a star collapsing into a black hole may emit a gamma-ray burst, which could be observable as well (Bromm & Loeb 2006; de Souza et al. 2011). Additional constraints can be provided by near-field cosmology or 'stellar archeology': examining chemical abundance patterns of extremely old, metal-poor stars in our local Universe. Such stars are presumed to be part of the second generation of stars and to have been enriched by Pop III SNe (e.g Karlsson et al. 2013).

More observational evidence regarding the shape of the IMF in different environments is also required, to pin down to which extent various processes are responsible for its shape, and to constrain variations with e.g. metallicity and radiation field. Particularly, we need more observations to determine the IMF in extreme environments, such as ULIRGs (Ultra-Luminous InfraRed Galaxies), which may be more common at high redshift, as well as in modest-metallicity dwarf galaxies.

*AND NOW FOR SOMETHING COMPLETELY DIFFERENT.*

— MONTY PYTHON

# THE END

# SUPPLEMENTAL MATTER

# SUMMARY (ENGLISH)

## COSMOLOGY

According to the Big Bang theory, about 13.8 billion years ago the Universe began as almost inconceivably hot and dense. And ever since then, it has expanded and cooled, eventually reaching the cold, tenuous state we see today. For a fraction of a second after the Big Bang, the Universe underwent a brief period of extremely fast expansion, known as inflation. During this process, tiny quantum mechanical fluctuations were enlarged to macroscopic scales and imprinted on the Universe as density fluctuations, which acted as seeds for the formation of structure. Over time, positive fluctuations in the matter distribution grew denser and more massive, giving rise to a highly sub-structured network of filaments, sheets and denser clusters of ordinary and dark matter, separated by immense voids. A few hundred million years after the Big Bang, the density in some of the nodes of this cosmic web had become high enough for the formation of stars and galaxies to become possible. The emergence of the first star brought the first source of light to the Universe and ended the previous Dark Ages.

The formation of structure in the Universe likely happened in a 'bottom-up' scenario, also called 'hierarchical' structure formation. Small overdensities are able to overcome the cosmological expansion and collapse first, and the resulting dark matter 'halos' then merge to form increasingly larger halos, which become the sites of galaxy formation.

At present, the complex physics associated with galaxy formation and evolution is still not well understood. Major ingredients to the recipe for galaxy assembly are the feedback processes from stars and black holes, which shape the interstellar medium (ISM) inside galaxies, or inside their smaller progenitor systems, out of which new generations of luminous objects will form.

## FIRST STARS

The very first generation of stars, formed out of the pristine primordial gas (consisting almost purely of hydrogen and helium; also called zero-metallicity gas[1]), are often called Population III or Pop III stars. Because the cooling processes in primordial gas are so different from solar metallicity gas (like most of the gas in the Milky Way disk today), the fragmentation behavior, and thus the star formation mode, may be different from what we see today. In particular, Pop III stars may have been typically more massive.

The first stars fundamentally transformed the early Universe through their radiative, chemical, and mechanical feedback. They were the first sources of hydrogen-ionizing photons, beginning the process of reionizing the Universe, and of photons capable of dissociating molecular hydrogen. They produced the first metals and also the first cosmic dust grains. Large amounts of energy were injected into the environment through stellar winds or when a star died in a supernova explosion (SN), creating shocks and turbulence, and possibly expelling gas out of the host halo. Additionally, the first stars may have been the sites where dynamically significant magnetic fields were created for the first time. And of course, following the epoch of the first stars, further generations of stars continue to shape galaxies through their feedback processes.

[1] The 'metallicity' of a gas is the fraction that is not in hydrogen or helium. Since most of the ordinary matter in the Universe is in the form of hydrogen and helium, astronomers use the word 'metals' as a shorthand to indicate all chemical elements heavier than hydrogen and helium.

## ORIGINS OF THE IMF

Stars form out of cold, molecular gas, which is found in molecular clouds in the interstellar medium, the gas and (in the non-primordial case) dust present in a (proto)galaxy that is not incorporated in stellar systems. Molecular clouds are embedded in regions of more tenuous and predominantly neutral, atomic gas, with a gradual transition in between. The presence of far-ultraviolet radiation (e.g. from stars) will significantly affect the chemistry and thermal balance of this gas, and such regions are then called photodissociation or photon-dominated regions (PDRs). In terms of physical and chemical processes, essentially most of the ISM is in PDRs. Far-UV radiation traveling into a PDR becomes gradually attenuated, and ionized species first give way to neutral atomic species, while deeper into the region molecules become stable, transitioning into a molecular cloud.

The initial mass function (IMF) and typical mass of a star are likely related to the thermodynamics of the gas. Numerical simulations have shown that the amount of fragmentation is very sensitive to the exact temperature-density relation in collapsing clouds. Additionally, systematic variations of the present-day IMF have been observed in extreme starburst environments and massive elliptical galaxies. Therefore, in order to determine the IMF of the first, as well as later generations of stars, it is vital to understand the chemical and thermodynamical properties of the feedback-regulated ISM.

## FIRST MASSIVE BLACK HOLES

Our current picture of galaxy formation and evolution is not complete without black holes. Evidence suggest that almost all massive galaxies contain a supermassive black hole (SMBH, with masses exceeding $10^5$ $M_\odot$[2]) at their center. Various correlations between the black hole mass and the properties of the host galaxy point to a common root or co-evolution between galaxies and their central black hole. Also black holes can be responsible for feedback effects, including radiative feedback, in particular their X-ray emission, and mechanical feedback in the form of winds and jets.

[2] $M_\odot$ or solar mass is a mass unit equal to the mass of the Sun.

Observational evidence suggests that some SMBHs with masses of $10^9\ M_\odot$ already existed less than 1 Gyr after the Big Bang, when the Universe was less than a tenth of its current age. Explaining how SMBHs with such large masses could have assembled so soon after the Big Bang presents quite a challenge. The main questions concern how and when the 'seeds' of these SMBHs formed and how their subsequent growth proceeded. Several pathways leading to the formation of seed black holes have been proposed.

Perhaps the most obvious scenario assumes that SMBHs grow from the remnants of the first stars. Whether or not a Pop III star will end its life as a black hole depends on its mass and angular momentum. However, it is not yet known if there would have been enough, if any, Pop III stars with masses above the threshold for the formation of suitably massive seeds. And even if very massive stars were able to form and collapse into seed black holes, it would be difficult for these (still relatively light) seeds to accrete sufficient mass and grow into a supermassive black hole in the available time.

A more promising pathway is therefore the 'direct collapse' scenario, in which primordial gas in a halo would collapse directly into a single central object, without fragmenting, resulting in massive seed black holes. For such a direct collapse to occur, it is necessary that fragmentation is limited, which is possible if the gas in the halo is kept hot. Hence, the formation of molecular hydrogen must be inhibited so that cooling can occur only through atomic hydrogen, as otherwise $H_2$ cooling will lower temperatures significantly and fragmentation may occur.

There are several plausible mechanisms that can suppress the formation of sufficient $H_2$, either through photodissociation of $H_2$ (by a supercritical ultraviolet (UV) radiation background), or through collisional dissociation of $H_2$ (by dissipation of a sufficiently strong magnetic field, trapping of Ly$\alpha$ photons, or the presence of strong shocks).

A likely outcome of the direct collapse scenario is the continued collapse of some gas to smaller scales in the galactic nucleus. As the gas contracts, it becomes optically thick and cooling photons can no longer escape. If the radiation pressure is strong enough to temporarily balance gravity, the resulting object will be a 'protostar'. Once the protostar has formed, it will accrete and eventually evolve into a seed black hole, which must still grow rapidly to explain the observed high-redshift SMBHs.

# THIS THESIS

## PART I: SUPERMASSIVE BLACK HOLE SEEDS

In the first part of this thesis, we investigate the formation of seeds of the first supermassive black holes through the direct collapse scenario.

Chapter 2 of the thesis explores the formation of a protostar resulting from the collapse of primordial gas in the presence of a strong UV radiation background. Particularly, we investigate the impact of turbulence and rotation on the fragmentation behavior of the gas cloud. As is known from present-day star formation, turbulence plays an important role in angular momentum transport and determining the fragmentation properties of collapsing gas clouds, since it can both locally compress the gas as well as provide additional support against collapse on larger scales. Previous numerical simulations of collapsing gas have shown that fragmentation may also depends on the amount of rotation. We have performed numerical 3D adaptive mesh refinement (AMR) simulations using the ENZO code, with the addition of the KROME package to improve modeling of the chemical and thermal processes. This enables us to simulate the formation of a protostar up to unprecedentedly high central densities of $10^{21}$ cm$^{-3}$ and spatial scales of a few solar radii. We find that the physical properties of the simulated gas clouds become similar on small scales, irrespective of the initial amount of turbulence and rotation. A single bound clump is formed at the end of each simulation, marking the onset of protostar formation. No strong fragmentation is observed by the end of the simulations, regardless of the initial amount of turbulence or rotation, and high accretion rates of a few solar masses per year are found. Given such high accretion rates, a quasistar of $10^5$ $M_\odot$ is expected to form within $10^5$ years, eventually leading to a massive seed black hole.

Other ways in which the gas may be able to stay hot and avoid fragmentation are explored in Appendix 2.B, using an analytical one-zone model. Particularly, we examine the interplay between magnetic fields, turbulence, and a UV radiation background during the gravitational collapse of primordial gas in a halo. We follow the evolution of a cloud of primordial gas from its initial cosmic expansion through turnaround, virialization, and collapse up to a density of $10^7$ cm$^{-3}$. It was found that in halos with no significant turbulence, the critical UV background intensity can be lowered significantly by the presence of either a magnetic field or turbulence (or both), with lower critical values for stronger fields and/or stronger turbulence. The reduction in critical UV intensity is particularly important, since it exponentially increases the number of halos exposed to a supercritical radiation background, and therefore the number of possible formation sites of massive seed black holes.

## PART II: ISM THERMODYNAMICS AND FRAGMENTATION

In the second part of the thesis, we focus on the equilibrium state of interstellar material (gas and dust) under various physical conditions, to better understand the chemo-thermal state and fragmentation behavior of gas in high-redshift galaxies.

In Chapter 3 of the thesis, we describe a photodissociation region (PDR) code. Our code, PDR-ZZ, is based on the PDR code by Meijerink & Spaans (2005), which we updated and significantly improved, and also extended for use in a wider range of physical conditions (e.g. much higher densities, metallicities from essentially 0 to solar, redshifts >0). This computational code can be used to study traditional PDRs, and can additionally be applied to many other situations where one is interested in steady-state gas at different optical depths, under various conditions; from diffuse and molecular clouds in the Milky Way to clouds near AGN and material in galaxies in the high-redshift Universe. The input parameters to the code are the total hydrogen density, the metallicity, the incident UV radiation field strength, the cosmic ray ionization rate, and the redshift; the combination of which defines the coupled chemical and thermal solution for the simulated region. In this chapter, we describe in detail which chemical species, reactions, and rates are used, as well as which cooling and heating processes are included in the model. Additionally, we describe our improvements to the convergence algorithms for both the chemical and thermal balance, which result in a large speedup and enable us to run many models in a relative short time.

As mentioned, feedback processes from stars and black holes shape the interstellar medium out of which new generations of luminous objects form. To understand the properties of these objects, e.g. the stellar initial mass function, it is vital to have a firm grasp of the chemical and thermodynamical properties of the feedback-regulated ISM. In Chapter 4 we explore in a systematic way the overall impact of various feedback effects, both radiative and chemical, on the chemical and thermal balance of the gas in different regimes, since the conditions in the ISM may vary greatly. A grid of models is run using our numerical code PDR-Zz, which provides 1D models using a steady-state approach. The grid covers a sizable range in total hydrogen density, metallicity, redshift, UV radiation field scaling parameter, and cosmic ray ionization rate. We provide insight in the most important processes and dominant chemical species in different regimes. We show the effects of radiative and chemical feedback on the effective gas temperature both in the outer layers and in the bulk of the cloud, and describe the differences. Finally, we examine the equations of state (EOS) in our parameter space, identify several regions of interest, and relate these to the IMF. We find a transition in EOS behavior around a metallicity of 0.1 solar, with an overall similar EOS for the bulk gas for metallicities below 0.1 solar due to the main heating and cooling channels scaling roughly with metallicity in gas optically thin to the UV continuum. For low metallicities, we find a relatively soft EOS, though coupled with the higher gas temperatures this may still lead to massive stars. Additionally, starburst-like conditions give rise to both a stiffer EOS and higher temperatures, resulting in a more top-heavy IMF.

# SAMENVATTING (NEDERLANDS)

## KOSMOLOGIE

Volgens de oerknaltheorie begon het universum ongeveer 13,8 miljard jaar geleden vanuit bijna onvoorstelbaar hoge temperatuur en dichtheid. En sindsdien is het continu geëxpandeerd en afgekoeld, tot het uiteindelijk zijn huidige koude, ijle staat bereikte. Een fractie van een seconde na de oerknal onderging het universum een korte periode van zeer snelle expansie, ook bekend als inflatie. Daarbij werden kleine kwantummechanische fluctuaties vergroot tot macroscopische schalen en ingeprent op het universum als dichtheidsfluctuaties, welke fungeerden als de kiemen voor het vormen van grote-schaalstructuur. In de loop van de tijd groeiden de positieve fluctuaties in de massadistributie in dichtheid en massa, waardoor een zeer gesubstructureerd netwerk van filamenten, wanden en dichte clusters van gewone en donkere materie ontstond, gescheiden door immense leegten. Een paar honderd miljoen jaar na de oerknal was de dichtheid van sommige knooppunten in dit kosmische web hoog genoeg geworden voor het vormen van sterren en sterrenstelsels. Het ontstaan van de eerste ster introduceerde de eerste bron van licht in het universum en beëindigde het voorgaande Donkere Tijdperk.

De vorming van structuur in het universum gebeurde waarschijnlijk in een 'bottom-up' scenario, ook wel 'hiërarchische' structuurvorming genoemd. Kleine overdichtheden zijn als eerste in staat om de kosmologische expansie te overwinnen en ineen te storten, en de resulterende donkere ma-terie 'halo's' smelten samen tot steeds grotere halo's, welke de sites worden voor de vorming van sterrenstelsels.

Momenteel wordt de complexe fysica geassocieerd met de vorming en evolutie van sterrenstelsels nog steeds niet goed begrepen. Belangrijke ingrediënten voor de totstandkoming van sterrenstelsels zijn de feedbackprocessen van sterren en zwarte gaten, die vorm geven aan het interstellaire medium (ISM) in sterrenstelsels, of in hun kleinere voorgangers, waaruit nieuwe generaties van stralende objecten zullen vormen.

## EERSTE STERREN

De allereerste generatie van sterren, gevormd uit het ongerepte primordiale gas (bijna uitsluitend bestaand uit waterstof en helium, ook wel nulmetalliciteit gas genoemd[3]), worden vaak Population III of Pop III sterren genoemd. Omdat de koelingsprocessen in primordiaal gas zo verschillend zijn van gas met zonsmetalliciteit (zoals het merendeel van het gas in de hedendaagse Melkwegschijf), kunnen het fragmentatiegedrag, en dus de stervormingsmodus, afwijken van wat we vandaag de dag zien. Voornamelijk de typische massa van Pop III sterren zou groter geweest kunnen zijn.

De eerste sterren transformeerden het vroege universum fundamenteel, via hun stralings-, chemische en mechanische feedback. Zij waren de eerste bronnen van waterstof-ioniserende fotonen, die het proces van herionisatie van het universum in gang zetten, en van fotonen in staat tot het dissociëren van moleculair waterstof. Zij produceerde de eerste metalen en de eerste kosmische stofdeeltjes. Grote hoeveelheden energie werden geïnjecteerd in het medium door middel van stellaire winden of wanneer een ster eindigde als supernova-explosie (SN), wat schokken en turbulentie creëert, en mogelijk het gas uit de omsluitende halo verdrijft. Bovendien waren de eerste sterren mogelijk de sites waar voor het eerst dynamisch significante magnetische velden ontstonden. En uiteraard, na het tijdperk van de eerste sterren gaan verdere generaties van sterren door met het vorm geven aan sterrenstelsels door middel van hun feedbackprocessen.

[3] De 'metalliciteit' van een gas is de fractie die zich niet in waterstof of helium bevindt. Aangezien de meeste gewone materie in het heelal voorkomt in de vorm van waterstof en helium, gebruiken sterrenkundigen de term 'metalen' als korte benaming om alle chemische elementen zwaarder dan waterstof en helium aan te duiden.

## OORSPRONG VAN DE IMF

Sterren vormen uit koud, moleculair gas, wat zich bevindt in moleculaire wolken in het interstellaire medium, het gas en (behalve in het zeer vroege heelal) stof dat aanwezig is in een (proto)sterrenstelsel en dat niet reeds in stellaire systemen geïncorporeerd is. Moleculaire wolken zijn ingebed in gebieden bestaand uit ijler en hoofdzakelijk neutraal atomair gas, met een geleidelijke overgang ertussen. De aanwezigheid van ver-ultraviolette straling (bijvoorbeeld van sterren) zal de chemie en de thermische balans van dit gas significant beïnvloeden, en deze gebieden worden dan fotodissociatie of foton-gedomineerde regio's (PDRs) genoemd. In termen van fysische en chemische processen bevindt in wezen het grootste deel van het ISM zich in PDRs. Ver-ultraviolette straling die zich verplaatst doorheen een PDR wordt geleidelijk afgezwakt, en geïoniseerde atomen maken eerst plaats voor neutrale atomen, terwijl dieper in de regio moleculen stabiel worden, de overgang makend naar een moleculaire wolk.

De initiële-massafunctie (IMF) en de typische massa van een ster zijn vermoedelijk gerelateerd aan de thermodynamica van het gas. Numerieke simulaties hebben aangetoond dat de hoeveelheid fragmentatie zeer gevoelig is voor het exacte verband tussen temperatuur en dichtheid in ineenstortende wolken. Daarnaast zijn er systematische variaties in de hedendaagse IMF waargenomen in extreme starburst-omgevingen en massieve elliptische sterrenstelsels. Om de IMF van zowel de eerste als van latere generaties sterren te bepalen, is het derhalve belangrijk om de chemische en thermodynamische eigenschappen van het feedback-gereguleerde ISM te begrijpen.

## EERSTE MASSIEVE ZWARTE GATEN

Ons huidig beeld van de vorming en evolutie van sterrenstelsels is niet compleet zonder zwarte gaten. Bewijsmateriaal suggereert dat bijna alle massieve sterrenstelsels een supermassief zwart gat (SMBH, met massa's van meer dan $10^5$ $M_\odot$[4]) bevatten in hun centrum. Verscheidene correlaties tussen de massa van het zwarte gat en de eigenschappen van het omsluitende sterrenstelsel wijzen op een gemeenschappelijke oorsprong of co-evolutie tussen sterrenstelsels en hun centrale zwarte gat. Ook zwarte gaten kunnen verantwoordelijk zijn voor feedbackeffecten, waaronder stralingsfeedback, in het bijzonder röntgenstraling (X-rays), en mechanische feedback in de vorm van wind en jets.

Observationeel bewijsmateriaal suggereert dat enkele SMBHs met massa's van $10^9$ $M_\odot$ reeds bestonden minder dan 1 Gyr na de oerknal, toen het universum nog geen tiende van zijn huidige leeftijd had. Verklaren hoe SMBHs met zulke grote massa's zo kort na de oerknal tot stand hebben kunnen komen is een hele uitdaging. De belangrijkste vragen betreffen hoe en wanneer de 'kiemen' ('seeds') van deze SMBHs gevormd zijn en hoe hun verdere groei verliep. Diverse mogelijke routes die leiden tot de vorming van kiem zwarte gaten zijn reeds geopperd.

Misschien wel het meest voor de hand liggende scenario gaat ervan uit dat SMBHs groeien vanuit de restanten van de eerste sterren. Of een Pop III ster al dan niet eindigt als een zwart gat is afhankelijk van massa en impulsmoment. Het is echter nog niet bekend of er genoeg, of zelfs maar enige, Pop III sterren geweest zijn met een massa groter dan de drempelwaarde bepalend voor de vorming van voldoende massieve zwarte gaten. En zelfs als zeer massieve sterren in staat waren om te vormen en ineen te storten tot kiem zwarte gaten, dan nog zou het moeilijk zijn voor deze (nog steeds relatief lichte) kiemen om voldoende massa te accreteren en te groeien tot een supermassief zwart gat in de beschikbare tijd.

Een veelbelovendere route is dan ook het 'direct collapse' scenario, waarbij het primordiale gas in een halo direct ineen zou storten tot een enkel centraal object, zonder te fragmenteren, wat resulteert in massieve kiem zwarte gaten. Voor het voorkomen van een dergelijke directe ineenstorting is het noodzakelijk dat fragmentatie beperkt is, wat mogelijk is als het gas in de halo heet kan worden gehouden. Derhalve moet de vorming van moleculair waterstof geremd worden, zodat koelen slechts mogelijk is via atomair waterstof, aangezien anders $H_2$ koeling de temperatuur aanzienlijk verlaagt en er fragmentatie kan optreden.

[4] $M_\odot$ of zonsmassa is een massa-eenheid die gelijk is aan de massa van onze Zon.

Er zijn verschillende plausibele mechanismen die de vorming van voldoende $H_2$ kunnen onderdrukken, hetzij via fotodissociatie van $H_2$ (door een superkritische ultraviolette (UV) stralingsachtergrond), hetzij via botsingsdissociatie van $H_2$ (door dissipatie van een voldoende sterk magnetisch veld, 'trapping' van Ly$\alpha$ fotonen, of de aanwezigheid van sterke schokken).

Een aannemelijke uitkomst van het *direct collapse* scenario is de verdere ineenstorting van een zekere hoeveelheid gas naar steeds kleinere schalen in de galactische kern. Op een gegeven moment tijdens de contractie wordt het gas optisch dik en de koelende straling kan dan niet meer ontsnappen. Als de stralingsdruk sterk genoeg is om de zwaartekracht tijdelijk te balanceren zal het resulterende object een 'protoster' zijn. Zodra de protoster gevormd is zal deze accreteren en uiteindelijk evolueren tot een kiem zwart gat, wat nog steeds snel zal moeten groeien om de waargenomen SMBHs op hoge roodverschuiving te verklaren.

## DIT PROEFSCHRIFT

### DEEL I: KIEMEN VAN SUPERMASSIEVE ZWARTE GATEN

In het eerste deel van dit proefschrift onderzoeken we de kiemen voor de eerste supermassieve zwarte gaten gevormd via directe ineenstorting.

In Hoofdstuk 2 van het proefschrift wordt de vorming van een protoster als gevolg van het ineenstorten van primordiaal gas in aanwezigheid van een sterke UV-stralingsachtergrond onderzocht. In het bijzonder onderzoeken we de impact van turbulentie en rotatie op het fragmentatiegedrag van de gaswolk. Zoals bekend van hedendaagse stervorming speelt turbulentie een belangrijke rol in het transport van impulsmoment en het bepalen van de fragmentatie-eigenschappen van ineenstortende gaswolken, omdat het in staat is om zowel het gas lokaal te comprimeren, als aanvullende steun te bieden tegen ineenstorting op grotere schalen. Eerdere numerieke simulaties van ineenstortend gas hebben aangetoond dat fragmentatie ook afhankelijk kan zijn van de mate van rotatie. We hebben numerieke 3D 'adaptive mesh refinement' (AMR) simulaties uitgevoerd gebruik makend van de ENZO code, met toevoeging van het KROME pakket om het modelleren van de chemische en thermische processen te verbeteren. Dit stelt ons in staat om de vorming van een protoster te simuleren tot ongekend hoge centrale dichtheden van $10^{21}$ $cm^{-3}$ en ruimtelijke schalen van een paar keer de straal van de zon. We zien dat de fysische eigenschappen van de gesimuleerde gaswolken gelijkaardig worden op kleine schalen, ongeacht de oorspronkelijke mate van turbulentie en rotatie. Een enkele gebonden massa heeft zich gevormd aan het einde van iedere simulatie, duidend op het begin van een protoster. Er wordt geen sterke fragmentatie waargenomen tegen het eind van de simulaties, ongeacht de oorspronkelijke mate van turbulentie of rotatie, en hoge groeisnelheden van enkele zonsmassa's per jaar worden vastgesteld. Gezien deze hoge groeisnelheden wordt verwacht dat een quasister van $10^5$ $M_\odot$ zich zal vormen binnen $10^5$ jaar, wat uiteindelijk zal leiden tot een massief kiem zwart gat.

Overige manieren waarop het gas in staat kan zijn om heet te blijven en fragmentatie te vermijden worden onderzocht in Appendix 2.B, met behulp van een analytisch één-zone-model. In het bijzonder onderzoeken we de wisselwerking tussen magnetische velden, turbulentie en een UV-stralingsachtergrond tijdens de gravitationele ineenstorting van primordiaal gas in een halo. We volgen de evolutie van een wolk bestaand uit primordiaal gas vanaf zijn initiële kosmische expansie doorheen 'turnaround', virialisatie en ineenstorting tot een dichtheid van $10^7$ $cm^{-3}$. Het blijkt dat in een halo zonder significante turbulentie de kritische UV-achtergrondintensiteit aanzienlijk kan worden verlaagd door de aanwezigheid van ofwel een magnetisch veld, ofwel turbulentie (of beide), met lagere kritische waarden voor sterkere velden en/of sterkere turbulentie. De vermindering in kritische UV-intensiteit is bijzonder van belang, omdat dit het aantal halo's blootgesteld aan een superkritische stralingsachtergrond exponentieel vergroot, en daarmee het aantal mogelijke vormingssites van massieve kiem zwarte gaten.

## DEEL II: ISM THERMODYNAMICA EN FRAGMENTATIE

In het tweede deel van het proefschrift richten we ons op de evenwichtstoestand van interstellair materiaal (gas en stof) onder verscheidene fysische omstandigheden, om een beter inzicht te krijgen in de chemo-thermische toestand en het fragmentatiegedrag van gas in sterrenstelsels op hoge roodverschuiving.

In Hoofdstuk 3 van het proefschrift beschrijven we een fotodissociatie-regio (PDR) code. Onze code, PDR-ZZ, is gebaseerd op de PDR code van Meijerink & Spaans (2005), welke wij geactualiseerd en aanzienlijk verbeterd hebben, en eveneens uitgebreid voor gebruik in een breder scala aan fysische omstandigheden (bv. veel hogere dichtheden, metalliciteiten van praktisch 0 tot zonsmetalliciteit, roodverschuivingen >0). Deze computationele code kan ingezet worden voor het bestuderen van traditionele PDRs, en kan bovendien toegepast worden op vele andere situaties waar men geïnteresseerd is in steady-state gas op verschillende optische dieptes, onder verschillende omstandigheden; van diffuse en moleculaire wolken in de Melkweg tot wolken in de buurt van AGN en materiaal in sterrenstelsels in het vroege heelal. De inputparameters voor de code zijn de totale waterstofdichtheid, de metalliciteit, de invallende UV-stralingsintensiteit, de kosmische-stralingsionisatiegraad, en de roodverschuiving; de combinatie hiervan bepaalt de gekoppelde chemische en thermische oplossing voor de gesimuleerde regio. In dit hoofdstuk beschrijven we in detail welke chemische species, reacties en reactiesnelheden worden gebruikt, en eveneens welke koelings- en verhittingsprocessen zijn inbegrepen in het model. Daarnaast beschrijven we onze verbeteringen aan de convergentiealgoritmen voor zowel de chemische als de thermische balans, welke leiden tot een grote snelheidswinst en maken het mogelijk om vele modellen te berekenen op relatief korte tijd.

Zoals eerder vermeld geven feedbackprocessen van sterren en zwarte gaten vorm aan het interstellaire medium, waaruit nieuwe generaties van stralende objecten zich vormen. Om de eigenschappen van deze objecten te begrijpen, onder andere de stellaire initiële-massafunctie, is het essentieel om een goed begrip van de chemische en thermodynamische eigenschappen van het feedback-gereguleerde ISM te hebben. In Hoofdstuk 4 onderzoeken we op een systematische manier de globale impact van verscheidene feedbackeffecten, zowel straling als chemisch, op de chemische en thermische balans van het gas in verschillende regimes, aangezien de omstandigheden in het ISM sterk kunnen variëren. Een raster van modellen is berekend met onze numerieke code PDR-ZZ, welke 1D modellen gebruik makend van een steady-state aanpak biedt. Het raster bestrijkt een omvangrijk gebied in totale waterstofdichtheid, metalliciteit, roodverschuiving, UV-stralingsintensiteit, en kosmische-stralingsionisatiegraad. We geven inzicht in de belangrijkste processen en dominante chemische species in verschillende regimes. We tonen de effecten van stralings- en chemische feedback op de effectieve gastemperatuur, zowel in de buitenste lagen als in de bulk van de wolk, en beschrijven de verschillen. Ten slotte onderzoeken we de toestandsvergelijkingen (EOS) in onze parameterruimte, identificeren een aantal interessante regio's, en relateren deze aan de IMF. We vinden een overgang in het gedrag van de EOS rond een metalliciteit van 0,1 keer de zonsmetalliciteit, met een globaal soortgelijke EOS voor het bulk van het gas met metalliciteiten beneden 0,1 zonsmetalliciteit, omdat de belangrijkste koelings- en verhittingskanalen ruwweg schalen met metalliciteit in gas dat optisch dun is voor het UV continuüm. Voor lage metalliciteiten vinden we een relatief zachte EOS, maar in combinatie met de hogere gastemperaturen kan dit nog steeds aanleiding geven tot massieve sterren. Verder leiden starburst-achtige omstandigheden tot zowel een stijvere EOS als hogere temperaturen, wat resulteert in een meer 'top-heavy' IMF.

# ZUSAMMENFASSUNG (DEUTSCH)

## KOSMOLOGIE

Der Urknalltheorie zufolge entstand das Universum vor ungefähr 13,8 Millionen Jahren in einem unvorstellbar heißen und dichten Zustand. Seitdem hat es sich zunehmend ausgedehnt und abgekühlt, um letztendlich den kalten, dünnen Zustand zu erreichen, den wir heute vorfinden. Für den Bruchteil einer Sekunde nach dem Urknall dehnte sich das Universum extrem schnell aus. Während dieser Phase, die auch als Inflation bezeichnet wird, wurden quantenmechanische Fluktuationen auf makroskopische Skalen vergrößert und dem Universum als Dichtefluktuationen aufgeprägt, die der Ursprung für weitere Strukturentstehung waren. Im weiteren Verlauf wurden positive Fluktuationen in der Masseverteilung dichter und massereicher, wodurch ein hochgradig strukturiertes Netzwerk aus Filamenten, Flächengebilden (Sheets) sowie dichteren Ansammlungen (Cluster) von gewöhnlicher und dunkler Materie, welche durch immense Leerräume (Voids) getrennt waren, entstand. Einige hundert Millionen Jahre nach dem Urknall war die Dichte in einigen Knotenpunkten dieses Kosmischen Netzes ausreichend hoch für die Bildung von Sternen und Galaxien. Die Entstehung des ersten Sterns brachte die erste Lichtquelle ins Universum und beendete damit das sogenannte dunkle Zeitalter (Dark Ages).

Die Entstehung von Strukturen im Universum erfolgte wahrscheinlich nach dem „Bottom-up“-Szenario, welches auch als „hierarchische Strukturbildung“ bezeichnet wird. Dabei schafften es kleine Überdichten, der kosmologischen Ausdehnung zu trotzen und zu kollabieren. Die daraus entstandenen Halos aus dunkler Materie verschmolzen zu immer größeren Halos, in denen wiederum Galaxien entstanden.

Auch heute ist die komplexe Physik der Galaxienentstehung und -entwicklung noch nicht umfassend verstanden. Wichtige Faktoren in der Galaxienentwicklung sind Rückkopplungsprozesse (Feedback) von Sternen und Schwarzen Löchern, die das interstellare Medium (ISM) in Galaxien und ihren Vorläufersystemen formen, in denen wiederum neue Generationen strahlender Objekte entstehen.

## ERSTE STERNE

Die ersten Sterne, die sich aus dem primordialen Gas (fast ausschließlich aus Wasserstoff und Helium bestehend Gas mit Metallizität null[5]) formten, werden oft als Population III oder Pop III Sterne bezeichnet. Da die Kühlungsprozesse in primordialem Gas so sehr von denen des Gases von der Metallizität der Sonne abweichen, können sich auch das Fragmentierungsverhalten und damit der Sternentstehungsvorgang von dem, was wir heute vorfinden, unterscheiden. Insbesondere sind Pop III Sterne typischerweise viel massereicher.

Die ersten Sterne haben das Universum wesentlich durch ihr Strahlungs-, chemisches und mechanisches Feedback verändert. Sie waren die erste Quellen für Wasserstoff-ionisierende Photonen, womit der Prozess der Reionisierung einsetzte, und für Photonen, die im Stande waren, molekularen Wasserstoff zu spalten. Sie produzierten die ersten Metalle und Staubkörner. Große Energiemengen wurden an ihre Umgebung abgegeben in Form von Sternwinden oder durch Supernova Explosionen, die wiederum Schocks und Turbulenzen verursachten und möglicherweise Gas aus dem Halo schleuderten. Die ersten Sterne könnten außerdem die ersten Orte gewesen sein, an denen sich zum ersten Mal dynamisch signifikante Magnetfelder bildeten.

[5] Die Metallizität eines Gases ist der Anteil der von Wasserstoff und Helium verschiedenen Elemente. Da die gewöhnliche Materie im Universum hauptsächlich in Form von Wasserstoff und Helium vorliegt, gebrauchen Astronomen den Begriff „Metalle“ zur Bezeichnung aller chemischen Elemente, die schwerer sind als Wasserstoff oder Helium.

Im Anschluss an die Periode der ersten Sterne haben weitere Generationen von Sternen Galaxien durch ihr Feedback geprägt.

## URSPRUNG DER IMF

Sterne bilden sich aus kaltem molekularem Gas, welches sich in Molekülwolken im interstellarem Medium befindet, also dem Gas und (mit Ausnahme des sehr frühen Weltalls) Staub, welche sich in (Proto-)Galaxien finden und nicht in Sternen gebunden sind. Molekülwolken sind in Regionen dichteren und überwiegend neutralem atomaren Gas eingebettet, wobei der Übergang dazwischen fließend ist. Das Vorhandensein von ferner ultravioletter (UV) Strahlung (z.B. von Sternen) beeinflusst das chemische und thermische Gleichgewicht dieses Gases. Solche Regionen werden auch als Photodissoziations- oder photonendominierte Regionen (PDR) bezeichnet. Im Hinblick auf physikalische und chemische Prozesse, befindet sich praktisch der größte Teil des ISMs in PDRs. Ferne ultraviolette Strahlung, die in eine PDR eindringt, wird sukzessive abgeschwächt. Mit zunehmender Eindringtiefe nimmt zunächst der Anteil von neutralen atomaren Spezies gegenüber den ionisierten Spezies zu, während weiter im Inneren der Region Moleküle stabil werden und sich zu einer Molekülwolke verbinden.

Die ursprüngliche Massenfunktion (IMF) und die typische Masse eines Sterns stehen wahrscheinlich im Zusammenhang mit den thermodynamischen Prozessen im Gas. Numerische Simulationen haben gezeigt, dass der Grad der Fragmentierung stark von der genauen Temperatur-Dichte-Beziehung in kollabierenden Wolken abhängt. Außerdem wurden in extremen Starburst Regionen (Gebiete mit ungewöhnlich hoher Sternentstehungsrate) und elliptischen Galaxien systematische Abweichungen von der heutigen IMF beobachtet. Aus diesem Grund ist es wichtig, die chemischen und thermodynamischen Eigenschaften des Feedback-regulierten ISMs zu verstehen, um die IMF der ersten Sterne und der Sterne der darauf folgenden Generationen zu bestimmen.

# MASSEREICHE SCHWARZE LÖCHER

Unser heutiges Verständnis von Galaxienentstehung und -entwicklung ist ohne Schwarze Löcher nicht vollständig. Es gibt Hinweise darauf, dass sich im Zentrum fast aller massiven Galaxien ein supermassereiches Schwarzes Loch (SMBH, mit Massen, die $10^5\ M_\odot$ überschreiten[6]) befindet. Zahlreiche Zusammenhänge zwischen der Masse des Schwarzen Loches und der umgebenden Galaxie weisen auf einen gemeinsamen Ursprung oder eine gemeinsame Entwicklung von Galaxien und ihren zentralen Schwarzen Löchern hin. Des weiteren können Schwarze Löcher für Feedbacprozesse, wie z.B. Strahlungs-Feedback, insbesondere durch Röntgenstrahlung, und mechanisches Feedback in Form von Winden und Jets, verantwortlich sein.

Beobachtungen deuten darauf hin, dass manche SMBHs mit Massen von $10^9\ M_\odot$ schon weniger als eine Milliarden Jahre nach dem Urknall existierten, als das Universum weniger als ein Zehntel seines heutigen Alters besaß. Die Beantwortung der Frage, wie sich SMBHs solch großer Masse so kurz nach dem Urknall bilden konnten, stellt eine große Herausforderung dar. Die wichtigsten Fragen beschäftigen sich damit, wie und wann sich die „Keimzellen" dieser SMBHs bildeten und wie sich ihr weiteres Wachstum gestaltete. Es existieren bereits mehrere Vorschläge für mögliche Wege, die zur Entstehung dieser Keimzellen für Schwarze Löcher geführt haben können.

Nach dem wahrscheinlich offensichtlichsten Szenario entstanden SMBHs aus den Überresten der ersten Sterne. Ob ein Pop III Stern sein Leben in einem Schwarzen Loch beendet, hängt von seiner Masse und seinem Drehimpuls ab. Es ist jedoch noch nicht klar, ob es, wenn überhaupt, genügend Pop III Sterne mit Massen oberhalb des Limits für die Entstehung von ausreichend massiven Keimzellen gab. Selbst wenn sich sehr massereiche Sterne formen und zu Keimzellen für Schwarze Löcher kollabieren konnten, wäre es für diese relativ leichten Keime noch stets äußerst schwierig, genügend Masse anzuhäufen, um sich in der verfügbaren Zeit zu einem Schwarzen Loch zu entwickeln.

Ein aussichtsreicherer Ansatz ist deshalb das „direct collapse" Szenario, bei dem primordiales Gas in einem Halo ohne vorherige Fragmentierung direkt in ein einziges zentrales Objekt kollabiert und dadurch ein massiver

[6] $M_\odot$ oder Sonnenmasse ist eine Maßeinheit, die der Masse der Sonne entspricht.

Keim für ein Schwarzes Loch entsteht. Damit so ein direkter Kollaps eintreten kann, ist eine Beschränkung der Fragmentierung nötig. Dies wiederum ist möglich, wenn das Gas im Halo heiß bleibt. Deshalb muss die Bildung von molekularem Wasserstoff gehemmt werden, damit eine Abkühlung nur durch atomaren Wasserstoff erfolgen kann. Eine $H_2$-Kühlung hingegen würde die Temperaturen zu schnell senken und eine Fragmentierung ermöglichen.

Es gibt mehrere plausible Mechanismen, die die Entstehung von genügend $H_2$ verhindern. Dies kann entweder durch Photodissoziation von $H_2$ (durch eine ausreichend starke (superkritische) ultraviolette Hintergrundstrahlung) oder durch Stoßdissoziation von $H_2$ (durch die Dissoziation eines hinreichend starken Magnetfeldes, das Einfangen von Ly$\alpha$-Photonen oder die Anwesenheit von starken Schocks).

Eine wahrscheinliche Folge des direct collapse Szenarios ist der kontinuierliche Kollaps von Gas auf kleinere Skalen im galaktischen Kern. Mit der Kontraktion des Gases wird es optisch dick und Photonen, die zur Kühlung beitragen, können nicht mehr entkommen. Wenn der Strahlungsdruck stark genug ist, um temporär die Schwerkraft auszugleichen, wird aus dem Objekt ein „Protostern“. Nachdem sich der Protostern gebildet hat, beginnt er zu akkretieren und entwickelt sich schließlich zu einem Keim für ein Schwarzes Loch, das wiederum sehr schnell wachsen muss, um die beobachteten SMBHs zu erklären.

## IN DIESER ARBEIT

### TEIL I: KEIMZELLEN SUPERMASSEREICHER SCHWARZER LÖCHER

Im ersten Teil dieser Arbeit untersuchen wir die Bildung von Keimzellen der ersten supermassereichen Schwarzen Löcher im direct collapse Szenario.

Kapitel 2 dieser Arbeit befasst sich mit der Bildung des Protosterns, der sich aus dem Kollaps von primordialem Gas in der Anwesenheit einer starker UV Hintergrundstrahlung ergibt. Wir untersuchen insbesondere den Einfluss von Turbulenz und Rotation auf das Fragmentierungsverhalten der Gaswolke. Aus der gegenwärtigen Sternentstehung ist bekannt, dass Turbulenz eine wichtige Rolle im Transport von Drehimpuls, sowie bei der Bestimmung von

Fragmentierungseigenschaften der kollabiereunden Gaswolke spielt, da sie sowohl lokal das Gas komprimieren als auch zusätzlich Unterstützung gegen den Kollaps auf größeren Skalen liefern kann. Frühere numerische Simulationen von kollabierendem Gas haben gezeigt, dass Fragmentierung auch von der Stärke der Rotation abhängen kann. Mit Hilfe des ENZO-Codes haben wir 3D adaptive mesh refinement (AMR) Simulationen durchgeführt. Für eine bessere Modellierung der chemischen und thermischen Prozesse nutzten wir zusätzlich das KROME-Paket. Dies machte es möglich, die Bildung eines Protosterns mit beispiellos hoher zentraler Dichte von $10^{21}$ $cm^{-3}$ und räumlichen Auflösungen von wenigen Sonnenradien zu simulieren. Wir haben festgestellt, dass, unabhängig von der anfänglichen Stärke der Turbulenz und Rotation, die physikalischen Eigenschaften der simulierten Gaswolken auf kleinen Skalen ähnlich sind. Am Ende jeder Simulation hat sich ein gebundener Klumpen gebildet, welcher den Anfang der Bildung eines Protosterns markiert. Weiterhin können wir unabhängig von der anfänglichen Stärke der Turbulenz und Rotation am Ende der Simulation keine starke Fragmentierung feststellen. Wir finden hohe Akkretionsraten von einigen Sonnenmassen pro Jahr. Folglich sollte sich innerhalb von $10^5$ Jahren ein Quasi-Stern von $10^5$ $M_\odot$ bilden, der letztendlich zur Keimzelle eines massiven Schwarzen Loches wird.

Weitere Möglichkeiten, wie das Gas heiß bleiben und damit einer Fragmentierung entgehen kann, werden mit Hilfe eines analytischen Ein-Zonen-Modells im Anhang 2.B untersucht. Insbesondere erforschen wir das Zusammenspiel zwischen Magnetfeldern, Turbulenz und UV Hintergrundstrahlung während des Kollaps von primordialem Gas im Halo. Wir folgen der Entwicklung einer Wolke aus primordialem Gas von ihrer initialen kosmischen Expansion, über den Umkehrpunkt und die Virialisierung bis zum Kollaps mit einer Dichte von bis zu $10^7$ $cm^{-3}$. Man hat herausgefunden, dass in Halos ohne signifikante Turbulenz die kritische Intensität der UV Hintergrundstrahlung deutlich reduziert werden kann durch die Anwesenheit von entweder einem Magnetfeld oder Turbulenz (oder beidem), mit geringeren kritischen Werten für stärkere Felder und/oder stärkere Turbulenz. Die Reduktion der kritischen UV Intensität ist von besonderer Bedeutung, da sie einen exponentiellen Anstieg der Anzahl der Halos zur Folge hat, die einem superkritischen Strahlungshintergrund ausgesetzt sind, was wiederum die Anzahl der möglichen Bildungsorte für Keimzellen von massiven Schwarzen Löchern erhöht.

## TEIL II: THERMODYNAMIK UND FRAGMENTIERUNG DES ISM

Im zweiten Teil dieser Arbeit konzentrieren wir uns auf den Gleichgewichtszustand des interstellaren Materials (Gas und Staub) unter verschiedenen physikalischen Bedingungen, um den chemo-thermischen Zustand und das Fragmentierungsverhalten des Gases in hoch rotverschobenen Galaxien besser zu verstehen.

In Kapitel 3 dieser Arbeit beschreiben wir einen Code für eine Photodissoziationsregion (PDR). Unser Code, PDR-Zz, basiert auf dem PDR code von Meijerink & Spaans (2005), welchen wir aktualisiert, bedeutend verbessert und für die Nutzung eines größeren Bereichs von physikalischen Anfangsbedingungen (z.B. höhere Dichten, Metallizitäten von 0 bis solar, Rotverschiebungen > 0) erweitert haben. Dieser Code kann zur Untersuchung von traditionellen PDRs genutzt werden. Zusätzlich kann er aber auch auf viele andere Situationen angewandt werden, in denen ein stationäres Gas unter verschiedenen Bedingungen bei unterschiedlichen optischen Tiefen untersucht werden soll; von diffusen Molekülwolke in der Milchstraße bis zu Wolken in der Nähe von AGNs und Material in Galaxien im hoch rotverschobenen Universum. Die Eingangsparameter des Codes sind die Gesamtwasserstoffdichte, die Metallizität, die Stärke des einfallenden UV Strahlungsfeldes, die Ionisationsrate kosmischer Strahlen und die Rotverschiebung. In Kombination definieren sie die gekoppelte chemische und thermische Lösung für die simulierte Region. In diesem Kapitel beschreiben wir im Detail, welche chemischen Spezies, Reaktionen und Raten genutzt wurden, sowie die im Modell enthaltenen Kühlungs- und Heizprozesse. Außerdem beschreiben wir unsere Optimierung der Konvergenzalgorithmen für das chemische und thermische Gleichgewicht, welche in einer großen Beschleunigung resultierten und es uns ermöglichen, viele Modelle in einer relativ kurzen Zeit zu rechnen.

Wie bereits erwähnt, formen Feedbackprozesse von Sternen und Schwarzen Löchern das interstellare Medium, aus dem sich wiederum neue Generationen von leuchtenden Objekten formen. Um die Eigenschaften dieser Objekte, wie z.B. die stellare IMF, zu verstehen, ist es von entscheidender Bedeutung, ein gutes Verständnis der chemischen und thermodynamischen Eigenschaften des Feedback-regulierten ISM zu haben. Da die Bedingungen im ISM sehr variieren können, untersuchen wir in Kapitel 4 systema-tisch die Gesamtwirkung von verschiedenen Feedbackeffekten (sowohl Strahlungs- als auch chemisches Feedback) auf das chemische und thermische Gleichgewicht des Gases in verschiedenen Regimes. Mit Hilfe unseres numerischen Codes PDR-Zz, welcher 1D Modelle unter der Nutzung eines stationären Ansatzes liefert, wurde ein Raster von Modellen gerechnet. Das Raster deckt einen beträchtlichen Bereich von Gesamtwasserstoffdichte, Metallizität, Rotverschiebung, Skalierungsparameter für das UV Strahlungsfeld und Ionisationsrate für kosmische Strahlen ab. Wir geben einen Einblick in die wichtigsten Prozesse und vorherrschenden chemischen Spezies in verschiedenen Regimen. Wir zeigen die Effekte von Strahlungs- und chemischem Feedback auf die effektive Gastemperatur sowohl in den äußeren Schichten als auch im Hauptteil der Wolke und beschreiben die Unterschiede. Letztendlich untersuchen wir die Zustandsgleichung (EOS) in unserem Parameterbereich, identifizieren mehrere interessante Regionen und setzen diese in Beziehung zur IMF. Wir finden einen Wechsel des Verhaltens der EOS bei einer Metallizität von ungefähr 0.1 solarer Metallizität mit einer insgesamt ähnlichen EOS für den Hauptteil des Gases für Metallizitäten unterhalb 0.1 solarer Metallizität. Der Grund hierfür ist, dass die Hauptkanäle für Heiz- und Kühlungsprozesse grob mit der Metallizität skalieren für Gas, das für das UV Kontinuum optisch dünn ist. Für niedrige Metallizitäten finden wir eine relativ schwache EOS. Gekoppelt mit der höheren Gastemperatur kann dies trotzdem zu massereichen Sternen führen. Außerdem fördern Starburst-ähnliche Bedingungen eine stärkere EOS und höhere Temperaturen, was zu einer eher „top-heavy“ IMF führt.

# POSTERS

The first poster illustrates Appendix 2.B, while the second poster shows results from the 3D simulations described in Chapter 2 (see following pages).

# The influence of magnetic fields, turbulence, and UV radiation on the formation of supermassive black holes

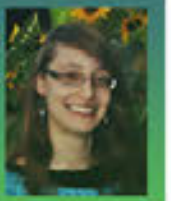

**Caroline Van Borm* and Marco Spaans**

**Van Borm, C., & Spaans, M. 2013, A&A, 553, L9**

Kapteyn Astronomical Institute, University of Groningen, PO Box 800, 9700 AV, Groningen, The Netherlands

*Currently at the Institut für Astrophysik, Georg-August-Universität Göttingen, Friedrich-Hund-Platz 1, 37077 Göttingen, Germany

rijksuniversiteit groningen

Georg-August-Universität Göttingen

## Context

Several very bright quasars have been detected as early as $z > 6 \Rightarrow$ some supermassive black holes (SMBHs) with masses of ~$10^9$ $M_\odot$ already existed < 1 Gyr after the Big Bang.

One possible scenario: SMBH seeds may have formed through the **direct collapse** of primordial gas in $T_{vir} \gtrsim 10^4$ K halos, whereby the gas must stay hot (~$10^4$ K) to avoid fragmentation[1-4]. Hence, **formation of $H_2$ must be inhibited**, otherwise $H_2$ cooling will lower the gas temperature to ~200 K.

$H_2$ can be **photo-dissociated** by a supercritical level of Lyman-Werner (UV) radiation, or destroyed by **collisional dissociation** as a result of the dissipation of a sufficiently strong magnetic field[5-6].

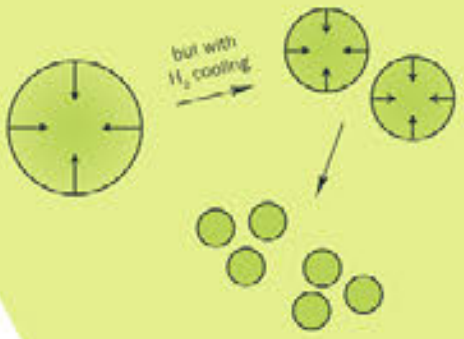

## Aim

To explore the interplay between **magnetic fields, turbulence**, and a **UV radiation background** in the post-recombination Universe and during the gravitational collapse of primordial gas in a halo. In particular, to examine the possibilities for avoiding fragmentation.

## Method

A **one-zone model** was used to follow the evolution of a cloud of primordial gas from its initial cosmic expansion through turnaround, virialization, and collapse up to a density of $10^7$ cm$^{-3}$.

(1) 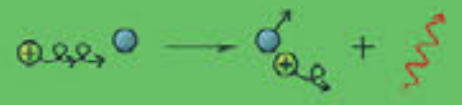

**Magnetic fields:**

Dissipated by (1) ambipolar diffusion (AD).

Magnified by:

- (2) gravitational compression,
- (3) small-scale dynamo: magnetic field lines are stretched and folded by turbulence until saturation[7].

(2) 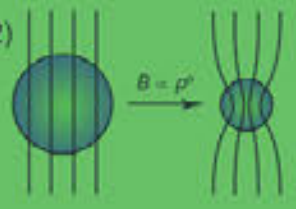

(3) 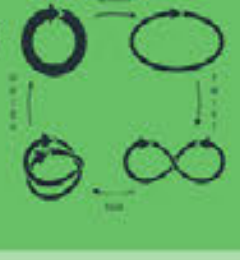

**Turbulence:**

Generated by infalling gas[8].

Turbulent dissipation = additional heating source.

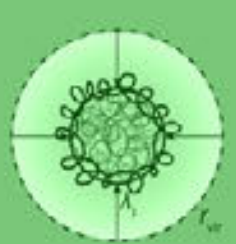

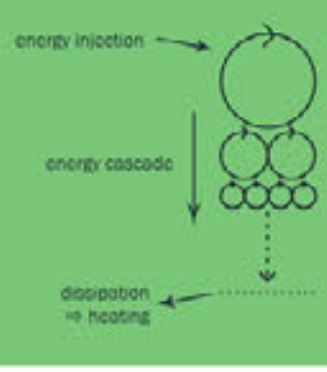

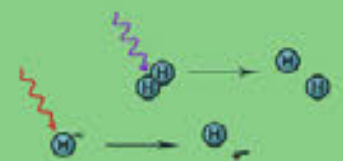

**UV background:**

Two different blackbody spectra considered:

- T4 (~ Pop II spectrum),
- T5 (~ Pop III spectrum).

## Results

**Without significant turbulence or magnetic field:**

- **T4:** $10 < J_{21}^{crit} \leq 10^2$
- **T5:** $10^4 < J_{21}^{crit} \leq 10^5$

Consistent with values found by [9] and lower by a factor ~10 than previous estimates[1,10] (due to different $H_2$ dissociation rates).

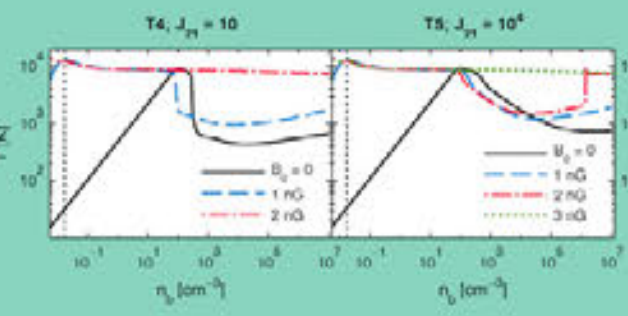

**With magnetic field:**

For $B_0$ ~ 2-3 nG: $J_{21}^{crit}$ is lowered by a factor ~10, and **the stronger the field, the lower $J_{21}^{crit}$**.

Without UV background: need $B_0$ ~ 13 nG for sufficient AD heating to overcome $H_2$ cooling.

Amount of AD heating depends on scaling of $B$ with $\rho \Rightarrow$ important to model this relation correctly.

Current upper limit on primordial magnetic field is ~1 nG comoving; a 2 nG field is then reached by ~2σ upward fluctuations.

**With turbulence ($M_{halo} = 10^9$ $M_\odot$):**

- **T4:** $1 < J_{21}^{crit} \leq 10$
- **T5:** $10^3 < J_{21}^{crit} \leq 10^4$

$\Rightarrow J_{21}^{crit}$ **is a factor ~10 lower** than without turbulence and magnetic fields.

This is due to turbulent heating, so $J_{21}^{crit}$ is even lower for larger halos and/or halos with stronger turbulent heating.

Turbulent halos with $M \gtrsim 10^{11}$ $M_\odot$ (depending on the strength of the turbulence) stay hot without any UV background.

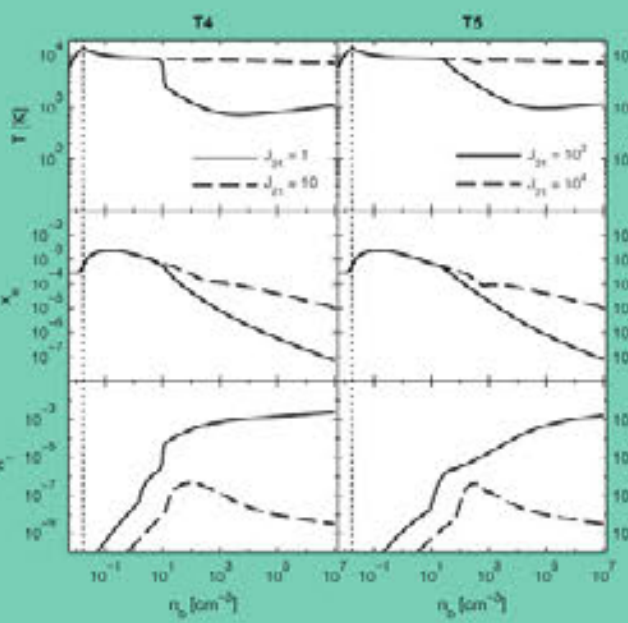

## Conclusion

The reduction in $J_{21}^{crit}$ as found here is particularly significant, since it exponentially increases the number of halos exposed to a supercritical radiation background, and thus the fraction of halos that are suitable candidates for direct SMBH formation[5].

With a sufficiently low $J_{21}^{crit}$, one could argue that this mechanism provided many, if not all, seeds for the SMBHs observed in galaxies today.

# ◉ SUPERMASSIVE BLACK HOLES – TOWARDS RESOLVING PROTOSTAR FORMATION ◉

**EFFECTS OF TURBULENCE AND ROTATION ON PROTOSTAR FORMATION AS A PRECURSOR TO SEED BLACK HOLES**

## ◉ INTRODUCTION ◉

Several very bright quasars have been detected as early as $z > 6$, meaning some supermassive black holes (SMBHs) with masses of $\sim 10^9$ $M_\odot$ already existed less than 1 Gyr after the Big Bang[1]. The seeds of these first SMBHs may have resulted from the direct collapse (with no or very little fragmentation) of hot primordial gas in $\geq 10^4$ K haloes[2-5], likely forming a supermassive star or a quasistar as an intermediate stage[6,7].

## ◉ AIM ◉

We explore the formation of a protostar resulting from the collapse of primordial gas in the presence of a strong Lyman-Werner radiation background. Particularly, we investigate the impact of turbulence and rotation on the fragmentation behaviour of the gas cloud.

## ◉ METHODS ◉

We performed 3D AMR simulations with a resolution of 64 cells per Jeans length using the ENZO code, combined with the KROME package to improve modelling of the chemical and thermal processes.
We follow the gravitational collapse of an isolated spherical gas cloud with a radius of 15 pc and top-hat density profile. The cloud is irradiated by a strong UV background which ensures a low $H_2$ abundance; the gas will cool mainly through atomic hydrogen and stay relatively hot[8,9].

**SIMULATIONS**

| Name | Turbulence (~% of $c_s$) | Rotation (% of $v_{Kep}$) |
|---|---|---|
| T4R0 | 40% | 0% |
| T4R1 | 40% | 10% |
| T4R2 | 40% | 20% |
| T2R1 | 20% | 10% |
| T8R1 | 80% | 10% |

## ◉ RESULTS ◉

((Want to know what happens when magnetic fields are added to this picture? Keep an eye out for our future research!))

**Stronger initial rotation**
⇒ Flatter, more extended disk and more pronounced spiral structures.

Number density projections along the z-axis (log N in cm⁻²)

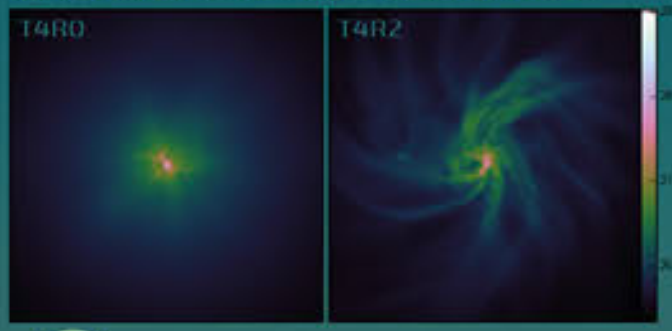

**Stronger initial turbulence**
⇒ More density structure, and more hot and cold patches.

Number density projections along the z-axis (log N in cm⁻²)

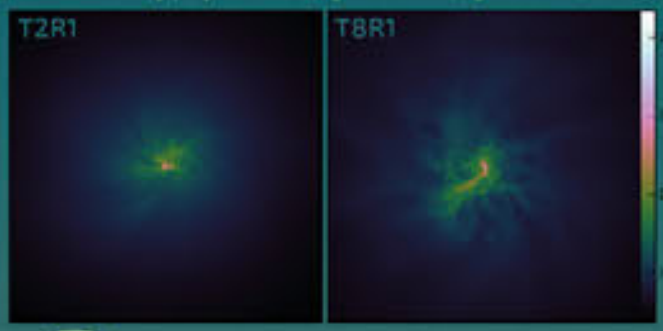

From here on: no longer possible to distinguish between different initial conditions.

**Physical properties become similar on small scales** (≲ 0.01 pc), irrespective of the initial amount of turbulence and rotation.

**Further evolution**: the clump and surrounding gas are expected to collapse into a **massive protostar**. Given the radial accretion rates of ~1-10 $M_\odot$/yr, a **quasistar**[7] of $10^5$ $M_\odot$ **will form within $10^5$ years**, evolving into a **seed black hole**, which will eventually grow into a **SMBH**.

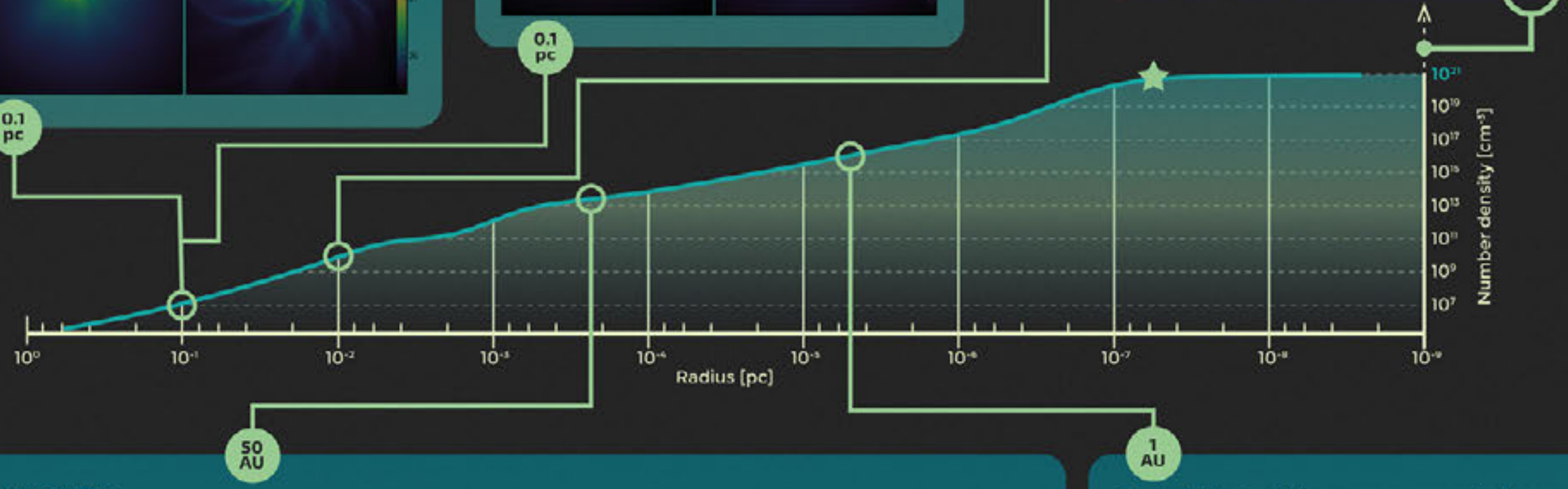

**Fragmentation**
In most cases just a **single** clump was found, though sometimes there were 1 or 2 additional (though less dense) clumps. These may still merge with the main clump later in time[10].

Number density projections along the z-axis (log N in cm⁻²)

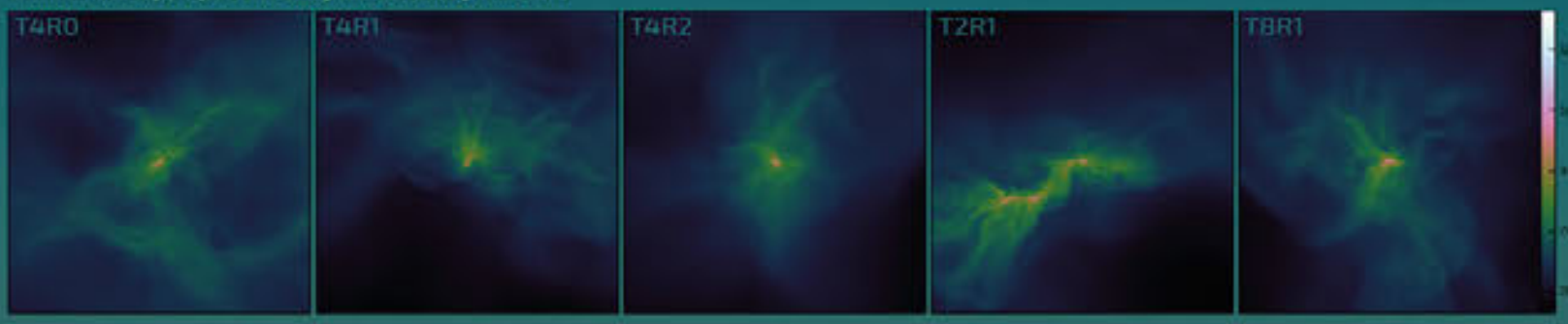

**Central object = hot, pressure-supported gas**
★ **R ~ 2x10⁻² AU**
**M ~ 7x10⁻² $M_\odot$**

Number density and temperature slices along the z-axis (log n in cm⁻³ and log T in K)

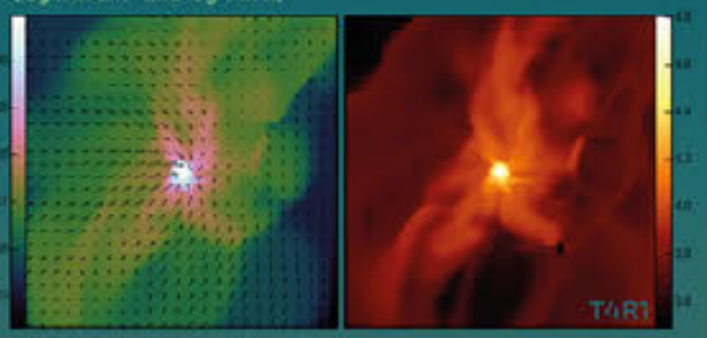

## ◉ REFERENCES ◉

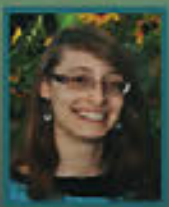

Contact Information
Caroline Van Borm
Kapteyn Astronomical Institute
University of Groningen
borm@astro.rug.nl

rijksuniversiteit groningen
GEORG-AUGUST-UNIVERSITÄT GÖTTINGEN

# CURRICULUM VITAE

Caroline Van Borm
December 7, 1989
Belgian
caroline.vanborm@gmail.com

## EDUCATION

| | |
|---|---|
| 2001 - 2007 | ASO (general secondary education)<br>Didasco Sciences - Mathematics<br>Koninklijk Atheneum Malle (Oostmalle), Belgium |
| 2007 - 2010 | **Bachelor of Physics** (great distinction, avg. 81%)<br>University of Antwerp, Belgium<br>*Thesis (theoretical): 'Effects of vortex core deformation on the vortex-vortex potential in Bose-Einstein condensates'*<br>*Thesis (experimental): 'Holography'* |
| 2010 - 2013 | **Master of Science in Astronomy** (cum laude, avg. 88%)<br>Kapteyn Institute, University of Groningen, The Netherlands<br>*Thesis: 'Origins of Supermassive Black Holes: The influence of magnetic fields and turbulence on the formation of seed black holes'* |

## PHD

| | |
|---|---|
| 03/2013 - 09/2016 | **PhD in Astronomy**<br>Kapteyn Institute, University of Groningen, The Netherlands<br>& Institut für Astrophysik, Georg-August Universität Göttingen, Germany |

*INFINITE DIVERSITY*
*IN INFINITE COMBINATIONS*

— GENE RODDENBERRY

# ACKNOWLEDGMENTS

You have reached the end of this book, while at the time of writing I have reached the end of my time as a PhD student. As I close this chapter of my life and turn the page to a new one, I'd like to thank the many people who have helped me along this journey in the past 3.5 years.

First and foremost, I want to thank my supervisor Marco Spaans for his unrelenting support, guidance and advice throughout the years. Thank you very much for being patient with me, and for the kind words and motivation you were somehow always able to provide. I feel privileged to have been your student and I couldn't have imagined a better supervisor and mentor.

I'd like to thank my collaborators at the Institute für Astrophysik in Göttingen for their advice and insights; Dominik, thank you for your guidance in the initial stages of my PhD project; Latif, thank you for helping me become familiar with ENZO and its many quirks, and your patience while doing so; and Stefano, thank you for introducing me to the KROME package and teaching me to use it.

Many thanks to Rowin, for making his PDR code available to me, and for taking the time to give me a crash course on how to use it.

I'm very grateful for the help of the secretaries at the Kapteyn Institute and the Institut für Astrophysik, for arranging many things 'behind the scenes', and for assistance with some German paperwork and regulations.

For providing some much needed help with the German translation of my summary, I'd like to thank my dad and Nadine; I owe you!

Furthermore, I want to thank the people in Göttingen who truly made it worthwhile: Laura, Dimitar, Sudeshna, Chris, Dennis, and Philipp. Thank you for all the fun and games!

Thanks to my office mates at Kapteyn: Robin, Mustafa, and Ming, for the pleasant work environment and fun chats we have had throughout those two years. Mustafa and Robin, I hope you'll both be able to finish your PhD the way you planned; and good luck with your future careers to all of you!

Thanks also to Sander, I was very glad you were co-tutoring the cosmology course with me (twice!). I enjoyed our discussions and I really appreciate your no-nonsense approach. It was nice knowing and interacting with you for all these years!

I also want to thank everyone else at Kapteyn, for contributing to the welcoming environment at the institute. I particularly enjoyed our lunchtime conversations, especially when the weather cooperated and we could enjoy the sunshine outside on the balcony! (I won't list names because I will inevitably forget someone, but if we've chatted, consider yourself included here!)

Many, many thanks go to Myrtle, my Göttingen roommate. Thank you so much for taking a chance and inviting me into your home, I've really enjoyed living with you and getting to know you. Thank you also for helping me navigate Germany as a foreigner, and for the many things you've taught me. And of course, thanks for all the lovely conversations and good times! I'll have you know I still, very uncharacteristically, watch football sometimes ;)

Thanks to my dear friend Charlotte, for regularly providing much-needed distractions and for not giving up on me, you're the best C:

I'm also grateful for my mom, dad, and sister, for their encouragement and support also in difficult times.

And last but not least, thanks to my partner Daniel, for putting up with my many long and technical descriptions of everything that went wrong on any given day. Thanks for being there for me :) And thanks also to our fluffy bundle of joy Marley, for being such an adorable menace.

*IT ALWAYS SEEMS IMPOSSIBLE*
*UNTIL IT'S DONE*

— NELSON MANDELA

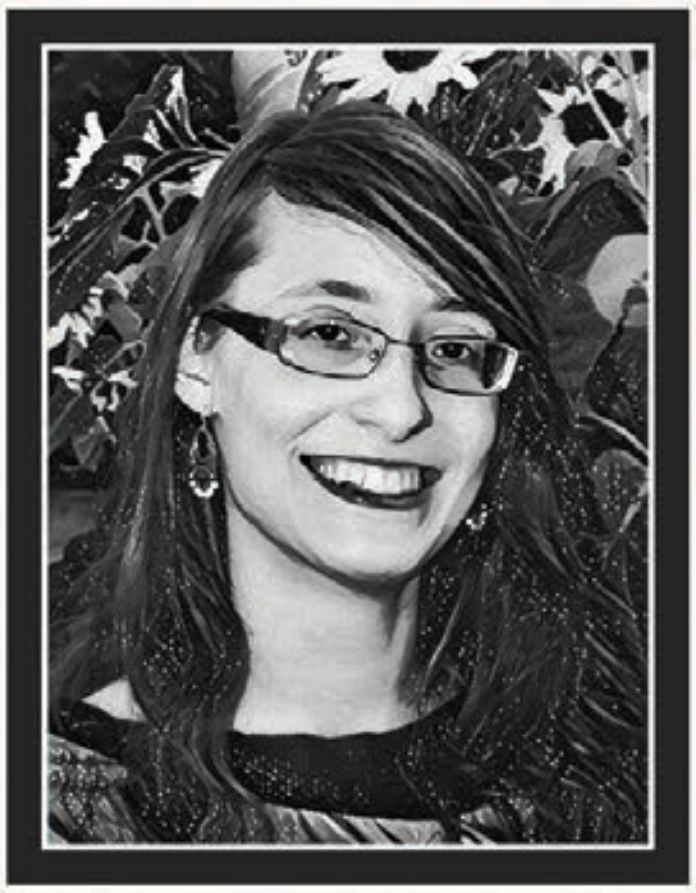

**CAROLINE VAN BORM** WAS BORN ON DECEMBER 7TH, 1989, IN HAGEN, GERMANY. IN 2007, SHE COMPLETED HER SECONDARY EDUCATION AT THE KONINKLIJK ATHENEUM MALLE (BELGIUM), WITH A SPECIALIZATION IN NATURAL SCIENCES AND MATHEMATICS AND A SPECIFIC FOCUS ON BOTH INDIVIDUAL LEARNING AND GROUP PROJECTS THROUGH THE DIDASCO PROGRAMME. SUBSEQUENTLY, SHE OBTAINED A BACHELOR'S DEGREE IN PHYSICS FROM THE UNIVERSITY OF ANTWERP (BELGIUM) IN 2010, WITH GREAT DISTINCTION, AND WENT ON TO FURTHER SPECIALIZE IN ASTROPHYSICS AT THE UNIVERSITY OF GRONINGEN (THE NETHERLANDS), WHERE SHE OBTAINED A MASTER'S DEGREE IN ASTRONOMY IN FEBRUARY 2013, CUM LAUDE. RIGHT AFTER GRADUATING, SHE STARTED A JOINT PHD PROJECT AT THE GEORG-AUGUST UNIVERSITÄT GÖTTINGEN (GERMANY) AND THE UNIVERSITY OF GRONINGEN; THE FRUITS OF WHICH CAN BE FOUND IN THIS BOOK.

CAROLINE.VANBORM@GMAIL.COM

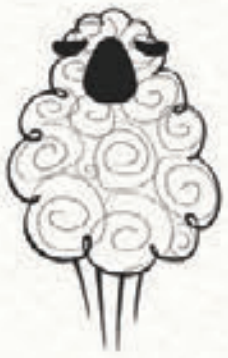

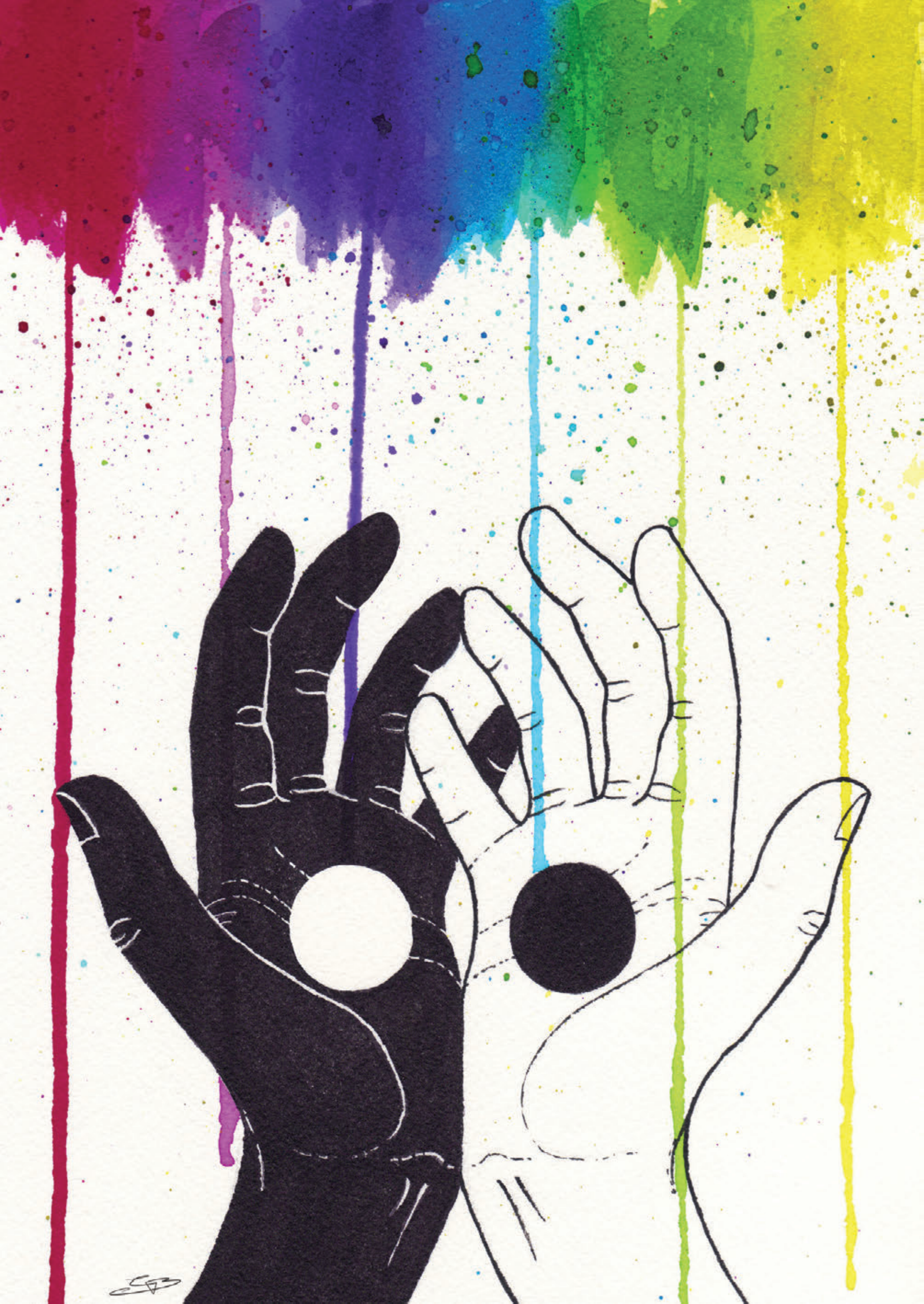